\documentclass[1p,times,showpacs]{elsarticle}
\pdfoutput=1
\usepackage{graphicx}
\usepackage{appendix}
\usepackage{amsmath,amssymb,latexsym}

\def\WIDTHC{1.2\textwidth}  
\def\WIDTHB{0.8\textwidth}  
\def\WIDTHA{1.0\textwidth}  

\begin{document}

\begin{frontmatter}

\title{Polymers with spatial or topological constraints: theoretical and computational results}

\author{Cristian Micheletti}
\address{SISSA, International School for Advanced Studies and \\ CNR-IOM Democritos and Italian Institute of technlogy (SISSA Unit)\\ via Bonomea 265, I-34136 Trieste, Italy}

\author{Davide Marenduzzo}
\address{SUPA, School of Physics and Astronomy, University of Edinburgh, \\ Mayfield Road, Edinburgh EH9 3JZ, Scotland}

\author{Enzo Orlandini}
\address{Dipartimento di Fisica and Sezione INFN, Universit\`a di Padova, I-35131 Padova, Italy}

\date{\today}

\begin{abstract}
In this review we provide an organized summary of the theoretical and computational results which are available for polymers subject to spatial or topological constraints. Because of the interdisciplinary character of the topic, we provide an accessible, non-specialist introduction to the main topological concepts, polymer models, and theoretical/computational methods used to investigate dense and entangled polymer systems. The main body of our review deals with: (i) the effect that spatial confinement has on the equilibrium topological entanglement of one or more polymer chains and (ii) the metric and entropic properties of polymer chains with fixed topological state. These problems have important technological applications and implications for the life-sciences. Both aspects, especially the latter, are amply covered. A number of selected open problems are finally highlighted.
\end{abstract} 


\end{frontmatter}

\tableofcontents

\section{Introduction}
\label{introduction}

Linear polymers are examples of soft-matter systems consisting of a
series of identical, or similar, covalently-bonded subunits
(monomers). The fundamental physical properties of a polymer, such as
its size and flexibility, strongly depend on its polymerization
degree, that is the number of constituting monomers.  Because of steric
hindrance and electrostatic self-repulsion, short polymer chains can
be aptly described as rigid molecules. On the other hand, when the
number of monomers in a chain is sufficiently large, polymers are
highly flexible and, in canonical equilibrium, explore a large number
of conformations that are typically highly self-entangled.

For individual open polymer chains this entanglement is purely
geometrical; in fact, a knot in the chain can be untied by a suitable
reptation of the polymer in space. On the other hand, if a closure
(cyclization) reaction occurs, or if the polymer termini are otherwise
constrained, then the geometrical self-entanglement is trapped in the
form of a knot (possibly a trivial knot, that is a ring), whose
topology cannot be changed by any manipulation of the polymer except
by cutting it.

Topological entanglement is a genuine characteristic of polymeric
systems that, in some circumstances, may severely affect not only the
physical properties of individual polymer chains, but also of polymer
melts. For example, entanglement can be trapped during the
crystallization of artificial polymers, and determine the properties
of the resulting crystal, such as its degree of
purity~\citep{deGennes:1984:Macromol,Saitta_et_al:1999:Nature,Saitta_Klein:2002:J_Chem_Phys}. It
is likely that the entanglement colocalizes with amorphous regions of
the crystal.

Artificial knotted polymer rings can be, nowadays, easily synthesized
in laboratory but their most ubiquitous and pervasive manifestation is
found in biopolymers such as DNA and RNA. These biomolecules can be
sufficiently long that they become entangled {\it in
vivo}~\citep{Wang:1996:AnnRevBio}, thus providing an ideal playground
to study the occurrence of knots and their biological
implications. For example, knots and other entanglements in DNA are
known to severely affect the efficiency of fundamental genetic events
such as replication and
transcription~\citep{Wassermann&Cozzarelli:1986:Science,Sumners:1990:MathInt,Sumners:1992,Sumners:1995}.
The biological necessity to maintain a control on the degree of
entanglement of these biomolecules {\em in vivo} is highlighted by the
fascinating fact that there exist specific cellular machineries, such
as topoisomerase enzymes, which are capable of simplifying the
topological complexity of the DNA entanglement by favouring the
selective cross-passage of pairs of DNA
strands~\citep{Rybenkov:1997:Science,Wang:1996:AnnRevBio,Sumners_et_al:1995,Liu_et_al:2006:J-Mol-Biol,Liu:2010:J-Mol-Biol:20460130}.

As illustrated by the above-mentioned examples, topological
entanglement is a ubiquitous aspect in polymer systems and has
important ramifications either in technological contexts or for the
understanding of complex physical and/or biological systems.

The objective of this review is to provide an overview of the salient
advancements that have been made in the past decades in the
theoretical and computational characterization of topological
entanglement in polymeric systems. In this regard, it should be
mentioned that most of the latest developments have been stimulated by
the advent of single-molecule manipulation techniques that have
redefined the ``hot'' open problems and applicative horizons of
polymer entanglement. We mention, in particular, the perspectives
opened by the possibility of confining one or more polymer chains in
nano-pores, nano-channels or through surface adsorption as well as
that of mechanically manipulating individual polymeric filaments.

The topics covered by the present review may broadly be collected
into four main sections which cover respectively: (i) the fundamental
notions of entanglement and polymer models, the state-of-the-art
characterization of (ii) self-entanglement of individual polymers and
(iii) several polymer chains in ``topological equilibrium'' and
finally (iv) the behaviour of one or more fluctuating polymers of
fixed topology.  For each section, we provide not only an up-to-date
account of available results, but also a full overview of the main
methodological (theoretical and computational) tools that can be
profitably used to attack the state-of-the-art problems. Finally, we
provide a perspective for future research by highlighting the key open
questions.

\section{First part: basic knot theory and polymer models}
\label{sec:2}

The first part of the review consists of
sections~\ref{knot_theory}--~\ref{pol_models} and aims at providing
the reader with an overview of:
\begin{itemize}
\item the main topological concepts,
\item the polymer models that are most-commonly studied theoretically
  or computationally,
\item the methods and algorithms used to characterize the degree of polymer entanglement.
\end{itemize}

In the literature there are several models of polymer chains in
solution and we will review some of them by pointing out advantages
and drawbacks.  A common, essential feature of these models is that
they neglect the chemical details of the monomers as much as possible
and describe the polymer in terms of a discrete (coarse-grained) chain
made by the repetition, in a string-like fashion, of the elementary
monomeric subunits. Ideally, viable models ought to be simple enough
to be theoretically tractable and yet sufficiently detailed to capture
the key characteristics of the macromolecule in equilibrium such as
its connectivity, its local rigidity and its flexibility at large
scales.

At the simplest level of model complexity one has the so called
\emph{ideal chains} where only the chain connectivity and local
stiffness are considered. In these models, the polymer chain is
typically embedded in the continuum three-dimensional space, and
self-avoidance is not accounted for. This leaves the chain free to
attain configurations that would be incompatible with steric
hindrance, such as those where distinct chain nodes or bonds are
overlapping. The minimalistic character of these models make them
amenable to extensive analytical characterization and to
straightforward numerical simulations, see Section~\ref{pol_models}.

More realistic descriptions of the polymer do take into account the
excluded-volume constraint in the chain, These constraints introduce
long-range (along the chemical sequence) correlations in the system
which make analytical progress very difficult. As a consequence, these
models are almost exclusively characterised by means of
computationally-intensive numerical simulations. The scope of the
latter is often extended by reducing the conformational degrees of
freedom of the model polymer, for example by embedding the chain on a
regular lattice.

In Section~\ref{knot_theory} we finally provide a concise summary of
the key concepts needed to define and characterize the topological
entanglement of a closed chain (and discuss to what extent it can be
extended to open linear chains) which is at the heart of the modern
branch of topology known as the \emph{knot theory}. Interested readers
can find a more systematic and in-depth coverage of the concepts and
results introduced below, in one of the several books that have
been published on the subject, see e.g. refs.~\citep{Cromwell_2004,Burde_Zieschang,Murasugi_1996,Adams:1994,Livingston:1993,Rolfsen:1976}

\section{Basic knot theory}
\label{knot_theory}

An operational definition of a knotted curve can be given by resorting
to our common experience of taking a piece of rope, tying a knot in it
and finally gluing the two ends together. The knotted rope can be
geometrically manipulated in three-dimensional space ($\mathbb R^3$)
in countless ways but, unless we use scissors and glue to cut and
rejoin the rope, it is impossible to turn it into a plain, unknotted
ring. The above example builds on our intuitive notion that a rope
has an approximately uniform thickness that forbids the rope
self-crossing during the manipulations.

Since all the configurations obtained by deforming the rope preserve
the initial knot type, it appears natural to define a knot as the
class of equivalence of configurations obtained by these manipulations
in the three-dimensional embedding space.

\begin{figure}[tbp]
\begin{center}
\includegraphics[width=\WIDTHB]{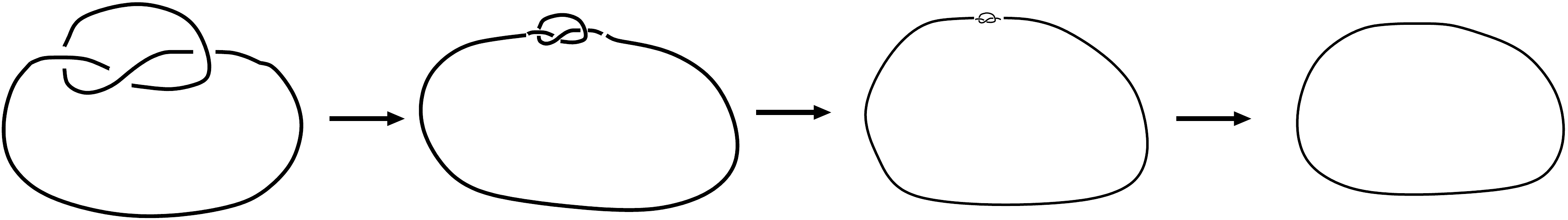}
\caption{Knot elimination by continuous deformation on a knotted loop.} \label{fig:singular_knot}
\end{center}
\end{figure}

Mathematically it is appealing to define a knot in a similar fashion,
that is as the set of continuous simple closed curves (i.e. $S^1$) that
are related by continuous deformations in $\mathbb R^3$
(isotopy). However, in contrast with our practical example, a curve in
$S^1$ has no thickness so that, based on the above definition, any
knot would be found equivalent to the unknot. This is readily seen by
considering the continuous deformation resulting from pulling the curve
tighter and tighter until the knot reduces to a point and hence
disappears (see Figure~\ref{fig:singular_knot}).

This difficulty can be overcome by basing the knot equivalence on the
notion of \emph{ambient isotopy}. An ambient isotopy is a
geometrical manipulation of the curve (isotopy) through the space in which the
curve is embedded (ambient). In other words, instead of deforming the knots
alone, one deforms the whole space in which they sit. These
deformations preserve the orientation of the curve and do not allow
the occurrence of instances such as the one depicted in
Figure~\ref{fig:singular_knot} in which the knot has been shrinked down 
to a point.

This restriction is sufficient to rule out the occurrence of
singularities, such as the one shown in
Figure~\ref{fig:singular_knot}, as the result of a continuous
deformation of a curve.

\begin{figure}[tbp]
\begin{center}
\includegraphics[width=\WIDTHB]{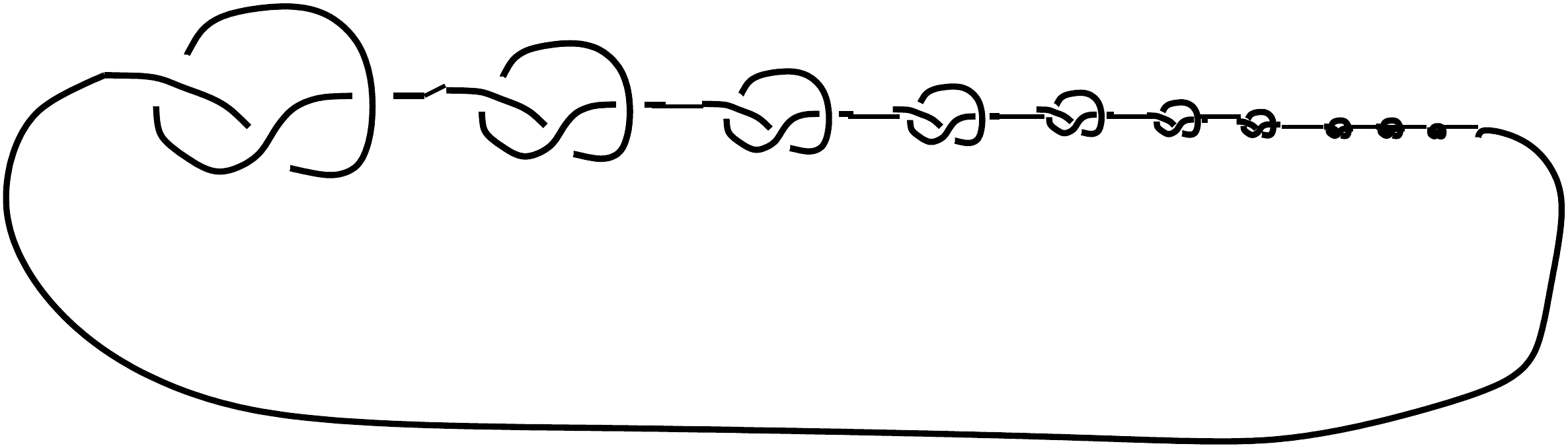}
\caption{A wild knot.} \label{fig:wild_knot}
\end{center}
\end{figure}

\begin{figure}[tbp]
\begin{center}
\includegraphics[width=\WIDTHB]{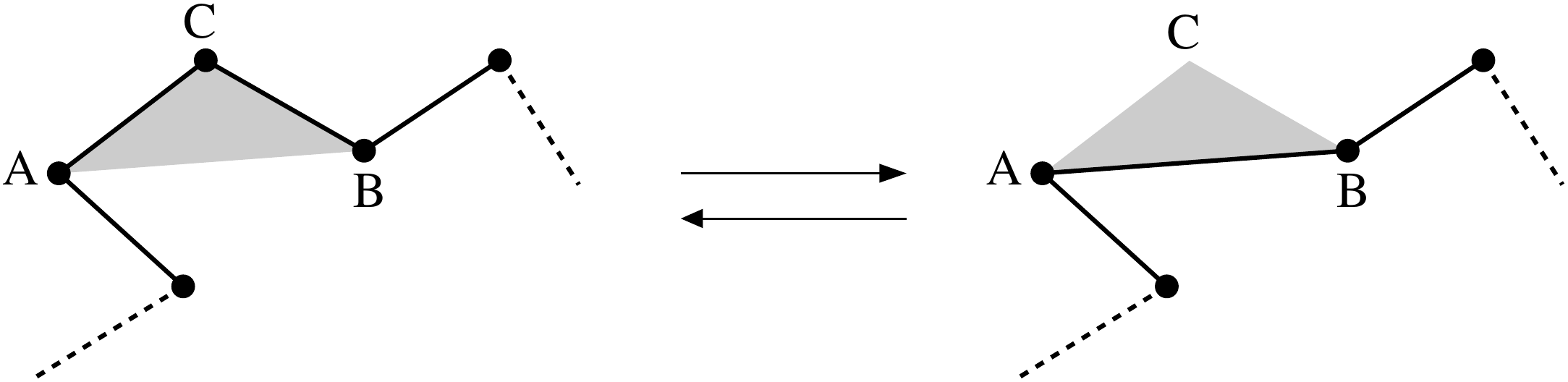}
\caption{Triangular moves: the removal or addition of the vertex C is allowed if the interior of the non-degenerate triangle with vertices ABC does not intersect any chain segment. We stress that the number of vertices and the length of the
  polygonal curve will increase or decrease after each triangolar move.} \label{fig:triangular_moves}
\end{center}
\end{figure}

Yet, it appears necessary to introduce restrictions also on the types
of admissible curves in order to disallow the occurrence of
``pathological'' cases such as the one depicted in
Figure~\ref{fig:wild_knot} and known as \emph{wild knots}. These
cases, that are very far from the intuitive, physical notions of a
proper knot, can be ruled out by requiring the curves to be
differentiable.  This severe constraint is impractical
in the discrete polymer models used in standard numerical
studies. In this situation the constraint is relaxed by defining a knot
as a simple, closed curve that is ambient isotopic to a  piecewise-linear 
(or polygonal) curve in $\mathbb R^3$ (\emph{tame knot})~\citep{Livingston:1993}.
This definition eliminates wild
knotting because polygonal curves are necessarily composed of a finite
number of bonds or edges. Within the space of polygonal curves, two
configurations are topologically equivalent (or ambient isotopic) if
one can be obtained from the other by a finite sequence of
\emph{triangular moves} (see Figure~\ref{fig:triangular_moves}). 

Detecting knot equivalences is at the heart of knot theory. For
instance, proving that it is impossible to deform one knot into
another is equivalent to proving that the two knots lie in different
equivalence classes. On the other hand proving that a knot is not
trivial i.e. that no ambient isotopy can move it into a perfectly
planar ring is equivalent to showing that it does not belong to the
equivalence class of the unknot.

\begin{figure}[tbp]
\begin{center}
\includegraphics[width=\WIDTHB]{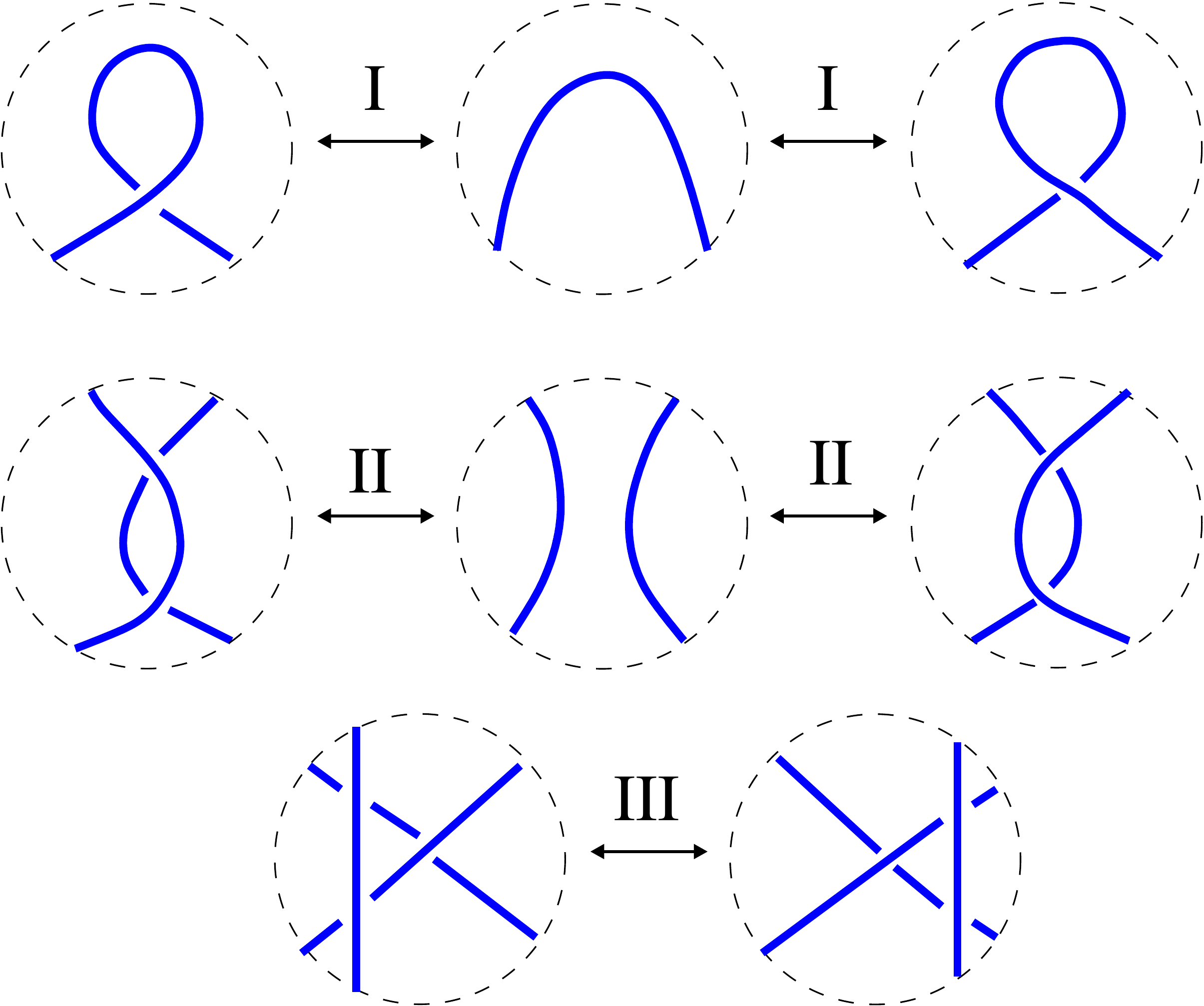}
\caption{Reideimeister moves.} \label{fig:reideimeister}
\end{center}
\end{figure}

Another way of detecting the knot type of a closed polygonal curve
with a definite arrangement (embedding) in $3d$ space consists of
looking at its planar representation, i.e. its \emph{knot
diagram}. Knot diagrams are planar projections of the
three-dimensional curve embedding whose only singularities are
transverse double points at which the \emph{underpass-overpass}
information is given by interrupting one of the branches of the
projected curves. In this mapping a movement of an embedding in
$\mathbb R^3$ corresponds to a change in its planar projection. For
this description there is a well established theorem stating that two
knots are equivalent if and only if (any of) their projections can be
made identical by a sequence of three possible \emph{Reidemeister
moves}~\citep{Reidemeister:1948} and planer isotopies (see Figure~\ref{fig:reideimeister}). 
Each knot diagram is characterized by a set
of transversely crossings whose number depends on the particular
projection direction and on the specific geometric realization of the
knot type. By using Reidemeister moves one can, in principle, minimise
the number of crossings compatible with the topology of that knot
type. This gives rise to a minimal knot diagram representation of
that knot type having the smallest possible number of crossings
$n_{cr}^{min}$ (\emph{crossing number}).

\begin{figure}[tbp]
\begin{center}
\includegraphics[width=\WIDTHB]{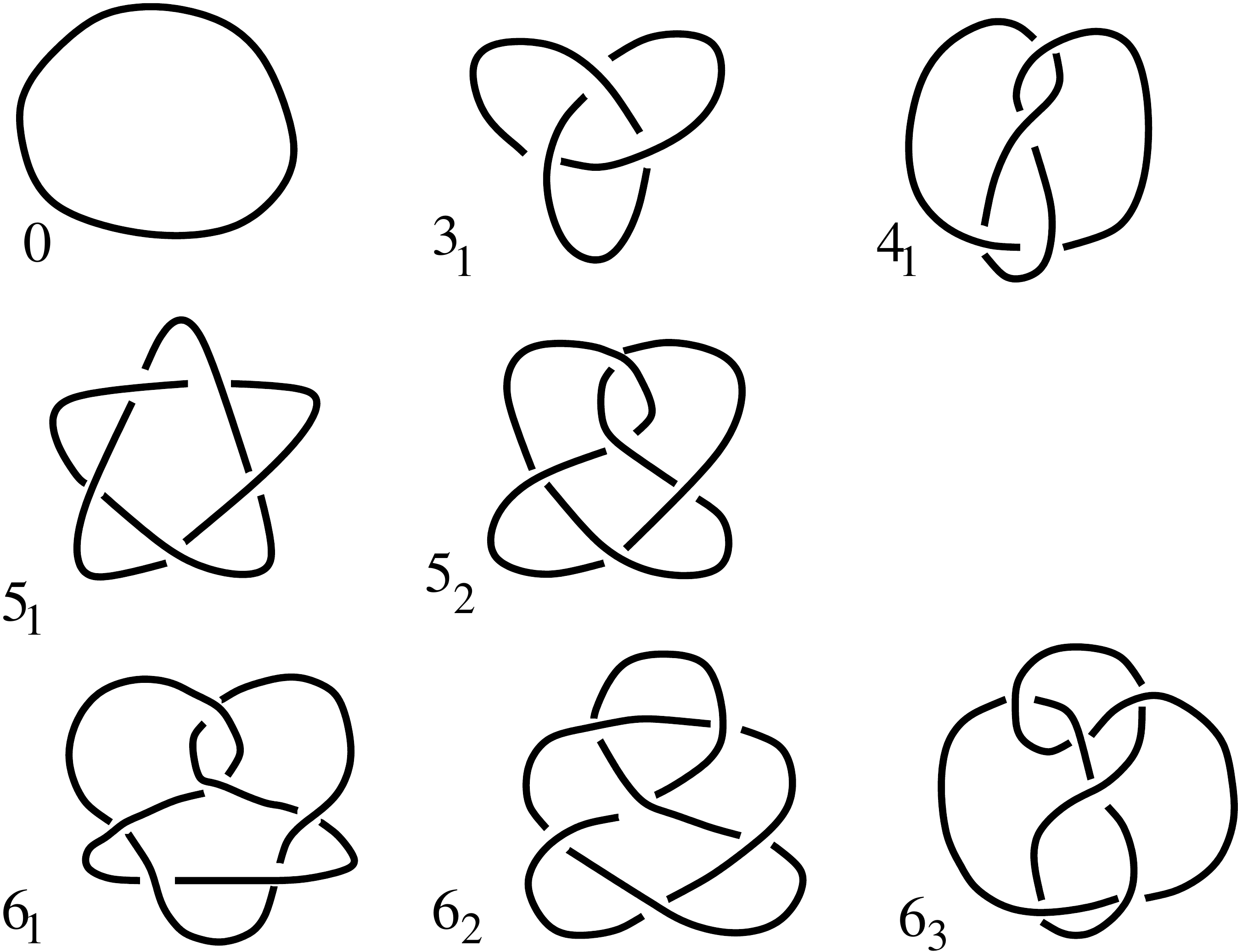}
\caption{Prime knots up to crossing number $6$.} \label{fig:knots_table}
\end{center}
\end{figure}

As a matter of fact, if we neglect chiral enantiomers (see
Section~\ref{chirality} below for a proper definition), 
the value of $n_{cr}^{min}$ provides
the most used criterion for classifying knots in groups of increasing
topological complexity.  An unknotted curve is, in fact, associated to
$n_{cr}^{min}=0$. At the next level of complexity one has the trefoil
knot, which has $n_{cr}^{min}=3$ etc., as shown in
Figure~\ref{fig:knots_table}.  As $n_{cr}^{min}$ increases one finds
topologically different knots having the same number of minimal
crossings which are conventionally distinguished by a different
numerical subscript appended to $n_{cr}^{min}$. The simplest example
is offered by the pair of 5-crossing knots, $5_1$ and $5_2$ sketched in
Figure~\ref{fig:knots_table}.

\subsection{Prime and composite knots}

As common experience teaches us, a non-trivial knot $\tau$ in a rope cannot be untied by introducing a second non-trivial knot $\tau'$ in different portion of the same rope (that is a portion that does not interfere with the first knot), no matter how one manipulates the doubly-knotted rope (with closed or otherwise constrained ends).
The notion that anti-knots do not exists can be actually proved rigorously
by using geometrical techniques (see Section~\ref{geometry} below). A
further result is that the resulting knot, $\tau \# \tau'$, belongs to
a knot type that is different from both $\tau$ and $\tau'$ and is
called the \emph{connected sum} or the \emph{composite knot} of the
two original knots. The two knots in the sum are called \emph{factors}
of $\tau \# \tau'$.

The notion of composite knots is very important as is leads to the
concept of a \emph{prime} knot i.e. of a non-trivial knot that cannot
be decomposed into a non-trivial connected sum. In other words, if a
prime knot $\tau$ is equivalent to the connected sum $\tau' \# \tau''$
this implies that either $\tau'$ or $\tau''$ are unknots.

The basic facts about connected sums are that: (i) the operation is
commutative (i.e. $\tau\#\tau'=\tau'\#\tau$) and (ii) that each non
trivial knot has a unique and finite factorization into prime
knots.

\begin{figure}[tbp]
\begin{center}
\includegraphics[width=\WIDTHB]{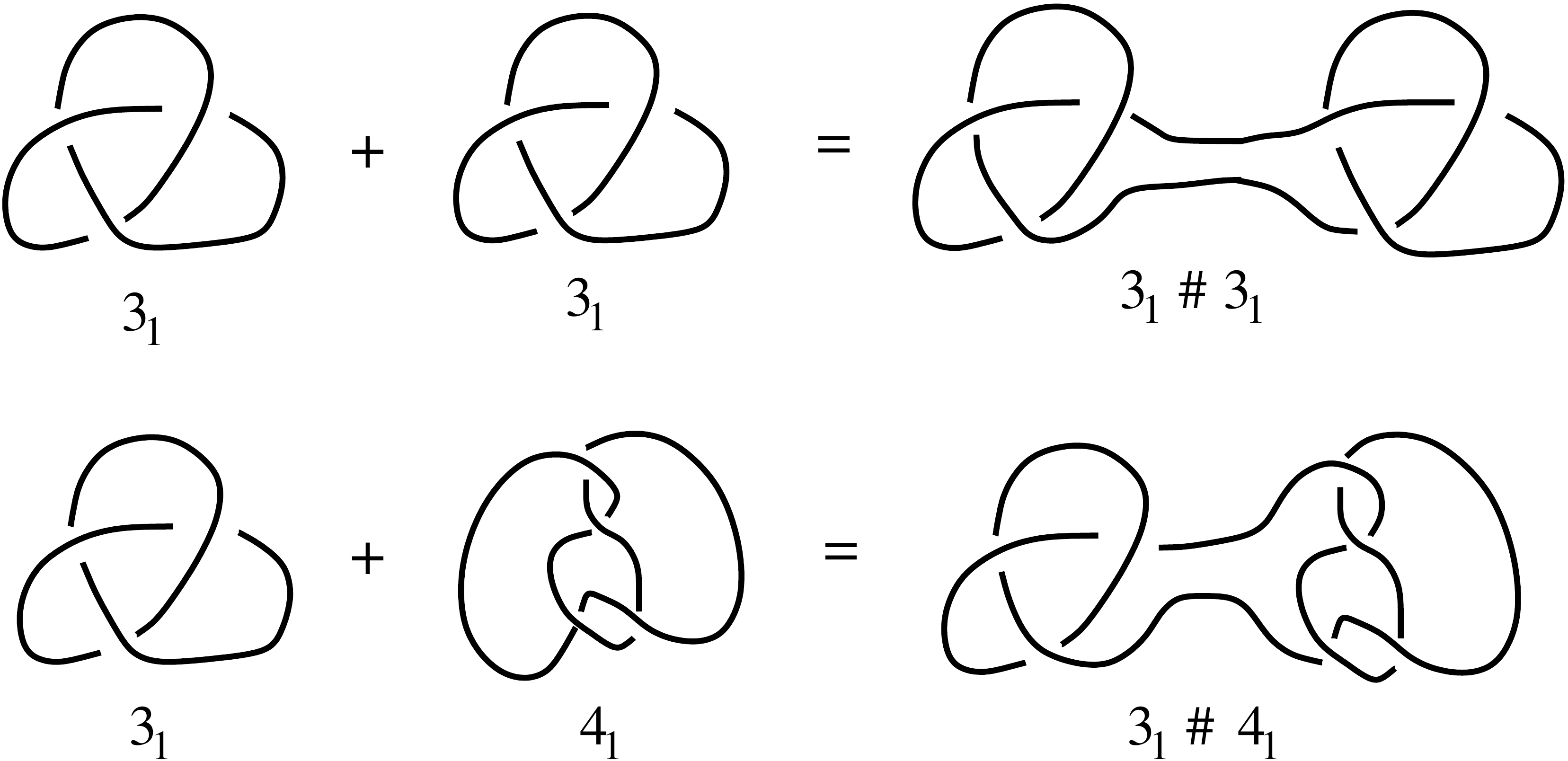}
\caption{Connected sum of simple knots. The composite knot in the upper panel is known as the granny knot and is obtained by connecting two trefoils with the same handedness.} \label{fig:connect_sum}
\end{center}
\end{figure}

For this reason standard knot tables, see for example the knot table
in~\citep{Rolfsen:1976}, lists only prime knots (some of which can
have two types of handedness, see next section).
The simplest prime knots, those with $n_{cr}^{min} \le
  6$ are shown in Figure~\ref{fig:knots_table} while the simplest
  composite knots, which are clearly generated by the connecting two
  trefoil knots are given in Figure~\ref{fig:connect_sum}.
  
An important fact is that the number of prime knots existing at a
given value of $n_{cr}^{min}$ grows exponentially with
$n_{cr}^{min}$~\citep{Ernst&Sumners:1987:MPCPS,Welsh:1991:CMSJB,Thistlethwaite1998}
This rapid growth is reflected in the fact that presently available
exhaustive tables of prime knots (minimal representations) exist only
for knots with up to $\sim 16$ crossings~\citep{prime_knots_16}. For
special knot families, such as the alternating knots described later
on, there exists tables of up to 22 crossings~\citep{alt_knots_22}.

\subsection{Knot handedness}
\label{chirality}

\begin{figure}[tbp]
\begin{center}
\includegraphics[width=\WIDTHB]{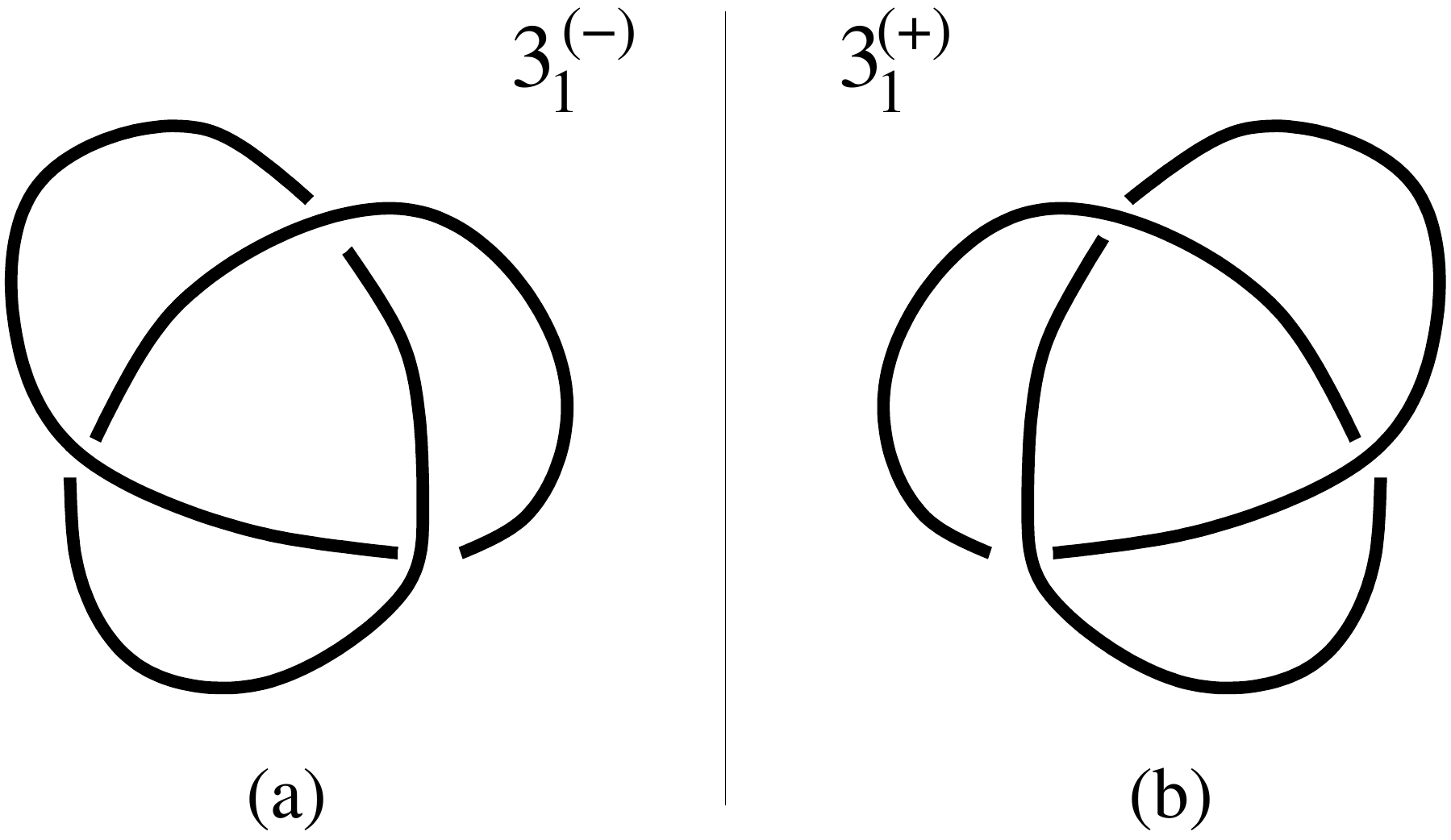}
\caption{The trefoil knot and its mirror image. The trefoil knot in panel (a) is left-handed and the one in panel (b) is right-handed}
 \label{fig:trefoil_mirror}
\end{center}
\end{figure}

\begin{figure}[tbp]
\begin{center}
\includegraphics[width=\WIDTHB]{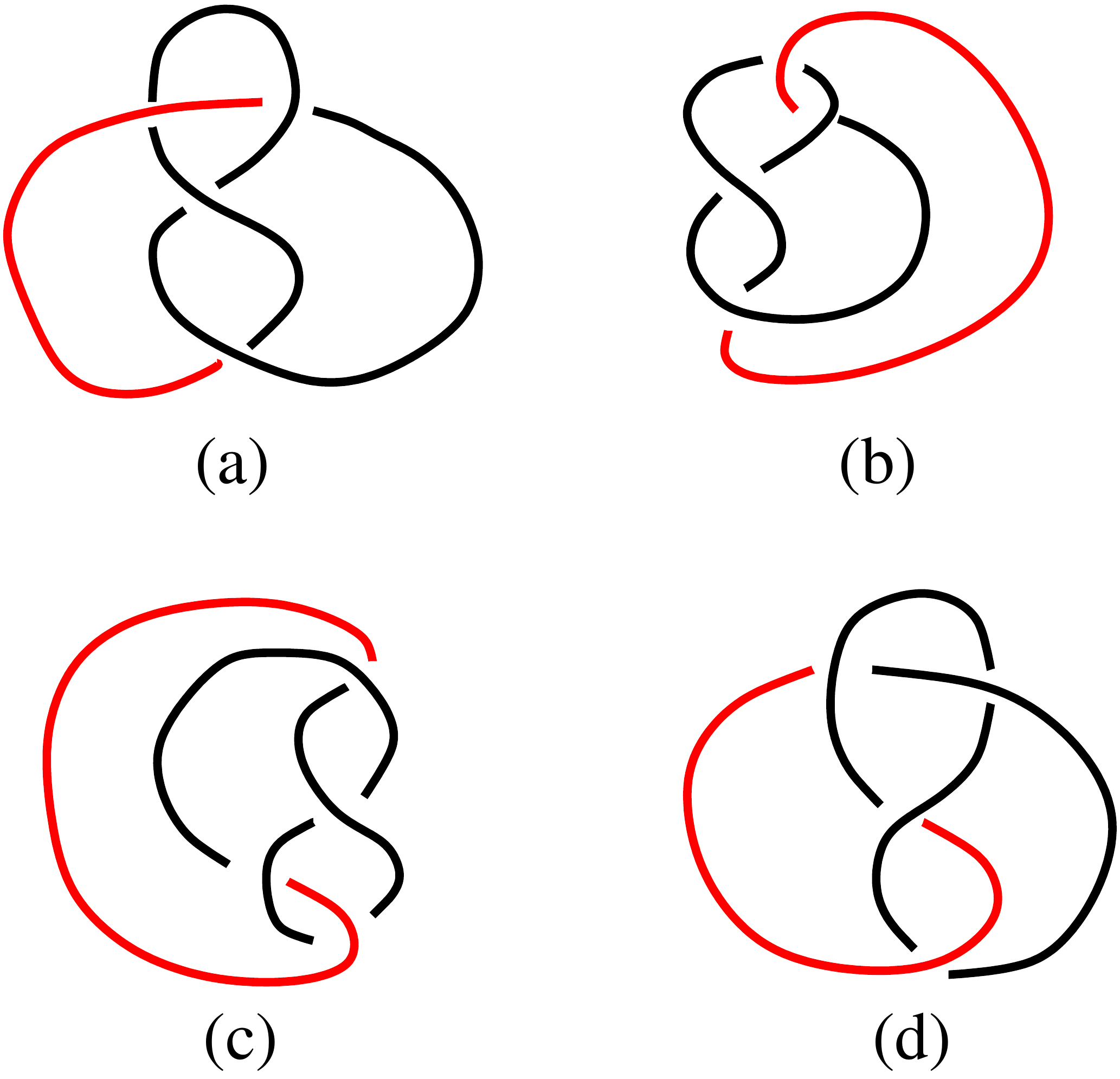}
\caption{Achiral knots: How to pass from the $4_1$ (a) to its mirror
  image (d) by smooth deformation. (b) Rearrange the red part of the
  curve. (c) Rotate the knot by $180^\circ$. }
\label{fig:figure_eight_mirror}
\end{center}
\end{figure}

An important notion in knot theory is \emph{chirality}, or
\emph{handedness}. Suppose we look at a knot in a mirror. Is the
mirror image equivalent to the original one or not ? As an example let
us consider the trefoil knot and its mirror image in
Figure~\ref{fig:trefoil_mirror}. The ``right handed" knot
  of (panel {\em b}) cannot be continuously deformed into the
  ``left-handed" version shown in panel {\em a}. For this reason, the
trefoil knot is said to be \emph{topologically chiral}. The homotopy
inequivalence of the two chiral versions of the trefoil was first
proved in 1914 by M. Dehn.  By contrast if a knot can be continuously
deformed into its mirror image we say that the knot is
\emph{topologically achiral}, or simply achiral (see for example the
$4_1$ knot in Figure~\ref{fig:figure_eight_mirror}).

\begin{figure}[tbp]
\begin{center}
\includegraphics[width=\WIDTHB]{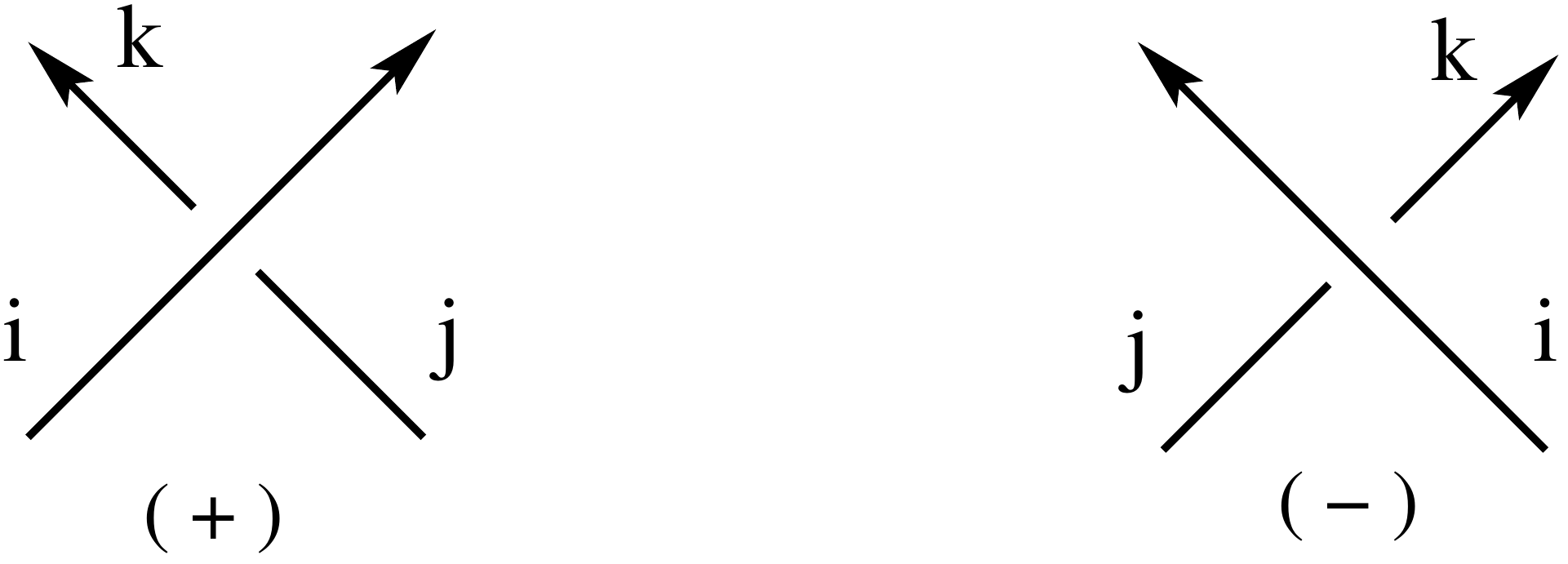}
\caption{The sign convention for signed crossings. The arc labelling
  is for the computation of the Alexander
  polynomial.} \label{fig:cross_sign}
\end{center}
\end{figure}

Achiral knots are relatively rare. Indeed among the first $35$ prime
knots in the knot table less than a $1/4$ are achiral $(\emptyset,
4_1,6_3,8_3,8_9,8_{12},8_{17},8_{18})$; all $49$ $9$-crossings knots
are chiral and only $13$ of a total of $165$ $10$-crossings knots are
achiral. The knot nomenclature introduced before is clearly incapable
of distinguishing enantiomers, as they have the same minimal crossing
number.  For prime knots with 7 or less crossings, the enantiomers can
be distinguished by computing the balance of left- and right-handed
crossings in the {\em minimal diagram}. To do so, it is first required to
assign an orientation to the knot diagram\footnote{We mention that there exist knots that, by using an appropriate ambient isotopy, can be transformed into themselves but with the reversed directionality. These knots are termed invertible and the simplest one is given by $8_{17}$}.
The handedness of each
crossing is assigned by the right-hand rule applied to the oriented
over- and under-passing segments as shown in
Figure~\ref{fig:cross_sign}. 

  The overall chirality is then
established by assigning a ``+1'' value to right-handed crossings and
``-1'' to the others and then taking the algebraic summation of the
crossing signs. A right- and left-handed trefoil would therefore be
associated to signed crossing numbers equal to +3 and -3,
respectively. The resulting +/- sign of the summation can be appended
to the knot name to distinguish the two types of enantiomers. For
example the right- and left-handed enantiomers of the trefoil are
denoted respectively by $3_1(+)$ and $3_1(-)$.

We stress that the above rule is formulated for the minimal diagram representation of prime knots with less than 8 minimal crossings and alternating knots (see next subsection) but can fail for an abritrary diagram of these knots. In general, the rule will fail even for minimal diagram representations of prime knots, as pointed
out by A. Perko~\citep{perkopair}. Bearing in mind these important limitations, the
chirality of connected sum of different prime knots can be suggested
by the sign of the summed crossings in the minimal diagram. For
example the composite knots $3_1(+)\#3_1(+)$ and $3_1(-)\#3_1(-)$ are
two enantiomers of the granny knot while the knot $3_1(+)\#3_1(-)$ is
achiral and is called the \emph{square knot}.

The concept of signed crossings can be also used to discuss the minimal
topological changes that can be applied to a given knot to turn it
into an unknot. Within a minimal knot diagram representation it is
easy to realize that one can convert a knot into the unknot by
reversing one or more crossings in the knot diagram. For example the
trefoil knot in its simplest form has only three crossings, and by
changing in one of these crossing the under-passing strand into the overpassing one and vice-versa, we obtain the unknot. In general, for a given knot diagram it is always
possible to find a set of crossings that can be switched over to
obtain the unknot.  On the other hand, for each knot diagram there can
be several possible choices of crossings that can lead to the
unknot. Moreover the number of crossing changes required might depend on the
diagram. It is then natural to define the \emph{unknotting number} as
the minimum (taken over all possible knot diagrams) number of crossing
reversals needed to turn a given knot into the
unknot~\citep{Rolfsen:1976,Livingston:1993}. 

There is no simple relation of the unknotting number to the minimal number of crossings of a knot. 
For example, the trefoil and figure eight knots both have unknotting number 1. 
Of the two five-crossing knots, $5_2$ has unknotting number 1 and $5_1$
has unknotting number 2. The granny knot has unknotting number 2
(essentially because each trefoil has to be unknotted).  It turns out
that if a knot has unknotting number 1 then it is automatically
prime~\citep{Scharlemann:1985}, but the converse is not true.

A natural generalization of the above scheme is the definition of the
minimum number of crossing changes required to convert one given knot
into another one. This number provides a distance in the space of
knots. General topological considerations can be used to provide
upper and lower bounds for knot distances which, in fact, are known
exactly only for a limited subsets of knot
pairs \citep{Darcy:2001:JKTR}.  A recent practical application of this
distance in knot space has been described in
ref.~\citep{Wang:1996:AnnRevBio,Flammini:2004:BPJ} in connection with
the topoisomerase action on circular DNA.

\subsection{Some knot families}

\begin{figure}[tbp]
\begin{center}
\includegraphics[width=\WIDTHB]{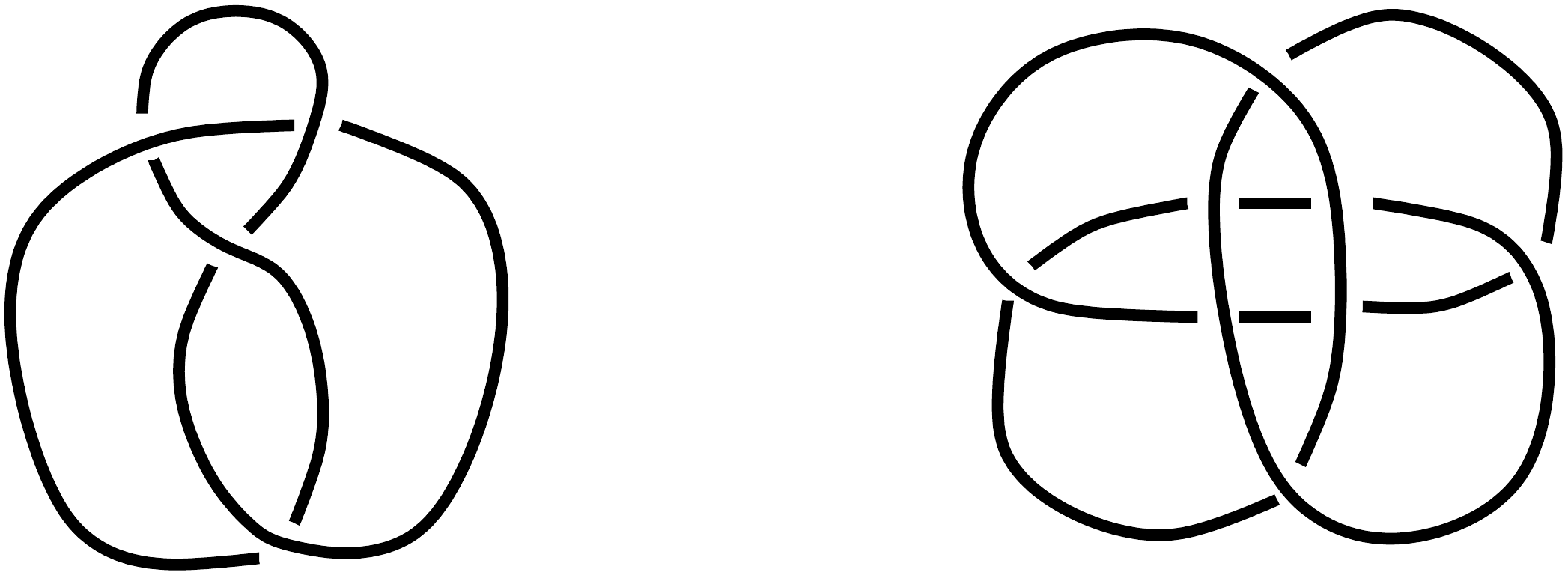}
\caption{A minimal (alternating) knot diagram for $4_1$ (left) and a
  diagram of the same knot that is not
  alternating.} \label{fig:alternating}
\end{center}
\end{figure}

As anticipated, prime knots can be partitioned in families according
to salient topological indicators, such as the minimal number of
crossings. An important one is the \emph{alternating knots}, that is
knots admitting a minimal diagrammatic representation where under and over
crossings alternate along the path.  Note that not all
diagrams of an alternating knot need to possess the alternating
properties described above, as illustrated in
Figure~\ref{fig:alternating}. Interestingly, most of
the simplest types of knots, that is those with sufficiently small
crossing number, are alternating.  In fact, all prime knots with crossing
number smaller than 8 are alternating, with the simplest
non-alternating instance being the $8_{19}$ knot.

\begin{figure}[tbp]
\begin{center}
(a)\includegraphics[width=0.45\textwidth]{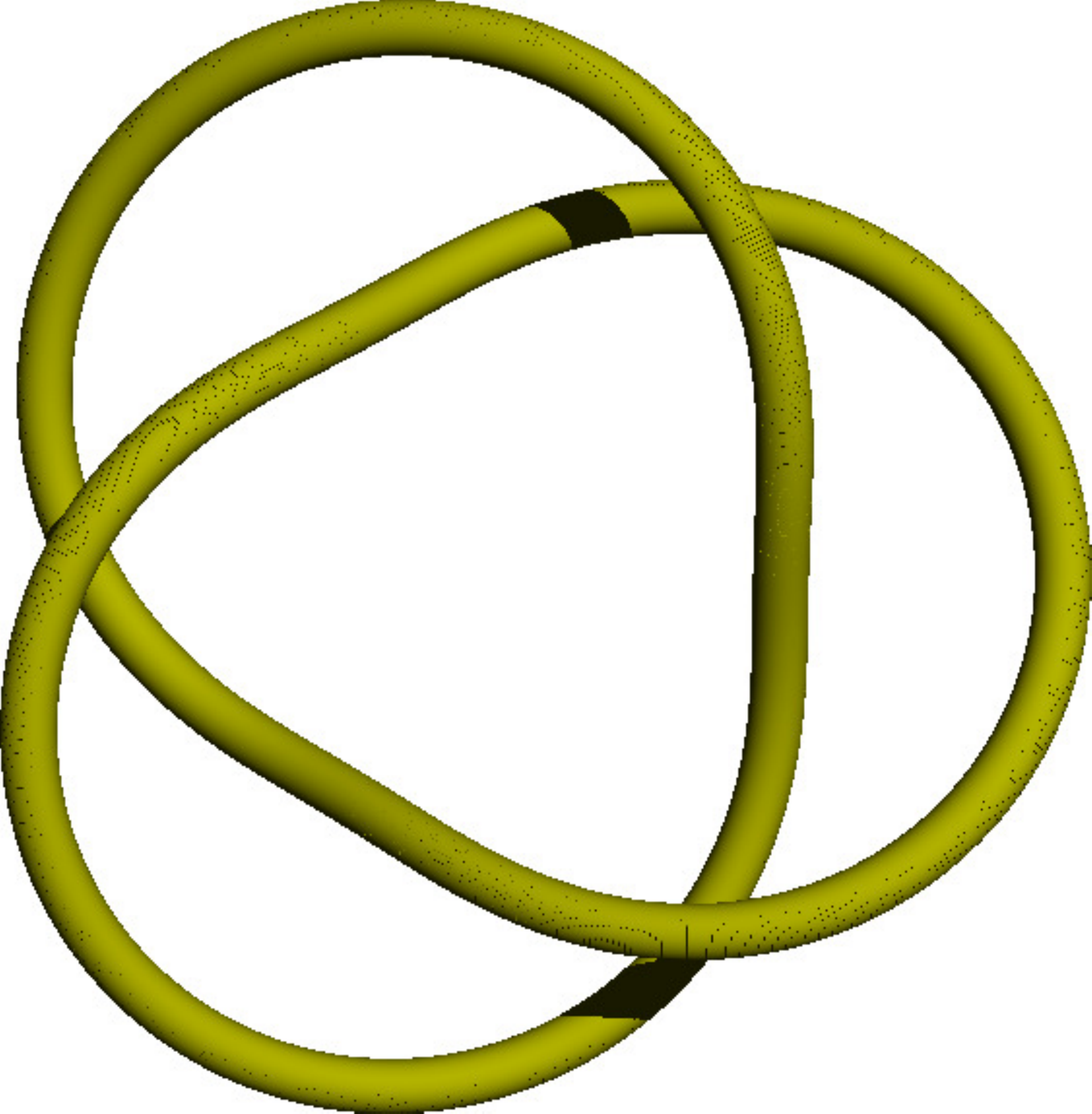} 
(b)\includegraphics[width=0.45\textwidth]{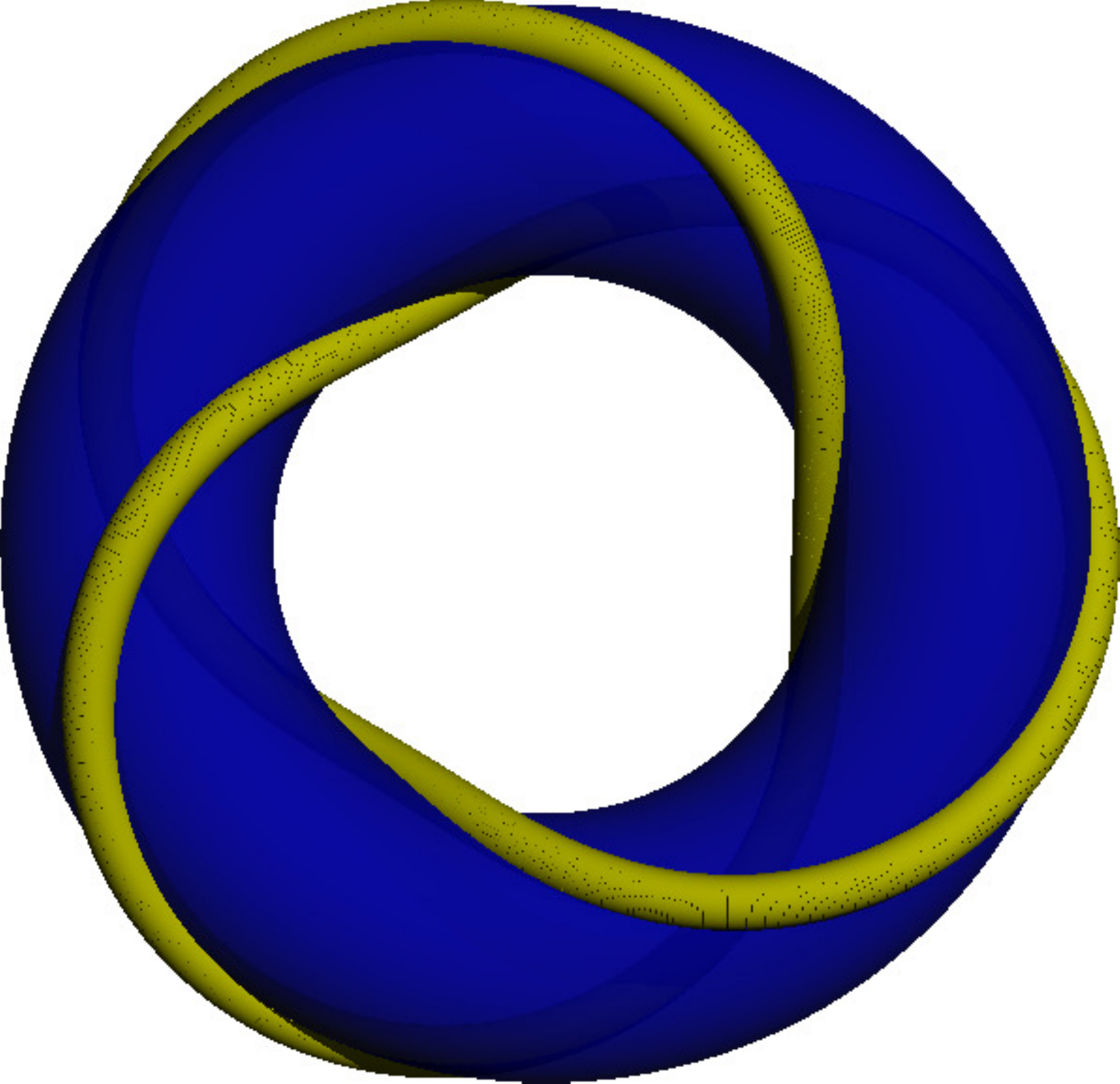}
\caption{The trefoil knot (a) drawn on the surface of a torus as a
  $(3,2)$ torus knot (b).} \label{fig:torus_knot}
\end{center}
\end{figure}

With the exception of the unknot no other knot can be drawn on the
surface of a sphere without self-intersections. Notably, there is a
class of knots that, nevertheless, can be drawn as a simple closed
curve on the surface of a standardly embedded (i.e. unknotted) torus. 
This is the family of \emph{torus knots}.

As an example we show in Figure~\ref{fig:torus_knot} the trefoil as a
(2,3)-torus knot: the knot is shown to wind around the torus for $p=2$
times longitudinally and $q=3$ times meridionally (a meridian being a
circle bounding the disc inside the torus). The $p$ and $q$ numbers
can be positive or negative depending on the orientation of the
winding directions and fully classify the $(p,q)$-torus knot. Basic
properties of such knots are: 
\begin{itemize}
\item $p$ and $q$ must be relatively prime;\footnote{if $p$ and $q$ are not relatively prime one obtains links (see subsection \ref{linking} for definition).}
\item a $(p,q)$-torus knot is topologically equivalent to a
  $(q,p)$-torus knot;
\item if $p$ or $q$ is $1$ or $-1$ the curve is
  unknotted;
 \item the mirror image of a $(p,q)$ torus knot is a
  $(p,-q)$ torus. In other words all the non-trivial torus knots are chiral.
\end{itemize}

A number of exact results have been proven for torus knots. Two
noteworthy ones regard the minimal crossing number and the unknotting
number. For torus knots $(p,q)$, the former is given by
to $\min \{p(q-1),q(p-1)\}$~\citep{Murasugi1991} while the uknotting number is is given by the simple formula $((p-1)(q-1))/2$ ~\citep{Kronheimer_Mrowka:1993:Topology}.

\subsection{Geometric techniques}
\label{geometry}

\begin{figure}[tbp]
\begin{center}
\includegraphics[width=\WIDTHB]{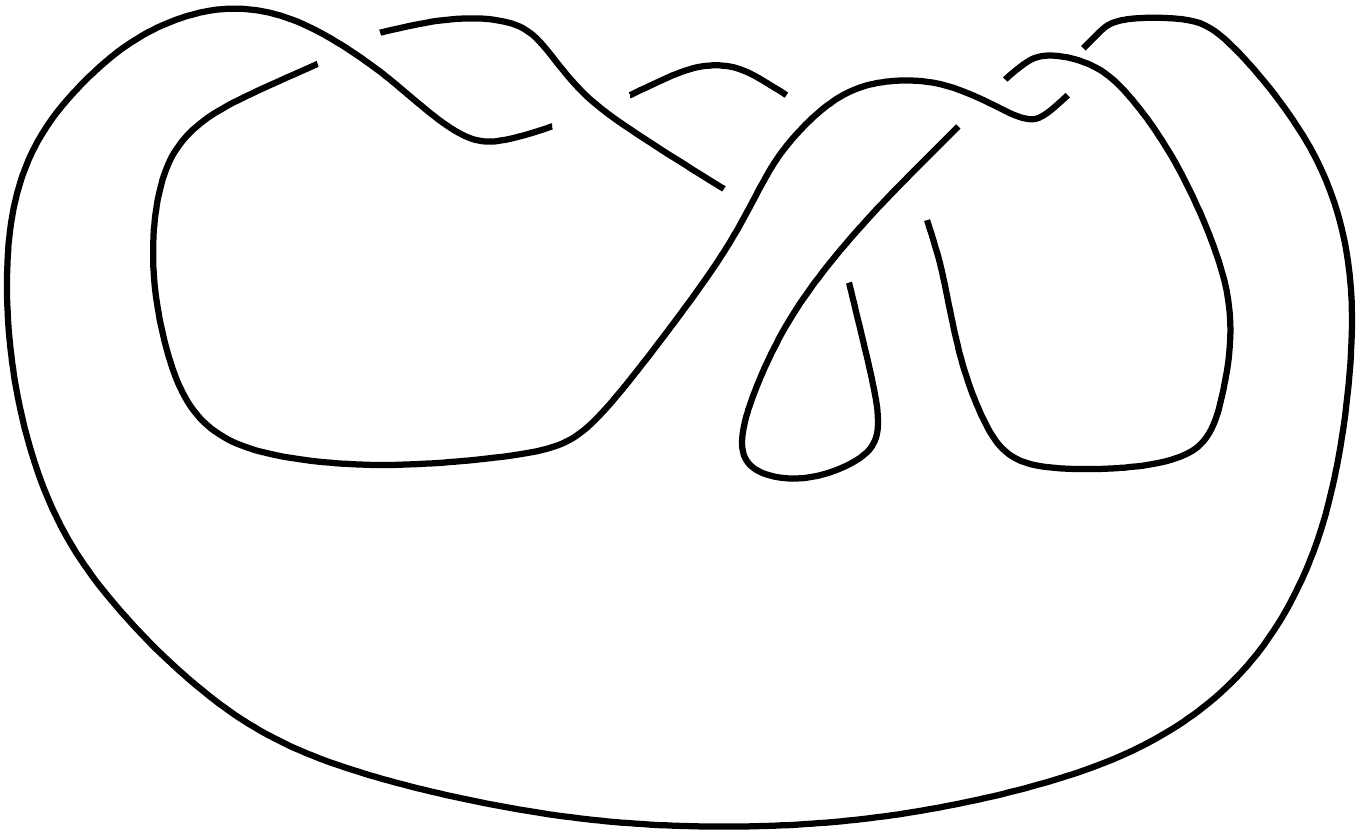}
\caption{The surface obtained from a disk by attaching two twisted
  bands is an orientable surface whose boundary is the trefoil
  knot.} \label{fig:seifert}
\end{center}
\end{figure}

One can also use geometrical quantities to characterize knots. The
boundary of a punctured sphere is necessarily the unknot. If, on the
other hand, we puncture a higher genus surface (e.g. a torus) the
boundary curve can be knotted.  In fact there is a well established
theorem saying that every knot is the boundary of an orientable
surface. The proof consists of an explicit construction first derived
by Seifert and the orientable surface is called a \emph{Seifert
  surface} of the knot (see Figure~\ref{fig:seifert} for a sketch of
the Seifert surface of the trefoil). A knot can have many Seifert
surfaces.
The lowest genus (orientable) Seifert surface of a knot
defines the \emph{genus} of that knot. For instance the trefoil and
figure eight knots both have genus 1 since they can each be the
boundary curve of a punctured torus.  It is not difficult to prove
that genus adds under the connected sum operation so the granny knot
has genus 2. As a by-product one has the important result that a knot
cannot be untied by connecting it to another non-trivial knot. In
other words anti-knots cannot exist.

It is important to distinguish between the genus $g$ of a knot
(related to be the boundary of a genus-$g$ surface) and its possibility
of lying on a standardly embedded surface of genus $G$. For example the
figure-eight knot has genus $1$ but it cannot be placed on a unknotted
torus (surface with $G=1$); it can lie however on a standardly
embedded surface with $G=2$.

\subsection{Polynomial invariants}
\label{subsec:polynomials}

One of the fundamental open problems in knot theory is to devise
general schemes for establishing whether two given knots are
equivalent or not. The use of sophisticated ``motion planning''
computational algorithms have proved useful in past years to aid the
geometrical simplification of knots to a minimal, or otherwise
unique~\citep{Pieranski:1998}, conformation which can then be
straightforwardly used to establish topological equivalence. However,
the geometrical embedding of even an unknotted closed curve in $\mathbb R^3$
can, in fact, be so entangled that the simplification to a planar loop is impractical. This may
indeed be impossible for unknotted polygonal curves deformed
by moves preserving the length of their edges \cite{Cantarella98}.

A powerful strategy, alternative to the geometrical simplification,
consists of associating to each knot type an \emph{invariant} quantity
whose value is unaffected by topology-preserving geometrical
manipulations of the curves.

In this way if the value of the invariant differs for two different
embeddings (or diagrams), then the two embeddings have different knot
types. Some invariants that have been obtained by studying the
three-dimensional manifold left after the knot is removed from
space, are \emph{the knot group}, the \emph{topology of branched and
unbranched covering spaces} and the properties of 
\emph{Seifert surfaces}~\citep{Rolfsen:1976,Adams:1994}. 

Nevertheless, these very precise invariants are not very useful
in applied numerical contexts, where it is typically necessary to
identify the knot type of thousands or millions of rings. For such purposes it is
customary to resort to invariants which are computed combinatorially
from knot diagrams and put in the algebraic form as polynomials. Perhaps
the best known knot polynomial invariant is the \emph{Alexander
  polynomial}. The latter is defined in terms of a single variable
$t$ and is computed starting from a given diagram according to the following
general algorithm:
\begin{enumerate}
\item Attach an orientation to the diagram and establish the sign of
  each crossing using the right-hand rule, as in Figure~\ref{fig:cross_sign}.

\item Assign a progressive numbering index to the $n$ arcs of the diagrams
  and (separately) to the $n$ crossings. 

\item Define an $(n\times n)$ matrix $M$. The elements of the $x$th
  row of $M$ are calculated by considering the $x^{\rm th}$ crossing in the
  diagram and the three arcs, $i$, $j$ and $k$ taking part to the
  crossing. For definiteness we shall assume that the $i^{\rm th}$ arc passes
  over arcs $j$ and $k$. All elements of the $x^{\rm th}$ row of $M$ are set
  to zero except for $M(x,i)$, $M(x,j)$, $M(x,k)$. These three entries
  are calculated as follows:
\begin{enumerate}
\item if the crossing $x$ is positive then $M(x,i)=1-t$, $M(x,j)=-1$ and $M(x,k)=t$.
\item if the crossing is negative then set
$M(x,i)=1-t$, $M(x,j)=t$ and $M(x,k)=-1$\ .
\end{enumerate}
Iterating the procedure for all crossings the matrix is completely defined.
\end{enumerate}
Deleting any one of these columns and any one row yields a
$(n-1)\times (n-1)$ matrix. This is the \emph{Alexander matrix}
associated to a given diagram. The determinant of the Alexander matrix
(which is therefore a minor of $M$) is the desired Alexander
polynomial, $\Delta(t)$ \citep{Rolfsen:1976}.

\begin{figure}[tbp]
\begin{center}
\includegraphics[width=\WIDTHB]{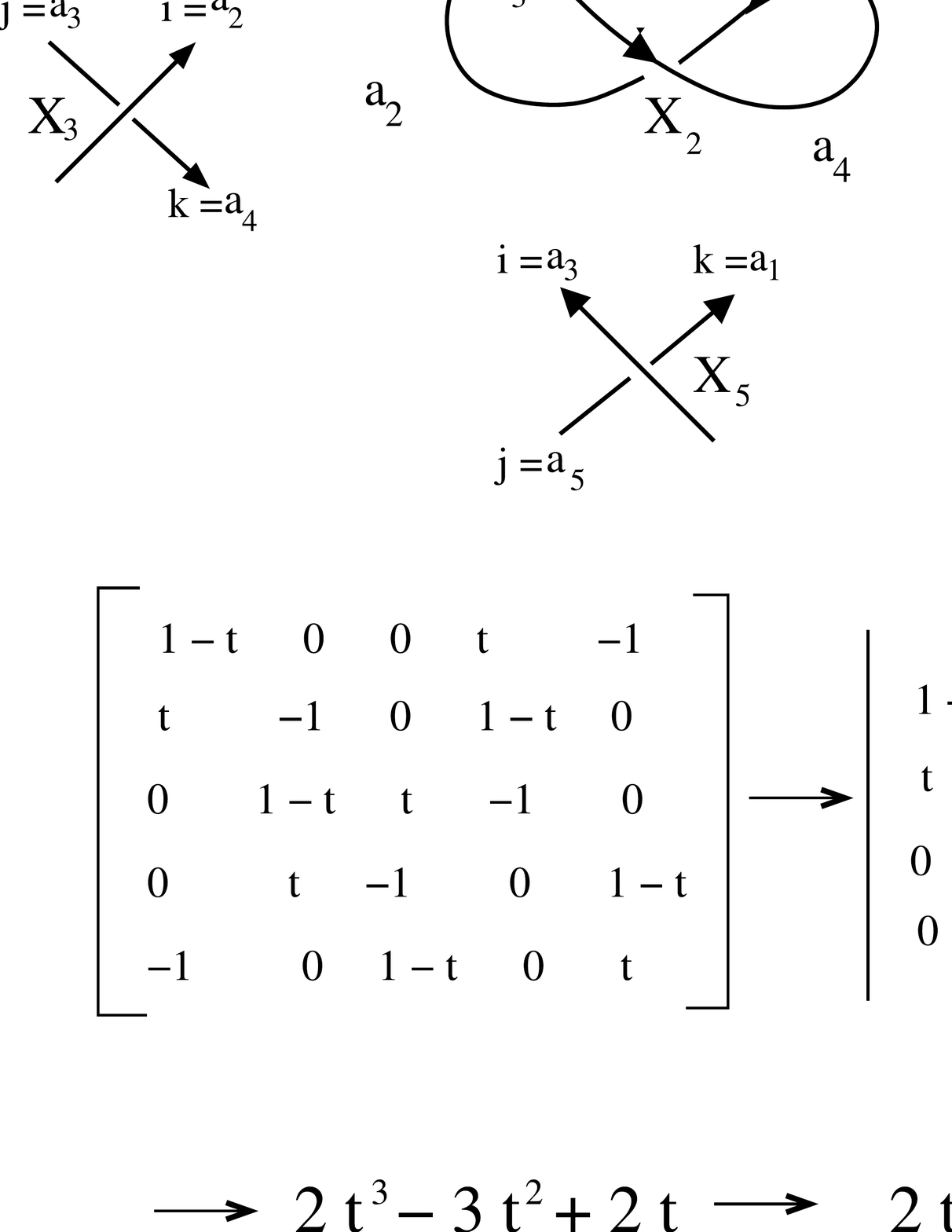}
\caption{Computation of the Alexander polynomial for knot $5_2$. The
  top panel shows the knot diagram with a given orientation: $X_i$
  denotes the $i^{\rm th}$ crossing (found as one goes around the diagram
  along the given orientation) whereas $a_i$ the arc between two
  consecutive undercrossings. Each crossing of the diagram is then
  drawn separately and coded according to the arcs involved taken with
  their orientation.  From these $n=5$ coded crossings it possible to
  calculate the elements of the matrix $5\times 5$ according to the
  algorithm described in the text. By deleting the last column and row
  one obtains the $4\times 4$ Alexander matrix associated to the
  starting oriented knot diagram.  The determinant of this matrix
  gives the Alexander polynomial. Note that in this example a term $t$
  can be factorized out of the diagram.}
\label{fig:alexander_5_2}
\end{center}
\end{figure}

The calculation of the Alexander polynomial for the five-crossing prime
knot $5_2$ is illustrated in Figure~\ref{fig:alexander_5_2}.

Strictly speaking, the Alexander polynomial is not uniquely defined
for a given knot type. This is evident because the size of the matrix,
$M$, and hence the determinant, depend of the number of crossings and
hence on the details of a given diagrammatic representation. 
It turns out that the Alexander polynomials of two
  different diagrammatic representations of the same knot type can
  differ only by a multiple of $\pm t^{m}$, $m\in \mathbb Z$.
  This ambiguity is immaterial if one computes $\Delta(-1)$, which is
  indeed a topological invariant.  This ambiguity, which is immaterial if one computes the topological 
invariant $\Delta(-1)$, is however present if the value of the polynomial is 
calculated for other values of $t$. By factoring out and removing the  $\pm t^m$ contribution one obtains a polynomial that is independent on the chosen projection 
of the knot.

  This gives the so called {\em irreducible} Alexander polynomial.
  For example in the typical case of $t=2$ the algorithm returns the
  number $\pm 2^{m} \Delta (-2)$, $m\in \mathbb Z$. To deal with
  this ambiguity one can find the largest value of $k\in \mathbb Z$
  such that the product $\pm 2^{m} \Delta (-2) 2^{-k}$ is an
  odd integer.  
With a slight abuse
of notation, it is customary to use the notation $\Delta(t)$ also for
the irreducible Alexander polynomial.

The Alexander polynomial satisfies a number of properties including:
\begin{enumerate}
\item $\Delta_{\tau}(t^{-1})$ and $\Delta_{\tau}(t)$ can differ only
  by factors of $\pm t^{\pm m}$; 

\item $\Delta_{\tau}(1)=-1$;

\item $\Delta_{\tau \# \tau'}(t)$ is equivalent to
  $\Delta_{\tau}(t)\Delta_{\tau'}(t)$. For example, since the
  Alexander polynomials of $3_1$ and $4_1$ are respectively
  $\Delta_{3_1}(t) = t^2-t+1$ and $\Delta_{4_1}(t) = t^2-3t+1$, we
  have for the granny knot $\Delta_{3_1\#3_1}(t) = (t^2-t+1)^2$ and
  $\Delta_{3_1\#4_1}(t) = (t^2-t+1)(t^2-3t+1)$.
\end{enumerate}

The first property can be used to check the correctness of a
calculated polynomial. For example, no correct calculation can return
the polynomial $\Delta_{\tau}(t) = t^3+t^2-1$, because
$\Delta_{\tau}(t^{-1})=t^{-3}+t^{-2}-1$ is not in the form $\pm
t^{p}\left ( t^3 + t^2 -1\right)$ and property $1$ is not satisfied.

The Alexander polynomial is relatively easy to implement on a computer
and it is often used in
detecting the knot type of a large number of configurations obtained by stochastic sampling, for example.  There are however some
shortcomings in using the Alexander polynomial as a knot
detector. First, and most importantly, the Alexander polynomial cannot
distinguish between knots and their mirror images.  Second, there
are several pairs of knot types having the same Alexander polynomial. For
instance, the (irreducible) Alexander polynomial of knot $8_{20}$ is
$\Delta(t)=(t^2-t+1)^2$, which is identical to the polynomial of the
granny knot $3_1 \# 3_1$ while $8_{21}$ has the same Alexander
polynomial as $3_1 \# 4_1$. The ambiguity can be even more dramatic in
the sense that non-trivial knots can have the same Alexander
determinant of the unknot. The simplest of such knots are the
Kinoshita-Terasaka Knot and its mutant, the Conway knot, which
have minimal crossing number equal to $n_{cr}^{min} =11$. 

\begin{figure}[tbp]
\begin{center}
\includegraphics[width=\WIDTHB]{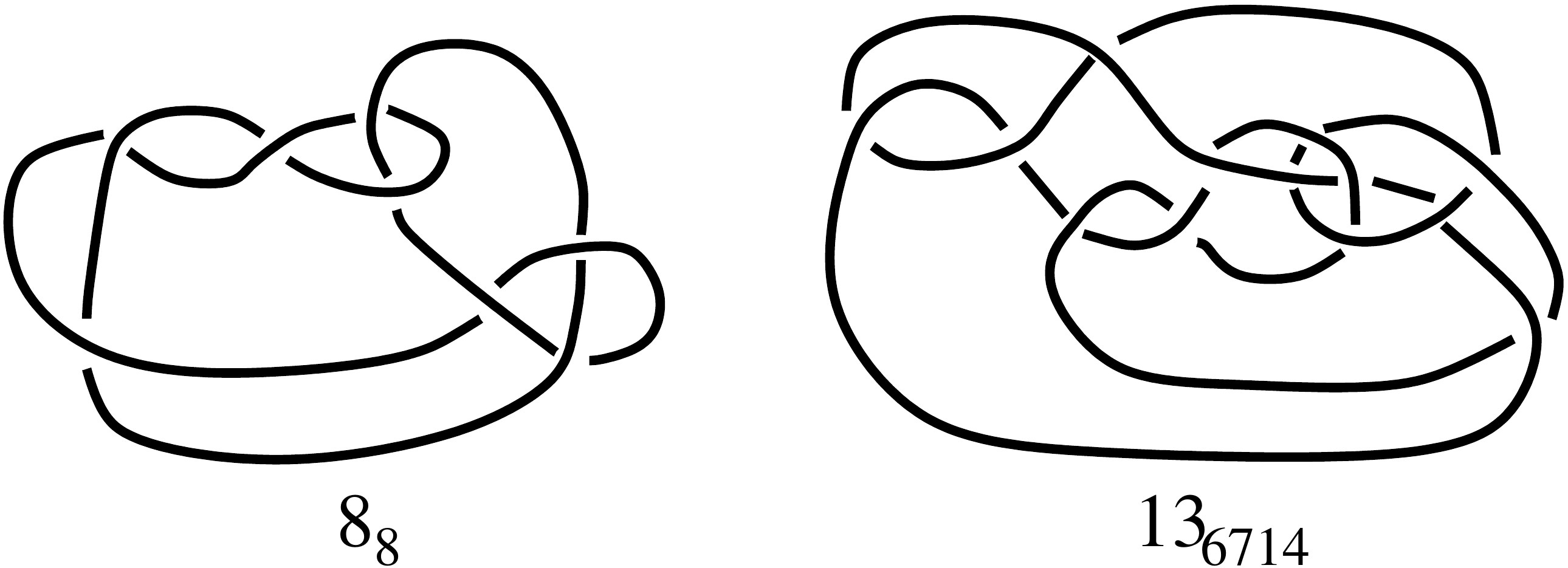}
\caption{Two different knot types having the same HOMFLY
  polynomial.} \label{fig:HOMFLY_fail}
\end{center}
\end{figure}

Other polynomials are known which are better discriminants of
knots. For example the Jones polynomial, the HOMFLY polynomial (see~\citep{Adams:1994} for a simple introduction) and the Vassiliev
invariants are known to be sensitive to knot chirality and to
distinguish a large class of knots that cannot be distinguished by the
Alexander polynomial. However, also for these polynomials there are
known examples of topologically inequivalent knots having the same
polynomials.  For example the two knots in
Figure~\ref{fig:HOMFLY_fail} have the same HOMFLY polynomial. Even the
proof that at least one of the known invariants is a perfect detector
of chirality or non-triviality is an open issue and there may exists
non trivial knots for which all the known polynomials are trivial.
\footnote{There is in fact a perfect algebraic knot and link
detector-two links are equivalent if and only if there is an
isomorphism of the fundamental group of the link complement that
matches up all (longitude, meridian) pairs--this is called the
peripheral structure of the link, and follows from a result of
Waldhausen~\citep{Waldhausen:1968:Ann_Math}. The problem
is that this invariant is not computable in any known way, so is not
very useful in computations.} An additional drawback of these more
sophisticated knot polynomials is that their computational cost
increases dramatically with the number of crossings in the
diagram. For this reason, if one is interested in performing a
detailed investigation of the knot complexity in random loops, a
strong reduction of $n_{cr}$ is recommended before computing these
invariants. The reduction of $n_{cr}$ for a given embedding is, in
general, a highly non trivial task to perform. Stochastic algorithms
have been recently introduced to tackle this problem and some of them
are reviewed in Section~\ref{knot_simp}.

\subsection{Open knots}
\label{openknots}

In various contexts, the notion of knottedness is loosely transferred
to open curves. Strictly speaking this is inappropriate because a
piece of curve is topologically equivalent to a linear segment, no
matter how many knots have been tied in. Indeed, an appropriate set of
manipulations of the open curve can lead to the untying of the knots.
It is, nevertheless, a common experience that these operations can be
lengthy and complicated and ultimately the characteristics of the
curve in terms of its spatial arrangement or fluctuation dynamics are very much
affected by its degree of entanglement.

To characterize the latter a ``virtual closure'' of the open chain is
often introduced so that the usual topological indicators of
knottedness can be used to specify the degree of
self-entanglement. Several closure procedures are possible. The
simplest one is the end-to-end closure obtained by bridging the two
ends with one segment. The amount of spurious entanglement introduced
by this closure depends on the proximity of the chain ends. If their
spatial separation is small compared to the chain radius of gyration
then the entanglement added by the end connector is typically
negligible. The opposite is true when the connector length is large.

To deal with the latter case it is preferable to prolong the chain
ends ``at infinity'' so that they can be connected by an arc not
intersecting the original chain. If the knots are localised in a small
portion of the chain, then the detected topological state will not
depend significantly on the specific closure procedure. Other
situations, for example when the chain attains a globular shape with
one or both ends tucked inside, the result will depend on the closure
scheme. These ambiguities have motivated the introduction of a
probabilistic concept of knottedness for an open curve. In this
situation, closure at infinity along all possible directions of
prolongation of the chain ends are considered and a probability
profile for the detected topological states is considered in place of
a single topological identifier~\citep{Millett:2005:Macromol}. In
several cases of practical interest these different schemes yield
consistent results. One notable example of this consensus is provided
by knotted proteins which constitute about 1\% of the set of
presently-known protein
structures~\citep{Potestio:2010:PLoS-Comput-Biol:20686683}. A very
high consistency exists among the different methods of closure
regarding the knotted/unknotted state of individual protein chains and
also regarding the occurring types of
knots~\citep{Virnau-Mallam-Jackson-knot-review:2011,Taylor:2000:Nature:10972297,Virnau:2006:PLoS-Comput-Biol:16978047,Lua:2006:PLoS-Comput-Biol:16710448,Lai:2007:Nucleic-Acids-Res:17526524,Khatib:2006:Bioinformatics:16873480,Bolinger:2010:PLoS-Comput-Biol:20369018,Potestio:2010:PLoS-Comput-Biol:20686683}.

Clearly, the cases of minimal ambiguity for the knotted state of an
open curve are associated to situations where knots are localised in a
small portion of the chain. Indeed a closely-related problem is the
determination of the length of a knot (in either a closed or open
chain)~\citep{Orlandini:2009:Phys-Bio}. This problem, which has been
recently tackled in ref. \cite{Min_entang_closure} by comparing
existing and new closure schemes, will be addressed in
Section~\ref{size_knots} of this review.

\section{Coarse-grained models of polymers in $\mathbb R^3$: from freely-jointed chains to semi-flexible self-avoiding ones.}
\label{pol_models}

In this section we shall introduce a number of standard polymer models
whose geometrical and topological properties in a variety of physical
conditions will be extensively discussed in the rest of this review.
The models will be described in the ``open chain'' formulation as it
is understood that the chain closure can be introduced by simply
enforcing the coincidence of the first and last monomer of the chain,
to obtain polymer rings.

Arguably, the simplest discrete polymer model is represented by the
\emph{freely-jointed chain}.  
The model  consists of a succession of  equally-long segments, usually
termed bonds,  that can take any arbitrary  relative orientation. This
model essentially captures the chain connectivity property while other
salient  aspects associated to  the energy  cost of  introducing local
bends in the chain, or disallowing self-avoidance, are not considered.

The oversimplified nature of the model brings the advantage of
allowing for a straightforward analytical or numerical
characterization and is a natural first step for characterizing the
behaviour of more realistic polymer models.

\begin{figure}[tbp]
\begin{center}
\includegraphics[width=\WIDTHA]{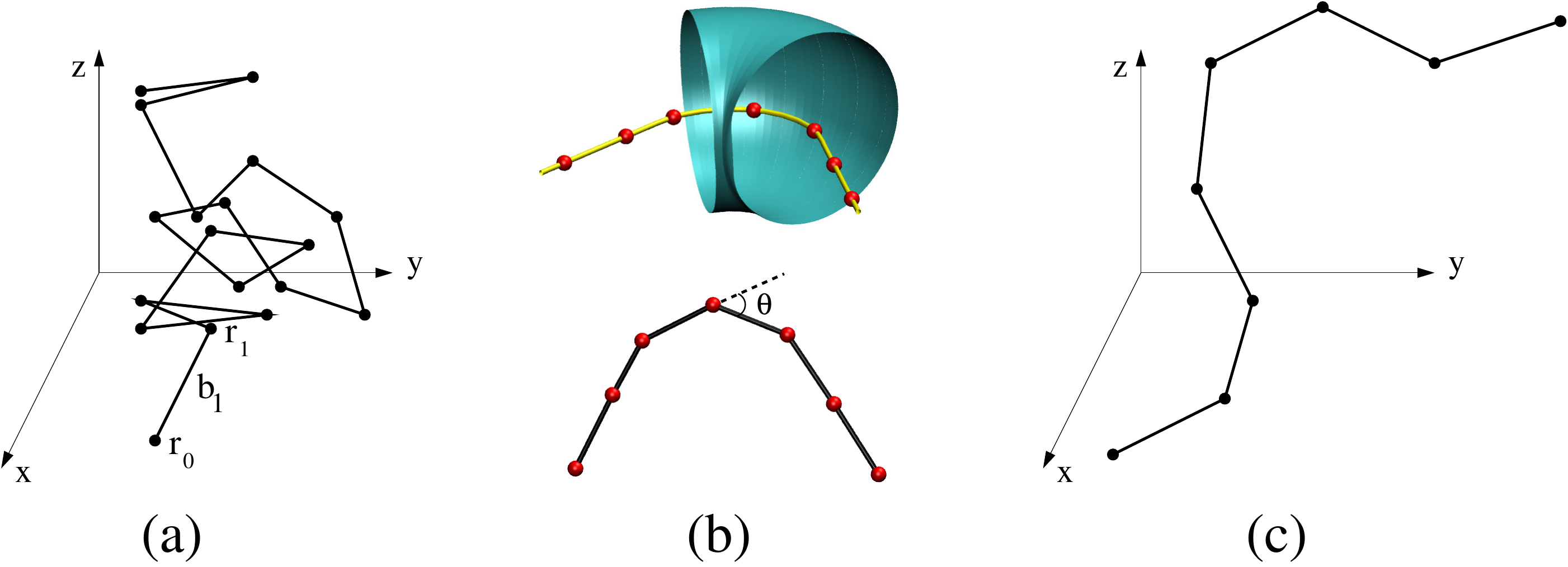}
\caption{(a) Possible configuration of a freely-jointed chain. (b) The
  yellow curve indicates the centerline of a continuous polymer with
  uniform thickness (tube). A portion of the tube hull is shown
  explicitly to illustrate that the polymer thickness cannot
  excessively have small local radii of curvature without
  self-intersecting. When the polymer is discretized (red
  beads), the curvature constraint reflects in a maximum allowed value
  for the angle, $\theta$, formed by consecutive bonds, as shown in the bottom panel. (c) Configuration of a discrete chain where the angle of two
  consecutive bonds is not allowed to exceed 45$^o$.}
\label{fig:fjc_tc}
\end{center}
\end{figure}

To illustrate this point, and for the purposes of introducing the
notation that will be used hereafter, we consider a freely-jointed chain configuration, $\Gamma$, with
nodes at positions $\{\vec{r}_0,\ \vec{r}_1,\ ...  \vec{r}_N\}$. All
bonds joining consecutive nodes, $\vec{b}_i \equiv \vec{r}_i -
\vec{r}_{i-1}$, have the same length, $b$, as illustrated in
Figure~\ref{fig:fjc_tc}a.

The mean square end-to-end distance has a linear dependence on the
chain length:

\begin{equation}
\langle R_{ee}^2 \rangle = \langle \sum_{i,j=1}^N \vec{b}_i \cdot
\vec{b}_j \rangle = \sum_{i,j=1}^N \langle \vec{b}_i \cdot \vec{b}_j
\rangle = \sum_{i=1}^N \langle \vec{b}_i \cdot \vec{b}_i \rangle = N b^2
\label{eqn:ree}
\end{equation}

\noindent where the statistical independence of the orientation of
different bonds was used.  Besides the end-to-end distance, a commonly
employed measure of the typical spatial size of the chain is given by
the radius of gyration, $R_g$:
\begin{equation}
\left \langle R_g^2 \right \rangle= \left \langle \frac{1}{N+1}  \sum_{i=0}^N | \vec{r}_i - \vec{r}_{CM}|^2 \right \rangle = {1 \over 2} 
\left (\frac{1}{N+1}\right )^2 \left \langle  \sum_{i,j=0}^N | \vec{r}_i - \vec{r}_{j}|^2 \right \rangle \ .
\label{radius_gyration}
\end{equation}

\noindent where $\vec{r}_{CM}$ denotes the location of the chain
center of mass, $\vec{r}_{CM} \equiv \sum_{i=0}^N \vec{r}_i/(N+1)$. In
the limit $N \gg 1$ one has that the average end-to-end distance and
the radius of gyration are tied by a simple linear relationship,
namely $\langle R_g^2 \rangle=N^2b =\langle R_{ee}^2 \rangle/6$ which
highlights that also $R_g^2$ depends linearly on $N$
~\citep{Flory:1969}.

\noindent As anticipated, the model is highly simplified and needs to
be suitably generalised before it can be used to describe key
physical aspects of a free polymer in solution. Arguably, the simplest
physical effect that can be introduced in the model to make contact
with the phenomenology of flexible chains is to consider the
correlations in the directionality of consecutive bonds.

These correlations arise, for example, when accounting for the polymer
bending rigidity.  By analogy with the case of an ideal elastic rod
slightly perturbed from the straight configuration, the bending energy
of a chain configuration, $\Gamma$, can be approximated by
the following energy function: 
\begin{equation}
E_{bend}(\Gamma) =  K_c \, \sum_i \left( 1 -  \frac{\vec{b}_i \cdot  \vec{b}_{i+1}}{b^2}\right)
\label{eqn:br}
\end{equation}

where $K_c$ is the chain bending rigidity.
Freely-jointed chains endowed with a bending rigidity term as above are
termed \emph{Kratky-Porod chains}.

Because of the simple form of the energy function of eq.~\ref{eqn:br} (which does not account for chain self-avoidance), it is possible to compute analytically several properties of Kratky-Porod chains in thermal equilibrium. Specifically, we shall assume that the statistical weight of a chain configuration, $\Gamma$, is given by the canonical Boltzmann weight:

\begin{equation}
{e^{-E_{bend}(\Gamma)/K_BT} \over \sum_{\Gamma^\prime} {e^{-E_{bend}(\Gamma^\prime)/K_BT}}}
\label{canonweight}
\end{equation}

where $K_B$ is the Boltzmann constant, $T$ is the temperature and the correct normalization of the statistical weight is ensured by the denominator where the summation stands for the integrated contribution 
over all possible chain configuration.
 
The effects that the local bending energy has on the
  correlation of the bonds directionality in the chain can be
  ascertained by calculating the average (equilibrium value) of the
  scalar product of two chain bonds at a separation $l$ along the
  chain. This quantity is straightforwardly calculating starting from
  $l=1$ and iterating to larger values of $l$ obtaining~\cite{Flory:1969}:

\begin{equation}
\langle \vec{b}_i \cdot \vec{b}_{i+l} \rangle = b^2\, e^{-l/l_p}\ 
\label{eqn:pers}
\end{equation}

 where $\langle \rangle$ denotes the equilibrium
average where configurations are weighted with the canonical weight of eq.~\ref{canonweight} and the quantity $l_p$, termed the persistence length, is given by
\begin{equation}
l_p = -b/\log\left [\coth\left ({K_c \over K_B T}\right ) - {K_B T
    \over K_c}\right ].
\end{equation}

\noindent The above expression clarifies that the bending rigidity
introduces an exponentially decaying correlation in the orientation of
bonds along the chain, as readily perceived by the comparative
inspection of panels (a) and (c) in Figure~\ref{fig:fjc_tc}.

Notice that, in the limit $K_c \gg K_B T$, that is when the average
value of the bending angle between consecutive bonds is small, the
expression of the persistence length simplifies to: $l_p = b {K_c
  \over K_B T}$, which is the well-known simple relationship tying the
persistence length and the bending rigidity in the continuum limit of
the Krakty-Porod chain (which leads to the so-called \emph{worm-like
  chain})~\citep{Marko_Siggia:1995}.

A further physical effect that introduces correlations in the bonds
orientations is the steric hindrance associated with a finite
thickness of the chain. When accounting for this effect, the model
chain should be viewed as the centerline of a discrete thick polymer
with cross-sectional radius equal to $\Delta$. At a global level, the
finite thickness of the chain introduces correlations in portions of
the chain that can be even far apart along the ``polymer
sequence''. In addition, this self-avoidance has an impact at a local
level as the chain cannot attain local radii of curvature smaller than
$\Delta$ ~\citep{Gonzalez:1999:Proc-Natl-Acad-Sci-U-S-A:10220368}, see
Figure~\ref{fig:fjc_tc}b . This implies that the angle $\theta$ formed
by two consecutive bonds must satisfy the following constraint:
\begin{eqnarray}\label{eq:angle}
\cos\theta &\ge& 1 - \frac{ b^2}{ 2\Delta^2}\ .
\end{eqnarray}

\noindent which, propagated along the chain, introduces an
exponentially-decaying correlation of the chain orientation that is
entirely analogous to the one in eq.~(\ref{eqn:pers}):

\begin{equation}
\frac{\vec{b}_{n+1} \cdot \vec{b}_1}{b^2} = \left[ 1- \frac{b^2}{4\Delta^2}\right]^n\ .
\end{equation} 

This correlation, which is readily perceived by the comparative
inspection of panels (a) and (c) in Figure~\ref{fig:fjc_tc} and the
decay length can be viewed as a thickness-related persistence
length given by~\citep{Marenduzzo:2003:J-Mol-Biol:12842465}:

\begin{equation}\label{eq:lp_TC_local}
l_p =  - \frac{b}{\ln\left(1-\frac{b^2}{4\Delta^2}\right)}\ .
\end{equation}

\noindent We stress that the above expression merely captures the {\em
  local} effects of the finite thickness of the chain.

We conclude this discussion by considering  how the interplay of the chain
contour length (that is the total chain length obtained by summing the length of all its bonds), $L_c=N b$, and of the persistence length, $l_p$,
controls the physical behaviour of the system.

To do so we shall consider a Kratky-Porod model of given contour
length, $L_c=Nb$, and given persistence length, $l_p$. The
discretization step, $b$, is assumed to be much smaller than the
persistence length (and therefore $K_c$ is suitably adjusted so to
reproduce the given persistence length). In this situation, the mean
square end-to-end distance is given by~\citep{Flory:1969}

\begin{eqnarray}
\langle R_{ee}^2 \rangle &=& \langle \sum_{i,j} \vec{b}_i \cdot
\vec{b}_j \rangle = b^2 \sum_{i,j} \exp\left[ - | i - j | b \over l_p\right]\\
&\approx& 2 l_p L_c \left [ 1 - {l_p \over L_c} \left (1 - e^{- L_c \over l_p} \right ) \right ]
\end{eqnarray}

\noindent From this expression it is readily seen that when the
contour length is much smaller than $l_p$, the polymer can be viewed
as a stiff chain. In fact, the mean square end-to-end distance is
approximately
\begin{equation}
\langle R_{ee}^2 \rangle \approx L_c^2 \left[ 1 - { L_c \over 3 l_p } + O\left({L_c^2 \over l_p^2}\right)\right]
\label{eqn:ree2}
\end{equation}
\noindent which, for $L_c \gg l_p$ reduces to the rigid rod value,
$\langle R_{ee}^2 \rangle \approx L_c^2$. For future reference it is
worth calculating separately the contributions to eq.~(\ref{eqn:ree2})
along the direction approximating the straight rod, $\langle
{R_{ee}^\parallel}^2 \rangle$ and in the plane perpendicular to it,
$\langle {R_{ee}^\perp}^2 \rangle$.  The first term, $\langle
{R_{ee}^\parallel}^2 \rangle$, is readily obtained by noting that the
average directionality of the chain is biased by the first bond,
$\vec{b}_1$, so that
\begin{equation}
\langle \vec{R}_{ee}^\parallel \rangle = {1 \over b} \sum_l \langle \vec{b}_l \cdot \vec{b}_1
\rangle = b \sum_l \exp\left({-l \over l_p}\right) = 
{ 1 - \exp\left ({ - L_c \over l_p}\right)\over  1 - \exp\left({ - b \over l_p}\right)}\ . 
\label{eqn:reepar}
\end{equation}
\noindent From expressions (\ref{eqn:reepar}) and (\ref{eqn:ree2}) and using $\langle
{R_{ee}^\perp}^2 \rangle = \langle {R_{ee}}^2 \rangle - \langle {R_{ee}^\parallel}^2 \rangle$ one
has, to leading order in $L_c/l_p$ and $b/L_c$:

\begin{equation}
\langle {{R_{ee}^\perp}^2} \rangle = {2 \over 3} {L_c^3 \over l_p}\ .
\label{eqn:rperp2}
\end{equation}

When $L_c \approx l_p$ the chain is said to be semi-flexible, in that
appreciable deviations from the straight configurations are possible
in thermal equilibrium. Instead, in the limit $L_c \gg l_p$, one has
\begin{equation}
\langle R_{ee}^2 \rangle \approx 2 l_p L_c
\end{equation}

\noindent By comparison with eq.~(\ref{eqn:ree}), the latter identity can be
used to say that the equilibrium size of the chain is equivalent to
that of a freely-jointed chain with same contour length, but where the
statistically independent segments have effective length equal to
twice the persistence length, $l_K=2 l_p$. The equivalent length, $l_k$, is
termed the \emph{Kuhn length}.

\subsection{Flexible and semi-flexible chains: the cylinder and   rod-bead models}
\label{cylinder_models}

It is important to stress that in the rigid and semi-flexible regimes
the non-local self-avoidance effects in unconstrained chains can be
largely neglected, as steric clashes will not be likely to occur
because of the limited contour length of the chain.

This is not true when $L_c$ largely exceeds $l_p$, which is the
condition in the flexible polymer regime. When self-avoidance is taken
into account, the end-to-end distance, $R_{ee}$ or, equivalently, the
radius of gyration, does not have a linear dependence on the contour
length anymore, rather it obeys the more general formula
\begin{equation}
\langle R_{ee}^2 \rangle \propto {N}^{2\nu} b^2
\end{equation}

\noindent where the self-avoiding exponent $\nu$ is about
0.588~\citep{Clisby:2010:Phys-Rev-Lett}, and is often approximated by
its mean-field estimate 3/5 ~\citep{Flory:1969,DeGennes:1979}).

The above discussion has been mostly focused on considerations of the
relative magnitude of the persistence and contour length.  For
modelling purposes a key element to consider is also the relative
magnitude of the polymer thickness, $\Delta$, and the persistence
length, $l_p$.

\begin{figure}[tbp]
\begin{center}
\includegraphics[width=\WIDTHB]{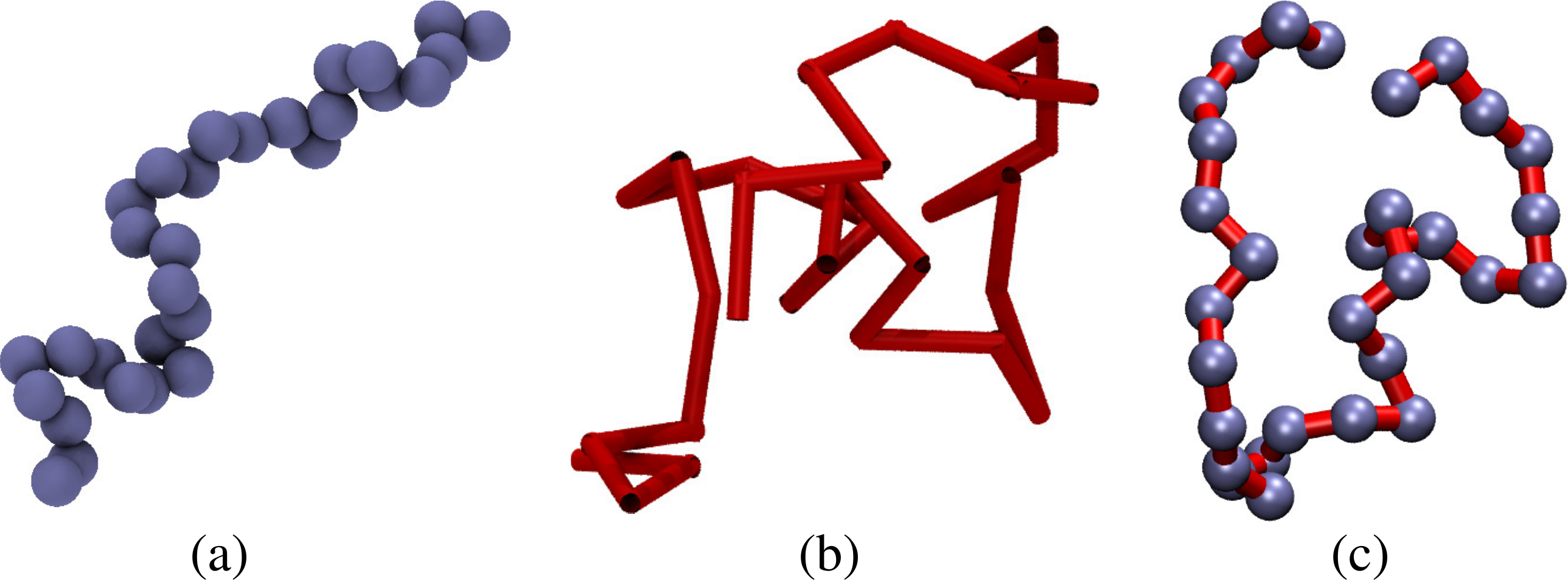}
\caption{Possible configurations of: (a) a chain of beads, (b) a chain of cylinders, (c) a chain of beads and cylinders.}
\label{fig:beads_cyls}
\end{center}
\end{figure}

For several biopolymers, such as polysaccharides, proteins and
single-stranded DNA, the persistence length is comparable with the
thickness. In such contexts, it is natural to model the polymers as
chains of spherical beads of radius $\Delta$, as shown in
Figure~\ref{fig:beads_cyls}a.  The so called \emph{bead model} can be
endowed with a bending rigidity which, together with constraints on
the minimum attainable local radius of curvature imposed by the
self-avoidance (see eq.~\ref{eq:angle}), determines the chain
persistence length\cite{Toan:2005:Biophys-J:15849251,Toan:2006:JPhysCondMat}.

Notice that when the persistence length largely exceeds the chain
thickness, then the bonds at a small arclength separation would tend to
be collinear. This observation suggests the possibility of modifying the
model so as to reduce the number of effective degrees of freedom in
the system. An effective way to do this is to describe the chain as a
sequence of cylinders with thickness $\Delta$ and axis length, $l$,
equal to a fraction of the persistence length (and yet much larger
than $\Delta$). A possible configuration of a flexible chain of
cylinders is shown in Figure~\ref{fig:beads_cyls}b.

Again, the correct persistence length would be reproduced by the
introduction of a suitable bending rigidity between consecutive
cylinders~\citep{Rybenkov:1993:Proc-Natl-Acad-Sci-U-S-A:8506378}. The
excluded volume interaction is simply enforced by disallowing pairs of
non-consecutive cylinders from overlapping. This is done by
calculating the distance of minimum approach of the centerlines of the
two cylinders and requiring to be larger than $2\Delta$, see
\ref{app:segdist}.  Note that the presence of a
  non-zero thickness, or excluded volume, renders an exact calculation
  of the tangent-tangent correlations impossible. Excluded volume interactions cause the tangent-tangent correlation to acquire appreciable deviations from the exponential decay above a certain arclength separation which depends on the interplay of $l$, $\Delta$ and $K_c l/k_BT$ (for more
  details see e.g. the comparison between the discretized Kratky-Porod
  model which disregards steric effects and molecular dynamics
  simulations above the $\theta$ point
  in~\citep{Rosa:2003:Macromolecules} which holds for the
  bead-and-spring polymer model defined above). 

The \emph{cylinder model} (with flexible joints) has been largely
adopted to model a variety of molecules and especially double-stranded
DNA (dsDNA).  In fact, the dsDNA persistence length is equal to 50 nm
while the bare DNA diameter has the much smaller value of 2.5 nm
(larger effective diameters are reported depending on the
concentration/type of counterions in solution
~\citep{Rybenkov:1993:Proc-Natl-Acad-Sci-U-S-A:8506378,Toan:2005:Biophys-J:15849251,Toan:2006:JPhysCondMat}).

A model that interpolates between the beads model and the cylinder
model may be described by a sequence of equally-long cylinders of
thickness $\Delta_1$; at the joints of the cylinders a sphere of
radius $\Delta_2$ is introduced, see Figure~\ref{fig:beads_cyls}c.
By varying $\Delta_1$ and $\Delta_2$ at fixed cylinder length, it is
possible to explore the interplay of various excluded volume terms on
the conformational statistics of the polymer rings of interest. In the
limiting case of unit length cylinders with $\Delta_1=0$ and
$\Delta_2$ finite, it is obtained the \emph{rod-bead
  model}~\citep{Chen:1981} which has been often used to study
the effect of self-avoidance on the knotting probability of polymer
rings (see Section~\ref{knot_prob})

We conclude by observing that, when the polymers are subjected to
spatial confinement, one needs to take into account further
length-scales in the problem, namely the typical width of the
confining geometry. In this case, the viability of cylinder models
depends on the possibility of identifying a cylinder length that is not
only much larger than $\Delta$, but also appreciably smaller than the
width of the confining region.

\subsection{Polymer models on regular lattices: combinatorial arguments and rigorous results}
\label{lattice_models}

The above mentioned models allow for a description of polymeric chains
through a limited number of continuous degrees of freedom.  Albeit
greatly simplified, the models are sufficiently complicated that the
characterization of their properties (especially when subject to
confinement) requires the use of advanced stochastic sampling
techniques such as those described in the next sections.

Occasionally, one deliberately resorts to over-simplified polymer
models to achieve an exhaustive enumeration of the configuration
space rather than sampling it with stochastic techniques. Almost
invariably, these approaches entail the embedding of the model polymer chain
on a discrete lattice, for example the cubic lattice.

\begin{figure}[tbp]
\begin{center}
\includegraphics[width=\WIDTHB]{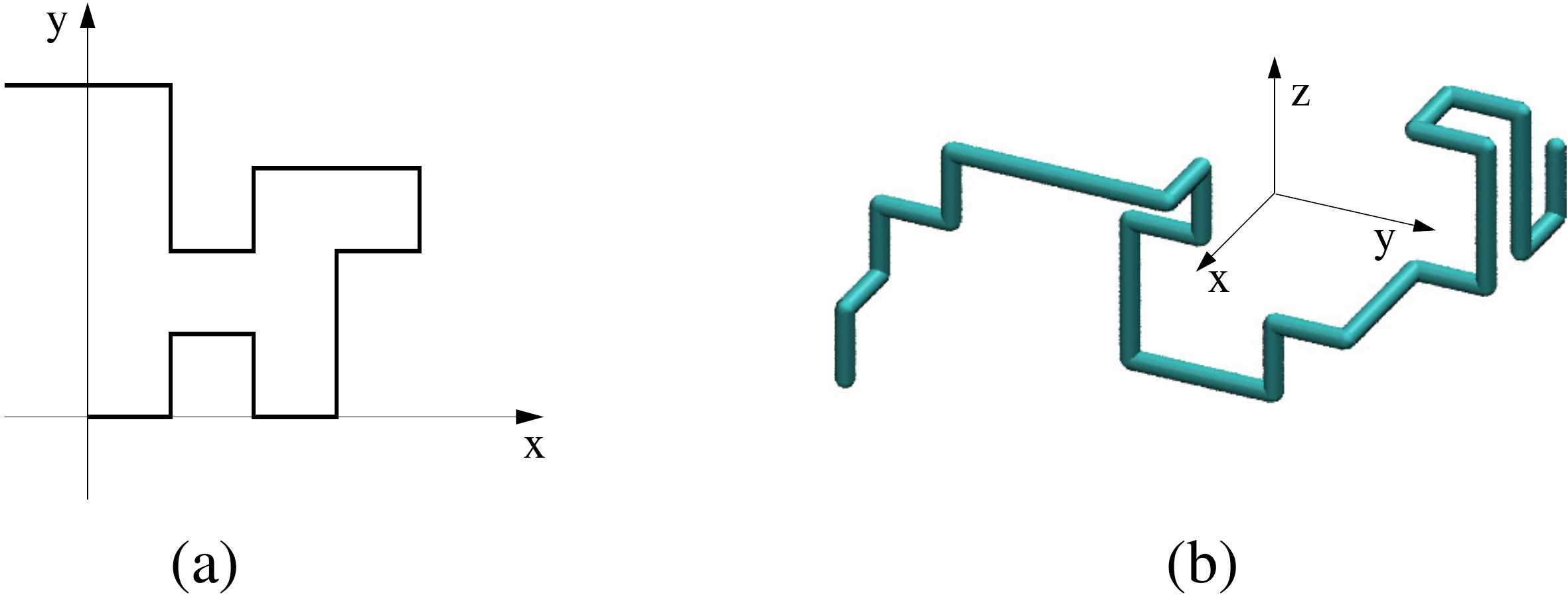}
\caption{Configurations of a self-avoiding walk embedded in: (a) a square lattice, (b) a cubic lattice.}
\label{fig:lattice}
\end{center}
\end{figure}

In the simplest case, the polymer is modeled as a walk which joins
nearest neighbours on the lattice. The self-avoidance is introduced by
preventing the chain from multiply-occupying any lattice edge or node,
as shown in Figure~\ref{fig:lattice}. A bending rigidity can also be
introduced by giving an energy penalty for non-collinear consecutive
segments of the walk. Other variants of this model are possible by
allowing the walk to bridge not only nearest neighbours but also, for
example, second-nearest neighbours. In this case chains are endowed
with a larger configuration freedom though the bond lengths can
fluctuate from configuration to configuration
\citep{Kremer_bond_fluctuation}.

Denoting by $c_N$ the number of distinct (up to translation)
self-avoiding walks with $N$ segments, hereafter referred to as edges
or bonds, on $3D$ lattice, one can use simple counting arguments to
compute by hand the values of $c_N$ for small $N$ values: $c_1=6$ ,
$c_2=30$, $c_3=150$, $c_4=4\times c_3 -24= 726$ etc.  Clearly, as $N$
increases the manual computation of hand $c_N$ is unfeasible so that
it is necessary to resort to efficient numerical techniques for the
exhaustive generation, and hence enumeration, of {\em all} possible
different walks or rings on a lattice up to a certain number of
segments.

These enumeration techniques have been an essential tool to establish
and verify several results regarding equilibrium properties of open
chains as well as rings. In particular, in the asymptotic limit of
large $N$, rigorous results can be established by combinatorial
arguments and simple mathematical analysis. Among the others we wish
to cite here the important theorem due to Hammersley
~\citep{Hammersley:1957:PCPS} .

\begin{equation}
\lim_{N\to \infty} {1 \over N}\, \log c_N \equiv \kappa \label{hamme}
\end{equation}
where $\kappa$ is the \emph{limiting entropy} per step of self-avoiding
walks. $\kappa$ depends on the $3D$ lattice and for the cubic lattice
the best estimate is $\kappa=1.544148\pm 0.000038$
~\citep{Rensburg&Rechnitzer:2008:JPA}. The above result establishes
that the number of self-avoiding walks increases exponentially rapidly
with the number of segments $N$. Few rigorous results are available
about the rate of approach to the limit. On the other hand a mapping
between the statistics of self-avoiding walks and the $O(n)$ model of
ferromagnets in the limit $n\to 0$,~\citep{DeGennes:1979} allows us to
establish the following scaling behaviour for $c_N$

\begin{equation}
c_N \sim e^{\kappa N} N^{\gamma-1}
\end{equation}
where $\gamma$ is the universal \emph{entropic exponent} whose value
depends on the dimensionality of the space in which the walk is
embedded. For three-dimensions, $\gamma \sim 1.566$
~\citep{Clisby_et_al:2007:JPA}. For ring polymers the corresponding
lattice model is a self-avoiding polygon (or simply polygon) i.e. a
self-avoiding walk with the termini that are one lattice spacing
apart. If we denote by $p_N$ the number of (undirected, unrooted)
self-avoiding polygons embeddable in a lattice, for the simple cubic
lattice we have $p_4=3$, $p_6=22$ and $p_8=207$. Again we count two
polygons as distinct if they cannot be superimposed by translation.
Also for polygons it is possible to prove that
~\citep{Hammersley:1961:PCPS}

\begin{equation}
\lim_{N\to \infty} {1 \over N}\, \log p_N \equiv \kappa \label{hamme_pol}.
\end{equation}
Notice that unconstrained polygons and self-avoiding walks have the
same limiting entropy. This is a not trivial result since, for
example, it does not hold for walks and polygons confined in prisms
(see Section~\ref{rig_prisms}).

\section{Second Part: Self-entanglement of single polymer chains}

In the second part of this review, spanning from sections
\ref{Monte_Carlo} to \ref{knot_prob_stress}, we summarize the
state-of-the-art of the characterization of the types and abundance of
knots formed by polymers which circularize either in unrestricted
environments or under spatial confinement. 

The interest in characterizing the probability that polymers,
fluctuating in equilibrium, form a knot upon circularization has grown
in parallel with the availability of experimental techniques capable
of detecting the knotted/unknotted state of circular polymers.

\begin{figure}[tbp]
\begin{center}
\includegraphics[width=\WIDTHB]{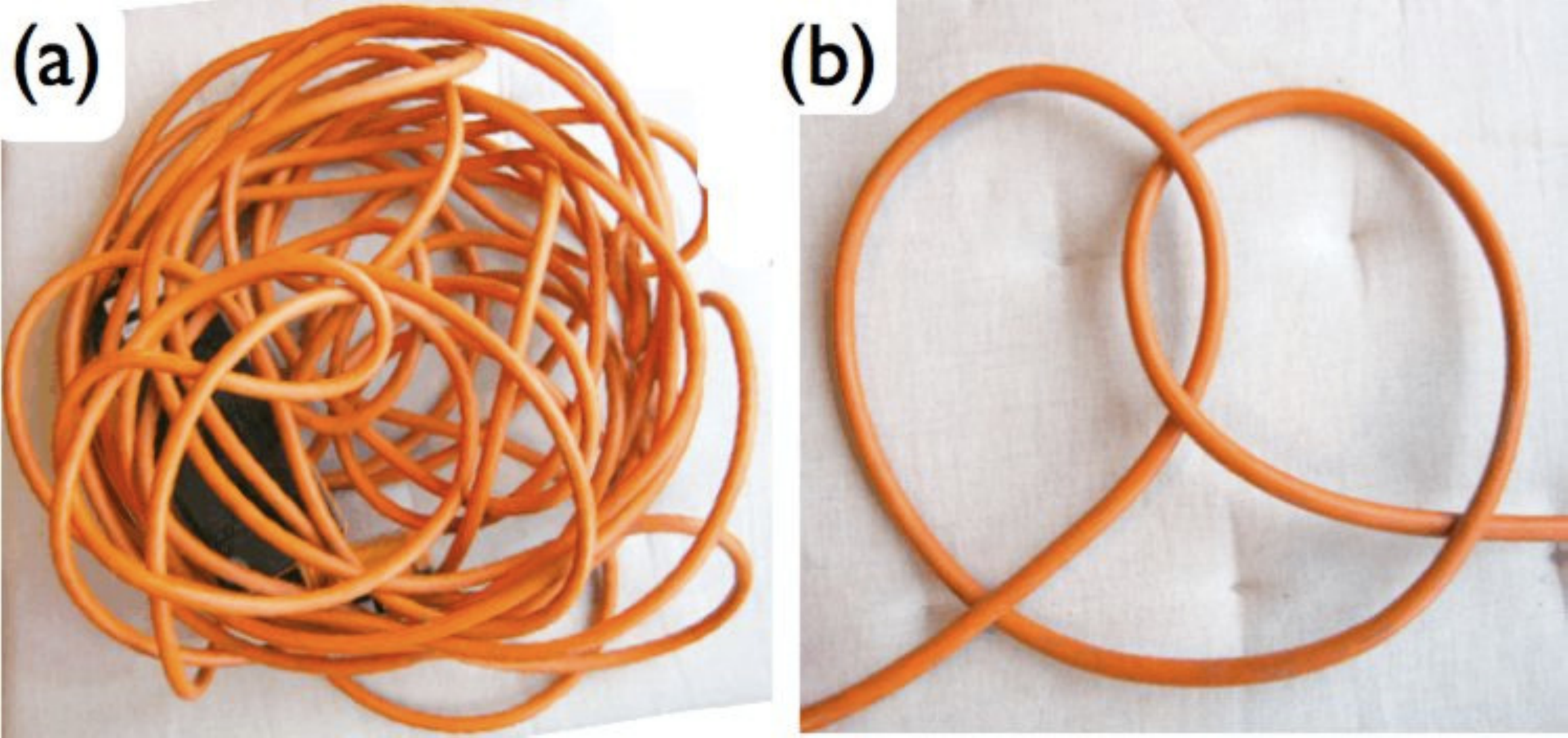}
\caption{(a) A 10m extension lead taken from the cupboard where it was
  stored. When the cable ends are attached to the power socket and to
  the hoover cable its entanglement is trapped in the form of a knot
  that can be of considerable complexity. The knot cannot be removed
  unless one of the cable ends is release and suitably passed through
  the tangle. By spreading out the cable in (a) so to simplify the
  geometry of the trapped knot, we recognise the $4_1$ knot, see panel
  b.  In our experience the knot complexity in the extension cord
  increases as it is repeatedly used and stored away (that is before
  our spouses force us to get up from the couch and engage the endless
  battle against the cable configuration entropy!).}
\label{fig:extension_cord}
\end{center}
\end{figure}

Common experience suggests that long ropes that are shaken or
shuffled, tend to become highly self-entangled. When their ends are
joined, the chances that the resulting arc loop is knotted are quite
high, just think of your garden hose or extension cord after it has
been repeatedly used and stored away, see Figure~\ref{fig:extension_cord}.

This observation leads to the question: what is the degree of knotting
of flexible polymers fluctuating in thermal equilibrium? A first
answer to this was already proposed in the early sixties by Frisch,
Wasserman~\citep{Frisch_Wasserman:1961:JACS} and
Delbruck~\citep{Delbruck:1962} (FWD) who conjectured that
infinitely-long rings are almost certainly knotted (their knotting
probability tends to 1 when their length goes to infinity). About
twenty years later this conjecture was turned into a rigorous theorem
for different models of polymer
rings~\citep{Sumners&Whittington:1988:J-Phys-A,Pippenger:1989:DAM}. The
key concepts underpinning the proof are given in
Section~\ref{knot_prob}.

Since the FWD conjecture several efforts have been spent on the precise characterization of how the knotting probability depends on the length of the ring polymer. More recently, the interest has shifted towards characterizing how the knot spectrum (that is the knot type and relative abundance/frequency) depends on the polymer length. Most of these investigations have been carried out by numerical means. The computational cost of such studies is significant and arises from two separate factors. On one hand, the sampling of a statistically-significant ensemble of independent polymer ring configurations can be very demanding, especially for confined rings. On the other hand, the precise assignment of the knotted state of each sampled configuration can be computationally intensive too as it entails the combinatorial calculation of global topological invariants, which rapidly grows as a function of the ring length (or degree of confinement).

Both these challenges are discussed in detail in the following
sections.  Particular attention is paid to the description of
efficient stochastic techniques which have proved to be useful in
sampling polymer ring conformations in a variety of physical situations
(see Section~\ref{Monte_Carlo}). Specifically, we first discuss and
illustrate the efficient sampling of unconstrained, flexible and
self-avoiding ring polymers. These results are used to illustrate the
viability of simplified polymers models in reproducing the
experimental knot spectrum, found via gel electrophoresis techniques,
for specific biomolecules, particularly
DNA~\citep{Shaw:1993:Science:8475384,Rybenkov:1993:Proc-Natl-Acad-Sci-U-S-A:8506378}.

Next, attention is turned to the case of polymer rings subject to a
variety of external physical constrains.  On one hand, this problem is
motivated by the advent of novel experiments based on micro- and
nano-fluidic devices, or on micromanipulation tools such as
AFM~\citep{Bustamante:1995:Physics-Today,Hansma:1996:J-Vac} and
optical tweezers~\citep{Moffit:2008:Ann-Rev-Biochem}. These techniques
allow an unprecedented quantitative characterization of the detailed
behaviour of long polymers in confined
conditions~\citep{Tegenfeldt:2004:PNAS,Reisner:2005:Phys-Rev-Lett,Stein_et_al:2006:PNAS,Bonthuis:2008:Phys-Rev-Lett,Reisner:2007:Phys-Rev-Lett}. This
may be achieved by inserting these molecules in narrow
channels~\citep{Tegenfeldt:2004:PNAS,Guo_et_al:2004:NanoLett}, by
adsorbing them onto smooth surfaces~\citep{Ercolini:2007:PRL}, or by
subjecting them to tensile stress~\citep{Strick_et_al:1996:Science}.
A detailed understanding of the statical and dynamical properties of
macromolecules in confined environment is crucial also for designing
novel applicative avenues for nano-confinement, such as the use of
nanochannels to pre-stretch and stabilize DNA molecules prior to their
threading through a nanopore~\citep{Austin:2003:Nat_Mat} or to make a
binding site along a long coiled DNA molecule accessible to external
probes~\citep{Zurla_et_al:2007}.  On the other hand, as noted
previously, genomic materials are also subject to spatial confinement
{\it in vivo}, and their entanglement properties are crucial to the
correct functioning of DNA replication, transcription etc.

The interest in polymer and knot adsorption is also motivated by the
fact that this is rather easy to realise in the lab. For instance,
when a polymer is rooted at the surface, by modulating (e.g. by means
of temperature) the effective attractive potential between the polymer
and the surface, it is possible to trigger an adsorption transition
from an extended regime (with high configuration entropy) to an
adsorbed one when it is mostly localised in proximity of the
surface~\citep{DeGennes:1979,Vanderzande:1998}. This phenomenon may
provide a way to visualise the impact of quasi-2D confinement on
topology. For a polymer embedded exactly in a 2D plane or a sphere it
is impossible to form knots -- how is this regime reached in practice
as a function of temperature? This question may now be answered
experimentally.

The interest in the behaviour under stretching is further sparked by
the recent use of modern micromanipulation
techniques~\citep{Svoboda&Block:1994:Ann-Rev-Bio,Bemis:1999:Langmuir,Moffit:2008:Ann-Rev-Biochem}
to characterize the mechanical response of entangled
polymers~\citep{Saitta_et_al:1999:Nature}.  Remarkably, it is nowadays
possible even to tie a knot in a piece of DNA or actin
fiber~\citep{Arai:1999:Nature} and then monitor its reptation dynamics
along the stretched filament~\citep{Bao_et_al:2003:PRL}.

The experimental difficulties in establishing the knotting probability
for ring polymers subject to geometrical/spatial constraints is
paralleled by the challenges found in theoretical characterizations of
analogous properties for simplified models. The difficulty arises from
the competition of several length scales, e.g. the polymer contour
length, the size of the confining geometry, and the polymer
persistence length (a measure of its flexibility). We shall present
the key theoretical concepts that have been introduced to tackle this
problem such as the de Gennes blob picture and the Odijk deflection
theory~\citep{Odijik:1983:Macromolecules} which are valid,
respectively, for weak and strong confinement. We shall also report on
the state-of-the-art computational techniques used to reduce the
impact on the sampling efficiency of the long relaxation times in
dense polymer phases. Finally, for lattice models, it is possible to
extend some of the rigorous results found in the unconstrained
situation to the confined case -- these results are reviewed in
Section~\ref{rigorous_confined}.

Specifically, here we discuss 3D confinement, adsorption and stretching in Sections~\ref{knot_prob_conf},~\ref{knot_prob_ads},~\ref{knot_prob_stress}
respectively.

For the aspect of 3D confinement, we here address the following key
questions:
\begin{itemize}
\item How does the knotting probability depend on the degree and the
  geometry of the confinement?
\item Does the crossover between the De Gennes and the Odijik regimes
  correspond to a significant change in the knotting probability?
\item What is the knot spectrum of linear polymers that circularize
  under spatial confinement?
\end{itemize}
\noindent As a biological application, we will review the studies of
the topological entanglement of viral DNA inside viral bacteriophage
capsids\cite{0953-8984-22-28-283102,Gelbart:2009:Science:19325104}. This is an extremely intriguing problem for a number of
reasons. For instance, until very recently, it was not possible to
account theoretically for the experimentally-observed viral knot
spectrum using simplified DNA models. Furthermore, one may ask how it
is possible that bacteriophages are still infective and can eject
their DNA into their bacterial host, given the high level of entanglement
of the DNA inside the capsid.  As we shall see, computer simulations
helped shed light on both these issues.

Finally, the three-dimensional confinement is complemented by the
discussion of the adsorption of polymer rings onto a two-dimensional
surface, and the behaviour of knots upon mechanical
stretching~\citep{Saitta_et_al:1999:Nature,Sulkowska12162008}.

In Section~\ref{knot_prob_ads} we discuss the adsorption of polymer rings onto a two-dimensional surface, 
 and discuss the following issues:
\begin{itemize}
\item How does the knotting probability depend on the degree of adsorption?
\item How are the universal properties of polymer adsorption affected
  by topology?
\item The critical point governing the adsorption
  transition of linear polymers has been well characterised
 ~\citep{DeBell&Lookman:1993:RMP,Vanderzande:1998}. Is this critical
  point governing also a topological transition separating a regime of
  high knotting probability from one in which knot formation is
  negligible?
\item What do we know about the knot spectrum of adsorbed polymers?
\end{itemize}
We close this Section with a list of unsolved open problems. It is
particularly long because there have been very few investigations on the topic
to date.

Finally, in Section~\ref{knot_prob_stress} we survey the few results
which have been obtained on the knotting probability and entanglement
complexity of polymer chains under stress at equilibrium.  Most of
them refer to lattice models and are based on combinatorial arguments
valid in the thermodynamic limit and on numerical simulations.

\section{Sampling at equilibrium by Monte Carlo methods}
\label{Monte_Carlo}

The analytic characterization of the behaviour of polymers subject to
geometrical or topological constraints is a challenging task even for
the simplest polymer models, such as freely-jointed chains~\citep{chandrasekhar43}.

Essential tools for making progress in such characterizations are
stochastic computational techniques such as Monte Carlo approaches.

In the present context, a Monte Carlo scheme consists of a set of
rules for the stochastic generation of a succession of polymer
configurations so that the steady state probability of visiting any
given conformation is given by its canonical weight.

In brief, the scheme can be formulated as follows.
We denote by $\Gamma_A$ the polymer configuration at a given Monte
Carlo step, and by $E_A$ its energy. At the next step a new polymer
configuration $\Gamma_B$ (with energy $E_B$) is selected with uniform
probability in the polymer configuration space. A stochastic criterion
is next used to decide whether, at the next Monte Carlo step, the
system configuration is given by $\Gamma_B$ or if the ``old'' one,
$\Gamma_A$, must be retained.  Virtually all employed selection
criteria are chosen so to satisfy the detailed balance 
condition~\citep{Hammersley_Handscomb:1964,Itzykson&Drouffe:1989}. 
In other words they ensure that the probability of
accepting the change in configuration, $w_{A\to B}$, satisfies,

\begin{equation}
{w_{A \to B} \over w_{B \to A}} = {e^{- \beta E_B} \over e^{- \beta E_A}}\ ,
\label{eqn:DB}
\end{equation}

\noindent where $w_{B \to A}$ is the probability with which one would
accept a change of configuration from $B$ to $A$ and $\beta$ is the
inverse temperature: $\beta^{-1} = K_B T$. Eq.~(\ref{eqn:DB}) can be
satisfied by many choices of the $w$'s. The one most commonly adopted
is known as the Metropolis criterion and states that the probability to accept the change from $A$ to $B$, $w_{A \to B}$ is equal to $1$ if $E_B < E_A$ otherwise, it is equal to $\exp^{-\beta(E_B -
  E_A)}$. In a definite MC simulation, this probabilistic acceptance of the move
is implemented by picking a random number in the [0,1] unit interval,
$r$, and accepting the change to configuration $B$ if $r <
\exp^{-\beta(E_B - E_A)}$. Notice that the condition $r
<\exp^{-\beta(E_B - E_A)}$ can be used even when $E_B < E_A$ as in
this case the inequality will always be satisfied. Because at each
Monte Carlo step, the probability of accepting the trial configuration
depends only on the configuration at the previous step (and not on
``older'' ones), the MC evolution realizes a Markov chain
~\citep{Hammersley_Handscomb:1964,Binder&Heermann:1988,Madras&Slade:1993}

Starting from an initial conformation, the repeated application of the
acceptance/rejection scheme ensures that in the long run, polymer
configurations $\Gamma_i$ are picked with probability proportional to
their canonical weight, $e^{-\beta
  E_i}$~\citep{Hammersley_Handscomb:1964,Binder&Heermann:1988,Itzykson&Drouffe:1989}.

In contexts of practical interest, the implementation of the Monte
Carlo algorithm differs from the above general formulation in that the
new conformation, $\Gamma_B$ is not picked randomly from the
conformational ensemble but is generated from $\Gamma_A$ by applying
some stochastic structural changes to it.

The choice of the set of stochastic deformations, commonly termed
``moves'', severely impacts the performance of the Monte Carlo
sampling. First, the moves must be compatible with the ergodicity
requirement. By this it is meant that any two possible conformations
might be connected by a finite number of deformation steps. If this is
so, then clearly the succession of generated states viewed with a
sufficiently large (but finite) time stride is equivalent to the most
general formulation where any configuration can be proposed at each
Monte Carlo step.  In addition, the moves can strongly affect the
autocorrelation time of the system. For a similar computational
investment, two sets of moves can, in fact, be associated with very
different breadths of visited phase space and hence number of
statistically-independent configurations.

\begin{figure}[tbp]
\begin{center}
\includegraphics[width=\WIDTHB]{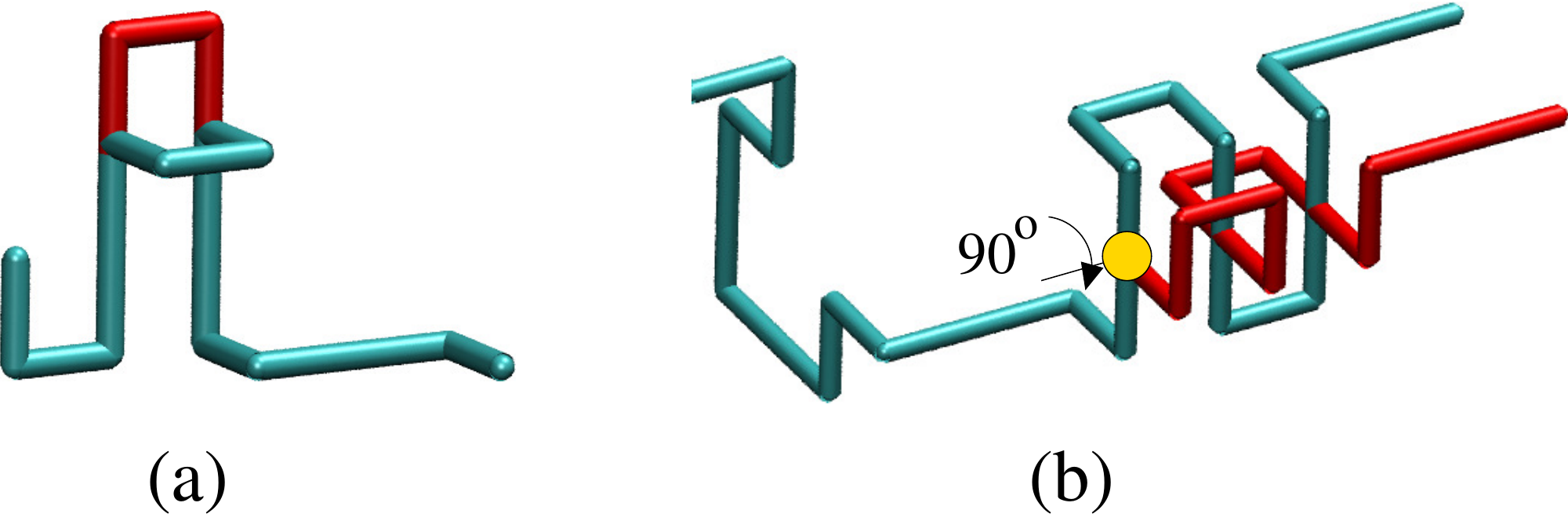}
\caption{Example of (a) crankshaft and (b) pivot moves on a
self-avoiding chain embedded in a cubic lattice. The portion of the
chain that is modified after the (a) local and (b) global move is
highlighted in red. For case (b) we have highlighted the bead,
rotation axis and rotation angle around which the pivot move is
performed.}
\label{fig:MCmoves}
\end{center}
\end{figure}

Customarily, the set of polymer moves usually consists of a
combination of local and global deformations. Local deformations are
those in which only a small number of consecutive monomers are displaced
from their current positions. In polymer models where the chain length
is not preserved (that is where the chain connectivity is enforced by
a suitable potential~\citep{FENE}), the local moves are simply
accomplished by displacing the monomers randomly within a cube or
sphere of preassigned size and centered in the current position.  In
polymer models where the chain length is preserved (as in the
freely-jointed or Kratky-Porod chain) the {\em local} displacement is
obtained through a crankshaft move as illustrated (in a cubic lattice
context~\citep{Madras&Slade:1993}) in Figure~\ref{fig:MCmoves}a.

These local deformations are associated with changes of the polymer
self-energy that are sufficiently small that the deformed structures
are accepted with good probability (say, about 50\%). This does not
necessarily imply a high efficiency of the scheme. In fact, the long
time required before the cumulated local deformations can globally
alter the structure, is reflected in large autocorrelation
times that scale proportionally  to $N^{2+2\nu}\sim N^{\ge 3}$ ($N$ being the polymer
length)~\citep{Caracciolo_Sokal:1986:JPA}.

The latter can be significantly reduced by introducing global moves
where a sizeable fraction of the polymer chain is moved. In an open
chain, a global deformation is obtained by a pivot move. The latter
consists of picking randomly a polymer bead and a random unit vector
through it (see appendix in~\citep{Allen:1987:CSL} for a discussion
of how to uniformly pick unit vectors in three-dimensions). The chain
portion following (or equivalently preceding) the picked bead is next
displaced by rotation through a random angle around the unit
vector. In closed chains (rings) the global move is instead realized
by operating a crankshaft move around an axis passing trough two
randomly-picked points of the chain. By analogy with the open chain
case this move is also termed ``two-point pivot'' move. Similar
procedures are performed also for lattice polymer models; the main
difference here is that the rotations in $\mathbb{R}^3$ are replaced
by symmetry operations of the underlying lattice
~\citep{Madras_Sokal:1988:JSP,Madras_et_al:1990:JSP}, as shown in
Figure~\ref{fig:MCmoves}b.

Global moves are obviously more computationally demanding than local
ones and the associated changes in polymer self-energy are usually
large and unfavourable\citep{Millett_2010_random_eq_polygons}. As
intuitively expected, they in fact, often lead to violations of
self-avoidance. Despite the fact that their acceptance probability is
small the extent of their associated structural changes is so large
that they can dramatically reduce the system autocorrelation time
$\tau$~\citep{Madras_Sokal:1988:JSP}. In particular, denoting by $f$
the acceptance fraction of pivot moves, it is possible to show that
$f\sim N^{-p}$ where $p\approx 0.11$ in
$d=3$~\citep{Madras_Sokal:1988:JSP}.  On the other hand after a time
of order $1/f \sim N^p$ the global conformation of the polymer should
have reached an ``essentially" new state. Consequently the
autocorrelation function \citep{Madras&Slade:1993} of global
properties of the configuration, such as mean-squared radius of
gyration, span along $x$, $y$, $z$ directions, should have a decay
time (autocorrelation time) of the order $1/f \sim N^p$. Local
observables typically evolve a factor $N$ more slowly than global one,
i.e. with $\tau \propto N^{1+p}$; topological properties, however, are
expected to be more slowly relaxing than global geometric
properties. A vivid illustration of this latter aspect
  in the contexts of open polymers undergoing a colapse transition is
  shown in ref. \citep{Mansfield:2007:J-Chem_Phys}.

Besides the above-mentioned local and global moves, a further type of
move, termed reptation, is adopted for polymer melts or confined
polymers. The use of this move is suggested by the observation that the
severe excluded volume constraints arising in dense polymer phases, force
each chain to move by slithering through the surrounding polymer
network~\citep{DeGennes:1979}. Algorithmically, the reptation move
consists of deleting the end bond from one of two chain extremities
and adding it, with a random orientation, to the other. The
autocorrelation time associated to this type of move scales like
$N^3$.

\subsection{Advanced sampling techniques: multiple Markov chains or parallel tempering}
\label{mmc}

The combination of the above mentioned local and global moves is, in
general, not sufficient to adequately sample the relevant
conformational space of polymers in dense solutions or in confined
geometries. In these situations, which are the main object of the
present review, even local moves are likely to incur in a high
ejection rate as a consequence of the high polymer density.

For a given computational investment the breadth of visited
conformational space can be enhanced by a generalised sampling scheme
where several Monte Carlo evolutions are run in parallel. In this
scheme, equivalently referred to as parallel tempering or more aptly
as multiple Markov chains, each Monte Carlo evolution is associated to
a different thermodynamic parameter which, for simplicity we shall
here take as an effective temperature
~\citep{Geyer:1991:CSS,Tesi_et_al:1996:JSP}. In contexts of polymer
confinement the parameter is more suitably chosen to be a confining
isotropic pressure. We shall also assume that the copies are labelled
and indexed so that the temperature systematically increases from the
first copy to the last.

Each copy of the system is evolved independently under a standard
Monte Carlo evolution for a preassigned number of steps. After this,
two neighbouring (in the sense of consecutive indexing) system copies,
are picked. We shall indicate with $E_A$ [$E_B$] and $\beta_A$
[$\beta_B$] the energy and inverse temperature of the copy $A$
[$B$]. A swap of the configurations between the
two copies is proposed. After the swap the energy of copy $A$, which is kept at
temperature $T_A$, would be $E_B$ while the energy of copy $B$ would
be $E_A$. The canonical weight of the system corresponding jointly to
copies A {\em and} B is $\exp({-\beta_A E_A - \beta_B E_B})$ before the swap
while it would be equal to $\exp({-\beta_A E_B - \beta_B E_A})$ after the
swap. By considering the ratio of these two canonical weights it is
possible to generalize the Metropolis criterion so to ensure that the
proposed swap is accepted/rejected respecting the combination of
canonical weights. By analogy with the single Monte Carlo evolution
this condition is ensured if the swap is accepted with probability
given by $\min(1,\exp({-\beta_A E_B - \beta_B E_A})/\exp({-\beta_A E_A -
  \beta_B E_B}))$.

By adopting this generalised Monte Carlo scheme the exploration of
phase space is greatly aided by the fact that different conformations
are passed between copies of the system kept at different
thermodynamic conditions~\citep{Tesi_et_al:1996:JSP}. Note that,
although the lag time separating the swap can be chosen arbitrarily,
the procedure becomes inefficient compared to the single Monte Carlo
evolution if the lag time is set to be longer than the autocorrelation
time of the individual Monte Carlo copies.

\subsection{Reweighting techniques}\label{reweight}

It is important to observe that the scheme described above does not
necessarily need be performed by assigning different temperatures to the
various copies of the system. It is very often convenient to
generalize the canonical weight of a configuration by introducing
suitable conjugate variables in addition to energy and temperature.

In the specific context of confined polymers, one such choice is
offered by volume and confining pressure, as discussed hereafter.

The simplest way by which one can generate an ensemble of
conformations under spatial confinement is to introduce a significant
energy penalty for configurations that do not respect the spatial
constraint. If, for example, one wishes to study a chain confined
in a spherical region one could add to the polymer potential energy a
very large energy penalty for each of its beads or segments protruding
from the spherical boundary. As explained later in section
\ref{rigorous_confined}, when the system density is large, most of the
Monte Carlo moves applied to a configuration that correctly lies within the allowed spherical region will likely cross the sphere boundary and
hence be rejected.  The Monte Carlo evolution under the spatial
constraint will therefore be very inefficient.

A practical solution to this problem is to substitute the constraint
of the ``hard'' boundary with a ``soft'' one resulting, for example,
from the application of an isotropic confining pressure
$P$. Accordingly, to the potential energy of a polymer configuration
is added the quantity $P\, V$, where $V$ is the volume of the smallest
sphere that enclosed the configuration. The multiple Markov chain
scheme can then be run with polymer copies kept at the same system
temperature, but at different confining pressures. In this scheme, for
a given computational investment, we will generate many more
independent configurations with very different degree of
confinement. Each of the system copies corresponds, in fact, to a
constant-pressure ensemble and not to a constant-volume one as in the
previously-described scheme. The data collected in the multiple-Markov
chains evolving at different pressures can nevertheless be suitably
processed to recover the equilibrium behaviour of the constant-volume
ensemble.  This method is known as thermodynamic
reweighting~\citep{Ferrenberg&Swendsen:1988:Phys-Rev-Lett} and its
principal aim is to optimally combine the data of the various Markov
chains so to compute the density of states as a function of the
relevant physical variable(s).

For the case considered here the key variable is the confining volume,
$V$. Let us focus on the set of independent configurations sampled
by $i^{\rm th}$ Markov chain kept at pressure $P_i$. The
histogram of how many configurations have been sampled for various
finely-discretized values, $V_0,\, V_1, ...$ of the confining volume is computed.
We shall indicate with $n_i(V_j)$ the number of hits in the $j$th
volume binning interval. When the total number of sampled
configurations in the $i^{\rm th}$ Markov chain is $N_i$, the
{\em expected} value of $n_i(V_j)$ is:
\begin{equation}
n_i(V_j) = N_i\, W(V_j)\, {e^{ - \beta P_i V_j} \over \sum_l W(V_l) e^{ - \beta P_i V_l}}
\end{equation}

\noindent where $W(V_j)$ is the desired weighted density of states of
the system which accounts not only for the large number of
configurations with volume $V_j$, but also for their canonical
weight. The density of states, $W(V_j)$ can therefore be recovered, up
to a multiplicative constant (which can be fixed by a suitable
normalization procedure) by inversion of the above relationship.  In
principle, an infinitely-large number of configurations sampled at a
given pressure could be used to recover the density of states $W(V_j)$
over the full range of allowed values of the enclosing volume. In
practice, in a finite simulation, most of the configurations will
cover a limited interval of volume values. The data collected in the
various Markov chains running at different pressures will cover
different volume ranges. If the ranges overlap appreciably, the
information can be combined into single determination of $W(V_j)$. The
optimal combination can be made using criteria which minimize the
statistical uncertainty on the $W$ profile
~\citep{Salzburg:1959:J-Chem-Phys,Ferrenberg&Swendsen:1988:Phys-Rev-Lett,Micheletti_et_al:2004:Phys-Rev-Lett}.

The method presented above can be generalised so that it is possible
to obtain the system density of states as a function of more than one
parameter. In particular, for the purpose of discussing how spatial
confinement affects the degree of knotting of polymers that
circularise in equilibrium it becomes necessary to recover the density
of states as a function of both the confining volume $V$ and the knot
type $\tau$: $W(V,\tau)$.  By doing so, the data collected in the various Markov
chains can be used to estimate the probability of occurrence of
various knot types as a function of confinement. Finally we mention
that, while the example considered here pertains to an isotropic
three-dimensional confinement, it can be straightforwardly applied also
when the chain is confined in, say, a slab of width $D$. In
this cases, in place of the isotropic confining pressure, $P$,
conjugate to the configuration volume, $V$, one introduces an
anisotropic pressure which conjugates with the calliper size of the
configuration measured in the direction perpendicular to the slab
planes (see Section~\ref{knot_prob_conf}).

\subsection{Knot simplifications}
\label{knot_simp}

One of the challenges that are faced when sampling spatially-confined
polymer rings, is that the configurations have a very high degree of
geometrical entanglement. For example, configurations of
freely-jointed rings of 100 segments confined in a sphere of radius
equal to 3 times the segment length produce planar projections with
about 150 crossings. The high degree of entanglement poses several
difficulties. For example, the calculation of Alexander determinants
for these projections is prone to round-off errors. In addition, if,
say, an unknotted configuration is associated to a badly-entangled
projection with hundreds of crossings, then even if its determinant is
correctly calculated, it will not be generally possible to assign
unambiguously its knotted state due to the fact that, e.g. the same
Alexander determinants may be shared by many knots with less crossings
than in the reference projection. 

These difficulties can be greatly reduced by subjecting the generated
configurations and their diagrammatic projections to
topology-preserving simplification procedures. To illustrate this
point suppose that the diagram of a certain closed curve has 11
crossings and that its (irreducible) Alexander polynomial is equal to
the one of the unknot. The knot identity cannot be established on the
basis of these informations alone because it could correspond either
to the unknot or to the Conway or Kinoshita-Terasaka knot (see section
\ref{subsec:polynomials}), which has minimal crossing number equal to
11.  However, if by a sequence of Reidemeister moves the diagram was
simplified to the point that it consists of 10 or less crossings, then
the previous ambiguity would be resolved in favour of the unknot.

For off-lattice chains, the topology-preserving simplification
algorithms involve the displacement, deletion and insertions of nodes
in the ring so to ultimately make the latter as smooth as possible and
with as few bonds as possible. In its simplest formulation, the method
consists of repeated attempted ``rectifications'' of the chain. At
each rectification step, a chain bead is chosen at random and is
tentatively displaced so to make it more collinear with its flanking
beads. The new bead position is accepted only if no bond crossing
occurs during the continuous chain deformation obtained by bridging
the previous bead position with the tentative one. Otherwise, the
previous position is retained, see \ref{app:topopres} for
details. Whenever a bead can be made exactly collinear with its
flanking ones, then the bead can be eliminated from the chain, so that
the flanking ones become joined by a single chain bond. The repeated
application of these moves typically results in dramatic
simplifications of the projected diagram \citep{Koniaris&Muthukumar:1991b,Taylor:2000:Nature:10972297}. Notice, however, that in general, this ``greedy''
approach cannot be expected to reach the minimally-complex diagram; in
fact, it is known that the geometrical complexity of particular
tangles needs to be increased first, before allowing for a full
simplification of the configuration.

The above algorithm is explicitly formulated for off-lattice chains
and does not lead naturally to an algorithm for chains embedded in a
lattice, for which other types of topology-preserving geometrical
simplifications have been introduced~\citep{Baiesi_et_al:2009:JCP}.  
For self-avoiding 
polygons on a cubic lattice a very effective scheme is provided by the so-called BFACF
algorithm, an acronym with the initials of the people who
first introduced it to sample walks with fixed extremities~\citep{Berg&Foester:1981:Phys-Lett-B,Aragao:1983:Nucl-Phys-B,Aragao&Caracciolo:1983:J-Physique}.

\begin{figure}
\includegraphics[width=\WIDTHB]{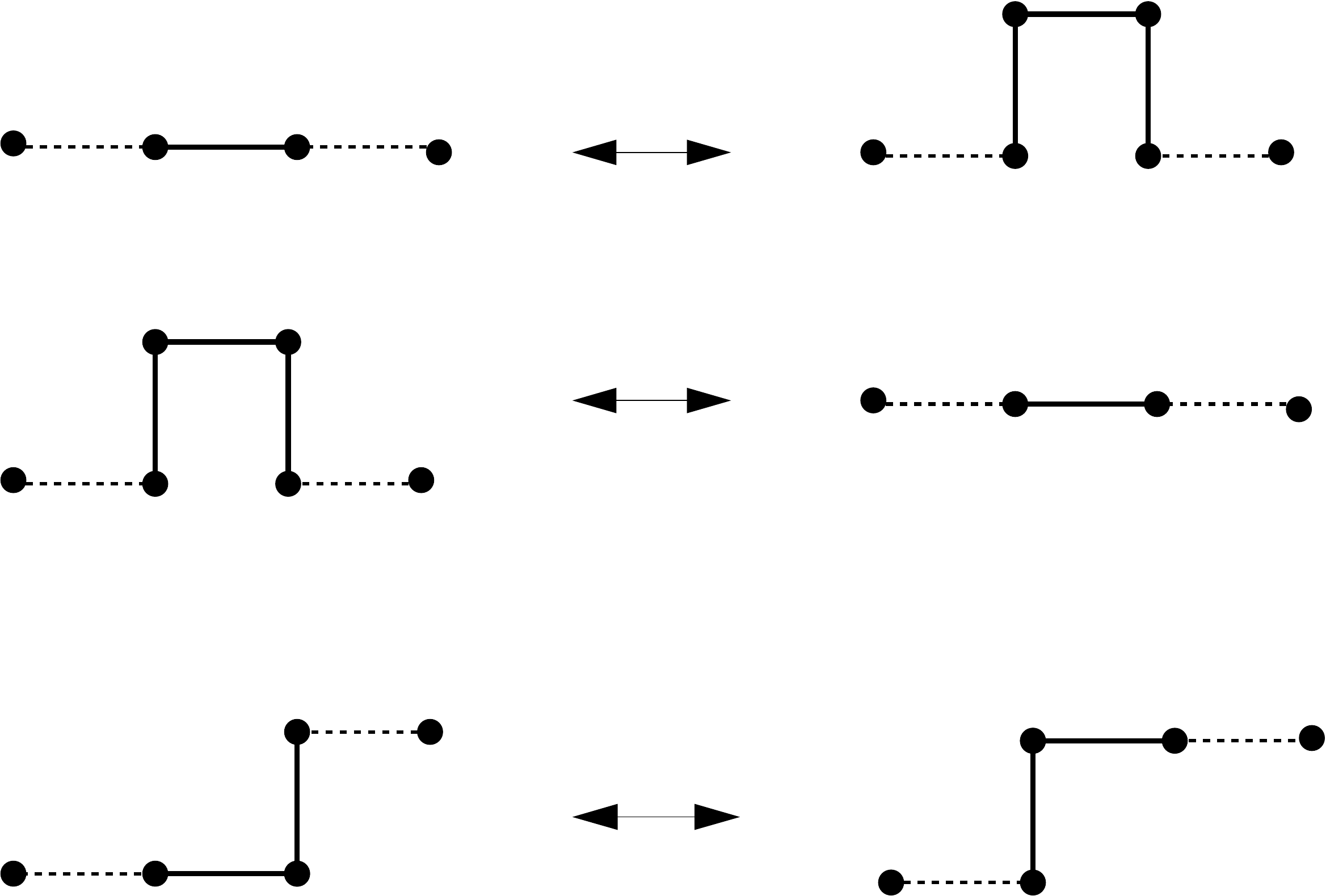}
\caption{ The BFACF moves.} \label{fig:bfacf}
\end{figure}

The standard BFACF can be defined as follows: (i) pick an edge of the
current polygon (say $\omega$) at random with uniform probability.
(ii) pick a unit vector $e$ perpendicular to that edge (iii) move the
chosen edge one lattice space along $e$ and insert two more edges at
its endpoints to keep the polygon connected. (iv) erase any double
edges (spikes) that may result from this operation. It is easy to see
that this procedure will result in one of the three moves illustrated
in Figure~\ref{fig:bfacf} with possible changes in the length of the
polygon $\Delta N \pm 2$ or $\Delta N = 0$.  If the new configuration,
$\omega'$, does not violate the self-avoidance constraints (i.e. no
steric clashes of chain beads) it then accepted with probability
proportional to a weight $w(K^2)$ where $w(K^2) = 1$ if $ \Delta N =
0, -2$ and $w(K^2) = K^2$ if $\Delta N = 2$.  The parameter $K$ is
called \emph{bond fugacity} and it is used to control the average length
of the sampled polygons: for $K$ approaching a critical value $K_c$
the average length of the sampled polygons diverges while for $K<<K_c$
the algorithm tends to sample polygons with the shortest length
allowed. It is easy to check that this implementation of the BFACF
algorithm is reversible.

Moreover, since neither BFACF move allows crossings of polygon bonds
the succession of BFACF moves manifestly preserves the knot type of
the original chain.

A more delicate question is whether by repeatedly applying BFACF moves
to an initial configuration one can {\em in principle} generate all
the possible polygon configurations with a certain knot type and
polygon length. The question is equivalent to asking whether the
ergodic classes of the Markov chain coincide with the knot types. A
positive answer to this question was given in
~\citep{Janse-van-Rensburg&Whittington:1991b:J-Phys-A} using an
argument based on Reidemeister moves. Since the BFACF is ergodic
within each knot class it has been often used to investigate
equilibrium properties of polygons with fixed knot type ( see section
\ref{fixed_knot}). Moreover, if the bond fugacity $K$ is slowly
decreased (annealing procedure) the algorithm would bring (in
principle) any given configuration to a minimal length that is
compatible with its knot type. This would drastically reduce the
number of crossings in the corresponding diagram allowing a
sufficiently precise knot detection also for geometrically-intricate
embeddings \citep{Baiesi_et_al:2009:JCP,ourselves}.

When the structure cannot be further simplified geometrically it is
still possible to manipulate the resulting projection removing further
non-essential crossings. One practical way to do this is to associate
the knot diagram with its Dowker code which essentially follows the
numbered succession of over and underpasses in the projection. 

Reidemeister moves can be performed very easily on the Dowker code by
algebraically manipulating it~\citep{Adams:1994}. Inspection of the
Dowker code can further reveal the factorisability of a knot. After
these simplifications, it could be the case that the simplified Dowker
code of a diagram (or one of its factors) matches the tabulated Dowker
code of a prime knot (there exist look-up tables of such codes for
knots of up to 16 crossings). In such situations the knot
identification is exact. In other cases such positive identification
cannot be achieved. This might reflect either the fact that the
knot is genuinely a very complicated one or that it has not be
simplified sufficiently to allow for its identification.

\section{Knotting probability and knot complexity in ring polymers}
\label{knot_prob}

A fundamental question regarding the knottedness of discretised closed
curves (polymer rings) in equilibrium is \emph{what is the knotting
  probability as a function of the ring length?}

Systematic attempts to address this issue date back to the early 1960s
when Frisch, Wasserman~\citep{Frisch_Wasserman:1961:JACS} and Delbruck
~\citep{Delbruck:1962} conjectured that \emph{sufficiently long ring
  polymers would be knotted with probability one}. While the statement
appears intuitive, it was not until 1988 that the validity of the
Frisch-Wasserman-Delbruck (FWD) conjecture was proved true for
self-avoiding polygons embedded in the cubic lattice
~\citep{Sumners&Whittington:1988:J-Phys-A,Pippenger:1989:DAM}.  The
difficulty of establishing the result rigorously is due to the fact
that knottedness, being a topological attribute, is a global property
of the chain and cannot be described in terms of local chain features
which are much simpler to characterize and control.

Here we give a qualitative argument of the proof (see
~\citep{Orlandini&Whittington:2007:Rev-Mod_Phys} or section
\ref{knot_prob_conf} for more details) which is based on the ingenious
idea of considering particular knot realizations having a localised
geometrical representation. In particular, considerations are
restricted to ``tight knots''.  A tight knot is the mathematical
version of tying a knot in a piece of string, and pulling it as tight
as possible.  For a self-avoiding walk on the simple cubic lattice,
each occupied vertex generates an excluded volume for itself, the dual
3-cell of the vertex. The key observation is that the tight knot
occupies a compact region of space ($3$-ball) which, owing to the
self-avoiding constraint, is inaccessible to the remainder of the
chain. Consequently, one has that a tight knot tied with a small
stretch of the chain cannot be ``undone'' by the remainder of the
chain. Because anti-knots do not exist, it is clear that if the {\em
local} inspection of the chain reveals the presence of a tight knot
then the chain is necessarily knotted. Accordingly, the FWD conjecture
would be proven if at least one tight knot is contained in
sufficiently long rings. 

The latter statement is proved resorting to the Kesten's Pattern
Theorem~\citep{Kesten:1963:J-Math-Phys} for self-avoiding walks where
the chosen pattern is the union of the curve portion taking part to
the tight knot and the 3-ball. Based on the pattern theorem, and
denoting by $P_N$ the probability that an equilateral polygon with $N$
bonds, or $N$-gon, is knotted, one has that the number of unknotted
polygons decreases exponentially fast for large values of $N$:

\begin{equation}
P_N=1-\exp{\left(-\alpha_0 N +o(N)\right)},
\label{knot_prob_free_eq1}
\end{equation}
\noindent where $\alpha_0$ is a strictly positive quantity. The
relationship in eq.~(\ref{knot_prob_free_eq1}) therefore proves the
FWD conjecture for polygons in the cubic lattice.

Similar results have been obtained for off-lattice polygons with
infinitely-thin bonds, 
such as Gaussian polygons~\citep{Diao:1994:JKTR} and
equilateral random polygons i.e. freely-jointed
rings~\citep{Diao:1995:JKTR}.  Despite the fact that these
non-self-avoiding polygons may thread several times through a knotted
region, it is still possible to prove that tight local knots occur
with high probability and that the topological $3$-ball surrounding
such knots is threaded through by the rest of the polygon with a
vanishing probability. In this case the FWD conjecture is proven in
the form
\begin{equation}
P_N > 1 - \exp{\left (-A\,N^{\epsilon}\right)}
\end{equation}
where $A$ and $\epsilon$ are strictly positive constants.

Extensions of these rigorous results to the case of off-lattice models
with finite thickness are not available yet.

From the above account, it is clear that the arguments employed to
prove the FWD conjecture are based on polygons with an arbitrarily
large number of bonds, $N$.  To present day, in fact, there are no
exact (analytic) results for the probability of occurrence of knots in
rings of a given finite length.  

The latter question has been consequently addresses by numerical
techniques such as exact enumeration or stochastic sampling
methods. In particular, Monte Carlo approaches such as the ones
described in Section~\ref{Monte_Carlo}, are commonly used to sample
the equilibrium conformations of a ring with $N$ bonds whose topology
is, in turn, established by the knot identification routines described
in Section~\ref{knot_theory}. The knotting probability is next
computed as a function of $N$ and the value of $\alpha_0$ is obtained
from a best-fit procedure with the functional form given by eq.~\ref{knot_prob_free_eq1}.

One of the earliest numerical attempts to characterise the knotting
probability was carried out in
ref.~\citep{Frank-Kamenetskii:1975:Nature} in the context of polygons
embedded on the body-centered cubic lattice. It was established that,
although infinitely-long lattice polygons are certainly knotted, the
constant $\alpha_0$ in eq.~(\ref{knot_prob_free_eq1}) is typically so
small that the occurrence of knots becomes detectable only for
polygons with several hundred bonds.

The first attempt to estimate $\alpha_0$ was done in
~\citep{Janse-van-Rensburg&Whittington:1990:J-Phys-A} where a two-point pivot algorithm was used to generate polygons of up to $N=1600$
bonds on the face-centered-cubic lattice. The study of their knotting
probability (based on the non-triviality of the Alexander polynomial)
lead to estimate that $\alpha_0$ was equal to $(7.6 \pm 0.9) \times
10^{-6}$.  Interestingly, even at the largest chain lengths, $N=1600$,
almost all the knots found were of the simplest type, namely trefoils.
An analogous calculation was later carried out for polygons with up to
$N=3000$ edges on the simple cubic lattice
~\citep{Yao:2001:J-Phys-A}. By using derivatives of the Jones
polynomial to detect knotting, it was established that $\alpha_0 =
(4.0 \pm 0.5 )\times 10^{-6}$.

Both these findings were later confirmed by detailed and extensive
simulations for polygons of up to $N=4000$ edges on the simple cubic
(SC), face centred cubic (FCC) and body centred cubic (BCC) lattices
~\citep{Janse-van-Rensburg:2002}.  It was shown that,
$\alpha_0=(4.15\pm 0.32)\times 10^{-6}$ (SC), $\alpha_0 = (5.91 \pm
0.32)\times 10^{-6}$ (FCC) and $\alpha_0 = (5.82 \pm 0.37)\times
10^{-6}$ (BCC). Note that these estimates indicate that $\alpha_0$ is
lattice dependent. If one interprets $\alpha_0$ as the inverse of the
characteristic polymer length, $N_0$, required to have an appreciable
fraction of knotted rings it turns out that, for each of these
lattices, the value of $N_0$ is a little larger than $10^5$. This
large value of $N_0$ is reflected in the computational difficulty of
collecting a statistically-significant ensemble of knotted polygons
through a simple stochastic sampling of the configuration space of
(long) polygons. If one is interested in characterizing the properties
of knotted polygons with a given topology then other sampling
strategies may prove much more efficient. For example, for polygons on
the cubic lattice one could resort to the topology-preserving BFACF
moves described in Section~\ref{knot_simp}.

The knotting probability has been also investigated numerically for
off-lattice models of rings such as the rod-bead model and the
cylinder models described in Section~\ref{cylinder_models}.

For the rod-bead model the
probability of being unknotted is still well represented by
(\ref{knot_prob_free_eq1}) but with $\alpha_0$ increasing as the
radius $\Delta_2 = r$ of the beads decreases, ranging from $3.7\times
10^{-3}$ for $r=0.05$ ~\citep{Deguchi&Tsurusaki:1997:PRE} down to
$1.25\times 10^{-6}$ for $r=0.499$
~\citep{Koniaris&Muthukumar:1991a,Koniaris&Muthukumar:1991b}, which is
rather similar to the value found for the lattice calculation.

The exponential decay of the unknotting probability of
eq.~\ref{knot_prob_free_eq1} appears to hold also for off-lattice
self-avoiding rings made by cylinders. This has been shown in
ref.~\citep{Shimamura&Deguchi:2000:Phys-Lett-A} where rings of up to
$N=150$ cylinders of unit length and different radii (see section
\ref{cylinder_models}) were considered. Interestingly, the study also
showed that the characteristic length $N_0$ increases with radius of
the cylinders which make up the ring; the dependence is approximately
exponential for small cylinder radii. As expected, thinner rings of cylinders
have a larger knotting probability at fixed number of cylinders.

The latter observation naturally prompts the consideration of the
knotting probability for rings of cylinders with vanishingly small
radius, that is freely-jointed rings. If the self-avoidance is neglected then
sizeable occurrence of knots in relatively-small
chains is expected~\citep{Vologodskii:1974:JETP,desCloizeaux:1979:J-Phys}.

An early example of such study is given by
ref. ~\citep{MichelsWiegel1986} who considered off-lattice ring
polymers consisting of a $N$ point masses tethered by harmonic
bonds. The pointwise character of the masses implies that the ring
polymer has no excluded volume. The ensemble of conformations explored
in thermal equilibrium by ring of up to $N=320$ beads was sampled by
means of a Langevin dynamics simulations. For this model it was
estimated that $\alpha_0 = (3.6 \pm 0.02) \times 10^{-3}$, which is a
much larger value than for self-avoiding polygons in a lattice.

The exponential decay of the unknotting probability is confirmed also
for Gaussian random polygons with $\alpha_0$ being about $2.9\times
10^{-3}$ (i.e. $n_0 \simeq 340$)
~\citep{Deguchi&Tsurusaki:1993:J-Phys-Soc-Jap,Deguchi&Tsurusaki:1997:PRE}.

\section{Polymers under geometrical confinement.}
\label{confined_polymers}

The studies reported above considered the knotting probability of a
single ring polymer that is unconstrained, i.e. not subject to spatial
conformational restriction. The introduction of spatial restraints is
expected to significantly increase the knotting probability,
consistently with our common experience that the careless packaging of
a long rope in a backpack will cause the rope to become highly
entangled.

In this section we shall report on the increasing number of studies
that in recent years have addressed the problem of knotting of chains
confined in restricted geometries\citep{0953-8984-22-28-283102}. As
an introduction the interplay between topological entanglement and
geometrical confinement in ring polymers we shall first brief review
the salient results obtained on the classical problem of how spatial
confinement affects the physical properties of {\em open} chains.

\begin{figure}[tbp]
\begin{center}
\includegraphics[width=\WIDTHB]{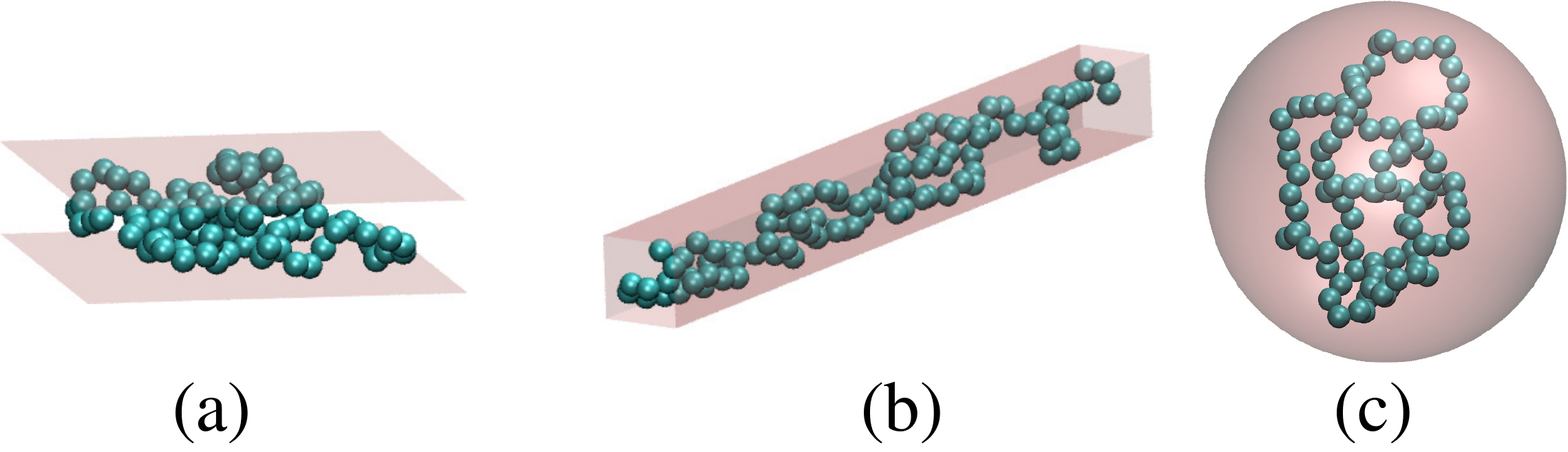}
\caption{Examples of spatial confinement of a chain of beads. Cases
(a), (b) and (c) correspond respectively to confinement in a slab, in
a square channel (prism), and in a sphere.}
\label{fig:confinement}
\end{center}
\end{figure}

Consider a polymer embedded in a convex subspace of $\mathbb R^3$. For
simplicity the subspace can be taken to correspond to a parallelepiped
where, in principle, any of the edges can be infinitely long.  The
polymer is considered geometrically confined if the length of at least
one of the parallelepiped edges is comparable or smaller than the
typical equilibrium size (radius of gyration or end-to-end distance)
of the chain in free, bulk, space. Let us consider three orthogonal
edges for the parallelepiped and indicate with $S_{||}$ the linear
subspace spanned by the unbound edge(s), and with $S_{\perp}$ the
co-space spanned by the bounded one(s). Denoting by $D$ the typical
length scale associated to $S_{\perp}$, we can classify the possible
geometrical confinements in the following tripartite scheme:

\begin{description}
\item[Slab-like confinement] In this case the polymer is confined
  along a single direction, i.e., $\mbox{dim}(S_{\perp})=1$ and
  $\mbox{dim}(S_{||}) = 2$. A simple geometry that describes this
  situation is the so- called \emph{slab} (or \emph{slit}) geometry where the chain is
  sandwiched between two parallel impenetrable walls set at a
  separation $D= L_z$ along the $z$ axis, see
  Figure~\ref{fig:confinement}a. 

  In practical contexts, this type of polymer confinement can be
  achieved inside slab-like micro-channels with $D \sim 1\mu m$ and
  $D_{||} >> D$ ($\sim 1 cm$)~\citep{Chen:2005:Macromolecules}. These
  devices can be realized for example by a soft lithography procedure
 ~\citep{Xia&Whitesides:1998:Angewandte:Chemie} on
  polydimethylsiloxane (PDMS).  Recently slab-like geometries at the
  nano-scales have been realized using electron beam lithography on
  silicon to define the patterns and wet-etched by inductive reactive
  ions (RIE). With this procedure nano-slabs with size $D\sim 33
  nm $ have been obtained~\citep{Bonthuis:2008:Phys-Rev-Lett}. Note
  that in experimental devices also $S_{||}$ is bounded but its
  typical length scales are much larger than the equilibrium polymer
  size and hence can be considered infinite for all practical
  purposes.

\item[Tube-like confinement] A tube-like confinement is realized when
  the polymer is bounded along two directions ($\mbox{dim}(S_{\perp})
  = 2$, $\mbox{dim}(S_{||})=1$), as in
  Figure~\ref{fig:confinement}b. The simplest realization one can think
  of is a tube-like geometry with cross section of diameter $D$. 

  Note that in biological systems this confinement is very common and
  occurs, for example, during the translocation of DNA through membrane
  or solid-state nano-pores~\citep{Smeets:2006:NanoLett}. In
  experiments micro- or nano-channels with
  rectangular cross-section are often employed. The size of the rectangle sides, $D_1$
  and $D_2$, usually ranges from $100-200 nm$
  ~\citep{Tegenfeldt:2004:PNAS} down to $35 nm$
  ~\citep{Reisner:2005:Phys-Rev-Lett}.

\item[Full 3D confinement] It is also possible to confine a polymer in
  a fully-bounded subspace of $\mathbb R^3$ such as a box or a sphere
  ~\citep{Sakaue&Raphael:2006:Macromolecules}. This problem has been
  studied theoretically from a variety of perspectives. For examples,
  scaling results are available for the free energy of compressed 
  polymer chains with and without excluded volume, see e.g.  section
  19.4 of ref. \cite{Grosberg1994}.  Here we shall be particularly
  interested in the impact of the full-3D confinement on polymer
  entanglement. In this respect we note that, the confinement of
  flexible polymers in spheres or ellipsoids is particularly
  noteworthy since it has direct bearing on the packaging of DNA
  inside viral capsids.
\end{description}

The above cases represent prototypical examples of how the
configuration space of a polymer chain can be limited by the
introduction of a physical constraint, namely the impenetrability of
the confining boundaries. It is important to note that polymer
confinement can also be realised without resorting to such physical
constraints. The simplest, and most important, example is offered by
\emph{surface confinement} where the polymer is constrained to
fluctuate in proximity of an attracting surface (polymer adsorption)
or around an interface between two immiscible fluids (polymer
localization). 

\subsection{Scaling arguments for confined polymers}

The equilibrium configurational properties of flexible self-avoiding
polymers subject to spatial confinement are governed by the
competition between two length scales: the average extension of the
polymer in the bulk, $R$, and the characteristic transverse or
calliper size of the bounded region $D$. Clearly, when $D$ is large
compared to $R$ then both $R_{||}$ and $R_{\perp}$ are equal to
$R$. When $D$ is smaller than $R$, then $R_{\perp}$ will be equal to
$D$, while the size of the squeezed-out polymer along the unbounded
direction, $R_{||}$, is expected to progressively grow for decreasing
$D$. The dependence of $R_{||}$ on $D$ can be established assuming the
validity of the following scaling relationship~\citep{DeGennes:1979}:

\begin{equation}
R_{||} = R\,  f \left ( \frac{R}{D}\right ), \label{scaling}
\end{equation}
where $f(x)$ is a dimensionless scaling function of the argument
$x=R/D$. 

For definiteness we shall first consider the case of a {\em slab
confinement} ($\mbox{dim} S_{||}=2$) of a self-avoiding polymer
consisting of a large number, $N$, of monomers. 

Consistent with the above observations, when $ D \gg R$, or
equivalently $x\to 0$, one must have $f(x\to 0) =1$, so that $R_{||}=R
\sim N^{\nu}$, where $\nu\approx 3/5$ is the Flory value of the
self-avoiding exponent in $d=3$. In the extreme case $R \gg D$ the
system essentially corresponds to a self-avoiding polymer in
two-dimensions. For such situation, the characteristic polymer size in
the slab plane is
\begin{equation}
R_\parallel \sim N^{\nu_{\parallel}}  \label{scalingbis}
\end{equation}
\noindent with $\nu_{\parallel}=3/4$ ~\citep{Vanderzande:1998}. In the
limit of large $N$, from the requirement of consistency between
eqns. (\ref{scaling}) and (\ref{scalingbis}) it follows that:
(i) for $x\gg1$, $f(x)$ must grow like $x^m$ and \\
(ii) $m= \frac{\nu_{||}}{\nu}-1$\ .

\noindent In conclusion, for a slab-like confinement with $R \gg D$ one has:
\begin{equation}
R_{||} \sim N^{\nu_{||}} D^{-(\frac{\nu_{||}}{\nu}-1) } \approx N^{3/4} D^{-1/4 }
\label{scaling2}
\end{equation}

\begin{figure}[tbp]
\begin{center}
\includegraphics[width=\WIDTHB]{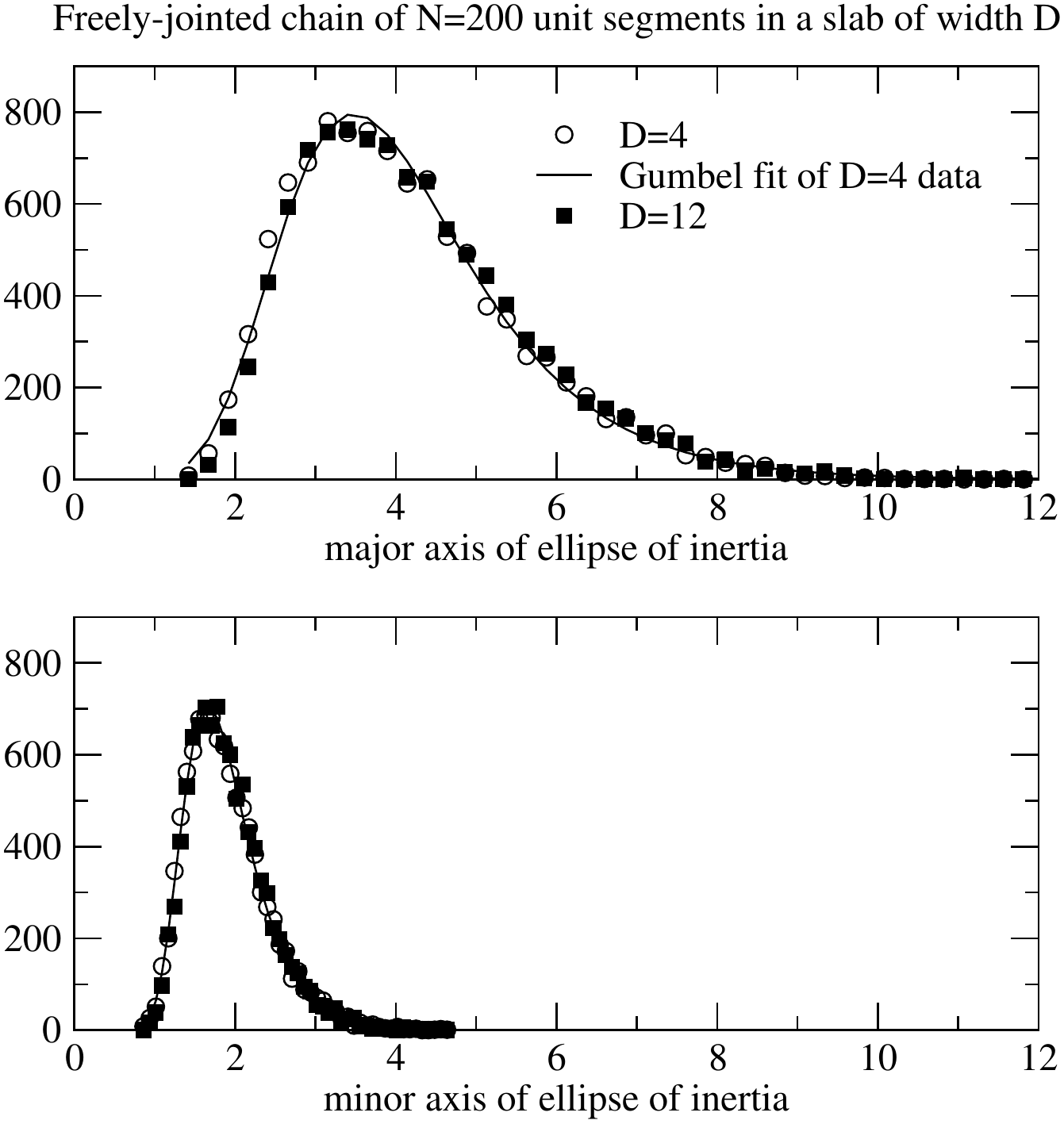}
\caption{Slab confinement of a freely-jointed chain of $N=200$ segments. A simple sampling technique was used to generate 10000 independent configurations of a FJC restricted in a slab of width $D=12$ (about equal to twice the value of the unconstrained radius of gyration) and $D=4$. The major and minor axis of the ellipse of inertia projected in the slab plane were calculated and their probability distributions are reported. No significant difference of the distributions for the two values of $D$ is observed. The continuous line represents a fit of the $D=4$ data using the Gumbel extremal statistics.}
\label{fig:slabfjc}
\end{center}
\end{figure}

\noindent Since $3/4 > 3/5$ it is clear that the net effect of the
confinement is to stretch the chain along the unconstrained
directions. Notice that this is not true for ideal chains since in
this case $\nu_{||}=\nu=1/2$, and therefore, the chain size in the
unconstrained directions remains of the same order of the free case.
This effect is illustrated in Fig.~\ref{fig:slabfjc}
  for a freely-jointed chain confined in two slabs of
  different width. The figure shows the distribution of the lengths of
  the major and minor axis of the chain inertial ellipse projected in
  the slab plane. It is seen that the distribution hardly differs in
  the two cases of confinement. Notice that even in the unconstrained
  case, the distributions of the major and minor axis lengths are
  centred on rather different values, consistently with the known
  anisotropy of self-avoiding walks \cite{Rudnik1987}.

 For self-avoiding chains the stretching effect is a genuine product of the
competition between the excluded volume interaction and the
geometrical confinement. It is worth to notice that
the law (\ref{scaling2}), obtained by a simple scaling argument, can
be shown to be valid for arbitrary values of slab-width $D$ and
extended to soft confining walls. These results have been obtained
by using variational arguments on the Edwards model of self-avoding
polymers ~\citep{Thirumalai:1997}. The same technique, applied specifically to polymers
in cylindrical pore, allows also to predict some more specific features of 
these confined systems such as the non-monotonic behaviour of the average extension
as a function of the confining size $D$~\citep{Vliet_Brinke:1990,Morrison_Thirumalai:2005} (see also Section 
\ref{sampling_confined}) 

The above argument can be transferred to the case of {\em tube-like
confinement}. Here, when $R\gg D$ the chain proceeds almost straight
along the axis of the tube, so that $\nu_{||}=1$. Therefore one has
\begin{equation}
R_{||} \sim N D^{-2/3 }\label{scaling_tube}.
\end{equation}
As one would expect, a stronger confinement induces a more pronounced
stretching effect on the polymer.

\subsubsection{Blob picture}
\label{blob_picture}

Considerable insight into the structural arrangement of confined
flexible chains is provided by the \emph{blob picture} which was first
proposed by de Gennes to facilitate the understanding of concentrated
polymer solutions and melts ~\citep{DeGennes:1979}. The key concepts
used to formulate the blob picture are as follows:

\begin{itemize}
\item the confined chain is viewed as a string of ``blobs''.  Each
  blob consists of $g$ monomers and has diameter $B$;

\item the chain portion in each blob does not experience the confining
constraints and therefore behaves as an unconstrained self-avoiding
chain. Thus one has $B \sim g^{\nu}\, b$ or, equivalently, $g \sim
(B\, b)^{1 \over \nu}$.

\item The blobs can be viewed as the effective monomers of a chain
  subject to the imposed confining constraints. Accordingly, the
  average chain extension in the unconstrained direction(s), $R_{||}$,
  would  scale as $R_{||} \sim (N/g)^{\nu_{||}} D \sim
  N^{\nu_{||}}\, D \, B^{- {\nu_\parallel \over \nu}}$\ .
\end{itemize}

\noindent The latter expression can be consistent with the result
given in the previous section, $ R_{||} \sim N^{\nu_{||}}
D^{-(\frac{\nu_{||}}{\nu}-1) }$ only if $B=D$. This fact establishes
an important structural property of confined self-avoiding chains. In
fact, the latter appear organised as a succession of blobs with
diameter equal to the characteristic size of the confining region.

\subsubsection{Odijk regime}

The considerations contained in the previous sections are limited to
cases where the characteristic size of the confining region, $D$, is
much larger than the chain persistence length. As a matter of fact,
the blob picture was formulated for fully flexible polymers and for
confinements where $D$ is much larger than the monomer size,
$b$. 

When $D\ll l_p$ the physics of confinement is not dominated
anymore by excluded-volume interactions but by the interplay of the
geometrical confinement and the intrinsic polymer elasticity.  In this
regime of very strong confinement De Gennes' blob theory is not valid
any more and it must be replaced by the Odijk's deflection theory
~\citep{Odijik:1983:Macromolecules,GiantMolecules2ndEd}.
Hereafter we give a brief account of the theory of Odijk which was originally formulated for
tube-like confinement.

The starting point is the observation that, in the limit of strong
confinement back-folding is energetically unfavorable and the
organization of the polymer in the geometry occurs by a sequence of
straight lines that deflect at the walls. Notably, recent theoretical
work by Odijk has pointed out a previously-overlooked effect, namely
the influence of the depletion attraction between the polymer and the
confining walls\cite{PhysRevE.77.060901}. This effective interaction is
expected to introduce new additional regimes in the polymer behaviour
because the localization at the confining walls makes the
formation of backturns in the open chain probable.

To estimate the typical contour length separation of consecutive chain
deflections, one considers first the statistics of an unrestricted
(i.e. not confined) stiff Kratky-Porod chain. In particular let us
consider a segment of the chain positioned in the middle of the tube
and oriented parallel to the tube axis. Using the result from eq.~(\ref{eqn:rperp2}) it is possible to estimate the contour length
separation, $\lambda$, between this segment and the one that first hits
boundary using the following equality~\citep{Yamakawa:1973:J-Chem-Phys}:
\begin{equation}
\langle {R_{ee}^\perp}^2 \rangle \approx 2 {\lambda^3  \over 3 l_p} = {D^2\over 4} \ .
\end{equation}

The distance between two deflection points along the chain is therefore
\begin{equation}
\lambda \simeq \left (D^2 l_p \right )^{1/3}\ .
\end{equation}
\noindent Notice that, since $D \ll l_p$ then $\lambda \ll l_p$. This
condition justifies the straight-line approximation for the chain
between two deflection points. The typical deflection angle $\theta$
formed by the (straight) chain and the tube boundary is given by $\tan
\theta = D / \lambda$. From this relationship it is easily obtained
the size of the chain projected along the tube axis:

\begin{equation}
R_{||} = L\cos \theta \simeq L \left [ 1-(D/\lambda)^2\right ] = L \left [1-\alpha_o \left (
\frac{D}{l_p}\right)^{2/3} \right ]. \label{scaling_odijk}
\end{equation}
The scaling (\ref{scaling_odijk})  has been later generalized \citep{Burkhardt:1997:JPA} for a
prism whose cross-section is a  rectangle with sides,  $D_1$ and $D_2$, both smaller than $l_p$:
\begin{equation}
R_{||} = L \left [1-\alpha_b \left [ \left ( \frac{D_1}{l_p}\right)^{2/3} +\left (
\frac{D_2}{l_p}\right)^{2/3} \right ] \right ]\ .
\label{scaling_burk}
\end{equation}

Nowadays, the condition $D < l_p$ can be achieved experimentally by
confining, e.g. ds-DNA molecules ($l_p\sim 50 nm$ ) into nanochannels
having typical sizes $D \sim 35 nm$ \citep{Reisner:2005:Phys-Rev-Lett,Reisner:2007:Phys-Rev-Lett,Persson:2009:Nano-Lett}. This
setup, combined with advanced imaging techniques, has made it possible to
test the validity of the strong-confinement scaling relationships
given by eqns. (\ref{scaling_odijk}) and (\ref{scaling_burk}).

\subsection{Rigorous results of lattice models of polymers in slabs and prisms}
\label{rigorous_confined}

We next report on the rigorous results that have been established
for the configuration entropy of confined self-avoiding walks and
polygons in a lattice.

\subsubsection{Walks and polygons in $D$-slabs}
In the cubic lattice a slab of width $D$ is defined by the subset of vertices
\begin{equation}
S_D = \{\vec{r}\in \mathbb{Z}^3 | 0 \le r_z \le D \}
\end{equation}
The boundaries of $S_D$ are the planes
\begin{equation}
\{\vec{r}\in \mathbb{Z}^3 | r_z=0 \}\quad \mbox{and} \quad \{\vec{r}\in \mathbb{Z}^3 | r_z = D \}.
\end{equation}
A walk or polygon ${\cal{C}}$ is contained in $S_D$ if every vertex
${\cal{C}}_i \in S_D$.  The system has translational invariance along
the unrestricted $x$ and $y$ directions. In the light of this
consideration it appears appropriate to remove the translational
symmetry by regarding two walks or two polygons confined to a $D$-slab
as distinct if they can not be superimposed by a suitable translation
in the $x$ or $y$ directions.

We start by considering the case of open self-avoiding walks (SAW) in a
$D$-slab. Let us denote by $c_N(D)$ the number of different such
walks made of  $N$ bonds. It can be shown that the limit
\begin{equation}
\lim_{N\to \infty} \frac{1}{N} \log c_N(D) = \kappa(D) \label{lim_walks_slab}
\end{equation}
\noindent exists for all slab widths, $D$. The limit $\kappa(D)$
defines the limiting entropy of SAWs confined in a $D$-slab. Moreover
$\kappa(D)$ is a strictly increasing function of $D$~\citep{Hammersley&Whittington:1985:JPA}: 
$\kappa(D) < \kappa(D+1) < \kappa$ where $\kappa$ is the
limiting entropy of SAWs in free space (see section
\ref{lattice_models}). In addition
\begin{equation}
\lim_{D\to\infty} \kappa(D) = \kappa. \label{eq_lim_slab}
\end{equation}

The limit (\ref{lim_walks_slab}) implies $c_N(D) = e^{\kappa(D)N +
  o(N)}$ and from the property $\kappa(D+1) < \kappa(D)$ we have the
interesting result that SAWs confined in a $D$ slab are
\emph{exponentially} few compared to SAWs confined in a $D+1$
slab.

Analogous considerations can be made for polygons in a
$D$-slab. Denoting by $p_N(D)$ the number of polygons in a $D$-slab we
have ~\citep{Whittington&Soteros:1991:IJC,Madras&Slade:1993}
\begin{equation}
\lim_{N\to \infty} \frac{1}{N} \log p_N(D) = \kappa_p(D) = \kappa(D)\label{lim_pol_slab}.
\end{equation}
Equation (\ref{lim_pol_slab}) not only proves the existence of the limiting
entropy for polygons in a $D$-slab but it generalizes the equality
between this entropy and the one of self-avoiding walks in free space
(see Section~\ref{lattice_models}), to the case of walks under
slab-like confinement. In other words, at least on a lattice, for
closed and linear chain in slab-like geometries the configuration
entropy per edge is the same.

\subsubsection{Walks and polygons in $(D_1,D_2)$-prisms}
\label{rig_prisms}

Since the equivalence between the configuration entropy of rings and
open chains holds both in free space and in slab-like geometries, it
would appear plausible that a similar property holds also for polygons
and walks confined within prisms.  Remarkably, this is not the case.
To see this, consider the case of a $(D_1,D_2)$ prism whose
cross-section is a rectangle with sides of length $D_1$ and
$D_2$. Similarly to $D$-slab we define a $(D_1,D_2)$-prism of average
size $D= \sqrt{D_1D_2}$ as the subset of vertices
\begin{equation}
S_{D_1, D_2} = \{\vec{r}\in \mathbb{Z}^3 | 0 \le r_y \le D_1 \quad \mbox{and} \quad 0 \le r_z \le
D_2\}
\end{equation}
The boundaries of $S_{D_1,D_2}$ are the planes
\begin{eqnarray}
\{\vec{r}\in \mathbb{Z}^3 | r_y=0 \}\quad &\mbox{and}& \quad \{\vec{r}\in \mathbb{Z}^3 | r_y = D_1
 \} \nonumber \\
\{\vec{r}\in \mathbb{Z}^3 | r_z=0 \}\quad &\mbox{and}& \quad \{\vec{r}\in \mathbb{Z}^3 | r_z = D_2
 \}
\end{eqnarray}
For self-avoiding walks confined in a $(D_1,D_2)$-prism a result similar to (\ref{lim_walks_slab})
holds \citep{Soteros&Whittington:1989:JPA}
\begin{equation}
\lim_{N\to \infty} \frac{1}{N} \log c_N(D_1,D_2) = \kappa(D_1,D_2) \qquad \forall
(D_1,D_2) \label{lim_walks_prism}
\end{equation}
Also for polygons, in a $(D_1,D_2)$-prism the limiting entropy is well-defined
\begin{equation}
\lim_{N\to \infty} \frac{1}{N} \log p_N(D_1,D_2) = \kappa_p(D_1,D_2) \qquad \forall (D_1,D_2)
,\label{lim_pol_prism}
\end{equation}
but now the following inequality between the limiting entropies of SAWs
and polygons holds~\citep{Soteros&Whittington:1989:JPA}:
\begin{equation}
\kappa_p(D_1,D_2) < \kappa(D_1,D_2). \label{free_ene_comp_prism}
\end{equation}
The non-trivial inequality of eq.~(\ref{free_ene_comp_prism}) 
can be obtained by applying the Kesten's pattern theorem as follows:
for a given $(D_1,D_2)$ prism, one can find a Kesten pattern which 
``fills" the entire prism and that can occur in a walk but not in a polygon.
On the other hand, by Kesten's pattern theorem, one knowns
that configurations  not containing such pattern
(in this case polygons) are exponentially rare and this gives the strict
inequality between the connnective constants.
The above result has notable implications.
In fact, it implies that polygons in a $(D_1,D_2)$-prism are
\emph{exponentially few} if compared to their self-avoiding walk
counterparts.  Hence, unlike the slab-like confinement, tube-like
confinement provides a more dramatic reduction of configuration space
for ring polymers than for linear ones. This is a first example of how
interplay between polymer topology and geometrical confinement may
lead to unexpected and notable phenomena.

\subsection{Sampling confined polymers at equilibrium}
\label{sampling_confined}

An efficient sampling of configurations of open or closed chains
fitting into a given convex region of $\mathbb R^3$ is, in general, a
challenging task. The simplest possible approach would be to rely on
unbiased sampling techniques, such as those described in section
\ref{Monte_Carlo}, to collect a large set of unrestricted polymer
configurations. The sampled conformations could next be processed {\em
a posteriori} to retain only those satisfying the given
constraint. The combination of unbiased sampling followed by this
\emph{rejection scheme} has two important advantages:
\begin{itemize}
\item the ergodicity of the sampling is satisfied whenever the unbiased Markov chain is ergodic,
\item for a given realization in free space the rejection technique
  can be easily applied to any desired geometries, making this method
  quite flexible and useful if one looks for a comparison between
  different types of confinement.
\end{itemize}
The major disadvantage of the method is that, although the unbiased
algorithm can explore in principle the whole configuration space, in
practice very few sampled configurations will satisfy the geometrical
constraint, particular when $D < R$. Indeed, the rigorous results
given in Section~\ref{rigorous_confined} indicate that number of SAWs
and self-avoiding polygons (SAP) in a slab or prism with typical size
$D$ are \emph{exponentially rare} compared to larger confinement size.
Consequently, the computational time required to generate the same
number of independent SAWs/SAPs grows exponentially as $D$ decreases.

As an example let us consider the problem of sampling self-avoiding
walks in the cubic lattice confined within slabs and prisms. We know
from Section~\ref{Monte_Carlo} that a very efficient method to sample
SAWs in free space is based on pivot moves
\citep{Madras&Slade:1993}. After collecting a large number, e.g. about $\sim
10^6$, of uncorrelated configurations and computing their metric
properties the rejection technique is used to partition the
configurations in subsets according to a given geometrical
constraint. Estimates of the canonical averages are then performed on
these subsets and plotted as a function of the typical size $D$ of the
confinement.

\begin{figure}[tbp]
\begin{center}
\includegraphics[width=\WIDTHB]{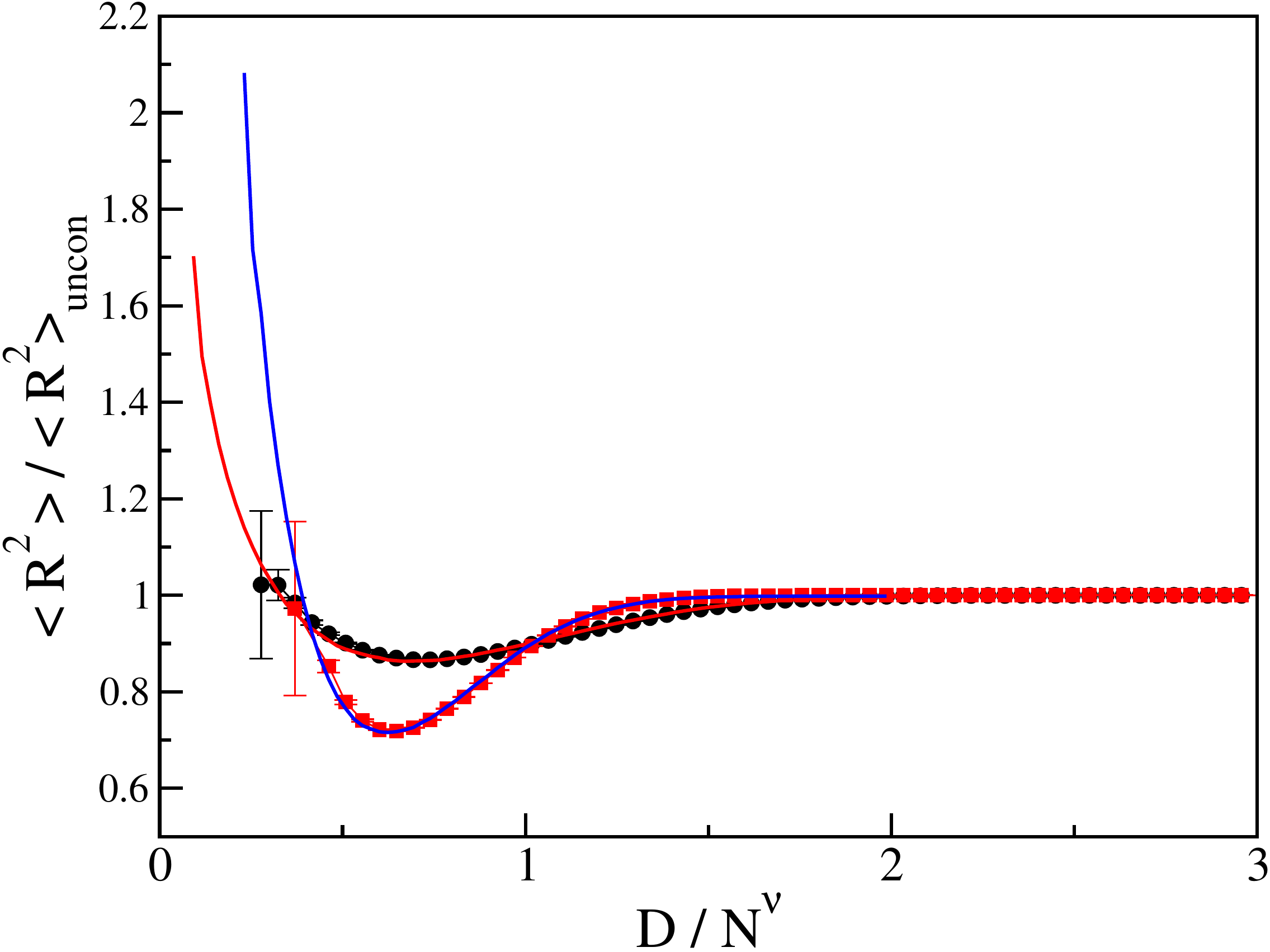}
\caption{Numerical data, from both simple- and
    advanced-sampling techniques, illustrating the dependence of the
    mean square radius of gyration of a self-avoiding walk as a
    function of the size of the confining slab and prism.  Data refer
    to self-avoiding walks of $N=600$ bonds on the cubic lattice
    confined in to a $D$-slab (circles) and into a $(D_1,D_2)$-prism
    (squares) of average size $D=\sqrt{D_1D_2}$. The mean squared
    radius of gyration $\langle R^2 \rangle$ was scaled by its value
    in free space, $\langle R^2 \rangle_{uncon}$ while the typical
    confinement size, $D$ was scaled by $N^{\nu}$, with
    $\nu=0.588$. The symbols refer to the data obtained by rejection
    techniques while the continuous lines refer to the importance
    sampling plus reweighting techniques.}
\label{fig:r2_comp_slab_chan_600}
\end{center}
\end{figure}

In Figure~\ref{fig:r2_comp_slab_chan_600} we show the mean squared
radius of gyration for self-avoiding walks on the cubic lattice
confined into a $D$-slab (black circles) and into a $(D_1,D_2)$-prism
(or rectangular channel) of average size $D=\sqrt{D_1D_2}$ (red squares). 
The length $D$ has been scaled by the mean extension $N^{\nu}$ ($\nu\approx
0.588$) of SAW's in free space while $\langle R^2 \rangle$ has been
scaled by the corresponding average square size of the walks in free
space.  Note that when $D > R$ the extension is independent on the
confinement, as intuitively expected. On the other hand for $D <
N^{\nu}$ the configurations start to experience the confinement and
the overall extension decreases down to a minimum value, suggesting an
overall mild compactification of the SAWs\cite{Vliet_Brinke:1990,Vliet_Brinke:1992,Thirumalai:1997,Morrison_Thirumalai:2005,FreyPRE2007}. This
average compactification is the result of the interplay of several
factors. For example it was pointed out by ten Brinke and co-workers
\cite{Vliet_Brinke:1990,Vliet_Brinke:1992} that upon progressive
confinement in a slab, the average (anisotropic) polymer conformations first aligns the major axes of the gyration matrix parallel to the plan, next it experiences a reduction in size along all directions and finally it spreads out along the unconstrained components. An analogous picture has been shown to hold for confinement in a channel\cite{Morrison_Thirumalai:2005}.
 
The results in Fig. \ref{fig:r2_comp_slab_chan_600} indicate that the
highest reduction of the chain overall size depends on the
dimensionality of the confinement, being more pronounced in the case
of prisms. Consistently with the picture reported above, when $D$ is
further decreased the SAWs start to stretch along the unrestricted
subspace $S_{||}$ and the overall mean square size, $\langle R^2
\rangle$ increases rapidly.  Note that for progressively smaller
values of $D$ the error bars increase in size due to the poor
efficiency of the unbiased sampling based on the rejection technique.

\begin{figure}[tbp]
\begin{center}
\includegraphics[width=\WIDTHB]{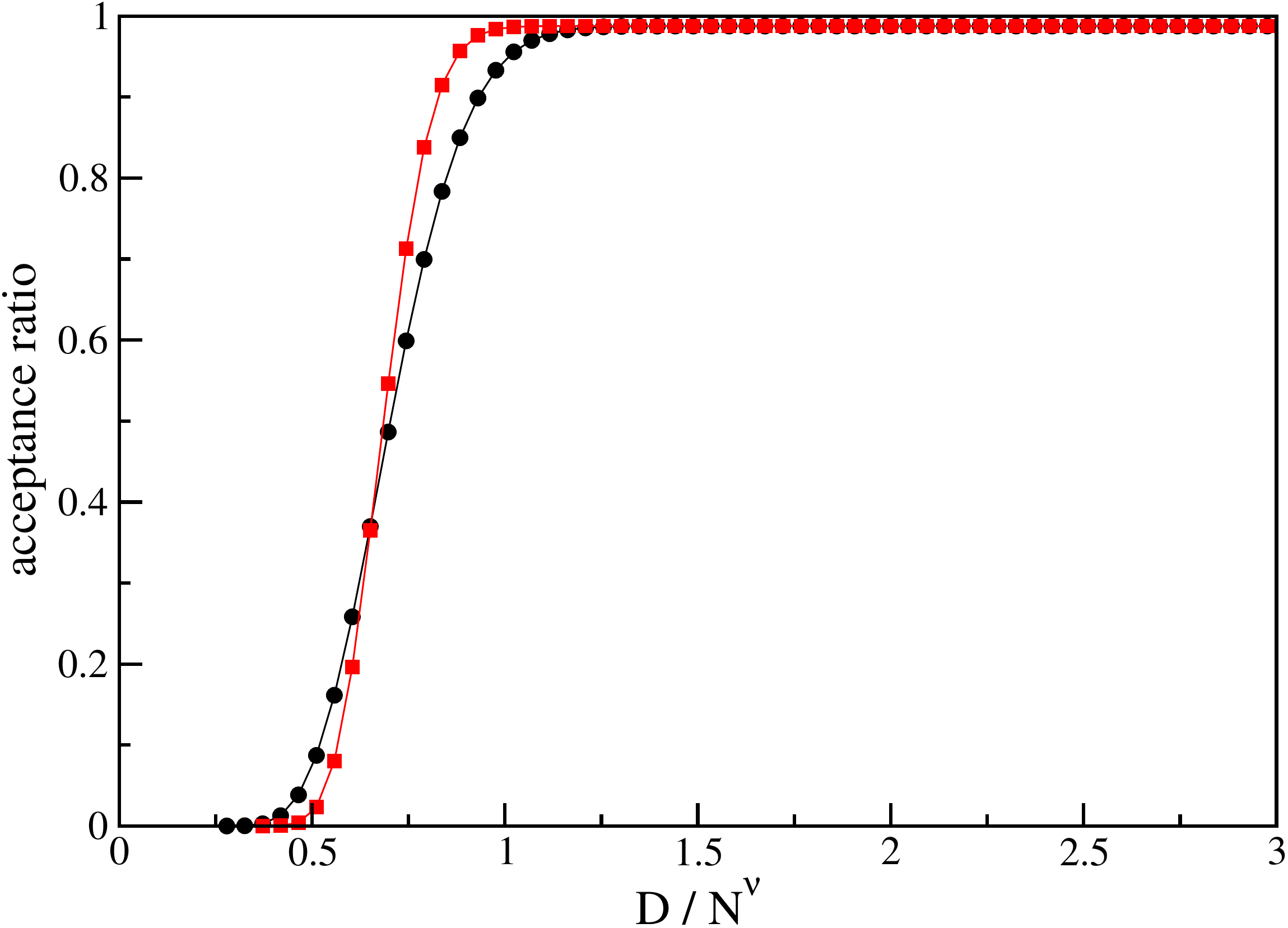}
\caption{Acceptance ratio of configurations of a self-avoiding walk
  that fits into a $D$-slab (circles) and into a $(D_1,D_2)$-prism of
  average size $D=\sqrt{D_1D_2}$ (squares). Data refer to walks of
  $N=600$ bonds.  The typical confinement size, $D$, was scaled width
  $D / N^{\nu}$ where $\nu=0.588$. Data have been obtained by sampling
  $10^6$ configurations and using the rejection technique.}
\label{fig:freq_conf_n600_slab_prism}
\end{center}
\end{figure}

This degradation of the efficiency of the method, for increasing
confinement, can be quantified by plotting the acceptance ratio of
configurations, with $N=600$ bonds, that fit into a subspace of
average size $D$ as a function of the scaled variable $x = D/ N^{\nu}$
(see Figure~\ref{fig:freq_conf_n600_slab_prism}) Notice that at $x=0.5$
only $5\%$ of the sampled configurations fit into the chosen subspace.

A possible way to overcome the sampling problem for small $D$ would be
to modify the Monte Carlo scheme so that it preferentially samples
configurations with the desired geometry. This goal, which clearly
alleviates, or entirely eliminates, the inefficiency of the rejection
scheme, can be achieved by e.g. mimicking the confining action of the boundaries of
$S_{\perp}$ through a repulsive potential $U({\cal{C}})$ that is zero if
the configuration ${\cal{C}}$ is in the interior of $S_{\perp}$ and
becomes strongly repulsive for configurations with $R_{\perp}$ close
to the size $D$ of $S_{\perp}$. One can then perform a standard
importance sampling -- Monte Carlo -- technique based on the generalised canonical
weight $w=\exp\left (U({\cal{C}})\right )$. 

In practice, at each Monte Carlo step the current configuration is
stochastically modified by using the moves of the unbiased method. The
usual Metropolis scheme is subsequently employed to reject or accept
the newly generated conformations based on their statistical weight,
$w$. In the simplest approach, the potential is set to infinity
outside the boundaries and the attempted move is rejected whenever
$R_{\perp}({\cal{C}}) > D$. This approach is often used in lattice
contexts.  Its major weakness is that a rejection of a proposed
configuration that does not satisfy the boundary condition could trap
the Markov chain into sub-regions of the conformational space and the
ergodicity of the algorithm in free space is no longer guaranteed.

Although a softening of the repulsive confining potential can
facilitate sampling, the method is prone to quasi-ergodicity problems.
In fact, as anticipated in Section~\ref{reweight}, for entropic
reasons, most of the sampled configurations will attain the maximum
allowed value of $R_{\perp}({\cal{C}})=D$ and any stochastic
modification is likely to produce a new configuration that violates
the geometry constraint leading to a rejection with high probability.

A very effective way to circumvent the above mentioned problems is to
use the multiple-Markov-chain scheme in combination with thermodynamic
reweighting techniques, as outlined in sections \ref{mmc} and
\ref{reweight}. To briefly recall the main points of the strategy, we
mention that various copies of the system are evolved at different
confining pressures. A suitable generalization of the Metropolis
criterion is used to accept/reject swaps of configurations at
different temperatures. The data collected at the various pressures are
next optimally combined to recover the desired equilibrium information
for different degree of confinement.

To understand why the method has major advantages over importance
sampling, first focus on a particular copy of the system kept at a
given pressure. Configurations sampled at this pressure will, in turn,
have a certain typical degree of confinement (the higher the pressure,
the stronger the confinement). Whenever the system copy is involved in
a successful swap, the configuration of the Markov chain will be
dramatically different from the previous one. The stochastic evolution
of the copy will then restart from a new point in configuration space
~\citep{Orlandini:1998:IMA}. 

It is interesting to follow the evolution from another perspective,
where a specific chain configuration (rather than a specific system
copy) evolves under the action of standard Monte Carlo moves
(crankshaft etc.) plus the swaps. When the latter are accepted the
evolving configuration will be subjected to the action of a different
pressure. When it eventually returns to its initial pressure it will
have evolved at other smaller and/or larger pressures which expectedly
will have left major changes to the configuration.  

The set of pressures used in the multiple Markov chain scheme should
be chosen such that the volume distribution of ``neighbouring'' copies
have a substantial overlap. If this condition is not met then swaps or
neighbouring copied would be accepted only rarely.  This powerful
method has become now a standard procedure in stochastic sampling
mechanics and it is in particular used in strongly interacting
systems.

Finally, as explained in Section~\ref{reweight}, the data coming from
the set of runs at different pressures are finally reweighted so as to
recover the key thermodynamic information, such as $c_N(s_z)$ and in
particular of $\sum_{s_z=0}^D c_N(s_z)\equiv c_N(D)$.

In Figure~\ref{fig:r2_comp_slab_chan_600} we compare the mean squared
radius of gyration of SAWs with $N=600$, confined in slabs and prisms,
obtained by the rejection technique (symbols) and with the importance sampling
method (solid lines).  Note that for all the values of $D$ for 
which the rejection
technique is still efficient the agreement between the two methods is
manifestly good. On the other hand, for sufficiently small values of
$D$, while the rejection estimates stops being  reliable, the importance
sampling data are still good enough to confirm the monotonic
increasing of $\langle R^2 \rangle$ after the minimum.

\section{Knotting probability in confined geometries}
\label{knot_prob_conf}
\subsection{Rigorous results for polygons in slabs and prisms}

The results reported in the previous section aptly illustrate the
dramatic reduction on the configuration entropy of open and closed
chains as a result of spatial confinement. Because the closed chains can
differ in topology, it is natural to pose the following series of
questions: (i) how is the balance between the populations of different
types of knots affected by the introduction of the confining
boundaries? (ii) Is it possible to have knots in rings confined in
slabs of prisms when $D$ is much smaller than the typical size of the
unconstrained polymer chain, i.e. when $D \ll N^\nu$ ?  (iii) Does the
Frisch-Wassermann-Delbruck conjecture, originally formulated for
unconstrained polygons, hold also for polygons in confined geometries?

We shall postpone to subsequent sections the answer to question (i),
which up to now has been tackled exclusively by computational
means, and focus on questions (ii) and (iii) which can be
addressed in rigorous terms for lattice polygons in a $D$-slab or a
$(D_1,D_2)$-prism. We shall provide a brief, non-technical
account of the results obtained originally in refs
~\citep{Tesi:1994:J-Phys-A,Soteros:1998:IMA}.

As in the unconstrained case, to prove the FWD in $D$-slabs and
$(D_1,D_2)$-prism it is sufficient to show that most sufficiently-long
($N \gg 1$) polygons contain at least a knot which is tied so tightly as
to prevent the remaining part of the polygon to thread through the
knotted region thus potentially unknotting the polygon. The proof is
described by three main properties of knotted curves.

The first property is that there exist no \emph{antiknots} that is
tangles that once connected (through the connect sum) to a given knot
will untie it. More precisely for any knot type $K$ there does not
exist a knot type $K'$ such as the connected sum $K \# K'$ is the
unknot. This properties can be proved by using the additivity of the
genus under connected sum, see Section~\ref{knot_theory}.

\begin{figure}[tbp]
\begin{center}
\includegraphics[width=\WIDTHB]{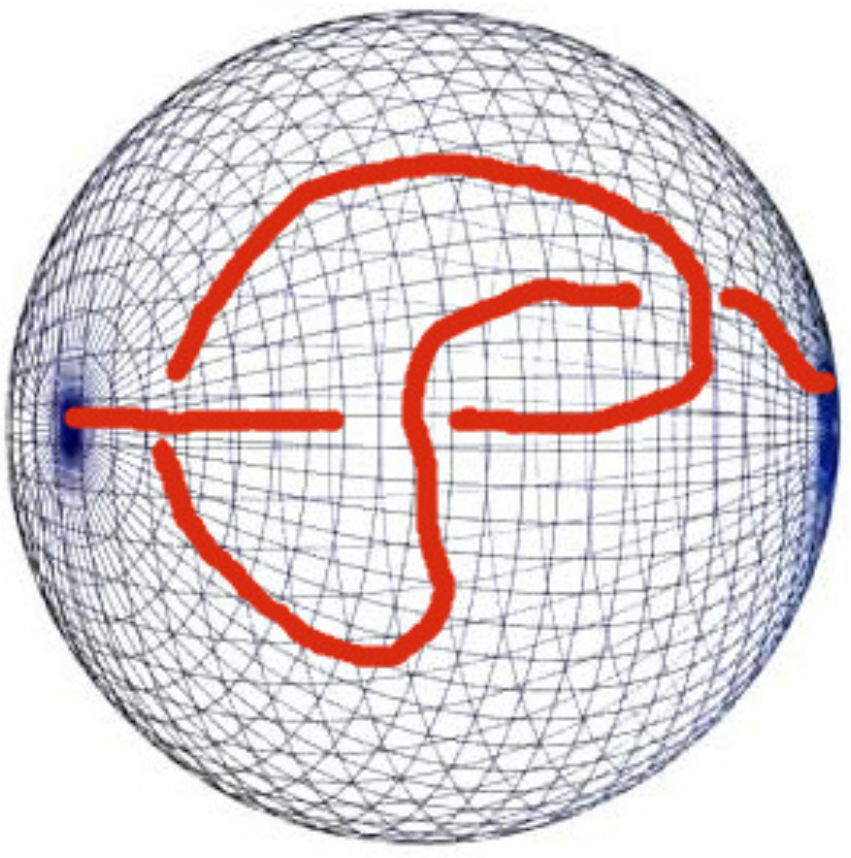}
\caption{A knotted ball pair.} \label{fig:knotted_ball_pair}
\end{center}
\end{figure}

\begin{figure}[tbp]
\begin{center}
\includegraphics[width=\WIDTHB]{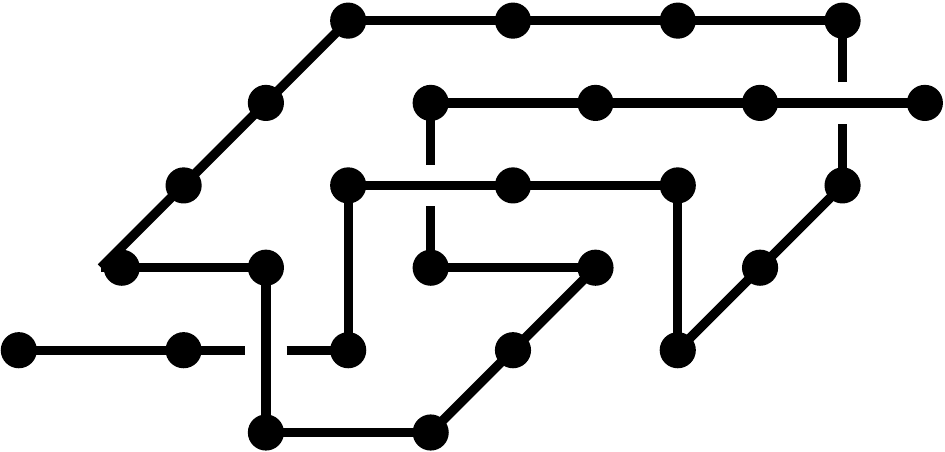}
\caption{A trefoil pattern that fits into a $1$-slab and $(1,3)$ prism.}
\label{fig:pattern_tref}
\end{center}
\end{figure}

Secondly, one has the notion of a knotted arc (see
~\citep{Sumners&Whittington:1988:J-Phys-A} for a technical
definition). One can capture the important part of a knot such as the
trefoil and tie it so tightly on the lattice that the rest of the walk
cannot pass through the neighbourhood of this subwalk and untie the
knot. As mentioned in the introductory part of the review, the subwalk
and its associated dual $3$-cells (i.e. the Wigner-Seitz cells of the
vertices of the subwalk) form a knotted ball-pair
~\citep{Sumners&Whittington:1988:J-Phys-A}, see
Figure~\ref{fig:knotted_ball_pair}. If the polygon contains a sub-walk
such that the walk and its dual $3$-cell neighbourhood form a knotted
ball pair then the polygon is certainly knotted. An
example of a knotted arc in a $1$-slab (i.e. slab with $D=1$) is shown 
in Figure~\ref{fig:pattern_tref}. Since this knotted ball pair fits into a
$1$-slab it will fit in any $D$-slab and the second ingredient is
satisfied for polygons in a $D$-slabs. Note that the same pattern of
Figure~\ref{fig:pattern_tref} fits into a $(1,4)$-prism.

Finally, one has the asymptotic property of lattice polygons that
knotted arcs corresponding to knotted ball pairs, such the ones in
Figure~\ref{fig:pattern_tref}, occur in most sufficiently-long
polygons. To establish this property it is necessary to generalize the
Kesten pattern theorem~\citep{Kesten:1963:J-Math-Phys}, originally
introduced for an unconstrained polygon, to the case of polygons in
$D$-slabs and $(D_1,D_2)$-prisms.

For polygons in a $D$-slab the theorem can be stated as follows:
Consider a subwalk $P$ which can occur at least three times in a
sufficiently long self-avoiding walk (Kesten pattern) confined into a
$D$-slab. Let $p_N(\bar{P},D)$ be the number of polygons which
\emph{do not contain} such a subwalk. Then
\begin{equation}
\lim\sup_{N\to\infty} \frac{1}{N} \log p_N(\bar{P},D) \le \kappa(\bar{P},D) < \kappa(D)
\label{pattern_slab}
\end{equation}
where $\kappa(D)$ is the limiting entropy of polygons in a
$D$-slab. This result implies that the fraction of polygons in a
$D$-slab which do not contain the given pattern, $P$, decreases
exponentially fast with the polygon length, $N$.  
Here we omit the details of the proof that can be found in
\citep{Tesi:1994:J-Phys-A}. It is however worthwhile to mention that
the idea behind this proof is to first establish Kesten's theorem
for SAW's in a $D$-slab and then use equality (\ref{eq_lim_slab}) to
extend the property to polygons confined in the same geometry.

Unfortunately, even if a pattern theorem is available for walks in a
$(D_1,D_2)$-prism~\citep{Soteros&Whittington:1989:JPA}, this result
cannot be straightforwardly extended to polygons because of inequality
(\ref{free_ene_comp_prism}). However, using transfer-matrix like
techniques, a Kesten's theorem for polygons has been proved in
~\citep{Soteros:1998:IMA}:
\begin{equation}
\lim\sup_{N\to\infty} \frac{1}{N} \log p_N(\bar{P};D_1,D_2)= \kappa_p(\bar{P};D_1,D_2)  <
\kappa_p(D_1,D_2) \label{pattern_prism}
\end{equation}
where $\kappa_p(D_1,D_2)$ is the limiting entropy of polygons in a
$(D_1,D_2)$-prism. This result implies that polygons in a
$(D_1,D_2)$-prism (with $D_1\ge 1 $ and $D_2 \ge 3$) which do not
contain the pattern $P$ are exponentially rare (as a function of chain
length) compared to the total number of polygons in a
$(D_1,D_2)$-prism.

The three properties can be used to prove the FWD conjecture for
polygons in $D$-slabs and $(D_1,D_2)$-prisms. To do so it is necessary
to identify a Kesten's pattern $T$ that is a tight trefoil arc. In
this way the ball pair corresponding to $T$ and its associated
$3$-ball neighbourhood is the knot $3_1$. A pattern $T$ fulfilling
these requirements and fitting into all $D$-slabs with $D\ge 1$ and
all $(D_1,D_2)$ prisms with $D_1 \ge 1 $ and $D_2\ge 4)$ is
provided by the example in Figure~\ref{fig:pattern_tref}.  Focussing
on $D$-slabs, if we now call $p_N(\bar{3_1},D)$ the number of
$N$-edges polygons which {\em do not} contain a trefoil as part of the
knot decomposition, we have the following set of inequalities
\begin{equation}
p_N^0(D) \le p_N(\bar{3_1},D) \le p_N(\bar{T},D).
\end{equation}
By using now the pattern theorem (\ref{pattern_slab}) we obtain
\begin{equation}
\lim_{N\to\infty} \frac{1}{N} \log p_N^0(D) \equiv \kappa^0(D) < \kappa(D) \qquad \forall D\ge 1.
\label{lim_un_slab}
\end{equation}
Denoting by
\begin{equation}
P_N(D) = 1-\frac{p_N^0(D)}{p_N(D)}
\end{equation}
the probability that a polygon with $N$ edges in a $D$-slab is knotted, one has that the limit
(\ref{lim_un_slab}) implies the following equality:
\begin{equation}
P_N(D) =1- e^{(\kappa(D)-\kappa^0(D))N +o(N)}= 1- e^{\alpha_0(D)N+o(N)}\qquad \forall D\ge 1
\label{FWD_slab}
\end{equation}
where $\alpha_0(D) >0$. 

The above equality proves the validity of the FWD conjecture
for lattice (cubic) polygons in a $D$-slab, i.e. that \emph{in a
  $D$-slab in $\mathbb Z^3$ all polygons are knotted except for a
  subset that decreases exponentially fast with the polygon length,
  $N$}. Note that
\begin{equation}
\lim_{D\to\infty} \alpha_0(D) =\alpha_0. \label{lim_alpha_0_lab}
\end{equation}

Similarly, for polygons in a $(D_1,D_2)$-prism we have
\begin{equation}
P_N(D_1,D_2) =1- e^{(\kappa_p(D_1,D_2)-\kappa_p^0(D_1,D_2))N +o(N)}= 1- e^{\alpha_0(D_1,D_2)N+o(N)}
\label{FWD_prism},
\end{equation}
$\forall D_1\ge 3, D_2 \ge 2$. Note that in (\ref{FWD_prism}) we
cannot replace the free energy for polygons $\kappa_p(D_1,D_2)$ with
the corresponding one for walks since $\kappa_p(D_1,D_2) <
\kappa(D_1,D_2)$. Embeddings of graphs in lattices can also be knotted
~\citep{Soteros:1992:Math-Proc-Camb} and there are extensions of these
results to knotted graphs~\citep{Soteros:1998:IMA}.

\subsection{Numerical results for polygons in slabs and  prisms}

The rigorous results described above concern the asymptotic behaviour
of confined polygons. Quantitative estimates of the knotting
probability or the knot spectrum at modest values of $N$, can be
obtained, at present, exclusively through stochastic sampling
techniques, such as Monte Carlo methods. The computational approach
can be used to estimate parameters such as $\alpha_0(D)$.  Although
the probability that a polygon of length $N$ approaches unity
exponentially rapidly with increasing $N$ (see eqs. (\ref{FWD_slab})
and (\ref{FWD_prism})), the values of $\alpha_0(D)$ can be so small
that to estimate them reliably it is necessary to sample extensively the
configuration space of confined rings of several hundred edges, and
this can be computationally very challenging and demanding (see
sections \ref{Monte_Carlo} and \ref{sampling_confined}).

\begin{figure}[tbp]
\begin{center}
\includegraphics[width=\WIDTHB]{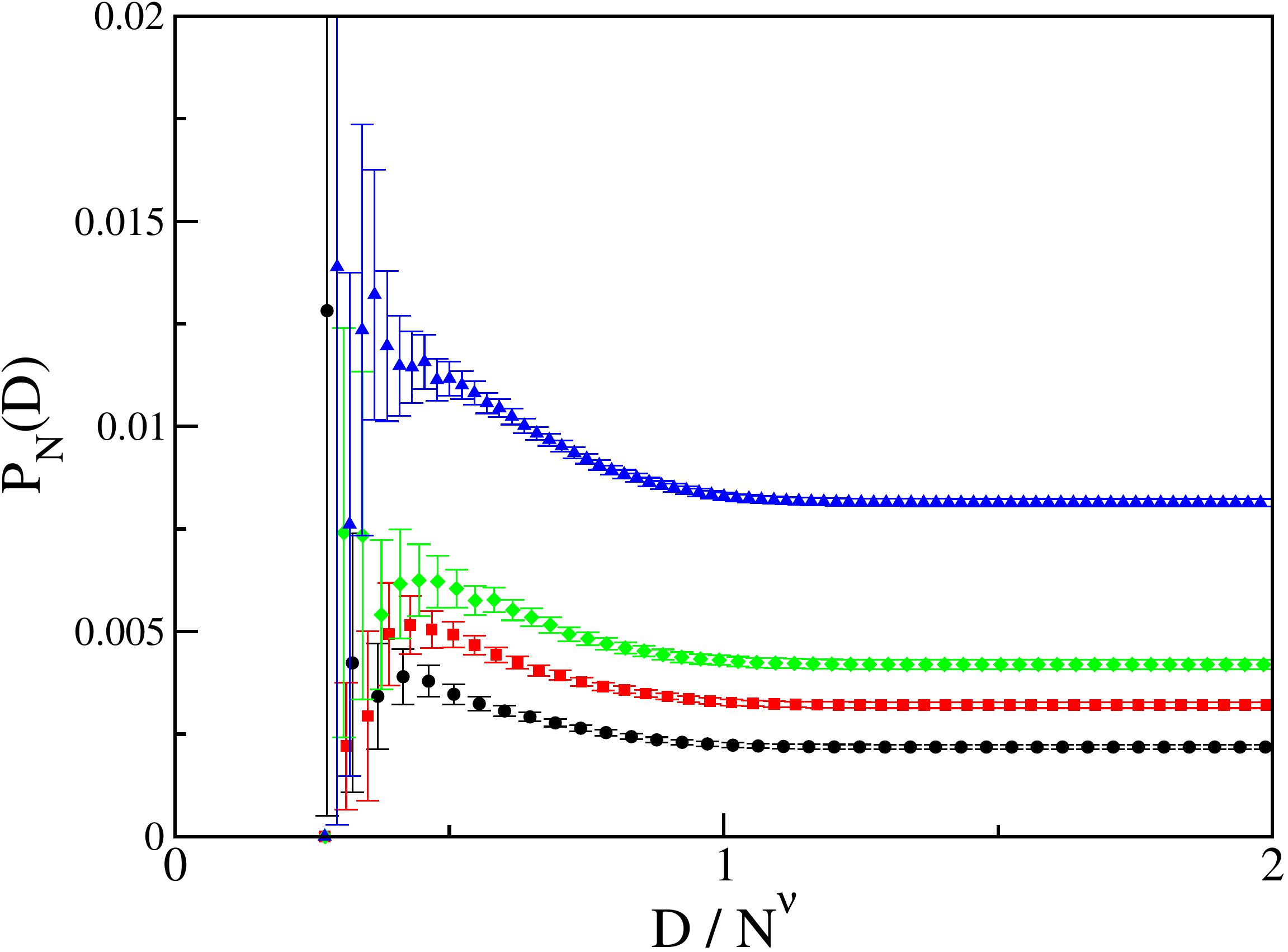}
\caption{Knotting probability vs $D / N^{\nu}$ for polygons of
  $N=600,800,1000,2000$ edges (bottom to top) on the cubic lattice confined inside a
  $D$-slab. The data are obtained by rejection technique.}
\label{fig:pknot_slab_all_rejection}
\end{center}
\end{figure}

One of the first attempts to study the knotting probability for
lattice polygons confined in $D$-slabs and $(D_1,D_2)$ prisms was made
in~\citep{Tesi:1994:J-Phys-A}. In this study a simple, unbiased
sampling method based on two-point pivot moves with rejection
techniques was used to select polygons respecting the confinement
constraint. The study
focused on the knotting probability of the confined rings (regardless
of the knot type) by computing the value of the Alexander polynomial
$\Delta(t)$ at $t=-1$ (see Section~\ref{knot_theory}).

We have repeated, with present-day computational resources, the
calculation of~\citep{Tesi:1994:J-Phys-A} and the results for the
knotting probability for polygons in a $D$-slab are shown 
in Figure~\ref{fig:pknot_slab_all_rejection}.  For each fixed value of
$N$ the knotting probability increases as $D$ is reduced. This is in
accord with the intuition that the increasing density of the
progressively confined polygons ought to reflect in a higher knotting
probability.  

However, it also appears that for sufficiently small values of $D$ the
knotting probability reaches a maximum at a particular value of $D$,
which will be denoted by $D^*=D^*(N)$, and then decreases as $D$ is
further decreased. Unfortunately in the region below $D^*$ the
rejection techniques become very inefficient and the error bars are
so large that it is not possible to establish with certainty the
decrease of the knotting probability as $D\to 1$. To confirm the
non-monotonic behaviour of the knotting probability it is therefore
necessary to improve the efficiency of the sampling scheme. In
ref.~\citep{Tesi:1994:J-Phys-A} this was accomplished by introducing
an ergodic $1$-slab two-point pivot move in the Monte Carlo algorithm. It was
accordingly established that for $D=1$ the knotting probability is
vanishingly small for every value of $N$. 

\begin{figure}[tbp]
\begin{center}
\includegraphics[width=\WIDTHB]{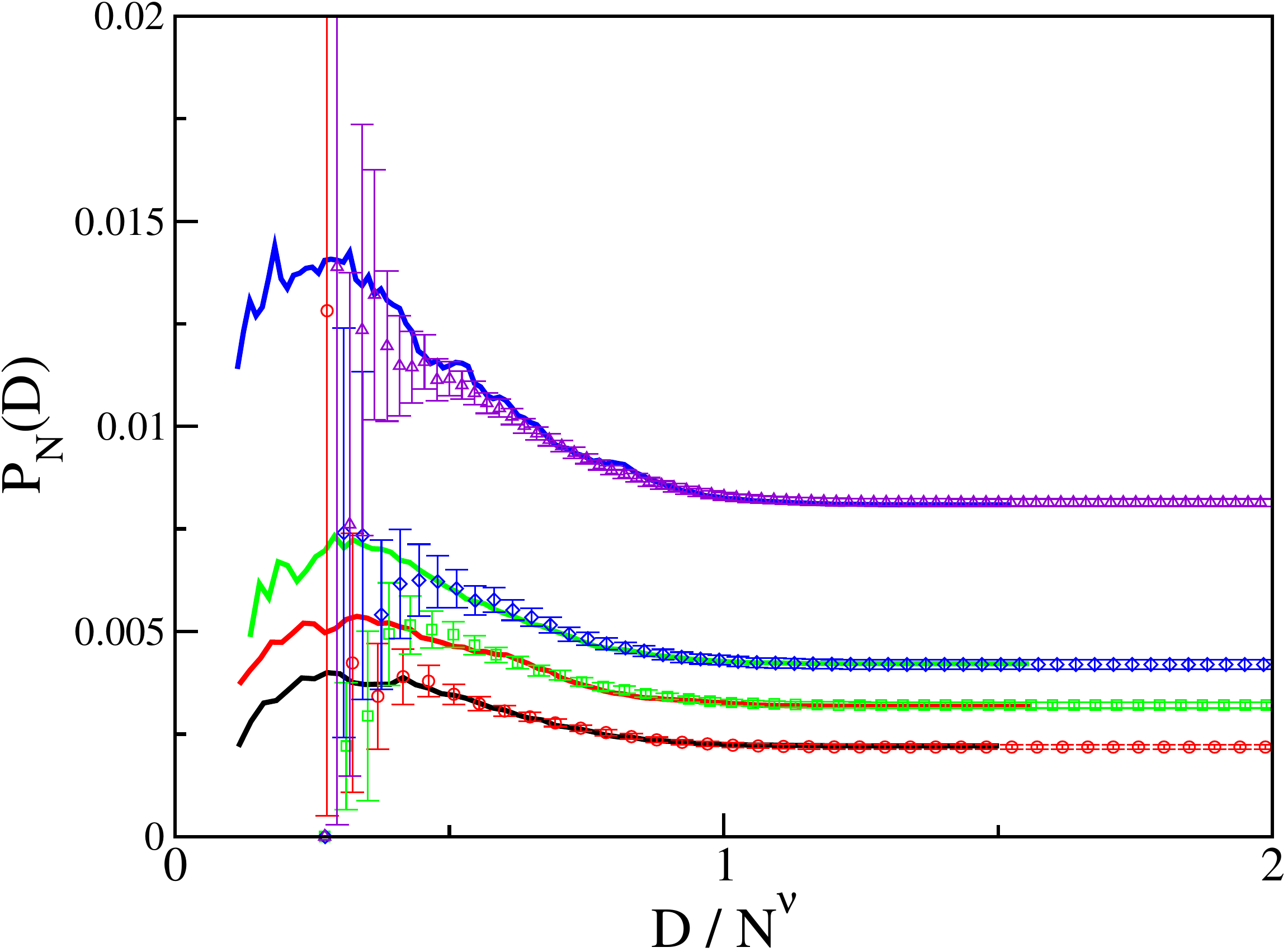}
\caption{Knotting probability vs $D / N^{\nu}$ for polygons of
  $N=600,800,1000$ and $2000$ bonds confined in to a slab. The symbols
  refer to the data of the previous Figure (i.e. obtained by rejection
  techniques) while the continuous lines refer to the importance
  sampling plus reweighting techniques.} \label{fig:pknot_slab_all}
\end{center}
\end{figure}

Importance sampling and reweighting techniques can be profitably used
to study knotting probability throughout the interval $ 1 \le D \le
D^*$ obtaining the results shown in Figure~\ref{fig:pknot_slab_all}. 
In this figure it is possible to compare the performance of the rejection technique (symbols) with the
ones (full curves) obtained by importance sampling. The differences of
the sampling efficiency become very pronounced for $D<D^*$, where the
importance sampling data start to decrease towards zero as expected
from the $D=1$ case.

\begin{figure}[tbp]
\begin{center}
\includegraphics[width=\WIDTHB]{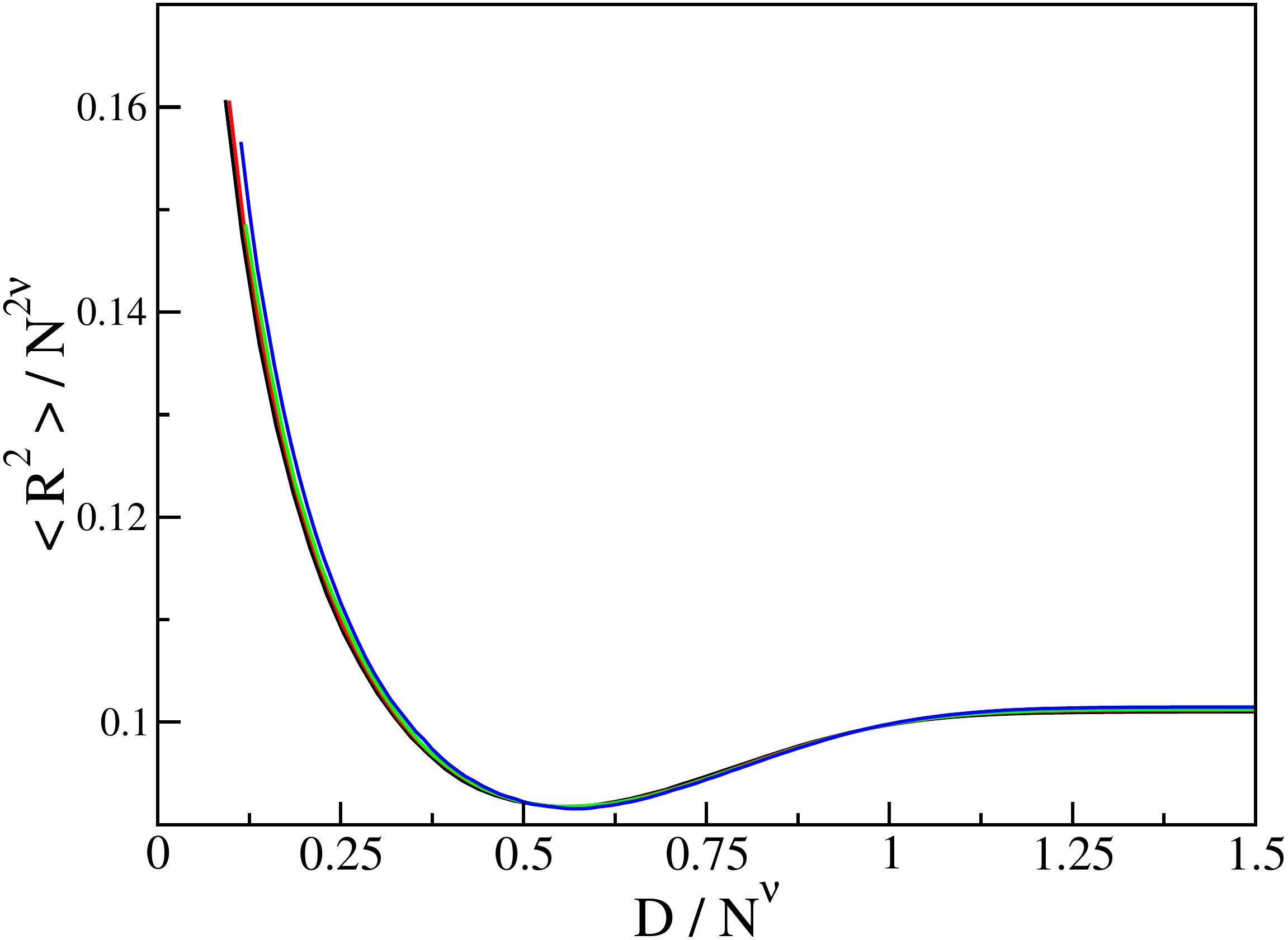}
\caption{Mean squared radius of gyration, scaled by $N^{2\nu}$, as a function of $D / N^{\nu}$.
Data refer to polygons of $N=600,800,1000$ and $2000$ bonds confined inside a $D$-slab.}
\label{fig:r2_slab_all_reweig}
\end{center}
\end{figure}

The presence of a maximum in the knotting probability may reflect the
competition between the higher probability for a polygon with smaller
average extension to be knotted, and the difficulty of forming knots
in confined geometries due to the polygon excluded-volume. If the
polygon is squeezed along a certain direction to the width $D$,
then it would spread out in the other directions leading to a decrease
in the fraction of knotted conformation. According to this argument,
the maximum of the knotting probability should correspond to a minimum in
the average extension of the polygons, or a maximum of its
density. This is indeed observed if one plots the mean squared radius
of gyration as a function of $D$ (see Figure~\ref{fig:r2_slab_all_reweig}).

At fixed value of $D$ one can compute the knotting probability as a
function of $N$ and estimate the value of $\alpha_0(D)$ using
(\ref{FWD_slab})~\citep{Tesi:1994:J-Phys-A}. These estimates were
performed for values of $D>D^*(N)$ and turned out to be consistent
with a functional the form $\alpha_0(D)=\alpha_0 + \beta /D$, where
$\beta$ is a positive constant. This suggests the following scaling
form for the unknotting probability
\begin{equation}
\frac{P^0_N(D)}{P^0_N} \simeq A e^{-N (\alpha_0(D)-\alpha_0)} =A e^{-N \beta /D}=f\left (
\frac{N}{D}\right ).
\end{equation}

A similar analysis can also be performed for polygons in a
$(D_1,D_2)$-prism. In the literature results are available for prisms
of square cross-section, $D_1 = D_2$~\citep{Tesi:1994:J-Phys-A}.  Here
we present results for a different ensemble where the rings are
enclosed in prisms where the area of the rectangular cross-section is
fixed to a given value, $D^2 \equiv D_1 D_2$, but the lengths of the
individual sides, $D_1$ and $D_2$, are not fixed. This is done because
the fluctuating geometry allows for a more efficient exploration of
the configuration space.

\begin{figure}[tbp]
\begin{center}
\includegraphics[width=\WIDTHB]{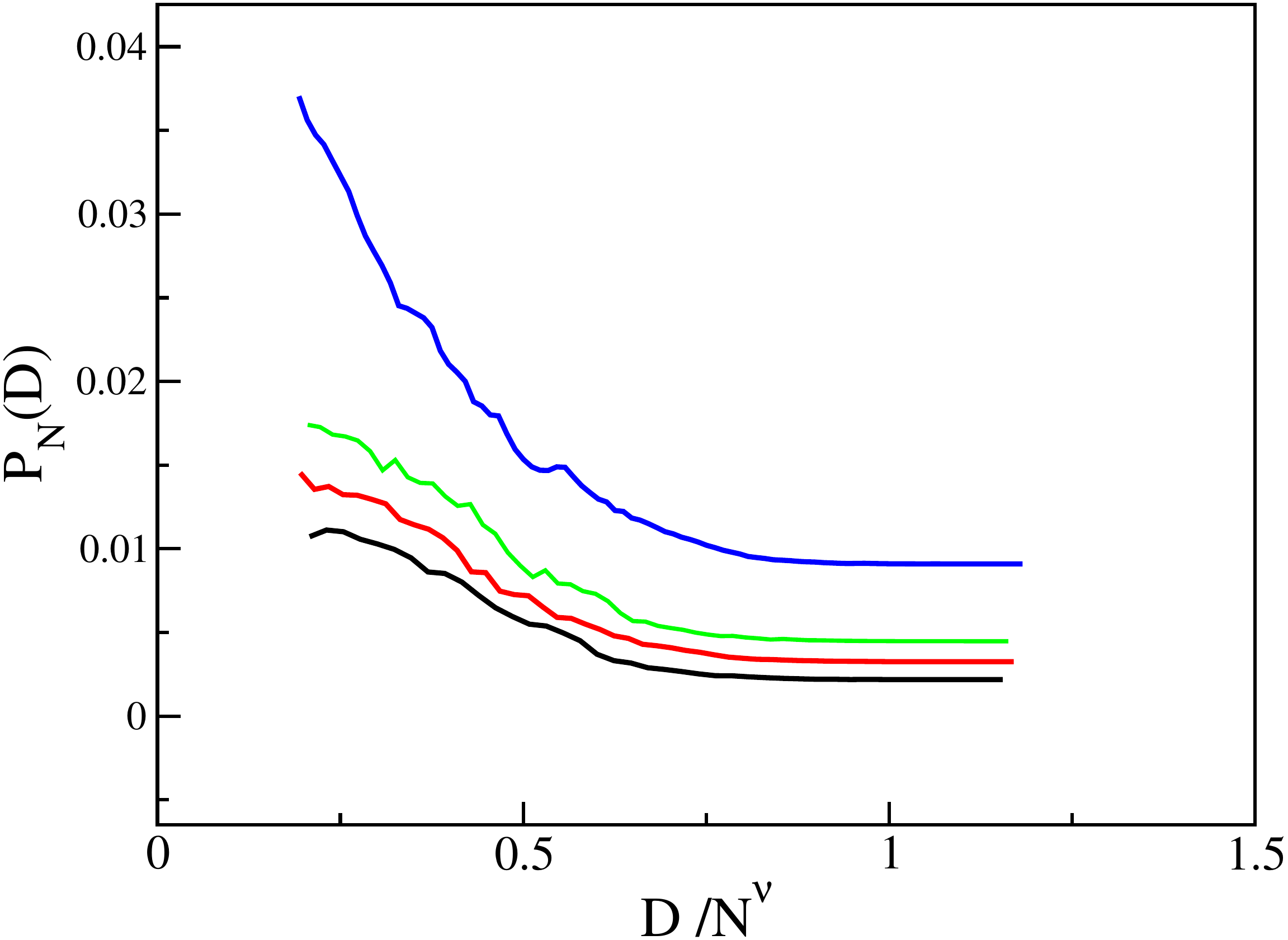}
\caption{Knotting probability $P_N(D)$ vs $D / N^{\nu}$ for polygons of $N=600,800,1000$ and $2000$ bonds (bottom to top) confined in to a $(D_1,D_2)$-prism of average size $D=\sqrt{D_1D_2}$.}
\label{fig:pknot_chan_all}
\end{center}
\end{figure}

In Figure~\ref{fig:pknot_chan_all} it is shown the knotting
probability for polygons in a $(D_1,D_2)$-prisms as a function of
$D=\sqrt{D_1,D_2}$ and for different values of $N$. The behaviour is
similar to the one found in the case of polygons in $D$-slabs. In particular,
for fixed $N$, the probability increases as $D$ decreases up to
$D^*(N)$ (maximum) and then start to decrease for $D<D^*(N)$.  

\begin{figure}[tbp]
\begin{center}
\includegraphics[width=\WIDTHB]{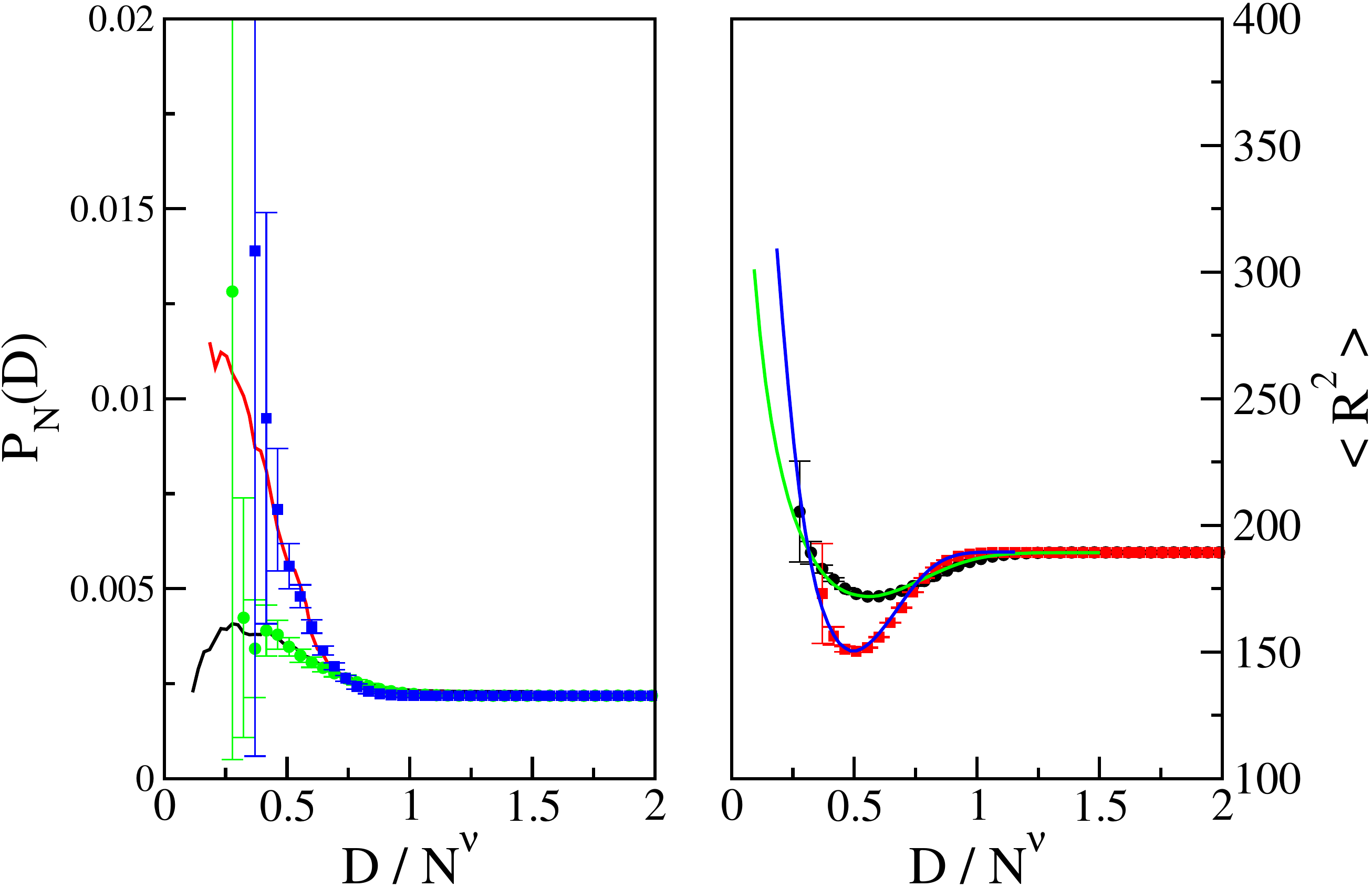}
\caption{Knotting probability (left panel)  and mean squared radius of gyration (right panel) vs $D
/ N^{\nu}$ for polygons of $N=600$ bonds confined inside a $D$-slab (circles) and
$(D_1,D_2)$-prism (squares). The symbols refer to the data obtained by rejection techniques while
the continuous lines refer to the importance sampling plus reweighting techniques. }
\label{fig:pknot_and_r2_comp_slab_chan_600}
\end{center}
\end{figure}

Note that in the limiting case $D_1=D_2=1$ no knots can possibly be
tied and the knotting probability must be equal to zero. The main
difference with the slab case is reflected by the much higher values
of the probabilities in the strongly confined regime (see
Figure~\ref{fig:pknot_and_r2_comp_slab_chan_600} left panel for a
comparison between slabs and prisms for $N=600$) and by the
difficulty, even for importance sampling techniques, to sample the
configuration space in where probability starts to decrease towards
zero. This is to be expected because confining a polygon in a prism
imposes more severe limitations to the polygon configuration space
than a slab-confinement. Indeed, for a given polygon length, the
minimum attainable radius of gyration is lower for the prism case than
the slab one. Consistently, the knotting probabilities are higher for
the prism confinement compared to the slab case, see Figure~\ref{fig:pknot_and_r2_comp_slab_chan_600}. 

\subsection{Knot spectrum in slabs and channels: an open problem.} 

All the results presented above concern the knotting probability
 of polygons confined in slabs and channels (of which prisms are a particular case). It would be
very interesting to look at the knot spectrum (i.e. the relative
frequency of occurrence of the different knot types) as a function of
the parameters characterizing the confinement. As far as we know there
are no results available in the literature on this argument.

\subsection{Polymer rings confined in spheres and cubes.}

Unlike the case of prisms and slabs, in a full three-dimensional
confinement there is no unbound co-space, $S_{||}$, available to the
confined ring. We accordingly expect important differences in the
knotting probability with respect to the slab and prism cases because
upon the progressive reduction of the confinement calliper size, $D$,
the polymer has no available co-space in which to expand. The result
is a progressive compactification of the conformation either as $N$
increases for fixed $D$ or as $D$ decreases at fixed $N$ (except when
the $D$ is so small to be comparable with the size of the polymer
monomeric units, e.g. a chain bead or a polygon edge).
In both cases one should expect a monotonic increasing of the knotting
probability unlike the cases of confinement in slabs and prisms where
a maximum in probability is found for intermediate values of $D$.

One of the earliest systematic attempts to characterize the occurrence
of knots in rings subject to spatial confinement was carried out by
Michels and Wiegel in 1986~\citep{MichelsWiegel1986}. These authors
used a simple-sampling scheme, based on stochastic molecular dynamics,
to explore the conformational space of a polygon with $N$ bonds, each
of length $b$. The polygon had no excluded volume and
  was treated as a {\em phantom ring} in that bond crossings were
  allowed during the MD evolution. The rings were subjected to an isotropic
spatial confinement by constraining them inside a sphere of radius
$R$.

The simple sampling scheme allowed Michels and Wiegel to sample ring
configurations constrained in a sphere with radius about $1/3$ of
the radius of gyration of the unconstrained polymer. From the analysis
of data of chains of up to $N\approx 200$ segments they concluded that
the unknotting probability could be expressed by a scaling formula:

\begin{equation}
P^0_N(R)=g({N^\alpha\,b^3 \over R^3}) \label{eqn:MichWiegel}
\end{equation}
\noindent Specifically, they suggested that $g$ was a decreasing exponential and that
$\alpha=2.28$.

\begin{figure}[tbp]
\begin{center}
\includegraphics[width=\WIDTHB]{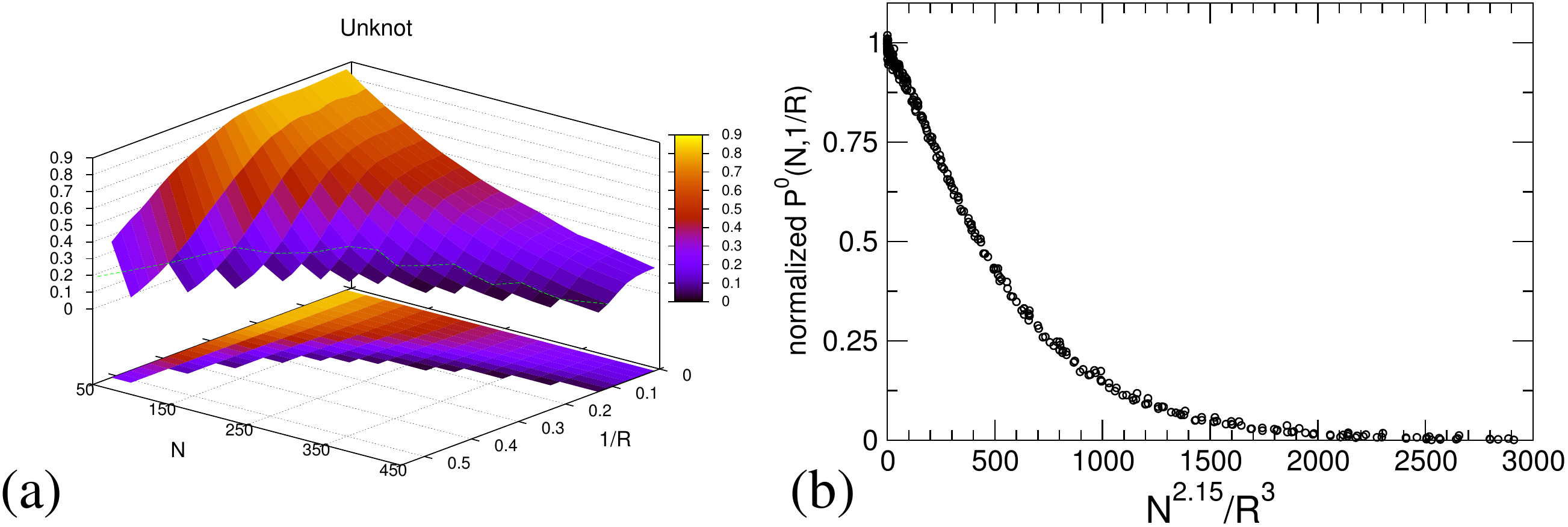}
\caption{(a) Unknotting probability, $P^0(N,1/R)$, for
equilateral freely-jointed rings of $N$ bonds (of unit length) confined inside a
sphere or radius $R$. Below the thin green line, the fraction of
unclassified knots is greater than 10\%. (b) All data points in the manifold in panel (a)
can be collapsed in a simple curve by plotting the normalised
unknotting probability, $P^0(N,R)/P^0(N,1/R=0)$ as a function of
$N^{2.15}/R^3$. Adapted from ref. \citep{Micheletti:2006:J-Chem-Phys:16483240}.}
\label{fig:confined_unknots}
\end{center}
\end{figure}

This pioneering study was recently revisited in ref.
~\citep{Micheletti:2006:J-Chem-Phys:16483240}. This latest
investigation was based on advanced sampling and thermodynamic
reweighting techniques to probe ring configurations with volume up to
ten times smaller than the one reached in the seminal studies of
ref.~\citep{MichelsWiegel1986}. In addition, the number of
independent configurations sampled at the highest density was also
increased by orders of magnitude. The dependence of the unknotting
probability as a function of the ring length and radius of confining
sphere is shown in Figure~\ref{fig:confined_unknots}a. The enhanced
thermodynamic sampling confirmed the validity of the general
functional form of eq.~(\ref{eqn:MichWiegel}) but with two important
differences: the scaling exponent $\alpha$ is 2.15 and the scaling
function is not a simple exponential, as shown in Figure~\ref{fig:confined_unknots}b.

\begin{figure}[tbp]
\begin{center}
\includegraphics[width=\WIDTHB]{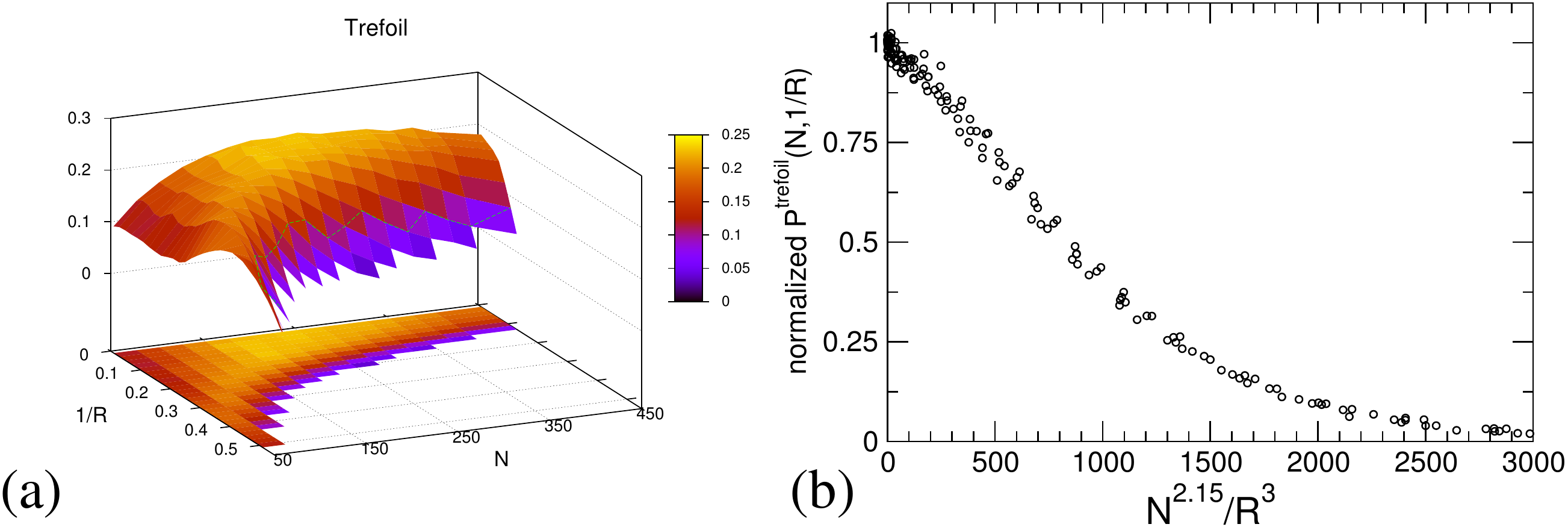}
\caption{(a) Trefoil probability, $P^3(N,1/R)$, for freely-jointed equilateral
rings of $N$ bonds (of unit length) confined inside a sphere or radius
$R$.  Below the thin green line, the fraction of unclassified knots is
greater than 10\%. (b) The data points for $N>250$ in the manifold in
panel (a) can be approximately collapsed in a simple curve by plotting
the normalised probability, $P^3(N,R)/P^3(N,1/R=0)$ as a function of
$N^{2.15}/R^3$. Adapted from ref. \citep{Micheletti:2006:J-Chem-Phys:16483240}.}
\label{fig:confined_trefoils}
\end{center}
\end{figure}

The enhanced sampling allowed us to investigate not only how the
probability of occurrence of unknotted polygons depends on $N$ and
$R$, but also to repeat the analysis for more complicated knot
types. Compared to the unknot case, the probability manifold of
non trivial knots presents major differences. While the unknot
manifold is always monotonically decreasing as a function of $1/R$ at
fixed $N$, (and vice-versa), for non trivial knots one observes a
"shoulder", as is readily appreciated in probability of trefoil
formation shown in Figure~\ref{fig:confined_trefoils}a. 

This peak can be explained with the following argument. For
non-trivial knots, the non-monotonic dependence of the probability
distributions as a function of $N$ in the unconstrained case, $1/R=0$,
is a well-known fact. It reflects the intuitive expectation that, as
longer and longer chains are considered, more complicated knots appear
at the expense of simpler ones. Consistent with this fact, the
location of the peaks at $1/R=0$, is shifted towards higher and higher
values of $N$ for knots of increasing complexity. The presence of the
shoulder at fixed small values of $N$ as a function of $1/R$ is
explained with an analogous intuitive argument. In fact, upon
increasing confinement, simpler knots are progressively superseded in
frequency of occurrence by more complicated knots.

The presence of the shoulder in the manifolds of Figure~\ref{fig:confined_trefoils}a 
suggests that the simple scaling function
of eq.~(\ref{eqn:MichWiegel}) cannot govern the behaviour of the
probability functions of non-trivial knots for all values of $N$. It
is noticed, however, that for sufficiently large $N$ (for example for
$N> 250$ for trefoils) the manifolds have a monotonic dependence on
both $N$ and $1/R$, analogously to the unknot case. 

It is therefore natural to pose the question of whether the same
scaling behaviour for unknots is valid for trefoils
too. Figure~\ref{fig:confined_trefoils}b shows indeed that the trefoil
probability data do collapse if they are plotted as a function of
$N^{2.15}/R^3$. Notice that the latter quantity is the argument of the
scaling function used for unknots, and is not fine-tuned to achieve a
good collapse of the trefoil data. The scaling function, $g$, see
eq.~(\ref{eqn:MichWiegel}), where the data collapse, does not have a
simple form; in particular it is not described by a single
exponential.

Due to the more limited statistics of trefoils compared to unknots,
the quality of the collapse is not as good as the one shown in
Figure~\ref{fig:confined_unknots}b, but it is nevertheless significant
and indicative that the scaling relation of eq.~(\ref{eqn:MichWiegel})
has a range of applicability wider than originally hypothesized by
Michels and Wiegel. At present, no simple argument can
  be offered for explaining the scaling relation and for tying the
  scaling exponent, $\alpha$, to the other relevant exponents
  governing the occurrence or knots in either unconstrained or
  confined rings or the length of the knotted region itself. In this
  respect, we mention that for unconstrained trefoil-knotted rings
  embedded in a cubic lattice, the length of the knotted region has
  been argued to scale as $N^{t}$, with $t \approx 0.75$. Based on
  this result (which is reported at length in section
  \ref{size_knots}) one could envisage that the exponent $\alpha$
  reflects the scaling of the knotted region volume; indeed the
  numerical value $3t=0.225$ is remarkably close to the
  previously-reported value of the Michels and Wiegel exponent. While
  this argument is appealing we mention two difficulties that deserve
  further investigations. In fact, on the one hand, the exponent
  $t\approx 0.75$ describes the scaling of the trefoil knot length in
  the unconstrained region only, while the Michels and Wiegel
  relationship appear to hold for various degrees of compactness. In
  addition, it appears difficult to extend the previous argument,
  which is tailored to rings with a given non-trivial topology, to
  unknots, for which the Michels and Wiegel relationship was first
  formulated. 

In recent years, the confinement-induced knotting of ring polymers was
studied not only in the context of freely-jointed chains, as discussed
above (and more recently in ref.\cite{ReithVirnau}) but also for
semi-flexible chains endowed with excluded-volume interactions and a
finite bending rigidity.

For example, the study of  ref.~\citep{Micheletti:2008:Biophys-J:18621819}
specifically addressed the knotting probability of a semi-flexible thick
chain subject to a progressive spatial confinement. 

This type of investigation can provide some insight into a
problem of high biological interest, namely the occurrence of knots in
DNA molecules that are tightly packed inside viral capsids, which is
discussed hereafter.

\section{DNA knotting inside viral capsids}
\label{dna_capsid}

The problem of characterizing the extent to which knots, and other
types of entanglement, arise in biomolecules such as DNA emerges quite
naturally if one considers that, in all types of organisms, the
genomic material is subject to a very high degree of 
confinement~\citep{0953-8984-22-28-283102,Jun:2010:Nat-Rev-Microbiol:20634810}.

For example, in eukaryotic cells, DNA molecules with a contour length
as large as $\sim$~1m are packed in a nucleus of diameter of about 1 $\mu$m. In
bacteria, mm-long DNA is typically hosted in $\mu$m size cells. At the
smallest scale, we have viruses, such as bacteriophages, whose
$\mu$m-long genome is packed inside capsids having diameter of about
$50$nm.

Upon considering the degree of packing common to all above types of
organisms, the consideration of the extent to which the spatial
confinement of the genome is accompanied by the self-entanglement, and
particular knotting, of the involved DNA filaments emerges naturally.

The understanding of the genome organization across eukaryotic,
bacterial and viral organisms is far from being complete.  Arguably,
one of the best characterised systems is represented by the smallest
types of organisms, namely bacteriophages. The latter, in fact, have
been probed in recent years by the most advanced experimental
techniques. This has provided a wealth of information about the
detailed structure of the capsids and, to a more limited extent, of
the organization of the genome held inside.

Prototypical examples are provided by bacteriophages such as $\epsilon15$, which have been resolved
by means of cryo-em techniques. These experiments, which did not enforce any {\em a priori}
symmetry on the resolved structure, have revealed a highly ordered organization of the DNA
contained in the capsid interior. In particular, the genome arrangement was compatible with that of
an inverse spool with a progressive increase of disorder moving towards the interior of the capsid.

The degree of order of the packaged DNA is affected by specific
details of the capsid and of the surrounding solvent. Computer models
of packaged DNA have shown, for example that the presence of an inner
core inside the capsid favours the increase of DNA
order~\citep{Forrey:2006:Biophys-J:16617089}. The degree of oblateness
or prolateness of the ellipsoid approximating the capsid, is also
known to influence the specific DNA
arrangement~\citep{Petrov:2007:J-Struct-Biol:17919923}. In addition, the
presence of counterions in solution, which can diffuse through the
capsid walls can affect the DNA self-repulsion and hence produce
different arrangements~\citep{Leforestier:2009:Proc-Natl-Acad-Sci-U-S-A:19470490}.

Upon considering the high degree of order observed in several
experimental measurements one might expect a low incidence of knots in
the packaged DNA. Model arrangement of well-ordered inverse spools
show, in fact, the virtual absence of knots after circularization
~\citep{Arsuaga:2002:Biophys-Chem:12488021}. The necessity to avoid
knots inside capsids might, at first sight, also appear to be a
biological necessity as the viral genome needs to be ejected through a
narrow channel (the virus tail) into the infected bacterial cell and
knots in the genome could obstruct the exit channel thus preventing
the infection~\citep{Matthews:2009:Phys-Rev-Lett:19257792}.

These two observations are challenged by a series of measurements
carried out on tailless mutants of the P4 virus. In these mutants, the
two complementary ends of DNA can anneal inside the capsid (in
wild-type viruses one end would be anchored to the tail) producing a
circular DNA molecule. The knotted state of the latter, which does not
change unless the molecule is broken, is revealed through gel
electrophoresis~\citep{Shaw:1993:Science:8475384,Stasiak:1996:Nature:8906784,Arsuaga:2002:Proc-Natl-Acad-Sci-U-S-A:11959991,Arsuaga:2005:Proc-Natl-Acad-Sci-U-S-A:15958528}. The experiments have
revealed a very high fraction (95\%) of knotted molecules, unlike what
was predicted on the basis of ordered spool models of packaged DNA. It is
important to remark that a smaller, but still large percentage of
knotted DNA molecules was observed in wild-type P4 genome where
circularisation occurs after the destruction of the capsid. The
knotted DNA molecules therefore reflect a fairly high degree of
entanglement already present inside the capsid. We remark here that
this aspect is not necessarily in contradiction with the fact that DNA
needs to be delivered efficiently through the phage head.  Indeed, it
is known that knots occurring in (unconstrained) chains tend not to be
localised~\citep{Orlandini:2009:Phys-Bio}, and a highly delocalised
knot may be fully compatible with the genome exiting from the tail
without occurrence of obstructions.

Before discussing the details of the knot spectrum of P4 DNA molecules
we shall first attempt to relate the findings with the results
discussed in the previous section about knots in confined rings.

For the purpose of this comparison, and following Vologodskii et al.
~\citep{Rybenkov:1993:Proc-Natl-Acad-Sci-U-S-A:8506378}, the
$10$Kb-long circularised P4 genome (corresponding to a contour length
of $3.4 \mu$m) was modeled as a flexibible ring of $N=200$
cylinders. The cross-sectional diameter, or thickness, of the
cylinders was set to 2.5 nm which corresponds to the diameter of
hydrated double-stranded DNA.  Between consecutive cylinders a bending rigidity was introduced appropriate to reproduce, within
Kratky-Porod models, the nominal dsDNA persistence length of 50nm.

The equilibrium properties of the model were studied with an advanced
sampling technique analogous to those described in the methods
sections (particularly multiple-Markov chains at various confining
pressures and reweighting
techniques)~\citep{Micheletti:2008:Biophys-J:18621819}.

\begin{figure}[tbp]
\begin{center}
\includegraphics[width=\WIDTHB]{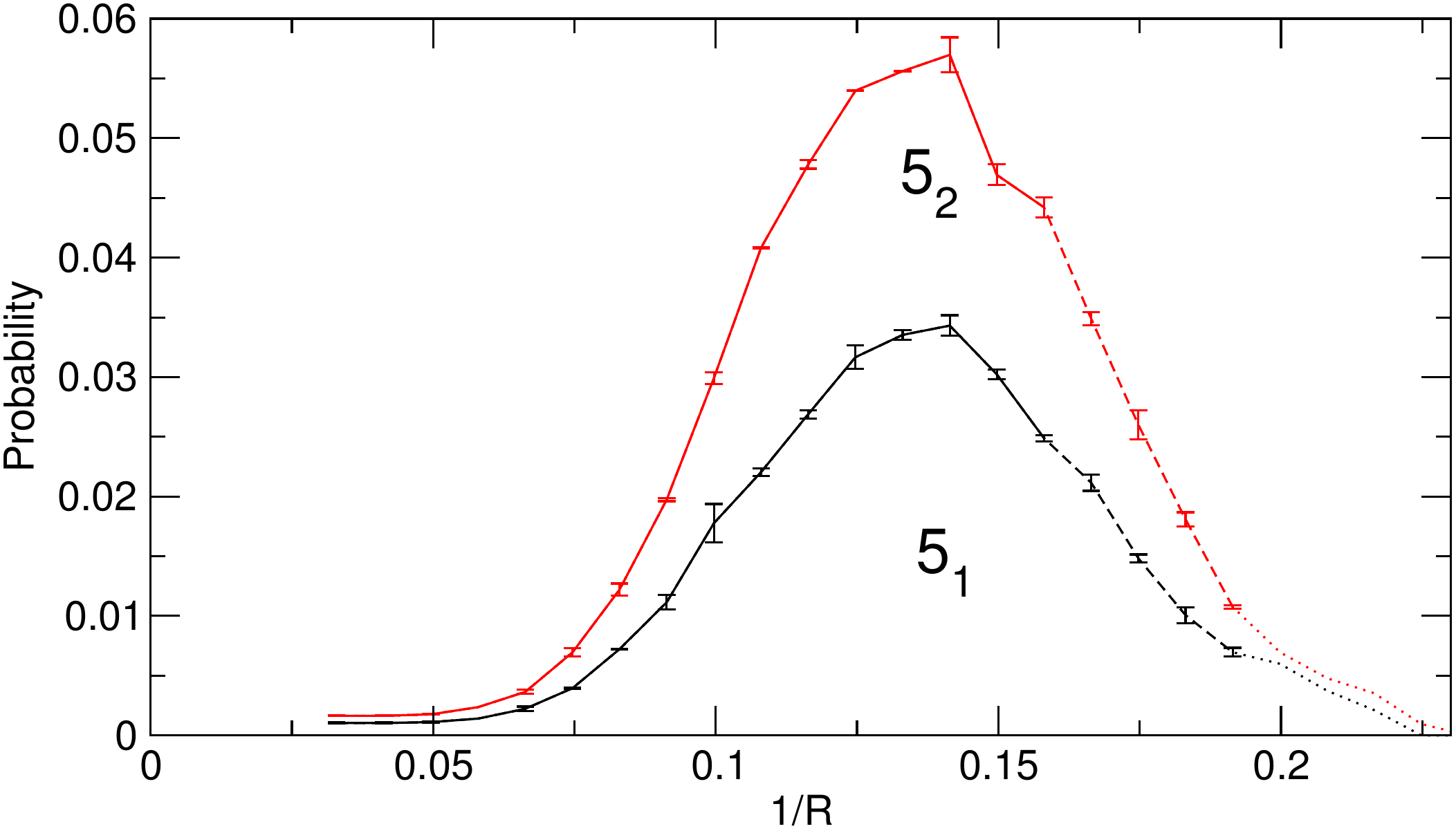}
\caption{Probability of formation of knots of type $5_1$ and $5_2$ in
flexible rings of cylinders of $N=200$ bonds (of unit length) confined
in a sphere of radius $R$. The bending rigidity of the rings and the
thickness of the cylinders has been set to match those of circular DNA
molecules of the P4 bacteriophage (see
ref.~\citep{Micheletti:2008:Biophys-J:18621819} for details). Dashed
[dotted] lines are used to indicate the regions where the knot
complexity becomes so high that 10 [50] \% of the rings form knots
that could not be classified after simplification. Figure adapted from
ref.~\citep{Micheletti:2008:Biophys-J:18621819}.}
\label{fig:5152}
\end{center}
\end{figure}

In accord with what was established for the confined freely-jointed rings, also
for the thick flexible ring of cylinders it was seen that, for
increasing spatial confinement, the knot spectrum is shifted towards
higher complexity with a similar progression of complexity. In
particular, it was seen (again consistent with the freely-jointed rings
case), that among the knots with five minimal crossings, the torus knot
$5_1$ is systematically about half as probable than knot $5_2$ over
a wide range of $1/R$ values, see Figure~\ref{fig:5152}.

This aspect appears profoundly different from what was established
experimentally for the knotting of the P4 genome. A series of
experiments, in fact, have shown that, among the simplest types of
knots, torus knots are much more abundant than twist ones.  In
particular, the knot of type $5_1$ was present in much higher
abundance than $5_2$, which, in fact, was virtually absent from the
measured spectrum. The bias in favour of torus knots is accompanied by
another apparent bias, namely the one in favour of chiral knots. It is
seen, in particular that knot $4_1$ appears much less frequently than
simpler or more complex knot types (e.g. $3_1$ and $5_1$).

This result then leads to the question of whether the knot spectrum of
DNA phage might at all be explained via a model of confined polymers
in thermodynamic equilibrium. One possibility is that the 
phage knot ``selection'' is a consequence of a certain degree of ordered spooling
in the DNA conformations, which, while not large enough to 
drive the formation of a perfect unknotted spool, may still provide a measurable
bias towards chiral and torus knots. 
For instance, Arsuaga and Diao recently showed that fluctuating, hence imperfect, DNA spools generate 
a knot spectrum closer to the experimental one~\citep{Arsuaga&Diao:2008:JCMMM}. 
However the generation of DNA spools came
from a mathematical algorithm rather than a force field or potential, so it
remained unclear what physical mechanism, kinetic or thermodynamic, provides the
ultimate driver for the bias towards torus and chiral knots. 

These biases are indirect clues about how DNA is organised inside the
capsid; in particular they are suggestive of certain degree of ordered
spooling, as this would be intuitively accompanied by a bias in chiral
and torus knots. The discrepancy of experiments and computational
predictions based on the simplified DNA models poses the interesting
question of what specific aspects of DNA self-interaction and/or the
genome packaging process need to be accounted for in order to explain
the observed measurements.

In a very recent contribution, we provided quantitative evidence 
supporting the scenario that the biases in the knot spectrum may be
explained via a more accurate force field which takes into account
DNA self-interactions in a more realistic way~\citep{Marenduzzo:2009:Proc-Natl-Acad-Sci-U-S-A:20018693}.
The main novelty in Ref.~\citep{Marenduzzo:2009:Proc-Natl-Acad-Sci-U-S-A:20018693}
is the introduction of a chiral potential which favours a small
twist angle between spatially contacting DNA segments. This potential would promote a 
cholesteric apolar ordering of DNA (with a pitch is in the $\sim \mu$m range, hence much larger than
the phage size). Such a chiral ordering {\em is} commonly observed in sufficiently 
concentrated DNA solutions {\it in vitro} (see e.g. Ref.~\citep{Grelet&Fraden:2003:PRL}). This additional chiral interaction
is also needed in models to discuss DNA in solutions at high enough densities to form a 
cholesteric phase, and is well documented in that literature~\citep{Kornyshvev:2002:EPJE,Tombolato:2005:J-Chem-Phys,Stanley:2005:Biophys-J,Tombolato_et_al:2006:PRL}. In the context of phage DNA, this interaction was never considered in simulations. 
Its effect is dramatic: the ordering increases so that DNA conformations now resemble much more 
fluctuating spools, and the main ``selection rules'' found in the experimental knot spectrum are
very well reproduced, as the $4_1$ and $5_2$ twist knots are now extremely rare, and knots are 
typically chiral. Torus knots outnumber twist knots in the simulations even for more complicated
knots, such as 7-crossing ones, which were not analysed in detail experimentally. Note that, for
computational reasons, instead of the P4 genome a mutant containing half the genome (and preserving the main features of the knot spectrum, as demonstrated in~\citep{Trigueros:2007:BMC-Biotechnol:18154674}).

A remaining intriguing question is how phages are able to efficiently eject their DNA, which is the first
step in the infection of a bacterial host, given that their genome is knotted with large probability, 
at least in the case of the P4 phage. The detailed simulations of Ref.~\citep{Marenduzzo:2009:Proc-Natl-Acad-Sci-U-S-A:20018693}
suggest a possible resolution of this seemingly paradoxical issue. By analysing the knotted configurations, 
it was found that the knots generated using the chiral interaction just described are highly delocalised
(see Section~\ref{size_knots} for a definition and discussion of knot localisation). This is very different to
what occurs in DNA knots in swollen, uncompacted configurations, which are (weakly) localised. The delocalisation
of the knot makes so that the local conformation which is found close to the opening of the phage where the DNA 
gets out is not appreciably affected by the knot state of the DNA, which is a global topological characteristic of 
the molecule. On the other hand, it is these local entanglements which are likely to impact on ejection dynamics, 
and hence knot delocalisation renders ejection relatively insensitive to the DNA topology. More computational studies 
would be needed to further elucidate the interesting problem of knot ejection. 

The latest theoretical and computational investigations provide
quantitative evidence in favour of the fact that the experimental
spectrum reflects specific aspects of DNA self-interaction (namely
cholesteric ordering) that emerge at high DNA packing
densities~\citep{Marenduzzo:2009:Proc-Natl-Acad-Sci-U-S-A:20018693}.

\section{Knotting of polymer rings undergoing adsorption transition}
\label{knot_prob_ads}

It is well known that if a long polymer chain is in proximity of an
attractive impenetrable surface, it may undergo a conformational
transition from a \emph{desorbed phase}, where the chain likes to be
extended in the bulk, to an \emph{adsorbed state}
~\citep{DeGennes:1979,Vanderzande:1998,Netz&Andelman:2003:Phys-Rep}
where it is essentially localized at the surface.

Therefore, the attractive substrate can be regarded as a means to
realize polymer confinement. The average degree of adsorption of a
long polymer is expected to depend significantly on the temperature of
the system~\citep{DeGennes:1979}. The latter, in fact, controls the
balance between the loss of configurational entropy when the polymer
adheres to the surface rather than being free in the bulk, and the
energy gain, $\epsilon$, for each monomer that is in contact with the
attractive substrate.

A crude estimate for this balance can be given by considering the
entropy per monomer of the free polymer in three-dimensions,
$\mu_{3d}$, and in two dimensions, $\mu_{2d}$ (clearly $\mu_{3d} >
\mu_{2d})$. The free energy per monomer of the fully desorbed and
adsorbed states are, respectively, $f_d \approx - T\, \ln \mu_{3d}$
and $f_a \approx -\epsilon - T\, \ln \mu_{2d}$.  Accordingly, at
temperatures lower than $T_c = \epsilon/ \ln(\mu_{3d}/\mu_{2d})$ the
polymer is expected to be fully adsorbed on the surface. The above
estimate for $T_c$ is admittedly crude as the adsorption is treated as
an {\em all or none} process as a function of temperature. As a matter
of fact, more accurate
theories~\citep{DeGennes:1979,Vanderzande:1998,Rensburg:2000} and
computational
studies~\citep{DeBell&Lookman:1993:RMP,Eisenriegler_et_al:1982:JCP}
indicate that below a critical temperature, $T_c$, the fraction of
polymer models adsorbed on the surface increases continuously from
zero to 1.

Extending considerations from the case of linear polymers to rings
with unconstrained topology, i.e. where ``virtual'' strand passages
are allowed so to establish an equilibrium population of knots (see
Section~\ref{knot_prob}) brings additional elements of interest.

In particular, it is most interesting to analyse how the fraction of
knotted rings varies with temperature. The salient physical mechanisms
at play in this problem are best formulated by considering a
partitioning of the ensemble of circular polymers into knotted and
unknotted rings.

The first observation is that for each of these two sub-ensembles one
can define a critical temperature for adsorption. In general these two
temperatures are not expected to coincide.  The second observation
regards the relative weight between the population of knotted and
unknotted rings. From what was discussed in Section~\ref{knot_prob} we
know that in bulk (above the adsorption temperature for both
sub-ensembles) the knotted population will outweight the unknotted one
for sufficiently long chains. At the same time, the lowest energy
configurations, which dominate the system thermodynamics at
sufficiently low temperatures, can only correspond to adsorbed {\em
  unknotted} rings. Only the latter, in fact, can be perfectly flat
and thus guarantee that each single monomer is in contact with the
surface. It is therefore to be expected that, as $T \to 0 $, (i.e.  well
below the adsorption temperature for both sub-ensemble), the fraction
of unknotted rings approaches 1.

Because of the complicated interplay of the two effects mentioned
above, no clear prediction can be made for how the knotting
probability depends on temperature in the neighborhood the adsorption
transition(s).

For polygons on the cubic lattice it is however
  possible to show that, for every finite value of $T$, the
  probability that a $N$-edge configuration is knotted goes to one as
  $N$ goes to infinity~\citep{Vanderzande:1995:J-Phys-A}. So except in
  the limiting case of zero temperature, when the polygon flattens
  completely by fully adhering to the surface, the polygon is knotted with high
  probability. 

Numerical calculations~\citep{Janse-van-Rensburg:2002:Cont-Math} indicate that, 
for a fixed value of
the adsorption energy, $\epsilon$, the knotting probability of sufficiently-long polymers is
described by the following functional form:
\begin{equation}
P_N(T) = 1 - e^{-\alpha_0(T) N + o(N)} \label{knot_prob_ads_eq1}
\end{equation}
\noindent where $N$ is the number of polymer monomers. 
 
It is readily realised that the above expression generalises the
expression for the knotting probability of unconstrained rings in
three dimensions, see eq.~\ref{knot_prob_free_eq1}.  Indeed, for
temperatures above the adsorption transition(s) $\alpha_0(T)$ is found
to be independent of temperature and equal to the bulk value.
Notably, a shoulder in proximity of the adsorption transition was
observed even if the data were not sufficiently asymptotic to
discriminate between the presence or the absence of a maximum around
the critical temperature $T_c$~\citep{Vanderzande:1995:J-Phys-A}. 
In particular it was not possible to
decide whether the adsorption transition of unknotted polygons differs
from the one of polygons with unrestricted topology.  Finally,
consistent with previous observations $\alpha_0(T)$ was found to
approach the zero value as $T\to 0$.

Further interesting observations, pointing to open problems could be
made about the model system. For example, a preliminary investigation on
the knot complexity shows that it decreases as $T$ decreases. No clear
results are however available for the knot distribution as a function
of $T$ in the adsorbed regime. The limited number of theoretical
investigations of the subject (which include the above mentioned works
and the study in~\citep{Michels&Wiegel:1989:J-Phys-A}) have left many
open questions deserving more in-depth investigations. A summary of
these open issues is provided below:

\begin{itemize}

\item it is unclear whether the adsorption transition depends on knot
type. More stringent numerical investigations are needed in this
respect.

\item up to now, the knotting probability in adsorbed polymers has
  been studied only for lattice polygons. It appears necessary to
  extend the investigation to off-lattice models of ring
  polymers. This would also allow a more stringent comparison with the
  novel experiments of knotted circular DNA adsorbed on mica
  surfaces~\citep{Ercolini:2007:PRL} or with future experiments based
  on single molecule imaging
  techniques~\citep{Rivetti:1996:JMB,Valle:2005:PRL,Witz:2008:PRL};

\item even for lattice polygons the knotting probability and the knot
spectrum have not been characterized extensively neither in proximity
of the adsorption transition nor in the fully-adsorbed phase, where a
possible comparison with the model of \emph{flat knots}
~\citep{Guitter&Orlandini:1999:J-Phys-A} can be pursued (see section
\ref{flatknots});

\item in the adsorption problem discussed so far it was assumed that
the surface exerts its attractive action over a lengthscale comparable
to the typical size $b$ of the monomers (short range potential). It
would be interesting to examine how the entanglement of the rings is
affected by extending the range of the attractive potential
~\citep{Netz&Andelman:2003:Phys-Rep};

\item so far very dilute solutions have been considered. In the case
of adsorption transition from semi-diluted solutions it would become
necessary to account for the mutual entanglement of polymer
rings. This problem leads to the largely-unexplored issue of how ring
linking probability depends on the polymer concentration in proximity
of the attractive surface, and on the strength of the adsorption
itself.
\end{itemize}

\section{Knotting of polymers under tension}
\label{knot_prob_stress}

A further notable means of imposing spatial constraints on polymers is
based on their direct {\em mechanical} manipulations. Owing to
technological advancements introduced in the past decade, it is
presently possible to apply a variety of controlled types of
confinement on few, or even single
biopolymers~\citep{Bustamante:1995:Physics-Today,Bemis:1999:Langmuir}. Examples
range from microfluidic
devices~\citep{Bakajin_et_al:1998:PRL,Guo_et_al:2004:NanoLett} which
are used to force polymers into channel or slab-like geometries to
experiments where one of the polymer ends (the other being anchored to
a surface) is pulled by an AFM tip or by means of optical or magnetic
tweezers~\citep{Svoboda&Block:1994:Ann-Rev-Bio,Strick_et_al:1996:Science,Askin:1997:PNAS,Moffit:2008:Ann-Rev-Biochem}.
The latter two types of probe are particularly interesting in the
present context as the ``topology'' of the stretched chain is not
fixed (at variance with the AFM case)~\citep{Arai:1999:Nature}.

The variety of available experimental setups provide concrete contexts
for validating or challenging the theories that had been developed 
for the response of polymers to mechanical probes either in 
the bulk~\citep{Pincus:1976:Macromol,DeGennes:1979} or
in confined geometries~\citep{Wang2001,FreyPRE2007,Jun_et_al_PRL:2008}.

Arguably, the most common type of mechanical probe is the application
of a stretching force at the two ends of a polymer. The application of
the force, which couples to the end-to-end distance of the chain,
reduces the chain configurational entropy. It is therefore interesting
to analyse how the equilibrium degree of entanglement of the chain is
affected by the application of the tensile force.

The latter question has been investigated for the case of lattice
polygons~\citep{Janse-van-Rensburg:2008:J-Phys-A,Janse-van-Rensburg:2008:J-Phys-A-B}. The
model was formulated so that the effects of both stretching and
compression could be treated on common grounds. Specifically, the
compressing/stretching force along the $z$ direction was coupled to
the calliper size, $s$, of the polygon measured along
the $z$ direction. In other words, because the concept
  of end-to-end separation is not applicable to closed polymers, the
  force is coupled to the largest distance projected by the ring
  polymer along the $z$ direction.  The calliper size therefore
  measures the width of the smallest slab (with plates perpendicular
  to the $z$ axis) which accommodates the polymer configuration.

In the \emph{constant force ensemble}, if we denote by $p_N(s)$ the
number of lattice polygons with calliper size $s$, the equilibrium
properties of the system are aptly  encoded in the canonical partition function
\begin{equation}
Z_N(\tilde{f}) = \sum_{s}p_N(s) e^{f s }
\end{equation}
where $f = \tilde{f}/k_B T$ is the \emph{reduced force} and
$\tilde{f}$ is the applied force (in this way reduced forces are
measured in units of inverse length). The unconstrained case is
recovered for $f=0$ where $Z_N(0) = p_N$. For $f<0$ (\emph{compressing
  regime}) the polygons are squeezed along the $z$ direction while for
$f>0$ (\emph{stretching regime}) conformations elongated along the
$z$-direction will be favored \footnote{Note that the statistical ensemble of the rings compressed by a constant negative force ($f <0$) is the conjugate of the statistical ensemble of rings confined in a slab of constant width. The properties of either ensembles can be recovered from the other through an appropriate reweighting tecnique (see Section \ref{reweight}).
This correspondence is exploited in Section \ref{sampling_confined} to achieve an efficient computational sampling of rings confined in a slab.}. 
By concatenation arguments the existence of the limiting free-energy
\begin{equation}
F(f) = \lim_{N\to\infty}\frac{1}{N} \log Z_N(\tilde{f}) 
\end{equation}
\noindent of this system for any (finite) values of the
force can be determined~\citep{Janse-van-Rensburg:2008:J-Phys-A}.

The metric properties of the stretched chains have been explored by
numerical simulations \citep{Janse-van-Rensburg:2008:J-Phys-A-B}. In
particular it is interesting to study how the average extension of the
polygons along the direction of application of the force:
\begin{equation}
\langle R_f \rangle = \frac{\partial}{\partial f} \log Z_N(f)
\end{equation}
depends either on the applied force $f$ or on $N$. 

A valuable theoretical framework for characterizing the behaviour of
linear (non closed) chains subject to a tensile force is provided by
the scaling analysis of
Pincus~\citep{Pincus:1976:Macromol,DeGennes:1979}. This theory
identifies two characteristic length scales in the problem: the
average modulus of the end-to-end distance in the unconstrained case,
$\bar{R}\propto N^{\nu}$, and the so called tensile screening length
$\xi=1 /f$.  The cases of weak and strong stretching forces correspond
respectively to the conditions $\bar{R}/\xi << 1$ and $\bar{R}/\xi
>>1$. According to Pincus~\citep{Pincus:1976:Macromol} three
specific stretching regimes should be observable for polymers in good solvent (see ref.~\cite{Morrison2007} for discussion of bad solvent effects).
\begin{itemize}
\item {\em Weak force regime}.

For small forces one should expect a linear increase in the extension $R_f$ as  $f$ increases
(Hooke's law). Assuming the validity of the scaling relationship:
\begin{equation}
\langle R_f \rangle = R\, \Phi ( \bar{R}/\xi),
\end{equation}
\noindent where $\Phi(x)$ is a dimensionless scaling function, one can recover Hooke's law if
$\Phi(x\to 0) \sim x$ so that
\begin{equation}
\langle R_f \rangle \propto R^2 f= N^{2\nu}f. \label{scaling_f}
\end{equation}
Note that for chains with no excluded volume interaction (ideal
chains) $\nu=1/2$ and $\langle R_f \rangle$ is linear in $N$ at low
forces. For self-avoiding walks $\nu \approx 0.588$  and the average extension is a
nonlinear function of $N$. In the ideal case the force is transmitted
along the backbone while for self-avoiding walks the transmission is
also through contacts between pairs of monomers due to excluded volume
interactions. Notice that scaling (\ref{scaling_f}) implies
\begin{equation}
f = K \langle R_f \rangle,
\end{equation}
\noindent so that the effective spring constant controlling the
Hookean behaviour, $K$, decreases with the polymer length as
$N^{-2\nu}$. It is important to notice that this Hookean regime should
not necessarily be interpreted as arising from an effective increase of the
average distance of the two chain ends. This point is best illustrated for a chain that is mildly pulled at both ends (which at low stretching forces can be slightly different from our case where the chain is pulled at the points with maximum and minimum values of $z$). The mild pulling at both chain ends leaves unaffected the end-to-end separation to
leading order in $f$. In fact, it merely causes the end-to-end distance vector to rotate towards the
stretching direction. This rotation produces in turn the linear (Hookean) increase of the projection of the end-to-end vector along the $z$ axis~\citep{Neumann1,Neumann2}.

\item {\em Intermediate force regime}. In the regime, which arises for
  intermediate forces, the value of $\langle R_f \rangle$ can be
  obtained according to a ``blob" interpretation similar to the one
  described for polymers in confined geometries. The application of
  the theory can be argued as follows. As we discussed above, the
  presence of small-enough stretching forces does not alter the shape
  of the polymers but causes a rotation towards the stretching
  direction. As the force is progressively increased, the shape will
  eventually distort so that the polymer will extend along the
  stretching direction and grow thinner perpendicularly to it. This
  thinning is expectedly uniform along the polymer producing a
  situation similar to the tube-like confinement where the blob
  picture was first introduced.

\begin{figure}[tbp]
\begin{center}
\includegraphics[width=\WIDTHC]{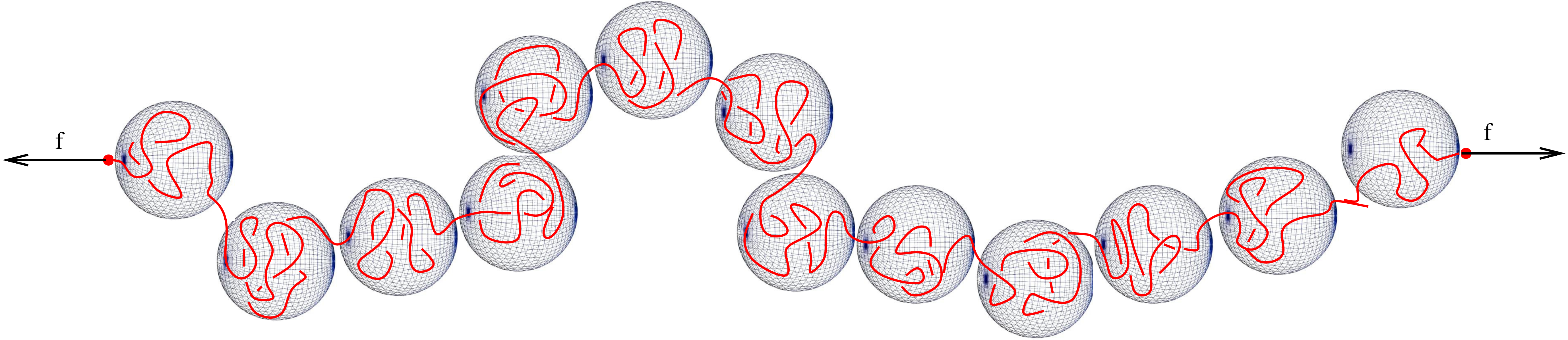}
\caption{Blob picture of stretched polymers in the intermediate force regime.} 
\label{fig:stretched_blobs}
\end{center}
\end{figure}

Analogously to the scheme outlined in Section~\ref{blob_picture} the
chain subjected to mechanical tension can be broken into a succession
of tensile blobs along the $z$ axis. The typical blob size is $\xi
\sim f^{-1}= (\tilde{f} /k_B T)^{-1}$ (see Figure~\ref{fig:stretched_blobs}). 
The blobs do not interact with one another
and the monomers contained in each blob behave as unperturbed
self-avoiding walks.  That is $\xi \sim g^{\nu}$ where $g$ is the
number of monomers in each blob. Since each chain is a stretched array
of tension blobs, their average extension along the direction of force
application is the product of the tensile screening length $\xi$ and
the number of these blobs, $N/g$ per chain
\begin{equation}
\langle R_f \rangle \sim \xi N/g \sim N f^{1/\nu -1} \sim N f^{2/3},
\end{equation}
where in the last term we have used the Flory value of the exponent
$\nu = 3/5$. This is the so-called Pincus regime. In this case
\begin{equation}
f = K \langle R_f \rangle^{3/2}
\end{equation}
with $K\sim N^{-3/2}$.

\item {\em Strong force regime}. For extremely large forces excluded
  volume effects become irrelevant since the bonds of the polymer are
  fully aligned along the stretching direction and monomers do not
  interact with one another.  In this case $\langle R_f \rangle $
  becomes comparable to $N$ and the force-extension relation for the
  chain is dominated by the bonding potential between consecutive
  monomers. As a result the force-extension curve would be model
  dependent.
\end{itemize}

\begin{figure}[tbp]
\begin{center}
\includegraphics[width=\WIDTHA]{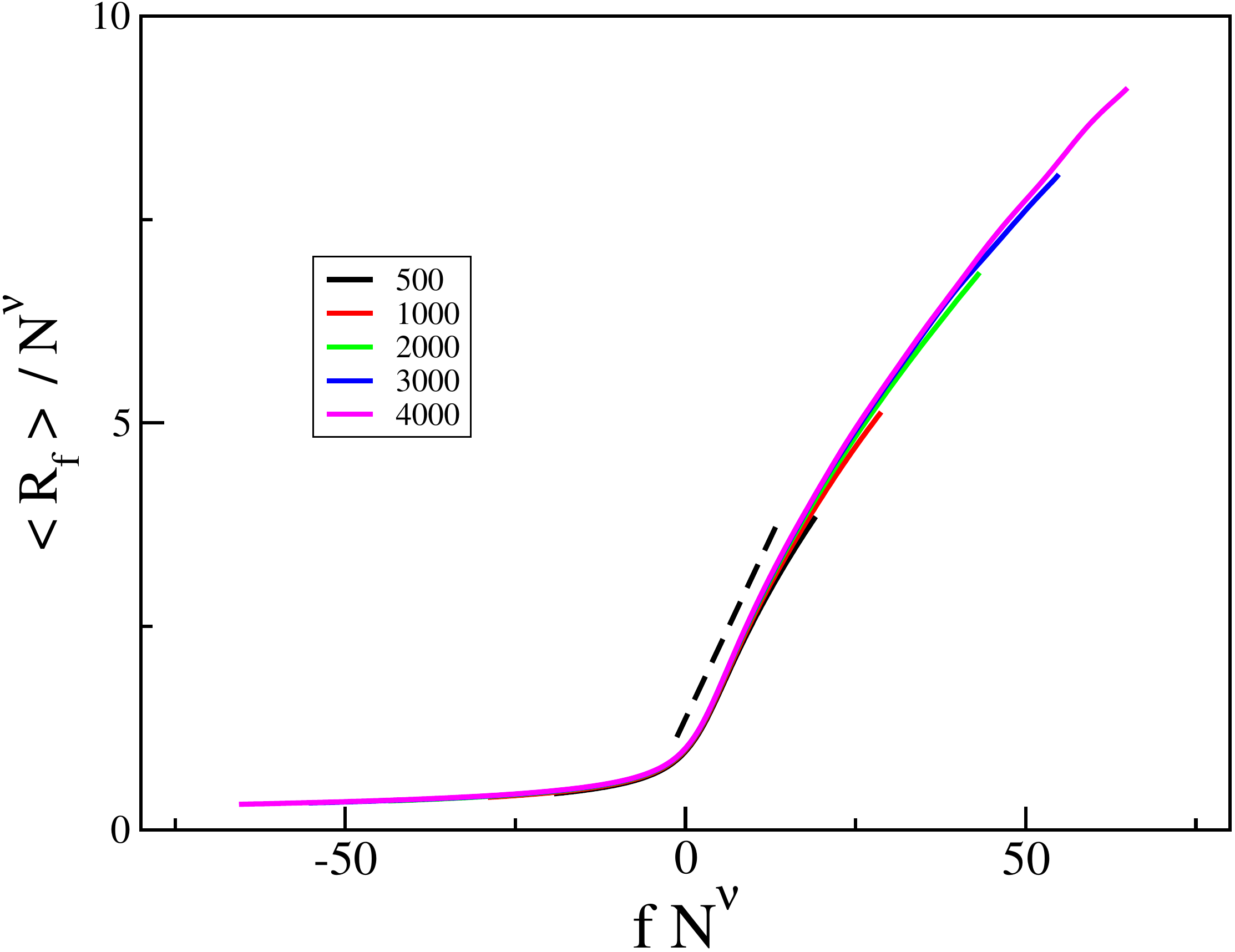}
\caption{Plots of $\langle R_f\rangle$ vs $f$ for stretched polygons of different values of $N$. To
underlying the weak adsorption regime the average extension along the force direction has been
scaled by $n^{\nu}$ whereas the force is multiplied by $n^{\nu}$. (The value of $\nu$ used is
$\nu\approx 0.588$). The dashed straight line indicates the expected linear behaviour for small values of
$f$.  } \label{fig:f_vs_Rf_scaled_weak_regime}
\end{center}
\end{figure}

\begin{figure}[tbp]
\begin{center}
\includegraphics[width=\WIDTHA]{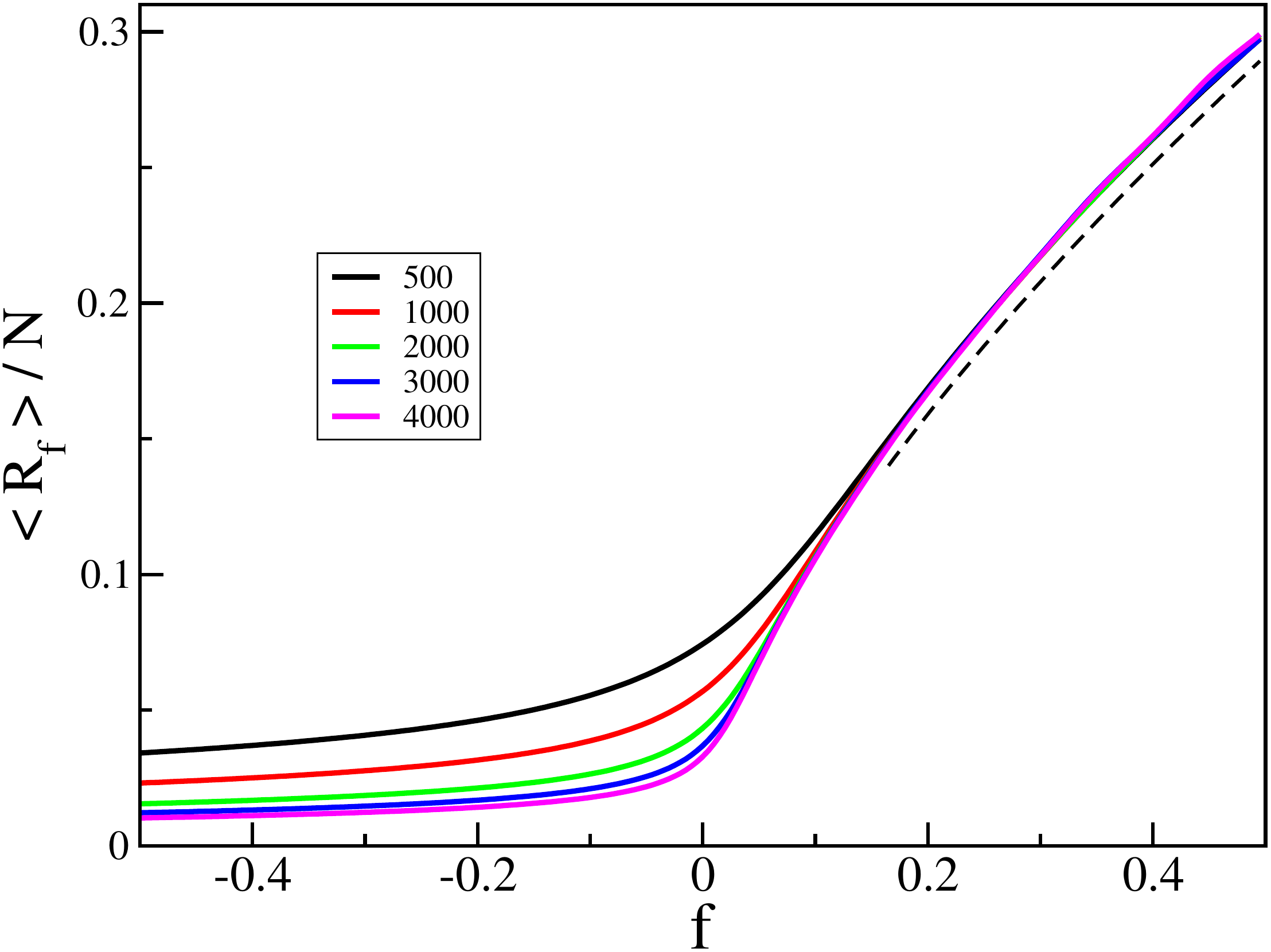}
\caption{Same data of the previous Figure. Now $\langle R_f\rangle$ is divided by $N$ to highlight
the Pincus regime. The dashed curve correspond to the fit $A\,N^{0.66}$ done on the $N=4000$ data.}
\label{fig:f_vs_Rf_scaled_pincus_regime}
\end{center}
\end{figure}

Figure~\ref{fig:f_vs_Rf_scaled_weak_regime} shows $\langle R_f \rangle
/ N^{\nu}$ vs $f N^{\nu}$ where $\nu = \nu_{saw} \approx 0.588$ for
positive forces. For relatively small positive forces, as $N$
increases, the data collapse onto a single curve. In this weak force
regime the behaviour is linear as suggested by the dashed straight
line. For larger values of $f$ the curves collapse nicely if we plot
$\langle R_f \rangle / N$ as a function of $f$ (see
Figure~\ref{fig:f_vs_Rf_scaled_pincus_regime}). The dashed curve
represents the best fit of the data to a power-law behaviour,
$A\,N^c$.  The estimate $c\approx 0.64$ is close to the value $2/3$
which characterizes the Pincus regime. In these data the values of $f$
are not large enough to see the strong force, non-universal,
regime. Notice that longer polygons reach the scaling regime at weaker
forces. Indeed the rigidity of a long chain is smaller than that of
short chains and this makes longer polygons more easily deformed. In
the contractile region, $f<0$, no scaling theory is available.

For the knotting probability some rigorous results are available
~\citep{Janse-van-Rensburg:2008:J-Phys-A}. In particular, for $f$
positive and sufficiently large it was possible to prove a pattern
theorem for stretched polygons. A key element of the proof is the
notion of a cut plane, that is a plane orthogonal to the stretching
direction that separates indepedent sub-parts of the polygon, in analogy
with the blobs in the Pincus scaling argument
~\citep{Pincus:1976:Macromol}. The proof relies on the fact that, for
typical polygon conformations, cut-planes occur with a finite density
along the stretching direction. 

The pattern theorem can be used to prove the FWD conjecture also for
sufficiently-stretched polygons. This result can be stated as follows:
for arbitrary large positive, though finite, values of the stretching
force, the probability that a stretched polygon is knotted approaches
$1$ as the length of the polygon (at fixed $f$) increases. In other
words, for $f> f_0$ the following large $N$ behaviour for the
unknotting probability holds
\begin{equation}
P^0_N(f) = e^{-\alpha_0(f)N + o(N)}. \label{unknot_f}
\end{equation}
Note that that eq.~(\ref{unknot_f}) can be proved for $f>f_0 = 4
\kappa$.  On the other hand a proof exists for $f=0$ (see section
\ref{knot_prob}) and it is quite intriguing that in the intermediate
region $0<f<f_0$ no proofs are at the moment available \footnote{A
  pattern theorem for biased self-avoiding walks has been proven
  in~\citep{Velenik:2008} for all values of $f>0$. If the technique
  in~\citep{Velenik:2008} were be applicable to models of pulled
  polygons one should be able to prove (\ref{unknot_f}) for all
  positive values of $f$} . This unknown region could correspond to
the weak stretching regime where the Pincus blob picture may not be
valid. For negative values of $f$ no rigorous results are available to
present day.

\begin{figure}[tbp]
\begin{center}
\includegraphics[width=\WIDTHA]{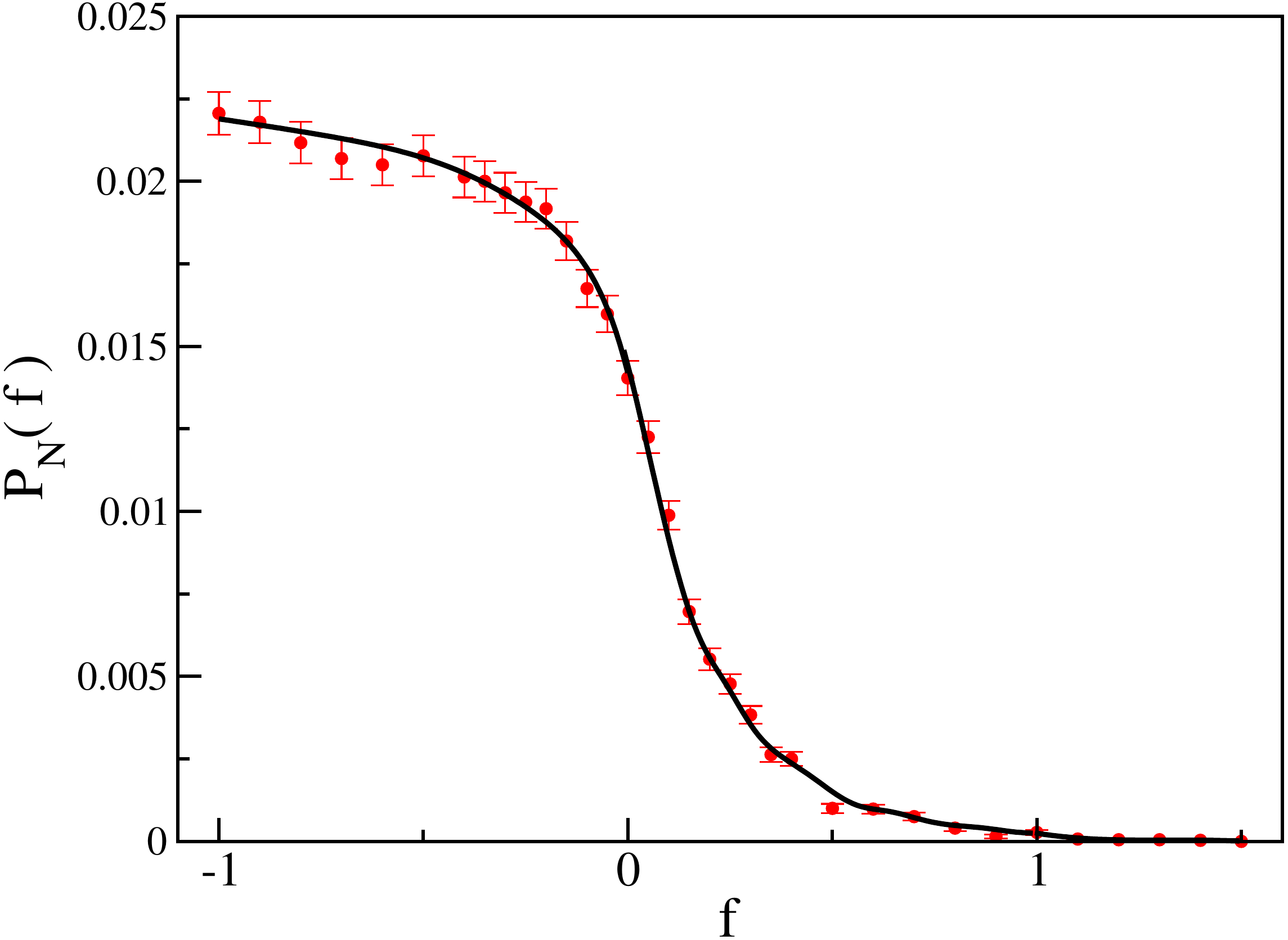}
\caption{Knotting probability for $N=3000$ bonds polygons subject to an external force $f$ coupled
to the span $s$ of the polygon along the z direction.} \label{fig:prob3000}
\end{center}
\end{figure}

The problem can, however, be investigated numerically by Monte Carlo
simulations~\citep{Janse-van-Rensburg:2008:J-Phys-A-B}. A typical
approach is based on the two-point pivot algorithm implemented on the importance
sampling techniques with weight proportional to $e^{fs}$. A MMC
sampling method is then performed on the space of $f$. Once a
reasonable number of uncorrelated configurations are sampled at a
given $f$, the knotting probability is computed by using the Alexander
polynomial (see Section~\ref{knot_theory}) as knot detector. The
results show a knotting probability that, at fixed $N$, decreases
rapidly as the stretching force increases while it increases (although
very slowly) in the compressive regime (see Figure~\ref{fig:prob3000}). 
For fixed values of $f$ the exponential decay
(\ref{unknot_f}) is confirmed in the \emph{whole range of $f$} and an
estimate of $\alpha_0(f)$ is obtained.

In addition to the case of rings mentioned above there are models of
linear chains under mechanical stress for which a non-ambiguous
definition of knottedness can be introduced. Examples are walks
spanning a slab and whose ends are anchored to the walls or walks in
which both extremities are anchored to an impenetrable surface (loops)
~\citep{Janse-van-Rensburg:2009:JSTAT}. For these models both the
thermodynamic and the entanglement properties have been explored
either by rigorous arguments or by Monte Carlo simulations. A
thermodynamic phase transition from swollen to ballistic (i.e. where the
end to end distance grows with $N$) phases) has been proved to exist
for $f=0^+$ and the unknotting probability has been confirmed,
numerically, to follow the law (\ref{unknot_f}) for the all range of
positive and negative values considered
~\citep{Janse-van-Rensburg:2009:JSTAT}. Polygons can also be stretched
by confining them into a tube and by applying a force along the axis
of the tube. In this case the geometrical constraint helps in proving
a pattern theorem \emph{for any positive values of $f$} and,
consequently, an extension of the FWD conjecture to polygons stretched
through a tube~\citep{Atapour:2009:J-Phys-A}.

With the exception of the studies just mentioned, the knotting
probability of polymers under tension and in general the dependence of
the entanglement complexity on the external force remains a rather
unexplored topic that deserves a more in-depth investigation. 

For example very little is known on the knot spectrum as a function of
$f$ and there are no results available (neither analytical nor
numerical) for the knotting probability for off-lattice models of
polymers under tension.

\section{Third part: Mutual entanglement of several polymer chains}
\label{sec:3rdpart}

The third part consists of Section~\ref{linking} and is devoted to
discussing the occurrence of a particular topological state in a polymer
melt, that is the {\em mutual entanglement} of several polymeric
chains (linking). Historically, this is arguably the oldest topic in
the topology of closed curves: in fact its origins can be traced back to
Gauss~\citep{Marathe:2006:MathInt}.

The mutual entanglement of chains and filaments that are densely
packed in the same region of space is known to affect their
rheological and elastic
properties~\citep{Edwards:1968:JPA,Kapnistos:2008:Nat-Mat}. The
topologically-entangled states of two or more circular polymers are
called links (or catenanes) \citep{Rolfsen:1976,Adams:1994}.  Links
are of interest to
chemists~\citep{Frisch_Wasserman:1961:JACS,Dietrich:1984:J-Am-Chem-Soc,Logemann:1993:J-Math-Chem},
but are also frequent in biology. For example, linked pairs of DNA
rings occur in the mitochondria of malignant cells and are
intermediates in the replications of circular
DNA~\citep{Adams:1992:Cell}. Linked rings are also present in
trypanosomes, microorganisms associated to leishmaniasis and Chagas
disease. In these organisms there are organelles, called kinetoplast,
which house thousands of DNA circles which can be interlinked to form
a network resembling chain mail~\citep{Chen_et_al:1995:Cell}.

In solutions with a sufficiently-high concentration of linear polymers
cyclization reactions will produce linked states of two or more
chains. In analogy with the case of knots, is natural to ask how
frequently, in thermodynamic equilibrium, a linked state would occur
compared to the unlinked one and how this linking probability and
linking complexity depends on polymer concentration, solvent condition
and degree of polymerization.

As for knots, linked states are not easily described in terms of local
variables and with the exception of analytical work based on the
Gaussian integral measure of links (linking
number)~\citep{Pohl:1981:Lectures-Notes,desClozeaux&Ball:1981:Comm-Math-Phys,Duplantier:1982:Comm-Math-Phys}
the linking probability of circular polymers in free solution has been
investigated mainly by numerical simulations and in the simplest case
in which only two rings are involved (two components
links)~\citep{Vologodskii:1975:JETP,Orlandini:1994:J-Phys-A,Hirayama:2009:J-Phys-A}.
Similar investigations have then been carried out for polymers in a
melt~\citep{Orlandini:2000:J-Phys-A,Brereton&Vilgis:1995:J-Phys-A}
but, with the exception of few studies on simplified
models~\citep{Tesi:1998:IMA,Arsuaga:2007:J-Phys-A,Soteros:2009:JKTR,Panagiotou_et_al:2010:J-Phys-A}
the linking probability and link spectrum of rings in confined
geometries is an issue that remains mostly unexplored. This subject is
discussed in Section~\ref{linking}.

\section{Mutual entanglement between polymer rings in solution: linked states}
\label{linking}

In solutions with sufficiently-high concentrations of linear polymers,
a circularization reaction will usually trap the chains in mutual
entanglement producing various topological states, known as
\emph{links} (see Section~\ref{knot_theory}).

Although the links arising from the ring-closure operation may involve
several chains, we will primarily focus on two-components links which
are made by a pair of chains. Such links are easier to define and
detect than generic multi-component links. Furthermore the latter can
be often characterized in terms of their pairwise subcomponents, as we
shall discuss later.

\begin{figure}[tbp]
\begin{center}
\includegraphics[width=\WIDTHB]{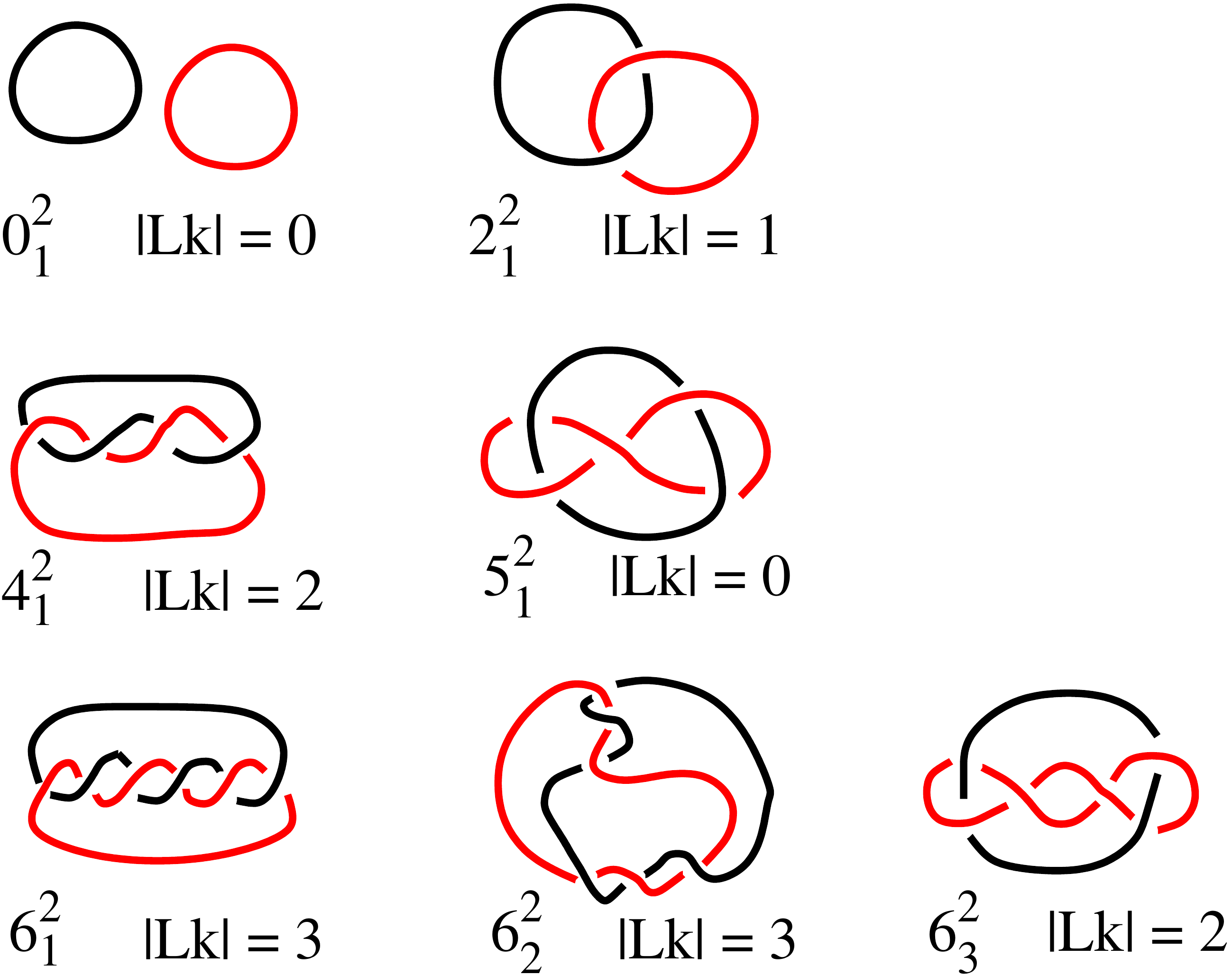}
\caption{The simplest two-components links with their symbol and linking number.} \label{fig:link_table}
\end{center}
\end{figure}

Consider two simple closed curves, ${\cal C}_1$ and ${\cal C}_2$ that are
disjoint, i.e. with no points in common, and embedded in
three-dimensional space. The curves are said to be \emph{topologically
  linked} or, more simply, linked if no smooth deformation exists by
which they can be pulled apart so that they lie in two different
half-spaces separated by a plane. In Figure~\ref{fig:link_table} we
show the diagrammatic representation of the simplest types of
$2$-components links. Notice that, while for isolated closed curves,
the salient topological information was contained in the {\em
  self}-crossings of their diagrammatic representations, for links,
the topological state is additionally specified by the {\em
  reciprocal} crossings between the curves. 
As reported in the figure, the standard nomenclature for links
parallels the one of prime knots with the addition of a superscript to
denote the number of link components so that, e.g. the unlink is
denoted as $0_1^2$. For future reference, we point to the reader's
attention the links $2_1^2$ and $5_1^2$ are known respectively as the
\emph{Hopf link} and the \emph{Whitehead link}.

Links between pairs of polymer chains are of paramount interest from
the chemico-engineering point of view, especially regarding the
synthesis of macromolecular compounds with so-called catenane units,
see refs.
~\citep{Wassermann:1960:J-Am-Chem-Soc,Frisch_Wasserman:1961:JACS}
~\citep{Dietrich:1984:J-Am-Chem-Soc,Logemann:1993:J-Math-Chem,Peinador:2009:J-Am-Chem-Soc}. Examples
of naturally-occurring links are common between nucleic acids
molecules and also protein complexes. For the latter case a remarkable
example is provided by the protein capsid of bacteriophage HK97 whose
resistance to degradation results from the mutual linking of its
tiling proteins~\citep{Wikoff:2000:Science}. Intriguingly, the
introduction of {\em de novo} designed links in proteins has now been
proposed as an effective method for stabilizing protein complexes
against thermal and chemical denaturation
~\citep{Yan:2001:Angew-Chem-Int-Ed,Blankenship&Dawson:2003:JMB}.

DNA molecules too offer several examples of topological linking.  For
example, linked pairs of DNA rings are found in mithochondria of
malignant cells~\citep{Hudson&Vinograd:1967:Nature}.  In bacteria,
DNA linking inevitably occurs during the replication process of the
circular
genome~\citep{Wassermann&Cozzarelli:1986:Science,Adams:1992:Cell}.
The removal of the resulting entanglement is essential for the
successful completion of the replication process and is provided by
enzymes, aptly named \emph{topoisomerases} because they are capable of
modifying the topological linked status of DNA rings by suitably
``cutting and rejoining'' DNA strands
~\citep{Wassermann&Cozzarelli:1986:Science,Adams:1992:Cell,Rybenkov:1997:Science,Alberts:2002,Liu_et_al:2006:J-Mol-Biol}.

As for the case of knots, there exist no general prescriptions or
algorithms that can be used to classifying exactly the topologically
state of a link starting from an arbitrary diagrammatic
representation. There are, however, several partial descriptors based
on \emph{link invariants} that can classify two-component links up to
a given degree of complexity. However, the most powerful descriptors
are also highly complex from the algorithmic point of
view. Consequently it is often necessary to strike a balance between
the power of the employed descriptor and the computational cost of its
evaluation.

The link invariant that requires the least amount of computation is
the \emph{linking number}.  Loosely speaking it corresponds to the
effective algebraic number of times the one curve winds around the other, and
viceversa. For two piecewise linear curves, ${\cal C}_1$ and ${\cal
  C}_2$, the linking number is conveniently evaluated by first
choosing an orientation for each curve and next projecting them onto a
plane avoiding intra- and inter-curve degenerate vertices. To each
crossing point where, say, ${\cal C}_1$ \emph{under}passes ${\cal
  C}_2$ we assign $+1$ or $-1$ value according to the crossing
orientation (see Figure~\ref{fig:cross_sign} where in this case the
strand labelled $k$ belongs to ${\cal C}_1$).  The sum of these signed
crossings is the linking number $Lk({\cal C}_1,{\cal C}_2)$ between
the two curves. Figure~\ref{fig:link_table} shows examples of
two-component link diagrams and the associated linking number. Note
that instead of focussing on the under-crossings of ${\cal C}_1$ one
can compute the sign of all the, say $K$, crossings between the two
curves. In this case the linking number is given by the formula
\begin{equation}
Lk({\cal C}_1,{\cal C}_2) = \frac{1}{2} \sum_{i=1}^K \sigma_i,
\end{equation}
\noindent where $\sigma_i$ is the sign of the $i$-th crossing. We emphasize
that, at variance with the case of single-component link (i.e. a
knotted closed curve) where standard topological invariants are
independent of the choice of orientation of the curve, the value of
$Lk$ does depend on how the two curves are oriented. For polymers that can be
oriented based on properties of their chemical sequence it is
meaningful to consider the signed value of $Lk$. In more abstract o
general contexts, where the choice of orientation is subjective, it is
more appropriate to  consider only the absolute value $|Lk|$ which is obviously
independent of the curves orientation.

For completeness we mention that there exist other algorithmic
formulations of the linking number which are equivalent to the one
given above~\citep{Rolfsen:1976}. One that is particularly useful for
analytic calculation on smooth curves is the definition introduced by
Gauss in terms of the double integral 
\begin{equation}
Lk_G({\cal C}_1,{\cal C}_2) = \frac{1}{4\pi} \int_{{\cal C}_1}\int_{{\cal C}_1} d\vec{r}_1 \times
d\vec{r}_2\frac{\vec{r}_1-\vec{r}_2}{|\vec{r}_1-\vec{r}_2|^3}
\end{equation}
where $\vec{r}_1$ ($\vec{r}_2$) defines a point on the curve ${\cal C}_1$ (${\cal C}_2$). The integral
$Lk_G({\cal C}_1,{\cal C}_2)$ is usually referred to as the \emph{Gauss linking number}, after
Gauss who introduced it to characterize the entanglement of the orbits of heavenly bodies.
Consistent with the former definition of linking number, the above integral returns an integer
value whose sign depends on the orientation of the two curves.

The limitations of the linking number as a descriptor of
topological-linking are illustrated in Figure~\ref{fig:link_table}
which portrays a Whitehead link $5_1^2$. Although this link has a
non-trivial topology, its linking number is zero as is the linking
number of the unlink $0_1^2$. This ambiguity has made necessary the
introduction of the concept of \emph{homologically linked} curves
which is used to denote two curves ${\cal C}_1$ and ${\cal C}_2$
having $|Lk({\cal C}_1,{\cal C}_2)| \ne 0$. Homologically-linked
curves are also referred to as being \emph{algebraically linked}, in
contrast to topological or geometrically linked ones.  Indeed, pairs
of curves that are homologically linked are guaranteed to be
topologically linked but, as shown by the previous example, the
converse is not true.

Despite the above limitations, the linking number is very commonly employed as a
detector of the concatenation of two curves. This is because it is
algorithmically simple to code and can be evaluated with a negligible
computational cost. Secondly, the linking number is often a complete
descriptor for the type of linking arising in certain contexts,
particularly biomolecular ones (e.g. DNA catenanes).  Finally, in the
context of double-stranded DNA circles, the
linking number of the two DNA strands is important. This is because the linking
number is a {\em topology-dependent} quantity which matches the sum of
two important {\em geometrical} quantities, namely local twist and the
writhe of the DNA duplex
~\citep{Fuller:1971:Proc-Natl-Acad-Sci-U-S-A:5279522,White:1971}.

More complete descriptors of topologically-linked states between two
curves are usually based on generalised polynomial invariants such as
the two-variable Alexander polynomial $\Delta(s,t)$
~\citep{Torres:1953:Ann-Math} or, better, the HOMFLY polynomial
reported in Section~\ref{knot_theory}. A detailed discussion of how to
compute the two variable Alexander polynomial from a given
$2$-component link diagram is provided in ref.
~\citep{Vologodskii:1975:JETP}.

\subsection{Linking statistics of two rings in unconstrained spaces}
\label{link_prob}

Consider the case where the linked state between two rings is not
fixed but can fluctuate in equilibrium or in steady-state conditions. A
concrete example of the latter situation can be provided by a solution
of circular DNA molecules in the presence of topoisomerase II enzymes
which continuously process the molecules cutting and rejoining DNA
strands. The action of the enzymes would therefore change the initial
topological linked state of the DNA circles producing a steady-state
distribution of linked or knotted molecules
~\citep{Alberts:2002}. Analogously, it is possible to envisage an
equilibrium linking probability for confined linear polymers whose
termini can attach to/detach from one of the bounding surfaces (which
provide a ``closure'' for the chains).

In these and similar cases it natural to ask how frequently the
equilibrated, or steady-state, configurations display a linked state
and how complex it is.  It is reasonable to expect that these
properties depend on several factors such as the degree of overlap
between rings, the quality of the solvent, the degree of confinement
etc. In the following we will review the salient available theoretical
results on this problem.

\subsubsection{Some analytical and rigorous results}

Because of the complexity of the problem, only a limited number of
analytical results are available for the statistics of two-component
links for rings in free space.  Early attempts were aimed at
characterizing the probability distribution of the linking number
between two closed rigid curves randomly displaced in $\mathbb{R}^3$
~\citep{Pohl:1981:Lectures-Notes,desClozeaux&Ball:1981:Comm-Math-Phys}. In
particular in~\citep{Pohl:1981:Lectures-Notes}, the Gauss integral
representation of the linking number was used to compute the
\emph{squared linking number} averaged over all relative rotations and
translations of the two curves and over all the possible reciprocal
orientation. The formula for the second moment of the Lk probability
distribution function was subsequently re-discovered and generalized
to space curves without a plane symmetry by using Fourier transform
techniques~\citep{desClozeaux&Ball:1981:Comm-Math-Phys}. A different
derivation and a further generalization to closed manifolds in
$\mathbb R^n$ was later provided
\citep{Duplantier:1982:Comm-Math-Phys}.

As intuitively expected, one of the difficulties of characterizing the
$Lk$ probability distribution for two curves in free space lies in the
fact that for the overwhelming majority of their possible relative
configurations, the two curves will be too far apart to be linked.

A more natural formulation of the linking problem is obtained by
introducing an external constraint limiting the phase space of
possible relative displacements of the two curves. A commonly-employed
scheme is to prevent the separation of the centers of mass of the two
curves from largely exceeding the average size of the rings.

An alternative, more rigorous formulation of the problem consists of
considering the ensemble of \emph{fixed linking type} and calculating
the limiting entropy of a system of two rings with a definite linked
(non trivial) state. This approach has been used for linked pairs of
polygons in the simple cubic lattice~\citep{Orlandini:1994:J-Phys-A}
and the results obtained are summarised hereafter. 

Let us denote by $p_{NN}^{(2)}(\tau)$ the number of polygons pairs,
each of $N$ bonds, forming a specific type of topological link (meaning their type of concatenation),
$\tau$. When no restriction is introduced for the knot type of each
component, it is expected that:
\begin{equation}
\lim_{N\to\infty}\frac{1}{2N} \log p_{NN}^{(2)}(\tau) = \kappa
\end{equation}
i.e. the limiting entropy of a two-component (non trivial) link equals
the one of the single polygon, independently on the link type.

Different results are obtained when the knot type of both components is kept fixed. Suppose, for
example, that neither of the two rings is knotted and let $p_{NN}^{(2)}(\tau|0,0)$ be the number of
such embeddings in the cubic lattice. In this case it is possible to prove that, independently of
the two-component link type $\tau$, the following inequality holds:
\begin{equation}
\lim_{N\to\infty}\frac{1}{2N} \log p_{NN}^{(2)}(\tau|0,0) = \kappa_0 < \kappa\ .
\label{lim_link_unknot}
\end{equation}
In other words, the key feature affecting the limiting entropy of a
two-component link is not the topological linked state between the two
rings but the requirement that \emph{both} components are
unknotted. One may \emph{conjecture} that this holds true also if both
components have a fixed non-trivial knot type $\sigma$:
\begin{equation}
\lim_{N\to\infty}\frac{1}{2N} \log p_{NN}^{(2)}(\tau|\sigma,\sigma) = \kappa_{\sigma}\ .
\end{equation}
This result, albeit being a plausible generalization of what was
established for unknotted polygons, has not been proved rigorously
because of the difficulty in proving the existence of the following
limit,
\begin{equation}
\lim_{N\to\infty}\frac{1}{N} \log p_{N}(\sigma) = \kappa_{\sigma}.
\end{equation}

Inequality (\ref{lim_link_unknot}) can be used to establish how the
knotting probability $P^{(2)}_{NN}(\tau)$ that {\em both} components
of the link of type $\tau$ are knotted grows as a function of $N$. One
has, in fact, that
\begin{equation}
P^{(2)}_{NN}(\tau) = 1 - e^{-\alpha_0 N +o(N)} \label{Prob_knot_link}
\end{equation}
\noindent when $N\to\infty$. This results is straightforwardly generalized to
$k$-components links as follows. If $k$ unknotted polygons, each of $N$ bonds, form a
$k$-component link of type $\tau$, then

\begin{equation}
\lim_{N\to\infty}\frac{1}{kN} \log p_{N,N,\cdots}^{(k)}(\tau|0,0,\cdots,0) = \kappa_0
\label{lim_link_unknot_k}
\end{equation}
\noindent irrespectively of the $k$-component link type,
$\tau$. Extension of these results to the links of higher dimensional
structures ($p$-spheres) can be found in~\citep{Soteros:1999:JKTR}.

Arguments similar to those in refs.~\citep{Orlandini:1994:J-Phys-A}
and~\citep{Soteros:1999:JKTR} can be used to establish that the
exponential growth rate of the number of topologically linked polygon
pairs is equal to that of topologically unlinked polygon
pairs. Therefore, unlike the knotting probability, it cannot be
concluded that all but exponentially few sufficiently-long pairs of
polygons are linked, even if the distance constraint is taken into
account. This leaves open the possibility that the linking probability
could still approach unity as $N\to\infty$ but in this case it would
not do so exponentially. It is important to stress once more that all
the results presented above have been obtained by assuming that both
rings have the same number of bonds, $N$.

\subsubsection{Numerical results on linking probability}
\label{link_prob_num}
The linking probability of two rings clearly depends on their spatial
separation. When the distance of the centers of mass of the rings is
much larger than the sum of their gyration radii, the linking
probability of the rings will be zero. As the distance is
progressively reduced it is intuitively expected that the linking
probability increases. It is therefore interesting to examine, by
computational means, whether the expected monotonic increase is indeed
observed within an equilibrated ensemble and whether the limiting
value of the linking probability is 1.

The first numerical investigation of these two aspects was carried out
by Vologodskii et al.~\citep{Vologodskii:1975:JETP} who considered
random (non self-avoiding) polygons embedded on a body-centred cubic
lattice. A Monte Carlo scheme, based on a chain-growth algorithm, was
first used to sample unrestricted conformations of a single polygon of
$N$ bonds.  Next, several pairs of uncorrelated polygons were picked
from the Monte Carlo generated ensemble. Each pair was embedded in the
lattice so that the centers of mass of the polygons were at a
pre-assigned distance, $d$, and their linked state was established by
using the two-variable Alexander polynomial $\Delta(-1,-1)$. By
repeating the process for each pair of rings 
the linking probability $P_{NN}^{(2)}(d)$ it was finally estimated. For ring lengths up to
$N=80$ the numerical data were well-consistent with
an exponential dependence of the linking probability on the
separation, $d$:

\begin{equation}
P_{NN}^{(2)}(d) \sim A e^{-\alpha_L d^3}\ .
\end{equation}

The effect of the excluded volume interaction on the pairwise linking
probability of self-and mutual-avoiding rings (on the simple cubic
lattice) was later studied in
~\citep{Orlandini:1994:J-Phys-A}. Because of the self-avoiding nature
of the rings, the efficiency of the simple sampling technique used in
~\citep{Vologodskii:1975:JETP} is poor.  An importance sampling method
was used instead where polygon pairs, deformed by two-point pivot moves,
were accepted according to a probability density that decreases
rapidly with the separation of the rings centers of mass. Suitable
reweighting techniques (see Section~\ref{Monte_Carlo}) were next used
to remove the effect of the external bias and compute the equilibrium
ensemble averages.

The linked state of each pair was probed by computing both the linking
number (homological linking) and the Alexander polynomial
$\Delta(-1,-1)$ (topological linking). For pairs of polygons, each
with up to $N=1800$ edges, the homological and
topological linking probability as a function of $d$ were calculated. In analogy with
the knotting of ring polymers in confining geometries, the degree of
the mutual entanglement turned out to be governed by the competition
between two length scales, namely the gyration radius, proportional to
$bN^{\nu}$, and the ring separation, $d$. The linking probabilities
calculated for various values of $d$ and $N$ are compatible with the
following simple scaling relationship:
\begin{equation}
P_{NN}^{(2)} \sim f\left ( \frac{d}{bN^{\nu}} \right ) \label{scaling_link_prob}
\end{equation}
\noindent which, in fact, yields a good collapse of the numerical data points
for $P_{NN}^{(2)}(d)$ obtained for various values of $N$ and $d$.

The scaling relation of eq.~(\ref{scaling_link_prob}) is consistent
with the vanishingly small probability of having two polygons
linked when their separation largely exceeds their typical size,
i.e. when $d/bN^{\nu} >> 1$, and also captures the fact that upon
bringing the rings progressively closer, the linking probability
increases and ultimately reaches 1. Using numerical means 
the conditional probability that two
topologically-linked rings (i.e. $\Delta(-1,-1)\ne 0$) are also
homologically linked (i.e. $|Lk| \ne 0$) was also analyzed. At large separations
($d/bN^{\nu} >> 1$) this conditional probability was found to be zero.
On the other hand, as $d/bN^{\nu}$ decreased, the fraction of
topologically linked polygons pairs that are also homologically linked
increasingly approached unity for $d/bN^{\nu}<<1$. This behaviour, which
is not evident \emph{a priori}, suggests that the linking number is a
very good indicator of topological linking for pairs of polygons that
strongly interpenetrate. By converse, the linking number is a poor
proxy for topological linking when the ring separation is large.

The impact of mutual excluded-volume interactions on the linking
probability between $2$ self-avoiding rings has been recently
re-examined in~\citep{Hirayama:2009:J-Phys-A} in off-lattice contexts
by using the rod-bead model (see Section~\ref{cylinder_models}). In
this study the linking probability was monitored as the radius of the
bead $\Delta_2$ was varied from zero (equilateral random
polygons with no excluded volume) to nearly 1.  In a fashion similar to the study of ref.
~\citep{Vologodskii:1975:JETP} single-ring configurations were first
generated using simple sampling techniques. The rings were not subject
to any spatial confinement but the sampling was limited to unknotted
rings. Next, pairs of single-ring configurations were randomly picked
from the sample and put at a given distance $d$. The
topologically-linked state of each pair was established using
$\Delta(-1,-1)$ and finally the linking probability,
$P_{NN}^{(0,0)}(d)$, was computed. The superscript $0,0$ is used to
remind of the unknotted topology of each component.

Throughout the explored range for $r_e$ it was confirmed that
$P_{NN}^{(0,0)}(d)$ is vanishingly small for
$d/bN^{\nu}>>1$. Remarkably, at variance with the lattice model of
ref. ~\citep{Orlandini:1994:J-Phys-A}, it was found that for $r_e \ge
0.2$, i.e. where excluded volume effects become relevant, the linking
probability is no longer a monotonic function of $d$, but has a
maximum at $d/bN^{\nu}$. Furthermore, Hirayama {\em
  et al.} ~\citep{Hirayama:2009:J-Phys-A} observe the breakdown of
the scaling assumption (\ref{scaling_link_prob}) for $r_e < 0.2$
i.e. in the regime where excluded volume interactions are negligible.

The comparison of these findings with those of
ref.~\citep{Orlandini:1994:J-Phys-A} suggests that salient aspects of
the linking probability are strongly affected by the introduction of
topological constraints, such as specific knotted states, on the
individual components. Future investigations of the problem ought to
shed light on this problem, possibly in connection with the expectedly
important role played by the size of the knotted region 
(see Section~\ref{size_knots}). 

\subsection{Linking probability in confined geometries}

\subsubsection{Statistics of two-components links confined in a cube}

As anticipated, it is particularly interesting to characterize the
linking probability of fluctuating rings that are spatially
constrained to lie in a region of width comparable to the average size
of unconstrained chains. 

To the best of our knowledge, the first investigation of the linking
probability in confined geometry was discussed
in~\citep{Orlandini:1994:J-Phys-A}. In this study, Monte Carlo
simulations were used to compute the linking probability of two
$N$-edged polygons embedded on a cubic lattice of lattice spacing $b$
and confined inside a cube with linear size equal to $D$.

Pairs of confined polygons were generated using the scheme similar to
ref.~\citep{Vologodskii:1975:JETP} that is by first generating
individual polygons in free space (using two-point pivot moves), then by
using rejection techniques to pick pairs of polygons which have total
calliper size (i.e. the calliper size of the pair) not larger than $D$
in all three Cartesian directions.

All configurations of the two polygons that could fit in the $D$-cube
were enumerated. As the embedding space is finite, two configurations
of the polygon pairs that differ for rigid translations along the
lattice axes are considered inequivalent.

The linked state of each polygon pair was probed using both the
homological and topological indicators, that is the linking number and
the Alexander polynomial $\Delta(-1,-1)$.  In analogy with the case of
knotting of individual polygons subject to geometrical confinement,
the occurrence of the two rings linking was found to depend on
the interplay of the average ring size, $b\, N^{\nu}$, and the average
confining length scale, $D$.

In particular, the linking probability was found to be a function of
adimensional ratio $D/(b N^{\nu})$. If $D >>b\, N^{\nu}$ (mild
confinement) the linking probability is essentially zero whereas as
$D/(b\, N^{\nu})$ decreases the linking probability increases
monotonically.

Regarding the viability of using linking number as a proxy for
topological linking, it was found that when the spatial confinement is
mild then almost all topologically-linked pairs (i.e. those with
$\Delta(-1,-1) \ne 0$) have linking number equal to zero. On the other
hand as $D/(bN^{\nu})$ decreases the two rings start to strongly
interpenetrate each other and, in analogy with the case of
unconstrained rings brought in close proximity, the resulting
topological linking is associated to a non-zero linking number.

This finding is not only interesting {\em per se}, but can be
exploited to speed up the computational detection of topological links
in systems with a high density of rings/polygons as it suggests that
the computational-effective linking can be reliably used.

By considering polygons without excluded volume interactions, Arsuaga
et al.~\citep{Arsuaga:2007:J-Phys-A} characterized the linking
probability for densities (polygon bonds per unit volume) much higher
than those examined in ref.~\citep{Orlandini:1994:J-Phys-A} for
self-avoiding polygons. More precisely, in
ref.~\citep{Arsuaga:2007:J-Phys-A} the uniform random polygon (URP)
model was used~\citep{Millett:2000:Hellas}.  In this model the polygon
configurations are obtained by joining $N$ points picked with uniform
probability in the unit cube (plus the closure condition).  The
polygons are clearly non-equilateral and the bonds are considered as
being infinitely thin. In the limit of
high densities, the probability of two URPs to be homologically linked
was found to approach 1 linearly, with slope proportional to $1/N$.

Because of its simplicity of formulation, the URP model lends to
analytical characterizations and, in~\citep{Arsuaga:2007:J-Phys-A} it
was rigorously shown that linking probability between a fixed closed
curve in the unit box and a uniform $N$-bonds random polygon
approaches $1$ at a rate at least of order $1/\sqrt{N}$.  We stress
that this result refers to a single uniform random polygon in presence
of a \emph{fixed} curve and, up to now, this has not been extended to
the more interesting case of two uniform random polygons mentioned
before.  Finally, we mention that in the same study, it was rigorously
shown that, for two $N$-bonds URP in the unit box, the second moment
of the linking number distribution is proportional to $N^2$.

\subsubsection{Statistics of two-components links in $D$-slabs and $(D_1,D_2)$-prisms}

Very few results are available for the linking probability of two
rings confined in $D$-slabs and $(D_1,D_2)$-prisms. In
~\citep{Tesi:1998:IMA} this problem was tackled numerically for
polygons embedded in the cubic lattice with lattice spacing $b$. To
overcome the sampling difficulties arising in confined geometries, a hybrid stochastic scheme was used. The method combined the strategy of
pairing unrestricted single-ring conformations and the method of
generating compact structures of individual rings described in section
\ref{reweight}. Specifically, polygon pairs are first generated using
importance sampling techniques according to the biasing weight
$w=\exp(-d^2/d_0^2)$, where $d$ is the distance of the rings centers
of mass and $d_0$ is a preassigned mean target distance. Next, from
the set of sampled polygon pairs, only the ones that fit into a
$D$-slab or $(D_1,D_2)$-prism were retained (rejection technique).

The global qualitative features observed in~\citep{Tesi:1998:IMA} for
the linking probability bear strong analogies to those observed for
the knotting probability of a single ring subject to spatial
confinement. In particular, the linking probability of polygon pairs
confined in a $D$-slab or $(D_1,D_2)$-prism does not have a monotonic
dependence on $D$. In facts, there is a maximum value for a specific
value of $D$ which depends on the polygon length, $N$, but is
independent of the scaling variable $D/(bN^{\nu})$. This indicates
that the knotting and linking probabilities (respectively for a single
ring and a pair of them) are influenced in analogous ways by the
introduction of geometrical constraints such as slabs or prisms.

One of the latest investigations of this subject were carried out by
Soteros {\em et al.} ~\citep{Soteros:2009:JKTR} who studied, by
rigorous arguments, the linking probability of pairs of
mutually-avoiding polygons confined within a tube and having the
same calliper size along the tube direction (tube spanning
constraint).

The additional spanning constraint causes each bond of one polygon to
be within a preassigned distance (related to the tube diameter) of
bonds belonging the other polygon. This is a more stringent distance
constraint than the one described before, which pertained to the
centres of mass of the two rings. In this confining geometry rigorous
results can be obtained for the linking probability of polygon pairs
in a $(D_1,D_2)$-prism. In particular it is obtained that the {\em
  homological} linking probability goes to one at a rate at least of
order $(1/\sqrt{N})$ (analogously to what was discussed before for the URP
model) while the {\em topological} linking probability goes to one
exponentially rapidly~\citep{Soteros:2009:JKTR}. Furthermore the
linking number grows (with probability one) faster than any function
$f \propto o(\sqrt{N})$ but it cannot grow faster than linearly in $N$
because of the prism constraint. It is interesting to notice that no
linear upper bound exists without the prism constraint since the
number of crossings can grow as fast as $O(N^{4/3})$ for a knot
~\citep{Diao:1998:Topology} or link~\citep{Diao:2002:JKTR} with
length $N$

\subsection{Mutual entanglements for polymers in concentrated solution}

All the results presented in the previous sections pertain to two
isolated rings brought in spatial proximity. These systems, while
interesting {\em per se} can be viewed as a first approximation to
more realistic systems where several rings can mutually entangle. In
practical contexts, these topological entanglements can occur when,
for example, linear polymers in a concentrated solution undergo a
closure reaction thus turning the geometrical entanglement of the open
chains into linked states of high topologically complexity.
The problem of mutual entanglement is manifestly important in the
equilibrium and dynamical properties of concentrated solutions of
polymers~\citep{DeGennes:1979,Doi&Edwards:1986,Everaers:2004:Science,Kapnistos:2008:Nat-Mat}.

However, a topological characterization of this entanglement in
equilibrium conditions is still largely unexplored and most of the
available results have been established either
numerically~\citep{Dickman&Hall:1988:J-Chem-Phys,Smith:1998}) or, in
the case of melts, in the context of the tube model
~\citep{DeGennes:1979,Doi&Edwards:1986}. This model postulates that
the mutual avoidance of the chains in a melt generates spatial
constraints which effectively force each chain to reptate through a
meandering tubelike
region~\citep{DeGennes:1979,Everaers:2004:Science}. One of the main
reasons for the widespread use of these concepts is that the
geometrical constraint given by the tubular neighborhood is much
easier to handle, and more amenable to calculations and simulations, than the topological constraints\cite{Wang2001}.  A
notable example is provided by the primitive-path method introduced a few years ago to 
analyze numerical simulations of dense polymer melts~\citep{Everaers:2005}. The  method allows to characterize the viscoelastic behaviour of the melts in terms of properties of the network of the coarse-grained paths traced by polymers chains.

\begin{figure}[tbp]
\begin{center}
\includegraphics[width=\WIDTHA]{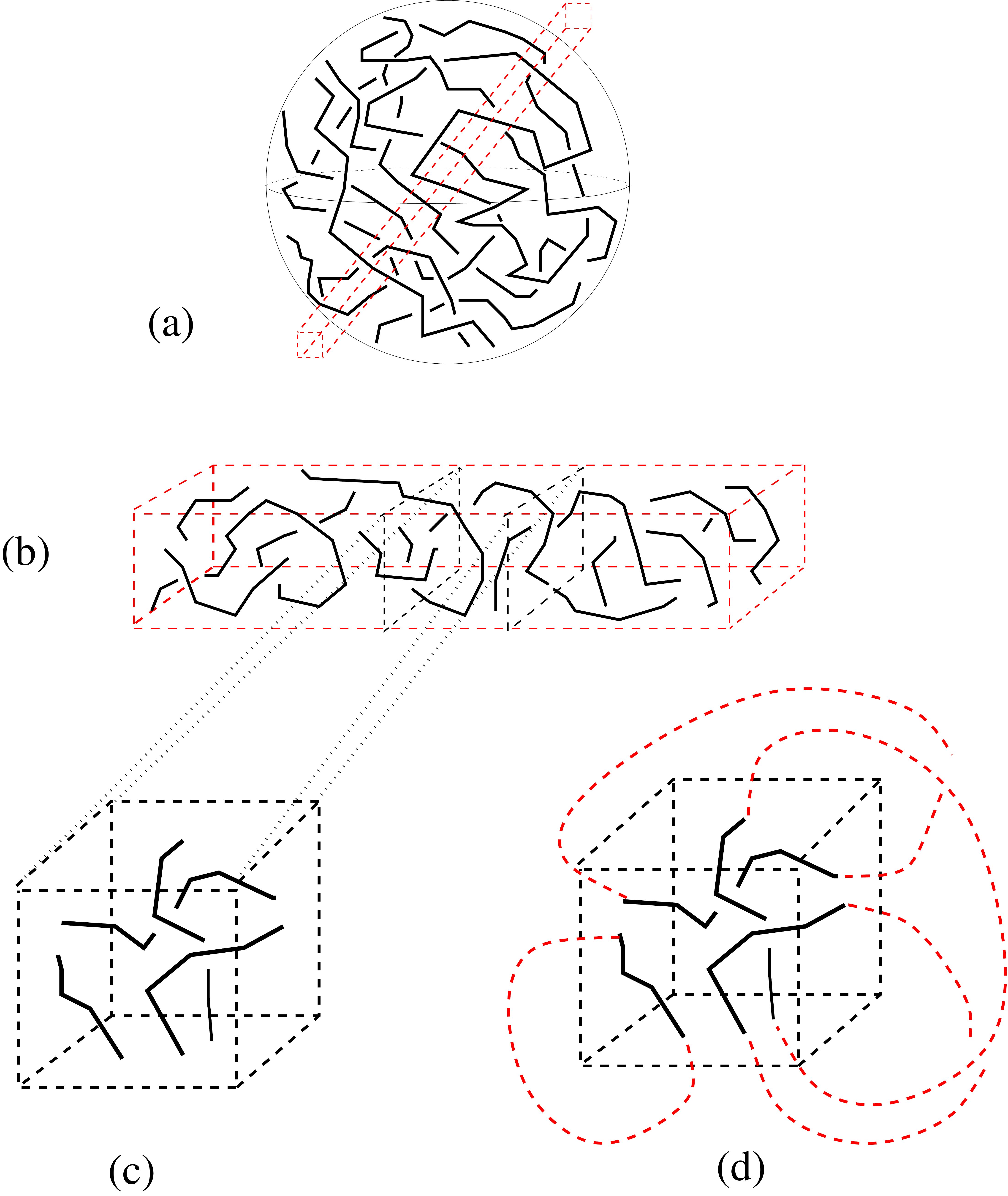}
\caption{(a) A carrot is extracted from a concentrated solution of polymers. (b) The carrot is
sectioned in smaller boxes (c). The entanglement of the sub-chains within an elementary box is
determined by computing the linking number between the rings obtained by closing outside the cube
the sub-chains.} \label{fig:drilling}
\end{center}
\end{figure}

An attempt to probe directly the topological complexity of
concentrated polymer solutions and melts was carried out in
refs.~\citep{Orlandini:2000:J-Phys-A,Orlandini&Whittington:2004:J-Chem-Phys}
The method consisted of considering a specific configuration of the
concentrated solution and to randomly ``drilling out'' from it a prism
(tube). Next, the chain portions trapped in the prism are
analysed. Most of these trapped chains have their ends at the prism
boundaries. These sub-chains can be both self-entangled and mutually
entangled and this entanglement can be properly defined in terms of
knotting and linking by bridging the two ends of each chain outside
the prism (see Figure~\ref{fig:drilling} panels (a) and
(b)). Sometimes it is necessary to introduce extraneous crossings
outside the prism but this number can be easily controlled (see
below).

By repeatedly applying this procedure it is possible, for example, to
compute the probability that a certain type of chain linking is found
in randomly-drilled prism etc. It is interesting to notice that,
within this approach, the complicated problem of characterizing the
equilibrium topological properties of concentrated solutions of
polymer chains can be, to some extent, reduced (or at least compared
to ) to the simpler problem of the linking probability of
$k$-components rings in $(D_1,D_2)$-prism.

The above approach was further simplified in
ref.~\citep{Orlandini:2000:J-Phys-A,Orlandini&Whittington:2004:J-Chem-Phys}
where the extracted prisms were subdivided into smaller pieces, such
as cubes. In this case, each cube and the $k$ sub-chains within it
form an entangled structure  of $k$ strings where knotted and linked state can be
rigorously defined (see Figure~\ref{fig:drilling}). The topological
properties of the entire prism can be inferred by properly identifying
the boundary faces of the elementary cubes.  A possible way of
describing this system, that is also amenable of numerical and
possibly analytical calculations, is in terms of a set of $k$ self-
and mutually-avoiding walks on a cubic lattice suitably embedded in a
cube of linear size $D$. This model has been explored in details by
Monte Carlo simulations and scaling arguments
in~\citep{Orlandini:2000:J-Phys-A,Orlandini&Whittington:2004:J-Chem-Phys}. In
these studies it was calculated the linking number between two chains
embedded in a cube and with their ends joined by arcs external to the
cube. As mentioned above it is possible that an additional,
extraneous, crossing is introduced during the closure outside the
box. If this is the case the sign of this extra crossing is chosen at
random with probability $1/2$.

This measure was later extended to $k$ chains by computing the
${k\choose 2}$ linking numbers between pairs of chains. The absolute
values of these ${k\choose 2}$ linking numbers were finally summed
algebraically so to obtained a single-valued measure, $L$, of the
degree of entanglement of a given arrangement of $k$ chains inside the
$D$-cube. Note that for $k=2$ the measure $L$ reduces to the linking
number between two rings.

The quantity $L$ was then averaged over the sampled configurations to
form the statistical average $\langle L \rangle$. It turns out that
for each $D$ and $k$ there is an initial set of values of the monomer
density $\rho$ at which $\langle L \rangle / {k\choose 2}$ is
approximately zero, and then increases roughly linearly with
$\rho$. As $D$ increases, the linear part of the curve becomes steeper
and sets in at smaller values of $\rho$. One expects that, in the
regime where $\langle L \rangle / {k\choose 2}$ is close to zero, the
$k$ chains will occupy essentially disjoint regions of space, while
when $\langle L \rangle / {k\choose 2}$ is linear in $\rho$ the
regions occupied by the chains will have substantial overlap. This
corresponds roughly to the distinction between the dilute and
semi-dilute regimes in polymer
solutions~\citep{DeGennes:1979,Doi&Edwards:1986}. In this situation,
two length scales are competing: one is the box dimension $D$ while
the second is the average extension $R$ of the single polymer chain in
free space, $R \sim bN^{\nu}$, where $N$ is the number of polymer
monomers. Linking is expected to occur rarely when $R << D$, and to be
frequent when $R \sim D$. Since $N = \rho D^3/k$, at fixed $D$ the
density $\rho^*$ at which linking is expectedly important is given by
\begin{equation}
D \sim bN^{\nu} \sim b\left ( \rho^* D^3 /k \right )^{\nu}
\end{equation}
so that
\begin{equation}
\rho^* \sim b^{-1}D^{(1-3\nu)/\nu} k.
\end{equation}

Note that $\rho^*$ is the critical value of the density above which
the system is in the semidiluted
regime~\citep{DeGennes:1979,Doi&Edwards:1986}. It is natural to
formulate the hypothesis that the mutual entanglement $\langle L \rangle /
{k\choose 2}$ is a function of the ratio of the two competing length
scales, $D$ and $R$:
\begin{equation}
\langle L \rangle / {k\choose 2} = f \left ( \frac{D}{bN^{\nu}}\right ).
\end{equation}
Using the fact that $N=\rho D^3/k$ one has $\frac{D}{bN^{\nu}} \sim
\left [ \rho b^{1/\nu}D^{(3\nu-1)/\nu}/k\right]^{-\nu}$ the dependence of $L$
on $D$ can be recasted in terms of a new function, $g$:
\begin{equation}
\langle L \rangle / {k\choose 2} = g \left (  \rho b^{-1}D^{(3\nu-1)/\nu}/k \right ) =g \left (  \rho
/\rho^* \right )
\end{equation}
Using Flory's value $\nu=3/5$ this implies
\begin{equation}
\langle L \rangle / {k\choose 2} = g \left (  \rho b^{-1}D^{4/3}/k \right )
\end{equation}
The validity of the above scaling relationship is confirmed by the
good collapse of the numerical data points obtained by plotting
$\langle L \rangle / {k\choose 2}$ as a function of $\rho D^{4/3}
/k$. The collapse is particularly good for densities $\rho$ up to
about $4\rho^*$. At higher densities, the collapse deteriorates,
probably due to the onset of the melt regime ($\rho >
\rho^{**}$)~\citep{Doi&Edwards:1986} where the efficiency of the Monte
Carlo algorithm starts to deteriorate. More sophisticated sampling
techniques are then necessary to look at the scaling regime of
$\langle L \rangle / {k\choose 2}$ in the, still unexplored, region
$\rho >> \rho^*$.

A further possible extension of the above-mentioned problem is the
study of the probability of occurrence of self-entanglement
(i.e. knotting) and how this depends on melt concentration. In
principle this could be done by joining the two ends of the same walk
by an arc outside the cube, and calculating (for instance) the
one-variable Alexander polynomial of the resulting simple closed
curve.

\section{Fourth part: properties of polymers with fixed topology}
\label{sec:4thpart}

The fourth part of this review, which comprises
sections~\ref{topological_constrains} and \ref{size_knots} departs
from the topics covered so far and which largely dealt with the
characterization of the topological state of polymer chains that
circularise in equilibrium. It is equally interesting, however, to
characterize the physical properties of polymer chains whose
topological state is fixed and cannot be altered during the
equilibrium fluctuation dynamics of the chain (e.g. because the
circularization occurs via a covalent bonding which is very resilient
at room temperature). In this case, it is clearly not meaningful to
investigate the knot spectrum. Instead it is interesting to examine
how the topological constraint affects the accessible configurational
space of the polymer and, in turn, how this affects the polymer
dynamics and equilibrium metric properties. These issues are covered in
this last part of the review. In particular, the following questions
will be addressed:
\begin{itemize}
\item To what extent does the presence of topological constraints influence the equilibrium properties of a single ring or a melt of rings? 
\item Are the limiting configuration entropy and the average extension of rings with fixed topology different  from the unconstrained counterparts? 
\item Is there any dependence on the type of topological constraint
  imposed on the system? In particular, do unknotted rings behave
  differently from rings with a nontrivial knot type?
\item More generally, what are the salient configurational features of a spatially confined molecule with a given topology?
\end{itemize}

A good prototypical example of how the salient physical properties of a
polymer ring can be influenced by fixing its topological state is
provided by the gel electrophoresis~\citep{Katritch_et_al:1996:Nature}
or stretching experiments~\citep{Saitta_et_al:1999:Nature} where a
detectably different response is associated to various knot types.

The presence of a topological constraint can have an even deeper
impact if the chain is geometrically confined into thin slabs, narrow
channels or small containers. This is because, in addition to the
relevant lengthscales of the problem, i.e. the contour length,
persistence length and width of the confining region, fixing the
topological state introduces a further lengthscale, namely the size of
the knotted region. For a knotted ring or a linear polymer the
competition between topology and confinement can dramatically affect
the translocation
ability~\citep{Alberts:2002,Kasianoviz:1996:PNAS}. This has profound
implications for the passage of a macromolecule through a narrow hole
or channel, a phenomenon that is ubiquitous in biological systems.

Furthermore, for a {\em melt} of rings in confined geometries the
topological constraints can strongly enhance the effective (entropic)
repulsion between rings which can dramatically affect their spatial
arrangement. Indeed, this mechanism is believed to be an important
ingredient in the spatial organization and segregation of highly confined DNA
molecules such as bacterial
chromosomes~\citep{Jun_and_Mulder:2006:PNAS} and eukaryotic
chromosomes during
interphase~\citep{Branco_and_Pombo:2006:PLOS,0953-8984-22-28-283102}
(see the discussion in Section~\ref{topologically-constrained-and-concentrated}).

The available results on the subject are accompanied by a general
overview of the main theoretical and computational challenges that
arise in systems of fixed topology. In brief, the challenges on the
analytical side arise from the fact that the (global) topological
constraints prevents exploitation of the mapping between the
statistics of rings with unrestricted topology and critical magnetic
systems. Field theory arguments cannot be applied any
more\citep{DeGennes:1979}. 

By necessity, most of the theoretical approaches to the problem
therefore start from the assumption that it is possible to transfer
scaling arguments developed in the unconstrained case to the
fixed-topology
one~\citep{Cates:1986:J-Phys,Grosberg_et_al:1996:PRE,Grosberg:2000:PRL}
and use numerical computation to support or verify {\em a posteriori}
the validity of this hypothesis
~\citep{Orlandini:1996:J-Phys-A,Orlandini:1998:J-Phys-A,Moore:2004:PNAS}. In
turn, from the numerical perspective the difficulty stems from the
fact that it is difficult to ensure {\em a priori}, that all the
relevant configurational phase space is explored by stochastic
modification of polymer rings which respects the given topological
constraint.

A conceptually-important feature of the fixed-topology problem answers
the question of how large a fraction of the ring is occupied, on
average, by a given knot type~\citep{Orlandini:2009:Phys-Bio}.

More precisely the following questions can be addressed:
\begin{itemize}
\item For knotted rings at equilibrium is the knot segregated into a
  small region in which all the topological details are confined or is
  it loosely spread over the entire chain ? \item Does the average
  length of the knotted region depends on physical parameters such as
  the polymerization degree, the stiffness of the chain, the quality
  of the solvent etc etc ? \item For knotted rings confined within
  thin slabs, narrow channels or closed small containers, is the
  typical size of the knot altered by the degree of confinement ?
\end{itemize}

All previous issues implicitly build on the notion that it is possible
to measure the size of the knotted region within a ring.  Actually,
this is a delicate conceptual point since the most natural way to
perform this measure consists of extracting arcs from the full ring
and looking for the presence of the knot within each of the excised
arcs. This scheme makes manifest the ambiguity of the definition of
the knot length because, as already mentioned, the notion of
knottedness of an open arc cannot be generally established except for
very special circumstances, such as when the termini are subjected to
suitable constraints, on in case of strong
adsorption~\citep{Ercolini:2007:PRL,Guitter&Orlandini:1999:J-Phys-A}. In
other circumstances, there is no unique way to establish the knotted
state of an open arc. As will be discussed in the review, this
ambiguity can actually be exploited to define a probabilistic notion
of knottedness of a given arc with a given (fixed) configuration, and
valuable insight into the fixed-topology problem can be thus
obtained. This is the subject of Section~\ref{size_knots}.

\section{Equilibrium properties of topologically constrained polymers}
\label{topological_constrains}

In this section we shall report on the results available for topologically
constrained rings. We shall first consider the case of isolated rings
whose knotted state is fixed, and next move on to the case of several
rings with a fixed type of mutual linking.  These two situations are
expected to arise from the polymer circularization in, respectively,
dilute and dense polymer solutions.

\subsection{Ring polymers in diluted solution: fixed knot type}
\label{fixed_knot}

We first consider the case the of isolated self-avoiding rings of fixed knot type.

\subsubsection{Limiting entropy and entropic exponents of rings at fixed knot type}

In principle, the equilibrium properties of a ring with a given knot
type, $\tau$, can be obtained by considering, in place of the full
ensemble of configurations (microstates), only those with the correct
knot type $\tau$. The introduction of this restriction usually makes
it impractical to characterize analytically the equilibrium
properties. The difficulty arises because the ``knottedness'' is a
global property of the chain and, for example, cannot be accounted for
by adding to the energy function terms that favour/disfavour specific
local geometries of the ring. For this reason there are very few
theoretical results based on analytical and/or rigorous approaches.

From the FWD theorem (see Section~\ref{knot_prob}) it is known that
the limiting entropy of unknotted self-avoiding polygons, $\kappa_0$, exists and
satisfies the inequality
\begin{equation}
\kappa_0  < \kappa, \label{ineq_1}
\end{equation}
\noindent where $\kappa$ is the limiting entropy of the whole set of
rings (i.e. summed over all the topologies). The question that rises
naturally from inequality (\ref{ineq_1}) is the following: is it
possible to prove rigorously the existence of the limiting entropy
$\kappa_\tau$ for rings with a given knot type $\tau$ ? Moreover: is
the strict inequality (\ref{ineq_1}) still valid if one replaces
$\kappa_0$ with $\kappa_\tau$ ? While a complete answer to the problem
is still lacking, the following rigorous results have been
established. Let $p_N(\tau)$ be the number of $N$-bonds polygons whose
knot type is $\tau$. Since all, but exponentially few polygons contain
at least one copy of, say, a knot $\tau^\prime$ different from $\tau$,
we have:
\begin{equation}
\limsup_{N\to\infty}\frac{1}{N} \log p_N(\tau) < \kappa \label{kappa_tau}\ .
\end{equation}
In other words, self-avoiding polygons with a particular knot type $\tau$ are
exponentially rare~\citep{Whittington:1992}. This number can be
compared to the number of unknots by noticing that a subset of
$N$-bond polygons with knot type $\tau$ can be obtained by
concatenating $M$-edges polygons with knot type $\tau$ and $N-M$ edge
unknotted polygons. This yields the following inequality
~\citep{Whittington:1992},
\begin{equation}
p_N(\tau) \ge \frac{1}{2}p_M(\tau)p_{N-M}(\emptyset)\ .
\end{equation}

By recalling the FWD result about the existence of the limiting
entropy of unknotted self-avoiding polygons, $\kappa_0$, one has that, for fixed $M$
and $N\to \infty$ the previous inequality can be recasted as 
\begin{equation}
\liminf_{N\to\infty} \frac{1}{N} \log p_{N}(\tau) \ge \kappa_0. \label{ineq_2}\  .
\end{equation}
The last inequality shows that, to exponential order, there are at
least as many polygons with knot type $\tau$ as unknots.  Notice that $\liminf$ and $\limsup$ are used
because the limit $\lim_{N\to\infty} N^{-1} \log p_N(\tau)$ is not
guaranteed to exist.

It would be interesting to establish whether eq.~(\ref{ineq_2}) holds
as an equality. At present, the only available rigorous result, based
on the existence of the limiting entropy $\kappa$ for the space of
\emph{all} unrooted polygons, is that
\begin{equation}
p_N \sim A e^{\kappa N +o(N)}.
\end{equation}
\noindent where $p_N$ is the number of all such polygons.

On the other hand, field-theory arguments, based on the mapping between
the statistics of self-avoiding rings and the $O(n)$ model for $n\to
0$~\citep{DeGennes:1979} give a more detailed asymptotic behaviour for
$p_N$,
\begin{equation}
p_N = A N^{\alpha -3} e^{\kappa N} \left ( 1 + \frac{B}{N^{\Delta}} + C N^{-1} +o(N^{-1}) \right )
\label{scaling_entropy_all}
\end{equation}
where $\alpha$ and $\Delta$ are respectively the \emph{entropic} and
the \emph{correction to scaling} exponents. They are known to be
universal quantities, in that they depend only on fundamental
properties of the system (such as the space dimensionality, $d$) and
not on model details. For unknotted  self-avoiding polygons we know rigorously the
existence of $\kappa_0$ but there is no field theory result available
in this case since, for the statistics of self-avoiding 
unknotted rings, no mapping to
the a $O(n)$-like model has been devised.

Nevertheless, it appears reasonable to conjecture that: 

\begin{equation}
p_N(\emptyset) = A(\emptyset) N^{\alpha(\emptyset) -3} e^{\kappa_0 N} \left ( 1 +
\frac{B(\emptyset)}{N^{\Delta(\emptyset)}} + C(\emptyset) N^{-1} +o(N^{-1}) \right ).
\label{scaling_entropy_un}
\end{equation}

For self-avoiding polygons with a given knot type $\tau$ no rigorous proof of the
existence of the limiting entropy $\kappa_\tau$ exists, though it
appears plausible that it obeys a scaling form analogous to the above
one,
\begin{equation}
p_N(\tau) = A(\tau) N^{\alpha(\tau) -3} e^{\kappa_\tau N} \left ( 1 +
\frac{B(\tau)}{N^{\Delta(\tau)}} + C(\tau) N^{-1} +o(N^{-1})\right ). \label{scaling_entropy_tau}
\end{equation}
Assuming the validity of eq.~(\ref{scaling_entropy_tau}) then
eq.~(\ref{kappa_tau}) implies that $\kappa_\tau < \kappa$ for every
knot type $\tau$.

The validity of the above assumptions can be tested using numerical
simulations. Two  different schemes (each with different
advantages and drawbacks) can be used to sample polygons (or in
general rings) with fixed knot type.

One possibility is to sample extensively the space of all polygons
(for example by using two-point pivot moves that change the knot
type). {\em A posteriori}, considerations can then be restricted to
only those conformations that have the desired knot type (identified,
for example, through the computation of knot polynomials). The
advantage of this approach is that algorithms based on  pivot moves are
ergodic in the whole class of polygons and are also quite
efficient. Aside from considerations about the use of imperfect
topological indicators, a major drawback of this scheme is that,
according to (\ref{kappa_tau}), the number of polygons with a fixed
knot type $\tau$ is, in the limit $N\to \infty$, exponentially rare
with respect to whole class so an exponentially-diverging (in polygon
length) simulation time is need to collect a reliable statistics for a
given $\tau$. On the other hand, for fixed (but large enough) $N$, a
reasonable amount of knots can be sampled obtaining a good statistics
for at least the simplest knot types.  Notice that for polygons on the
cubic lattice the population of trefoil knots is non-negligible for $N
\sim 10^5$ whereas for off-lattice equilateral rings, the maximum
incidence of trefoils (corresponding to a probability of about 25 \%)
is observed for $N \approx 300$. 

The second approach consists of sampling configurations always
remaining in the region of phase space associated with the given knot
type, $\tau$. The spirit of these (generalizable
~\citep{Farago:2002:EPL}) approaches is aptly illustrated in lattice
contexts. Polygons at fixed knot type on the cubic lattice can be
sampled using the BFACF algorithm
~\citep{Berg&Foester:1981:Phys-Lett-B,Aragao:1983:Nucl-Phys-B,Aragao&Caracciolo:1983:J-Physique}
described in Section~\ref{knot_simp}. 

The BFACF algorithm is, indeed, a viable tool for sampling polygons
with a fixed knot type though it can lead to long correlation
times. The severity of the latter problem can be largely reduced using
the Multiple Markov chain method (see Section~\ref{Monte_Carlo}) where
various copies of the system are run at different values of the bond
fugacity~\citep{Orlandini:1998:IMA}. In
~\citep{Orlandini:1996:J-Phys-A,Orlandini:1998:J-Phys-A}, by using the
BFACF algorithm, the scaling hypothesis (\ref{scaling_entropy_tau})
has been tested for different knot types. Within numerical
uncertainty, it was found that
\begin{eqnarray}
\kappa_\tau=\kappa_0 \quad {\rm and}\\
\alpha(\emptyset) = \alpha. \label{entr_exp_0}
\end{eqnarray}
Moreover
\begin{equation}
\alpha(\tau) = \alpha(\emptyset) + N_f \label{entr_exp}
\end{equation}
where $N_f$ is the number of prime factors in the knot decomposition
of $\tau$. Equation (\ref{entr_exp}) implies that $\alpha(\tau)$ is
independent of $\tau$ if $\tau$ is a prime knot.  Similar results have
been later found for off-lattice models of knotted rings as Gaussian
random polygons and rod-bead
rings~\citep{Deguchi&Tsurusaki:1997:PRE}. In this study the sampling
of rings with fixed knot type was based on the first approach
described above, i.e. a stochastic generation of rings with free
topology and subsequent selection of configurations of a given knot
type detected by topological invariants such as the Alexander and the
Vassiliev polynomials.

\begin{figure}[tbp]
\begin{center}
\includegraphics[width=\WIDTHA]{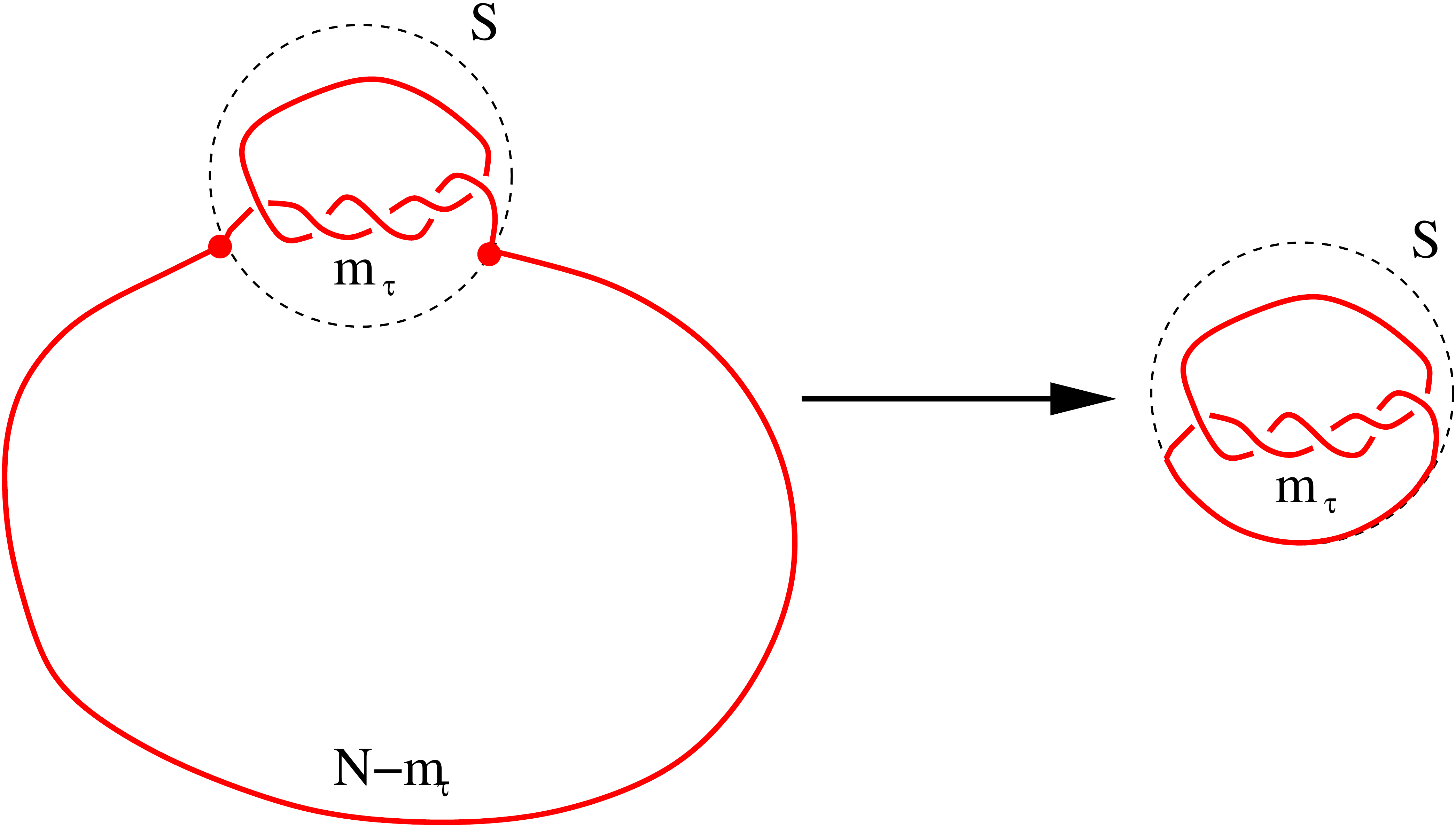}
\caption{The sphere $S$ separates the knotted region into two arcs, one of which
will be knotted with knot type $\tau$ ($5_1$ in the figure) if it is closed by an arc on the sphere.}
\label{fig:5_1_ball}
\end{center}
\end{figure}

The results of eq.~(\ref{entr_exp}) can be understood if one assumes
that knots are tied relatively tightly in the polygon. To turn this
observation into a more quantitative statement let us consider a
simple closed curve, $\omega$, in three-dimensional space, having the
topology of prime knot, $\tau$. Consider a geometric sphere, $S$,
intersecting $\omega$ in exactly two points, dividing $\omega$ into
two 1-balls, $\omega_1$ and $\omega_2$.  Suppose that $\omega_1$ meets
$S$ at its two boundary points but is otherwise inside $S$ and that
$\omega_2$ meets $S$ at its boundary points but is otherwise outside
$S$, see Figure~\ref{fig:5_1_ball}. Next, convert $\omega_1$ into a
simple closed curve $\omega_1'$ by adding a curve on the sphere $S$ so
to join the end points of $\omega_1$.  The same construction converts
$\omega_2$ to a simple closed curve $\omega_2'$. 

\noindent If $\omega_1'$ has knot type $\tau$ and $\omega_2'$ is
unknotted then we can say that the knot is \emph{localized} inside
$S$.  Let the total length of the knotted segment $\omega_1$ be
$m_\tau$, and assume that $\omega$ has a total contour length equal to
$N$, as sketched in Figure~\ref{fig:5_1_ball}. Define $M_\tau$ to be
the infimum of $m_\tau$ over all possible intersections with geometric
spheres which cut $\omega$ in exactly two points.  Define $N_\tau$ to
be the expected value of $M_\tau$ taken uniformly over all polygons of
knot type $\tau$. If $N_\tau = o(N)$ , i.e. $\lim_{N \to \infty}
N_\tau/N$, then in the large $N$ limit, the ``average" knotted polygon
looks like an unknotted polygon with a small sphere containing a
knotted arc (of known knot type!) attached to it. In this case the
knot $\tau$ is said to be \emph{tight} (or \emph{localized} or
\emph{segregated}). An extreme case would be when $N_\tau$ is a
constant independent on $N$. If the knot is tight we
  expect that the number of inequivalent places where it can be
  accommodated along the ring grows proportionally to the ring length
  itself. In other words there exists a positive number $\epsilon <
  1$, such that there are of order $\epsilon N$ positions in the ring
  where the knot can be created by concatenating a small polygon of
  knot type $\tau$ and $N_\tau$ edges and an unknotted polygon with
  $N$ edges. This gives
\begin{equation}
p_{N_\tau+N}(\tau) \sim \epsilon N  p_N(\emptyset) 
\end{equation}
and from the scaling form~(\ref{scaling_entropy_tau}) we have
\begin{equation}
\alpha(\tau) = \alpha(\emptyset)+1,
\end{equation}
if $\tau$ is a prime knot. Note that we have
implicitly assumed $\kappa_\tau=\kappa_0$ but, rigorously, we only
know that $\kappa_{\tau}\ge \kappa_0$ \footnote{A. Y. Grosberg has
  given a qualitative argument for which the equality $\kappa_0=
  \kappa_\tau$ holds: see A.Y. Grosberg, in {\it Ideal Knots}, 
Series of Knots and Everything, vol. 19, World Scientific 1998}. 
This argument can be similarly generalized to cases where $\tau$ is a knot
with $N_f$ prime factors. It is indeed sufficient to choose $N_f$ out of
$\epsilon N$ locations in about $\left ( \epsilon N \right )^{N_f}$
ways and the rest of the argument is the same giving the result in
(\ref{entr_exp}).

Because in the large $N$ limit the statistics of topologically
unconstrained polygons is dominated by composite knots then, if
eq.~(\ref{entr_exp}) holds one would be expect to see a significant
difference between the values $\alpha$ and $\alpha_(\emptyset)$.  This
is, however, not the case; indeed there is strong numerical evidence
supporting the validity of relation (\ref{entr_exp_0}). 

A full comprehension of this result is presently
lacking, although some notable progress in this direction was made in
the context of subclasses of lattice animals partitioned with respect
to their cyclomatic index (i.e. the number of independent cycles),
which, although having different entropic exponents altogether make up
the class of animals with entropic exponents equal to those of
lattice trees \citep{Soteros&Whittington:1988:JPA}.

\subsubsection{Metric properties of rings at fixed knot type.}
\label{metr_fixed_knot}

The above argument on the localization, or segregation, of the knotted
portion of the chain not only provides a basis to rationalise eq.~(\ref{entr_exp}) but has also important implications on the metric
properties of self-avoiding polygons of fixed knot type, in particular regarding the
mean square radius of gyration, $\langle R_N^2\rangle$ (see
eq.~\ref{radius_gyration}). For the whole
(i.e. topologically-unconstrained) class of self-avoiding rings, field theory
arguments suggest the asymptotic form 
\begin{equation}
\langle R_N^2\rangle = A_{\nu} N^{2\nu} \left ( 1 +B_{\nu} N^{-\Delta} +C_{\nu} N^{-1}
+o(N^{-1})\right), \label{r2scal}
\end{equation}
where the subscript $\nu$ is used to distinguish the $A$, $B$ and $C$ coefficients of eq.~(\ref{r2scal}) from those in eq.~(\ref{scaling_entropy_all}). Available theoretical
estimates~\citep{Guida&Zinn-Justin:1997:JPA} for the exponents $\nu$ and $\Delta$ yield:
\begin{eqnarray}
\nu &=& 0.5882 \pm 0.0010 \nonumber\\
\Delta &=& 0.478 \pm 0.010 \label{Zinn}
\end{eqnarray}
whereas the best available numerical estimates for the self-avoiding walk are
as follows
\begin{eqnarray}
\nu &=& 0.587597 \pm 0.000007 \qquad \mbox{\citep{Clisby:2010:Phys-Rev-Lett}} \nonumber\\
\Delta &=& 0.56 \pm 0.03 \qquad \mbox{     \citep{Li:1995:JSP}}. \label{Li}
\end{eqnarray}

As for the case of $p_N(\tau)$, no field theory argument is available
for the scaling behaviour of the mean squared radius of gyration
$\langle R_N^2(\tau) \rangle$ for self-avoiding rings at given knot type
$\tau$. However, analogously to eq.~(\ref{scaling_entropy_tau}) it is
plausible to expect that
\begin{equation}
\langle R_N ^2 (\tau)\rangle = A_{\nu}(\tau)N^{2\nu(\tau)}\left [ 1 + B_{\nu}(\tau)
N^{-\Delta(\tau)} +C_{\nu}(\tau)N^{-1} + o(N^{-1})\right] \label{r2scalknot} \ .
\end{equation}
The validity of the above scaling hypothesis can be ultimately checked
using numerical simulations. Specific aspects of the above
relationship can, however, be checked for consistency against the
localization assumption discussed before.

In particular, if knots were subject to the discussed
  localization, then, in the large $N$ limit, a polygon with knot type
  $\tau$ consisting of $N_f$ prime knot components should be
  statistically similar to an unknotted ring with $N_f$ roots
  (i.e. distinguishable vertices).  On the other hand, since rooted
  polygons have a different entropy with respect to un-rooted ones and
  yet have the same average extension, we should expect
  $\nu(\tau)=\nu(\emptyset)$ and $A_{\nu}(\tau)=A_{\nu}(\emptyset)$
  for all $\tau$.  Note that neither the metric exponent $\nu$
  (universal quantity) nor the amplitude $A_{\nu}(\tau)$ should depend
  on the knot type. It is however possible that the topological
  constraint affects the higher order correction to the scaling
  relation of eq.~(\ref{r2scalknot}) but, in order to get a more
  definite answer on this question a direct measure of the size of the
  knot within the ring is needed. The numerical investigations
  performed up to now by different groups and on different models of
  self-avoiding rings have shown a good agreement on the independence
  of $\nu$ on the knot type
  $\tau$~\citep{Janse-van-Rensburg&Whittington:1991a:J-Phys-A,Quake:1994:Phys-Rev-Lett,Orlandini:1998:J-Phys-A,Matsuda:2003:PRE,Brown:2001:Phys-Rev-E,Mansfield&Douglas:2010:J-Chem-Phys}.
For the amplitude the numerical evidence
$A_{\nu}(\tau)=A_{\nu}(\emptyset)$ is less
sharp~\citep{Quake:1994:Phys-Rev-Lett} but results obtained with more
asymptotic values of $N$ seems to confirm that the amplitude $A_{\nu}$
is constant over $\tau$.

Compared to available theoretical results, there are very few
experiments that, up to now, have explored the influence of the
topological constraints on the equilibrium properties of isolated
rings.  The main reason resides in the extreme difficulty in
synthetizing highly purified unknotted and unconcatenated
rings~\citep{McKenna:1987:Macromol}. In the few cases when these
conditions could be met it was indeed found by viscosity
measurements~\citep{Roovers:1985:Macromol}, that the average size of
isolated unknotted rings is the same as for (isolated) linear
chains. While the topological constraint of being unknotted does not
seem to change the average extension of these rings there are
indications that it entails a decrease of the $\theta$ temperature,
which is the temperature at which the second virial coefficient
becomes zero signalling the occurrence of a coil-collapse
transition~\citep{DeGennes:1979,Vanderzande:1998}.  The decrease of
$\theta$ was found by measuring the rheological properties of
unknotted ring polystyrenes in
cyclohexane~\citep{Roovers&Toporowski:1983:Macromol,Roovers:1985:J-Pol-Sci}
and was later confirmed by light scattering measurements of the second
virial coefficient of highly-purified ring polystyrenes in
cyclohexane~\citep{Takano:2009:Polymer}.

The fact that the unknottedness constraint affects the $\theta$
temperature, where the equilibrium statistics of the rings is
compatible with that of ideal chains, suggests that the
effective topologically-dependent interactions are very weak
compared to excluded volume interactions and that they may manifest
themselves when the latter can be neglected.

At the $\theta$ point, in fact, the metric exponent $\nu$ for all
rings is $1/2$ since the ensemble essentially corresponds to random
walks conditioned to return to the origin.

For infinitely-thin rings with trivial topology it was
conjectured that the average size scales as $N^{\nu(\emptyset)}$ with
$\nu(\emptyset)=\nu_{SAW} \approx 0.588 \simeq 3/5$. This conjecture
builds on the seminal work of des Cloizeaux who studied the conformational properties of linked rings and concluded that topological constraints on a long flexible ring have a similar effects as excluded volume interactions~\citep{desCloizeux:1981:J-Phys-Lett}. 
On the basis of this observation, for rings with no excluded volume interactions, the
exclusion of all non-trivial knots from the statistics is expected to be equivalent to the introduction of an effective
\emph{topological repulsion} that swells the
infinitely-thin unknotted rings up to an average
size that scales as the one of self-avoiding rings (although corrections to the asymptotic scaling point at a different universality class for self-avoiding rings and infinitely-thin unknotted ones\cite{Moore:2005:Phys-Rev-E-Stat-Nonlin-Soft-Matter-Phys:16485967}).

This conjecture
was later supported by scaling arguments \citep{Grosberg:2000:PRL}
leading to a more specific prediction for
infinitely-thin unknotted rings: 
\begin{equation}
\langle R_N ^2 (\emptyset)\rangle = \left \{  \begin{array}{cccc} &(b^2/12)N  &\mbox{if}  &N << N_0
\\ &A(b^2/12) N^{2\nu_{SAW}} & \mbox{if}  &N >>  N_0\end{array}\right.
\label{scaling_grosb}
\end{equation}
where $N_0=1/\alpha$ is the characteristic length of the unknot,
i.e. it appears in the exponential decay of the unknotting
probability, $\exp(-N/N_0+o(N))$. Expression (\ref{scaling_grosb})
suggests a crossover from a small $N$ regime, where the knotted
infinitely-thin ring is characterised by the random
walk exponent $\nu=1/2$, to a large $N$ regime in which the average
size scales as the one of self-avoiding walks. Evidence in favour of
the scaling relation of eq.~(\ref{scaling_grosb}) was subsequently
provided by extensive
simulations~\citep{Shimamura&Deguchi:2002:PRE,Dobay:2003:PNAS,Moore:2004:PNAS}.

Notably, these studies suggest that the above mentioned-relationship
also holds for any set of infinitely-thin rings with fixed knot type. At present, no exact result is available to confirm the asymptotic
validity of the numerical findings. A number of arguments have however
been suggested, ranging from the relative strength of topological and
excluded volume
interactions~\citep{PhysRevE.59.R2539,Shimamura&Deguchi:2001:PRE,Shimamura&Deguchi:2002:PRE}
to considerations made on the fraction (and size) of rings that,
compared to the average case, are under- or
over-knotted~\citep{Moore:2004:PNAS}. Interestingly, the latter
argument predicts that the $\nu$ exponent for sufficiently long rings
of a given knot type would be directly inherited from the asymptotic
one of the unknotted rings, $\nu(\emptyset)$.

\subsection{Knotted rings in proximity of an impenetrable surface}

As previously discussed, the configurational entropy of a
spatially-confined chain is clearly smaller than when it is
unconstrained in the bulk. A ubiquitous type of polymer confinement is
obtained in proximity of a surface. As the chain approaches the
surface, the chain configurational entropy decreases producing a soft
entropic force pushing the polymer away from the
surface~\citep{Marenduzzo&Orlandini:2006:EPL}.

The case when the ring approaching the surface is knotted, and cannot
change the knot type, is a notable example of the interplay that can
arise between geometrical and topological constraints. In particular,
one may ask what free energy loss, or equivalently entropic repulsive
force, is experience by surface-adsorbed polymers with different
topologies. Is this force large enough to be detectable in
single-molecule experiments (which would hence provide a new means of
detecting the ring topologies)?

To quantify the impact that topology can have on the free energy cost
of placing a polymer close to a wall, consider the particularly simple
case of a freely-jointed chain, made up of $N$
statistically-independent links, each of length $b$, and rooted at a
point at a distance $x_0$ from an impenetrable wall, which for
defineteness we may take parallel to the $yz$ plane and placed at
$x=0$. It is a simple polymer physics exercise to find the partition
functions of a linear open and a ring polymer rooted, e.g. via the
so-called image method~\citep{Doi&Edwards:1986,Eisenriegler:1993}, in
the continuum limit (valid for $N\gg 1$). These are respectively given
by
\begin{eqnarray}
Z_{\rm open}(N,x_0)
& \propto & {\rm erf}\left(\frac{\sqrt{3}x_0}{b\sqrt{2N}}\right)\\
Z_{\rm loop}(N,x_0)
& \propto &
\left[1-\exp\left({-\frac{6x_0^2}{Nb^2}}\right)\right]
\end{eqnarray}
where ${\rm erf}$ denotes the error function, and we have omitted a
constant which does not depend on $x_0$.

\begin{figure}
\centerline{
\includegraphics[width=4.1in]{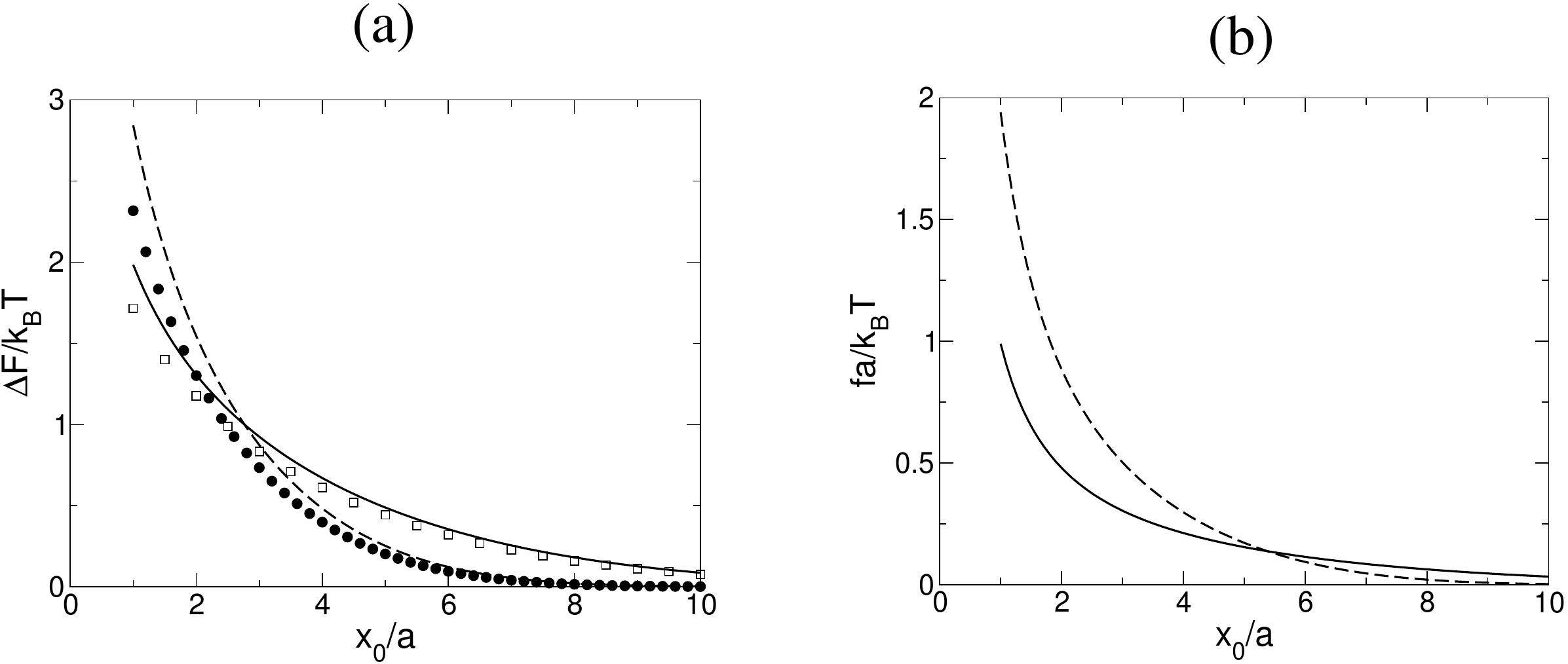}
}
\caption{Plot of (a) the free energy which is lost, and (b) the force
  required to keep the polymer in place, when a linear (solid lines)
  and a looped (long dashed lines) freely-jointed chain with $N=100$
  are placed at a distance $x_0$ (in units of the Kuhn length) from a
  solid wall, according to Eqs.~\ref{open-approach},\ref{loop-approach}.  In
  (a) we also show the numerical estimate for the free energy loss for
  a linear (open squares) and a looped (filled circles) chain with
  same contour length. The difference between these and the analytical
  calculations is due to finite size effects.}
\label{entropic_approach_curves_open_loop}
\end{figure}

From the partition function one can then find both the free energy
loss due to the decrease in configurational entropy, and the repulsive
entropic forces (this is the derivative of the free energy with
respect to distance from the wall). After some algebra it is possible
to show that the repulsive force, for open and looped configurations,
is respectively given by (note that we keep the dependence on $N$ and
$x_0$)
\begin{eqnarray}
\label{open-approach}
  f_{\rm open}(N,x_0) =  k_B T{\partial \ln Z_{\rm open}(N,x_0) \over \partial x_0} & = & \frac{2\sqrt{3}k_BT}{\sqrt{2\pi N}b}
  \frac{\exp\left(-\frac{3x_0^2}{2Nb^2}\right)}
  {{\rm erf}\left(\frac{\sqrt{3}x_0}{\sqrt{2N}b}\right)} \\ 
  f_{\rm looped}(N,x_0) = k_B T{\partial \ln Z_{\rm looped}(N,x_0) \over \partial x_0}  & = & \frac{12x_0k_BT}{Nb^2}
  \frac{\exp\left(-\frac{6x_0^2}{Nb^2}\right)}
  {1-\exp\left(-\frac{6x_0^2}{Nb^2}\right)}
\label{loop-approach}
\end{eqnarray}
These forces are plotted in
Figure~\ref{entropic_approach_curves_open_loop}. It can be seen that
the repulsive force builds up first for the open chain, whose
statistical size is larger. However, upon close approach it is the
ring which feels a stronger repulsive force. This different trend may
be understood by comparing the change in the entropic exponent,
$\gamma$, for a ring and an open chain when they are tied to a
wall~\citep{Marenduzzo&Orlandini:2006:EPL}. It is actually possible to
show that, at least in the thermodynamic limit, the more complex the
topological constraints are, the steeper the entropic repulsive force
will rise for $x_0\to 0$. Thus, for a network of freely-jointed chains
locally made of $n_L$ rings and $n_M$ linear branches tied together at
a common root held at distance $x_0$ from the surface, we find, for
$x_0\to 0$, $f(N,x_0) \sim \phi k_BT/x_0$, with $\phi=2 n_L+n_M$
through an analysis similar to the one above.

Therefore entropic forces for different network topologies are
different. What about the case of a fixed knot type approaching a
solid wall? In order to answer this question, it is necessary to
compute the free energy loss, or equivalently the partition function, of a ring
with fixed knot type as a function of $x_0$ (and $N$). This may sound
like a difficult task, but in this case it can be realised via a small
change in a standard knot generation algorithm. In particular, in
Ref.~\citep{Marenduzzo&Orlandini:2006:EPL} we have generated several
fluctuating freely-jointed rings with Monte-Carlo simulations,
and we have computed how the probability of having the topology of an
unknot, or a simple prime knot changes with the distance to the wall,
$x_0$. The free energy of a polymer chain with knot type $\tau$ as a
function of $x_0$ is then equal, up to an additive constant, to minus
the logarithm of the probability of forming knot type $\tau$ if the root
is a distance $x_0$ from the solid wall. The constant can be adjusted
so that the free energy loss is 0 as $x_0\to \infty$. It is then
straightforward to compute the entropic forces and in particular the
difference between forces felt by different knot types. In order to
reduce the statistical errors in the force curve and enhance the
signal to noise ratio it is useful to employ additional tricks, which
are however not important conceptually -- the reader interested in or
needing these may consult Ref.~\citep{Marenduzzo&Orlandini:2006:EPL}
for more technical details.

\begin{figure}[tbp]
\begin{center}
\includegraphics[width=\WIDTHA]{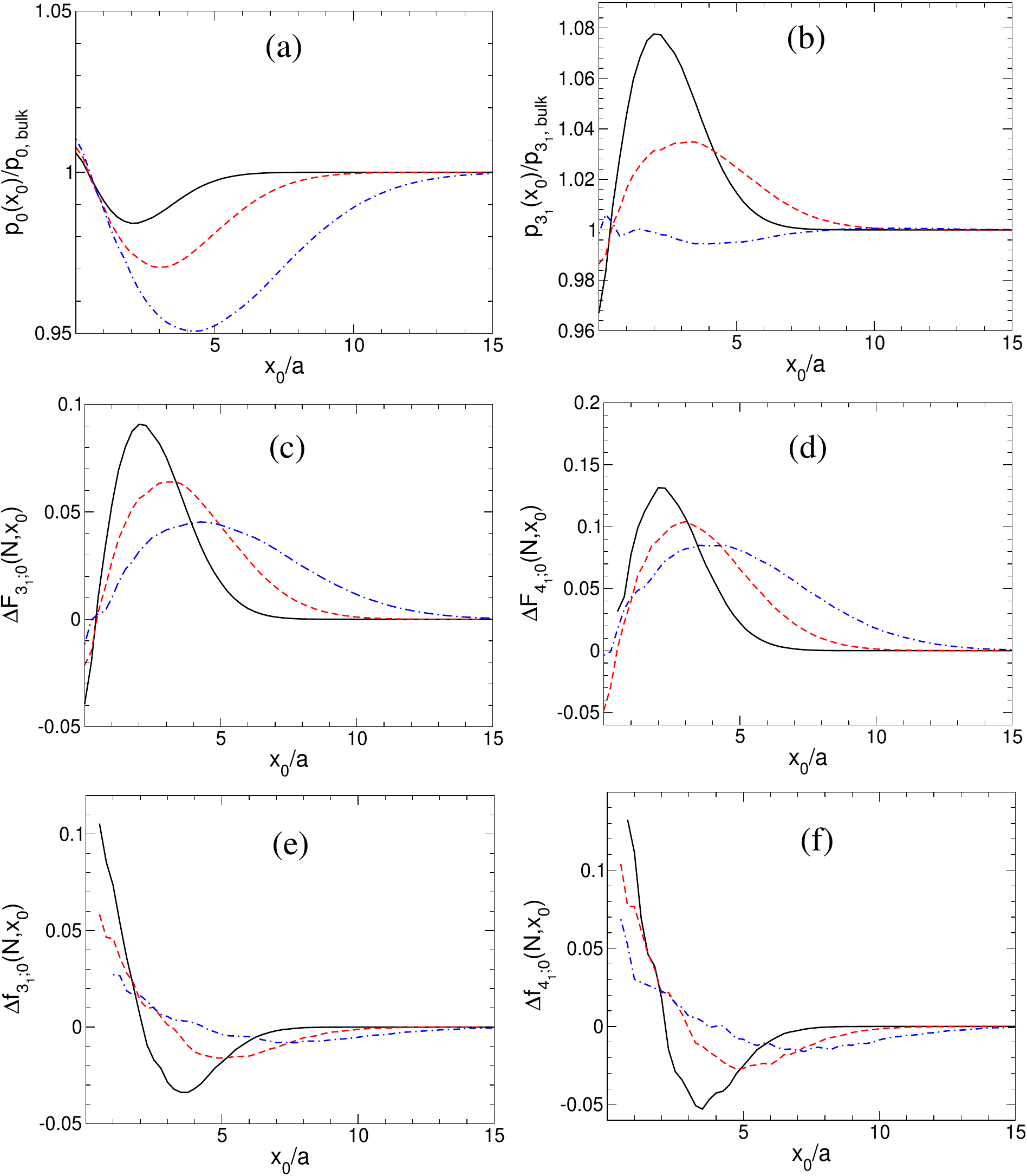}
\caption{Panels (a) and (b): plots of the knotting probabilities
  (normalised to their bulk value) for polymers of given knot type
  approaching a wall ((a) refers to the unknot and (b) to the
  trefoil).  Panels (c) and (d): plots of the differential free energy
  for a trefoil (c) and a $4_1$ knot (d) approaching a wall. The free
  energy differences are computed with respect to the free energy
  spent for the case of the unknot.  Panels (e) and (f): differences
  between the force felt by a trefoil and unknot when they are
  approaching the wall (e), and between a $4_1$ knot and an unknot.
  Solid, dashed and dot-dashed lines respectively correspond to chains
  with $N=50$, $N=100$, and $N=200$ Kuhn lengths. Figure reproduced
  from Ref.~\cite{Marenduzzo&Orlandini:2006:EPL}.}
\label{knot-approach-curves}
\end{center}
\end{figure}

Figure~\ref{knot-approach-curves} shows an example of the results which
can be obtained. This figure shows how the unknotting and trefoil
formation probability change with respect to their bulk value (panels
a and b respectively), and also shows the difference in the free
energy (panels c and d) and in the force (panels e and f) ``approach
curves'' for two polymers of different given knot types. From the
figure it is apparent that all the computed probabilities are
non-monotonic with respect to $x_0$. The unknot probability (panel a)
decreases for intermediate $x_0$ and then increases for very small
$x_0$, to slightly higher values compared to the bulk one. The
deviation from the bulk value increases with the size of the ring. For
small chains ($N=50$ and $N=100$), the qualitative behaviour of the
trefoil (panel b) and the $4_1$ knot is the same, and is opposite to
that of the unknot -- these probabilities display a {\it maximum} for
intermediate $x_0$ (or even oscillations for larger $N$).  From the
knotting probabilities one can extract the difference in free energy
losses for different knot types. It can be seen that these have a
maximum value of $0.1$ $K_BT$, and decreases with $N$, so that they
would vanish in the thermodynamic limit. However real polymers are
often less than $100$ Kuhn lengths long and for those the difference
in free energy losses might be picked up by single molecule
experiments.

There are several open problems having to do with knots in restricted
geometries which could be studied within a similar framework. For
instance, it would be interesting to analyze the free energy profile
as a function of position in suitably-patterned microchannels and
microdevices. For instance, in a sinusoidal or ratchet-shaped channel
the free energies of different knot types will be different from each
other and depend on height. Is it possible to sort knots in this way?
Eventually, it would be also interesting to study the dynamical
properties of knots in these geometries, for instance when they are
subjected to a uniform stretching flow. Clearly in these cases the
Monte-Carlo algorithm may only be used to generate initial conditions,
which then need to be evolved via a dynamic algorithm (e.g. Brownian
dynamics or Lattice Boltzmann according to whether hydrodynamic
interactions can be neglected as a first approximation or need to be
included). While it may be that these more complicated microfluidic
geometries will actually be more readily studied or more interesting
for experiments, understanding the case of an impenetrable surface, as
discussed above, is a necessary first step.

\subsection{Topological and volume exclusion interaction between pairs of rings}
\label{virial_top}

We now turn to the case of a semidilute solution of rings. It is
interesting to characterize the behaviour of two or more nearby rings with
fixed topology, that is of given knotted and linked state.

Let us first consider a system consisting of only two unlinked rings
which fluctuate in equilibrium subject to the constraint that the
distance of their centers of mass, $r$ is held fixed (see section
\ref{link_prob}). 

When the distance $r$ is larger than the average size of each ring
the two rings are statistically independent. Upon decreasing $r$,
the two rings start to feel each other and mutual self-avoidance
interactions become important. In addition the constraint of keeping
the two rings unlinked further decreases the number of configurations
that would be accessible to each of the isolated rings. This
\emph{topological entropy} (or free energy) loss $\Delta F_{topo}(r) =
F_{SAW}(r) - F_{unlink}(r)$ gives rise to an effective repulsive
potential $U_{topo}= \exp(-\Delta F_{topo}(r)/k_BT)$ and,
consequently, to a \emph{topological second virial coefficient},
$B_{topo}^(2)$, which adds to the second virial coefficient arising from
excluded volume:
\begin{equation}
B_{topo}^{(2)} = 2\pi\int_{\mathbb{R}} r^2 \left [ 1 - \exp(-\Delta F_{topo}(r)/k_BT) \right ] d r.
\end{equation}
The quantity $B_{topo}^{(2)}$ can be estimated by considering the
linking probability between two rings \cite{Iwata1985,tanaka:4201,Deguchi-tsurusaki:1997}.
Starting from the fact
that the linking probability between two rings whose centers of mass
are held at a distance $r$, $P_{link}(r)$ follows the Mayer relation,
$P_{link}(r)=1 - \exp(-\Delta F_{topo}(r)/k_BT)$.

This scheme was first explored in ref.~\citep{Vologodskii:1975:JETP}
where $P_{link}(r)$ was numerically investigated for random polygons
on a centred cubic lattice (see Section~\ref{link_prob_num}). The
numerical data suggested that $P_{link}(r)$ had the following
functional dependence on $r$,
\begin{equation}
P_{link}(r) = 1 - A_0 \exp\left (-\alpha_0 r^3 \right)
\end{equation}
from which the following topological second virial coefficient was obtained:
\begin{equation}
B_{topo}^{(2)} = \frac{2}{3} \pi \frac{A_0}{\alpha_0}.
\end{equation}
\noindent Note that for random polygons the second virial coefficient
due to excluded volume interactions is zero and only $B_{topo}^{(2)}$
is present. If, on the other hand, real polymer rings are considered
(i.e. with excluded volume) the effect of $B_{topo}^{(2)}$ would be to
shift the value of the temperature at which the total second virial
coefficient $B_{SAW}^{(2)}+B_{topo}^{(2)}$ is zero\cite{Iwata1985,tanaka:4201}.  This value would
correspond to the new $\theta$ point value for unlinked and unknotted
rings~\citep{Hirayama:2009:J-Phys-A} discussed at the end of Section~\ref{metr_fixed_knot}.

A related problem was considered in
ref.~\citep{Marenduzzo&Orlandini:2009:JSTAT}, which focused on two
approaching {\it rooted} lattice rings.  In this case, in place of the
distance between the rings centres of mass, the distance between the
two rooting points, $r$, was used.  By comparing the free energy
losses upon close approach of both linked and unlinked rings, it was
possible to separate the topological and excluded-volume contributions
to the repulsive entropic force between the rings. Interestingly, we
found that the effect due to mutual avoidance dominates the
topological contribution by about an order of
magnitude.

The system considered has a loose analogy with a
  situation that may be realised experimentally, namely by
  allowing/disallowing topoisomerase enzymes, such as topo-II, to act
  {\em in vitro} on circularized DNA molecules. In the presence of
ATP, the topo-II strand-passage action can link/unlink and knot/unknot
the DNA
rings\cite{Rybenkov:1997:Science}. \footnote{The
    topo-II action therefore produces a steady-state probabilities for
    linking and knotting circular DNA molecules that are significantly
    lower than those measured when DNA circularization occurs in
    equilibrium. } When topo-II does not act, then
  the linking state of the circular DNA molecules cannot change. Based
  on the previous theoretical consideration (but bearing in mind that
  they pertain to an equilibrium situation, which is unlike the steady
  state action of topo-II) it could be anticipated that the
  free energy/forces recorded when the two loops are placed close
  together should not differ much whether or not topo-II or other
  similar enzymes are included in the experiments. 

Interestingly, the rooted polymer simulations (unlike the ones in
which the centres of mass are controlled) can be used to test
field-theoretical predictions on the entropic exponents of polymeric
networks, which rely on renormalisation group estimates. Thus,
observing that when the two roots approach each other the two rings
locally look like a 4-leg branch point, one can use the formulas in
Ref.~\citep{Duplantier:1986:Phys-Rev-Lett,Duplantier:1989:J-Stat-Phys}
to predict the following scaling for the free energy loss $\Delta F$
in the limit $x\to 0$:
\begin{equation}\label{fieldtheory}
\Delta F \propto -\frac{2\gamma_2-\gamma_4-1}{\nu}\log{(x)}
\end{equation}
\noindent where $\gamma_2=1$ and $\gamma_4=1/2$ are the
$\epsilon$-expansion estimates of the entropic exponents for 2- and
4-leg branch points (see Ref.~\citep{Duplantier:1989:J-Stat-Phys} for
details). Remarkably, simulations confirm the field theory scaling
close to $x_0\to 0$ to a very good approximation.  This is
particularly notable, as the theory does not fix either the knot type
of each of the rings, or their linking number.
 
At the same time, the field theory predicts $\Delta F \propto
\log{N}$, so it only costs a few $k_B T$ to position the two roots
very close together, and there is only a very weak logarithmic
dependence on polymer size. While this value is experimentally
detectable, it does not ensure ring segregation in semidilute
equilibrium solution (in other words thermal fluctuations lead to ring
interpenetration, as they do for linear
polymers~\citep{Grosber_et_al:1982:Makr_Chem_Comm})).

It is also interesting to consider the approach of ring pairs when they 
are in the globular phase~\citep{Chuang_et_al_JCP_2000, Marenduzzo&Orlandini:2009:JSTAT}
and intrachain interactions dominate
over interchain ones. In that case the range of the
entropic/topological repulsion is smaller compared to the case of two
polymers in the swollen phase, but the force is ``harder'' and
increases much more steeply. This case may be more relevant for nearby
interphase chromosomes segregating into territories during interphase
(see below for more details on this application of linking and
entanglement in semidilute suspensions)
\citep{Blackstone:2010:J-Math-Biol:20379719}. Finally, the case of several rings
coming together, possibly by entropic forces
\citep{Marenduzzo:2006:Biophys-J:16500976,Toan:2006:Phys-Rev-Lett:17155512},
to form a rosette is also considered -- even in this more complicated
case field-theoretical estimates work semiquantitatively.

\subsection{Topological constraints in concentrated solutions.}
\label{topologically-constrained-and-concentrated}

We have seen that the topological entropic repulsion arising between
two unlinked rings brought in spatial proximity, does not appreciably
alter the salient equilibrium properties, such as the $\nu$ exponent,
of the two-rings system. The situation may, however, dramatically
change in concentrated solutions or melts of
rings~\citep{Everaers:2004:Science,Kapnistos:2008:Nat-Mat}.  

The effects of topological constraints where many chains are close
together is difficult to investigate analytically and, with the
exception of few findings based on perturbation analysis on an
effective field theory of unlinked
rings~\citep{Brereton&Vilgis:1995:J-Phys-A}, all the known results
have been obtained by numerical investigations and scaling
analysis~\citep{Muller:1996:Phys-Rev-E:9964837,Cates:1986:J-Phys,Brown&Szamel:1998:J-Chem-Phys,Brown:2001:Phys-Rev-E}. Within
these approaches it was found that unknotted and unlinked rings in
concentrated solutions are more compact than their linear
counterparts. In particular the mean squared radius of gyration of a
ring scales as $N^{\nu}$ with $\nu$ between $0.40$ and $0.42$
\citep{Muller:1996:Phys-Rev-E:9964837}. This value is significantly
smaller than the Gaussian value $1/2$ expected in concentrated
solutions and has a theoretical explanation based on the following
Flory-like argument~\citep{Cates:1986:J-Phys}.

\noindent In Section~\ref{virial_top} we discussed the entropy
loss of two unlinked rings brought progressively close. This entropy
costs would give rise to an effective topological repulsion that adds
to the mutual excluded volume interaction. While in dilute solutions
this repulsion is overwhelmed by self-avoidance, in concentrated
solutions it becomes more important, presumably because the crowded
environment limits severely the amplitude of the conformational
fluctuations of the rings.

For an $N$-monomer ring of size $R$, the intensity of this effect
should be a function of the average number of overlapping rings $k$
that, in arbitrary dimension $d$, should scale as $R^d \frac{\rho}{N}$.
The $k$ neighbours act as obstacles for the given ring decreasing its
conformational entropy.  This gives a contribution to the free energy
proportional to $k_BT R^d /N$. This term favours a compact spatial
organization of the ring.  On the other hand an entropic penalty has
to be paid if the ring is confined within a spherical cavity of radius
$R< bN^{1/2}$. This penalty was estimated in the context of the Flory
theory for the coil-globule transition and scales as $k_BT
\frac{Nb^2}{R^2}$.  By minimizing the sum of these two terms with
respect to $R$ one obtains $R\sim N^{\nu}$ with $\nu =
\frac{2}{d+2}$. Note that in $d=3$ this gives $\nu = 2/5$, i.e. a
value that closely agrees with the numerical estimates found in
simulations~\citep{Muller:1996:Phys-Rev-E:9964837,Brown&Szamel:1998:J-Chem-Phys,Brown:2001:Phys-Rev-E}
and with recent small-angle neutron scattering experiments carried
out on cyclic polydimethylsiloxane
(PDMS)~\citep{Gagliardi:2002:Appl-Phys-A}. Later numerical
investigations with higher values of $N$ and more detailed scaling
arguments \citep{Muller:2000:PRE,Vettorel:2009:Phys-Biol} have revealed
that actually the value $\nu = \frac{2}{d+2}$ describes the data in a
crossover region of $N$ values separating the short $N$ regime, in
which the topological interactions are weak and rings appear to be
Gaussian (as their linear counterpart in concentrated solution), and
the large $N$ regime in which the rings assume a more globular
structure with $\nu = 1/3$. Note that this large $N$ regime can be
described in terms of a ring squeezed within a network of topological
obstacles created in a self-consistent way by the surrounding
rings~\citep{Obukhov:1994:PRL,Khokhlov&Nechaev:1985:Phys-Lett}.

\subsubsection{Out-of-equilibrium and entropic effects in chromosome territories}
\label{chromterr}

A notable system where the above mentioned effects could be at play is
given by chromosome territories.

Most of a cell's life is spent to perform a variety of functions, including preparation for cell division. This phase of the cell cycle is known as {\it interphase} -- whereas division occurs during {\it mitosis}~\citep{Alberts:2002}. Interphase and mitosis are arguably the most important of the phases in the cell's cycle. Most of us associate chromosomes with elongated X-like structures: however these shapes are only observed during mitosis, when chromosomes condense and become more readily visible. 
On the other hand, during interphase, chromosomes are better described 
described as swollen polymers with thickness of approximately 30 nm
(the diameter of the chromatin fiber) and persistence length in the
40-200 nm range, confined in a quasi-spherical volume (the eukaryotic
nucleus)~\citep{Rosa:2008:PLOS,Cook:2009:J-Cell-Biol}.While theory predicts that linear polymers should strongly
intermingle in confinement if in thermodynamic equilibrium 
(see e.g. the discussions in ~\cite{Rosa:2008:PLOS,0953-8984-22-28-283102,Grosberg:2011:inpress,Mirny:2011:Chromosome-Res:21274616,Liebermann:2009})
chromosomes instead segregate into spatially distinct regions, or {\em territories}.

A possible explanation to this effect is that the physics of
interphase chromosomes is strongly dominated by out-of-equilibrium
effects.  This is at the basis of the proposal by Rosa and Everaers
in~\citep{Rosa:2008:PLOS}, who suggest that the observed segregation
is kinetically driven. The key idea behind the
  simulations in~\citep{Rosa:2008:PLOS} is that chromosomes enter
  interphase (the ``ordinary'' cellular phase and hence the one with
  the longest time duration) after the mitotic chromosome
  separation. Therefore, the starting mitotic chromosome
  configurations are compact and well segregated. Consistently with
  this fact, Rosa and Everaers chose the initial configurations of the
  model chromosome as solenoidal structures in which loops are stacked
  up in a cylindrical state. By means of large scale parallel
  simulations, the authors managed to follow the dynamics of one or a
  few polymers as long as a human chromosome starting from a mitotic
  state. Periodic boundary conditions were introduced to model what
  happens in the nucleus where chromosomes are close to each
  other. Such long polymers were found to not be able to equilibrate
  and intermingle, so that they remain segregated as in the initial
  mitotic configuration, at least when interstrand crossing was
  disallowed. Interestingly, yeast chromosomes are about two orders of
  magnitude shorter than human ones, thus equilibrate much faster --
  indeed neither molecular dynamics simulations nor experiments show
  territories in yeast chromosomes.

Chromosome territories, however, may also originate from {\em
  thermodynamic} forces. In this case, it is topological and
entropic repulsion which can lead to
segregation~\citep{Dorier:2009:Nucleic-Acid-Res,deNooijer:2009:Nucleic-Acid-Res,Cook:2009:J-Cell-Biol,Vettorel:2009:Phys-Biol,Nicodemi:2009:Biophys-J:19289043}. According
to this explanation, there are scaffolding DNA binding proteins which
act as ties~\citep{Marenduzzo:2007:Trends-in-Gen} forcing a looped
topology on chromatin in interphase chromosomes. Simple steric
repulsion between rosettes on different chromosomes made up by joining
several loops can then be estimated to lead effectively to
segregation~\citep{deNooijer:2009:Nucleic-Acid-Res,Cook:2009:J-Cell-Biol,Marenduzzo&Orlandini:2009:JSTAT}. In
particular, Ref.~\citep{Dorier:2009:Nucleic-Acid-Res} also considered
an ensemble of equilibrated unlinked freely-jointed loops and
demonstrated that these too can segregate, which shows that
segregation can be purely topologically driven. Lattice simulations of
very long ring polymers also support this
view~\citep{Muller:1996:Phys-Rev-E:9964837,Vettorel:2009:Phys-Biol}. Of
course, the kinetic and thermodynamic explanation are not necessarily
mutually exclusive and in reality they could complement each other.

In any case, our understanding of the physical and topological
properties of interphase chromosomes is still far from conclusive, and
both theoretical and experimental breakthroughs are in order before a
satisfactory picture can be reached. For instance, the role of the
looping proteins, or ``ties'' mentioned above might lead to
interchromosomal, as well as intrachromosomal, contacts. How is this
avoided in practice? Apart from the segregation into territories, what
is the local and global organization of one interphase chromosome? It
is likely that in order to answer questions like these a serious
revision of the force field used to characterise confined DNA will
be needed. Once more, large scale simulations may safely be
anticipated to be the primary method of investigation.

It is also interesting to compare the problem of chromosome
segregation in eukaryotes to that in prokaryotes. In bacteria like
{\it E. coli}, the genome is circular, and replication is concurrent
with the segregation of the sister
chromosomes~\citep{Jun:2010:Nat-Rev-Microbiol:20634810}. Apart from the topology
of the genome, looped versus linear, there is also a fundamental
generic difference between chromosome segregation in bacteria and in
eukaryotes: the geometry of confinement. While the nucleus is
essentially spherical, bacterial cells which contain the DNA are
typically elongated: for instance {\it E. coli} may be modelled as an
ellipsoid with one major axis and two nearly degenerate minor
ones. The ratio between major and minor axes varies between 2 and 4 in
the wild-type (the variability is due to the fact that the cells grow
while replicating). These two differences both help segregation and it
has recently been proposed
in~\citep{Jun_and_Mulder:2006:PNAS,Jun:2010:Nat-Rev-Microbiol:20634810}, on the
basis of scaling arguments and numerical simulations, that entropic
forces alone (i.e. self-avoidance) are enough to lead to chromosome
segregation in a slowly dividing {\it E. coli}. This is mainly because
an elongated geometry entropically favours configurations where the
two sister chromosomes do not mix. Although the scenario may thus seem
simpler in bacteria, the number of open questions is just as
impressive as for the eukaryotic case~\citep{PhysRevE.76.031901}. For
instance, in the lab {\it E. coli} cells usually divide much more
rapidly than in a minimal medium, with the ``firing'' of multiple
replication origins, which leads to the coexistence of more than two
chromosomes in a cell. What happens in this case, which is
topologically more complex than the one considered in
Ref.~\citep{Jun_and_Mulder:2006:PNAS}?

Furthermore, most
experimental evidence points to the fact that the DNA in bacteria is
strongly supercoiled, and it would be extremely interesting to
understand how this impacts the structure of the bacterial DNA
during replication. The development of ever more accurate bacterial
cell biology experiments is also likely to shape our understanding of
the biophysics of the bacterial genome. For example, in an interesting
recent study~\citep{White:2008:Nature} White {\em et al.} have shown
experimental evidence which suggests that chromosome segregation in
{\em E. coli} is not random but is driven in a manner that results in
the leading and lagging strands being addressed to particular cellular
destinations. It may therefore be that entropic forces are
complemented by active ones (e.g. due to a segregation protein
machinery) in order to ensure an error-free cell replication.

\section{The size of knots}
\label{size_knots}

In Section~\ref{fixed_knot}, in order to understand the scaling
behaviour of the entropic and metric properties of polymer rings with
fixed knot type, we used the working hypothesis that knots are tightly
tied in the chain. This assumption could be verified {\em a
  posteriori} by looking at the statistics of the effective size of
the knotted region within the chain. Questions that may be relevant in
this respect are for example the following:

\begin{itemize}
\item What is the average size of the knotted portion of a ring in
  equilibrium?

\item What is the fraction of the chain contour length that is taken up by the knot?

\item How do the above properties depend on external conditions such
  as the quality of the solvent, the degree of a spatial confinement,
  the intensity of an applied stress etc?
\end{itemize}

The above questions can be made more precise if formulated through the
notion of the \emph{degree of localization of a knot}. In the
following we shall denote with $N$ the total number of bonds in the
ring, and with $\langle N_{\tau}\rangle $ the average length (in unit of
chain bonds) of the region where the knot of type $\tau$ resides.
Three possible scenarios can occur:

\begin{enumerate}

\item when, upon increasing $N$, $\langle N_{\tau}\rangle $ increases
  more slowly than any power of $N$, then the knot is said to be
  \emph{strongly localized} within the ring. Note that this definition
  includes either the situation where $\langle N_{\tau}\rangle $ grows
  as any power of $\log N$ or the case in which is $\langle
  N_{\tau}\rangle $ is independent on $N$.

  \item When $\langle N_{\tau}\rangle $ is $o(N)$ but yet grows as
  $N^{t}$, with $0<t<1$, then the the knot is said to be \emph{weakly
    localized}. In this situation it is possible that short chains
  display knotted region that are comparable in size with the whole
  chain and only for very large (depending on $t$) values of $N$
  the knotted region becomes negligible relatively to the chain contour length.

\item Finally, when $\langle N_{\tau}\rangle $ is proportional to $N$
  then the knot is said to be \emph{delocalized}.

\end{enumerate}


\subsection{Knotted arc}
\label{knotted_arc}

The major difficulty one encounters in estimating $\langle
N_{\tau}\rangle$ relies on the notion of \emph{knotted arc} that, in
general, cannot be defined rigorously (see Section~\ref{openknots}).
Indeed, if we consider open equilateral chain the standard definition
of knot type (see Section~\ref{knot_theory}) will tell us that such
curves are unknotted. This is because an open piecewise chain with a
knot tied in it can be deformed, possibly with moves that do not
preserve edge lengths, into a straight line without passing the curve
through itself. Notice, however, that by using only edge-preserving
moves it is not presently known whether an arbitrary configuration of
an equilateral piece-wise linear chain can be deformed into a straight
segment (and such deformations have been proven to be impossible for
certain non-equilateral chain configurations)~\citep{Toussaint01,Cantarella98}.

There are cases, however, where the notion of knotted arc can be well
defined. Simple examples are open chains in which the two extremities
lay at the surface of a three-ball (i.e. homeomorphic to a the
standard three ball $\{(x,y,z), x^2+y^2+z^2\le 1 \}$) that contains
all the rest of chain.  Indeed, in this case the definition of knotted
arc can be made rigorous through the idea of knotted ball pair
introduced in Section~\ref{knot_prob_conf} and more carefully defined
in~\citep{Sumners&Whittington:1990:J-Phys-A,Orlandini&Whittington:2007:Rev-Mod_Phys}. This
can be easily seen for the class of \emph{unfolded} self-avoiding
walks since they can be uniquely completed to $(3,1)$ ball-pairs. To
see how the construction works let us consider for example the set of
$x$-\emph{unfolded} walks~\citep{JansevanRensburg:1992:J-Phys-A}. They
are characterized by having the $x$ coordinates with the property
$x_0< x_i <x_N$ for $1\le i \le N-1$.  We add half-edges in the
negative and positive $x$-direction to the first and last vertices of
the walk and construct a $3$-ball with two parallel faces
perpendicular to the $x$-axis, and four more faces (perpendicular to
the remaining coordinate directions) so that the walk (and the added
half-edges) is properly embedded in the $3$-ball. The walk and the
$3$-ball containing it form uniquely a $(3,1)$ ball-pair for which the
notion of knot is well defined. Unfortunately, although unfolded walks
are not exponentially rare in the set of all
SAWs~\citep{Hammersley&Welsh:1962:QJMO}, they are still too rare to
search for them within a long walk as possible candidates of knotted
regions.  Even if the above scheme based on unfolded walk is not going
to solve the problem of how define a knotted arc \emph{for any arc} it
does suggest some useful extensions
~\citep{JansevanRensburg:1992:J-Phys-A}.

\begin{enumerate}

\item One idea is to choose a direction at random and construct two
  parallel rays, whose origins are the two extremities of a given walk
  and which are parallel to the chosen direction. Since all such rays
  will have irrational direction cosines they will not pass through
  any of the vertices of the cubic lattice. If we regard the two rays
  to join themselves at the point at infinity the resulting curve is a
  closed curve that will be, almost surely, simple. For this closed
  curve the topological entanglement is well defined and can be
  detected for example by computing some invariants. Note that in
  general the value of the invariant will depend on the chosen
  direction so it will be convenient to average over all directions.
  The main problem of this construction is that it can create or
  destroy entanglements. For instance, a knotted arc present in the
  chain may not be necessarily detected as as being knotted upon
  closure, because one of the rays might pass through the $3$-ball
  associated with the knotted arc thus unknotting the curve (see
  Figure~\ref{fig:false_knot}). Depite these potential pittfalls,
  closing over all possible directions has been shown to yield robust
  results for both compact and unconstrained ring configurations. Its
  main limitation resides in the heavy computational cst associated
  with the repeated calculation of knot invariants. A robust and more
  computationally-effective alternative is offered by the
  minimal-entanglement closure scheme introduced recently by some of
  us\cite{Min_entang_closure}.

\begin{figure}[tbp]
\begin{center}
\includegraphics[width=\WIDTHA]{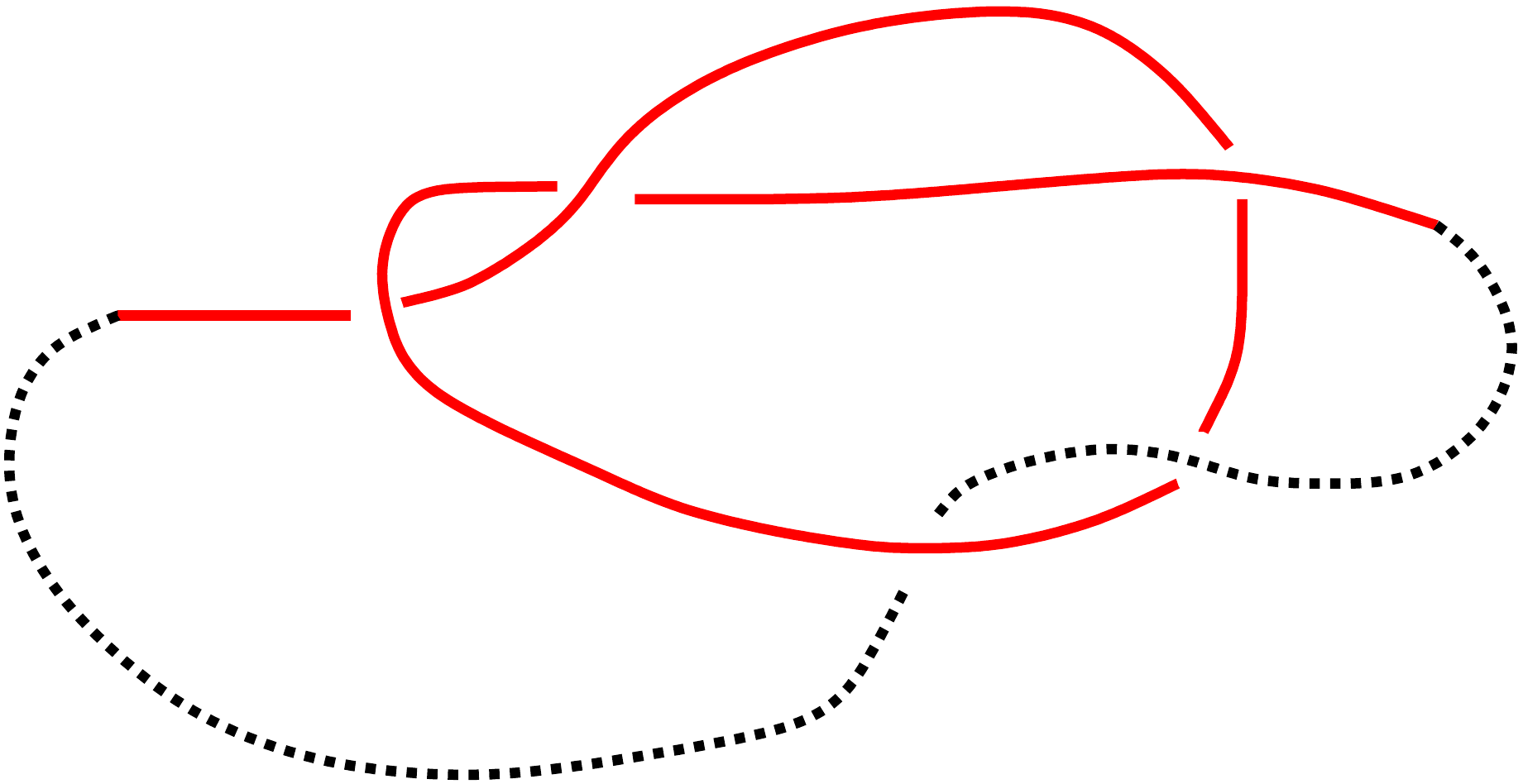}
\caption{Example of an unknotted ring that contains a knotted region (solid line)}
\label{fig:false_knot}
\end{center}
\end{figure}

\item An alternative and apparently simpler closing scheme is to join
  the ends of the arc with a line segment. The immediate problem is
  that the added line segment will in general pass through vertices of
  the lattice and the resulting closed curve might not be a simple
  curve (i.e. not self-avoiding). One way to get around this problem
  is to add to the ends of the walk two parallel line segments (of
  length say $\epsilon < b$ ) in a direction with irrational direction
  cosines, and then join up the end points of the resulting object to
  form a closed curve that is almost always a simple closed curve.  As
  for the previous scheme the knot detection may depend on the
  direction cosines chosen so that an averaging over them might be
  necessary. Clearly, also in this case the addition of the segment
  closing the curve may change the entanglement status of the starting
  arc.
\end{enumerate}

It is clear from the schemes described above that the closing
construction is not unique and so is the detection of the topological
entanglement in the arc.  This uncertainty poses severe problems for a
"deterministic" definition of knotted arc. One may rely instead on a
more probabilistic approach as suggested
in~\citep{Millett:2005:Macromol} and outlined in section
\ref{openknots}.  The idea consists in analyzing the spectrum of
knots generated by multiple closures of the same open chain. Each
closure is obtained by connecting the ends of the chain to a point
chosen (randomly) on the surface of a sphere that contains the
walk. Different closures correspond to different chosen points on the
sphere. For each fixed open chain a distribution of knots is estimated
and the \emph{most probable knot} gives the knot type of the open
chain. Interestingly enough it turns out that the knot obtained by a
simple direct end-to-end closure usually coincides with the knot type
that dominates the random closure
spectrum~\citep{Millett:2005:Macromol}.

\subsection{Flat knots}
\label{flatknots}

\begin{figure}
(a)\includegraphics[width=0.48 \textwidth]{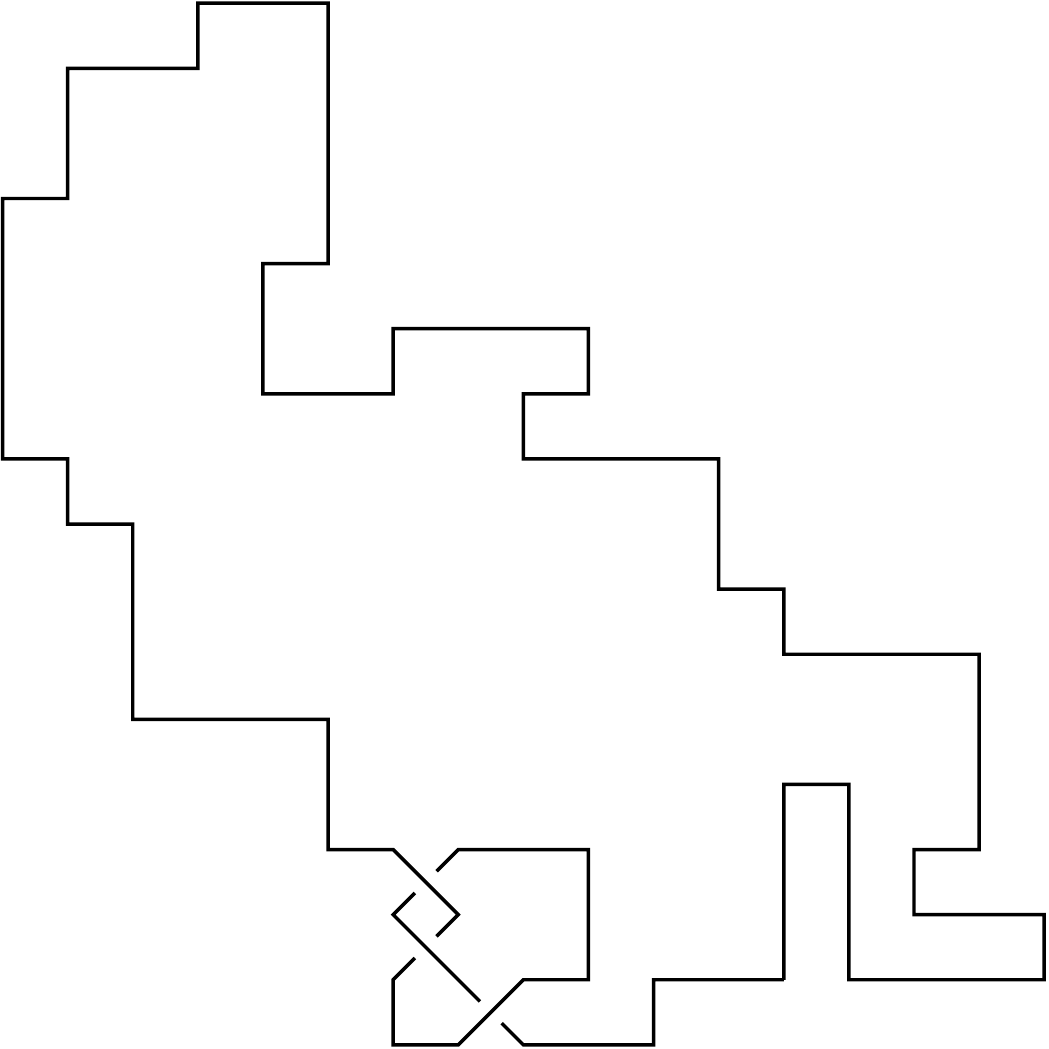}
(b)\includegraphics[width=0.48 \textwidth]{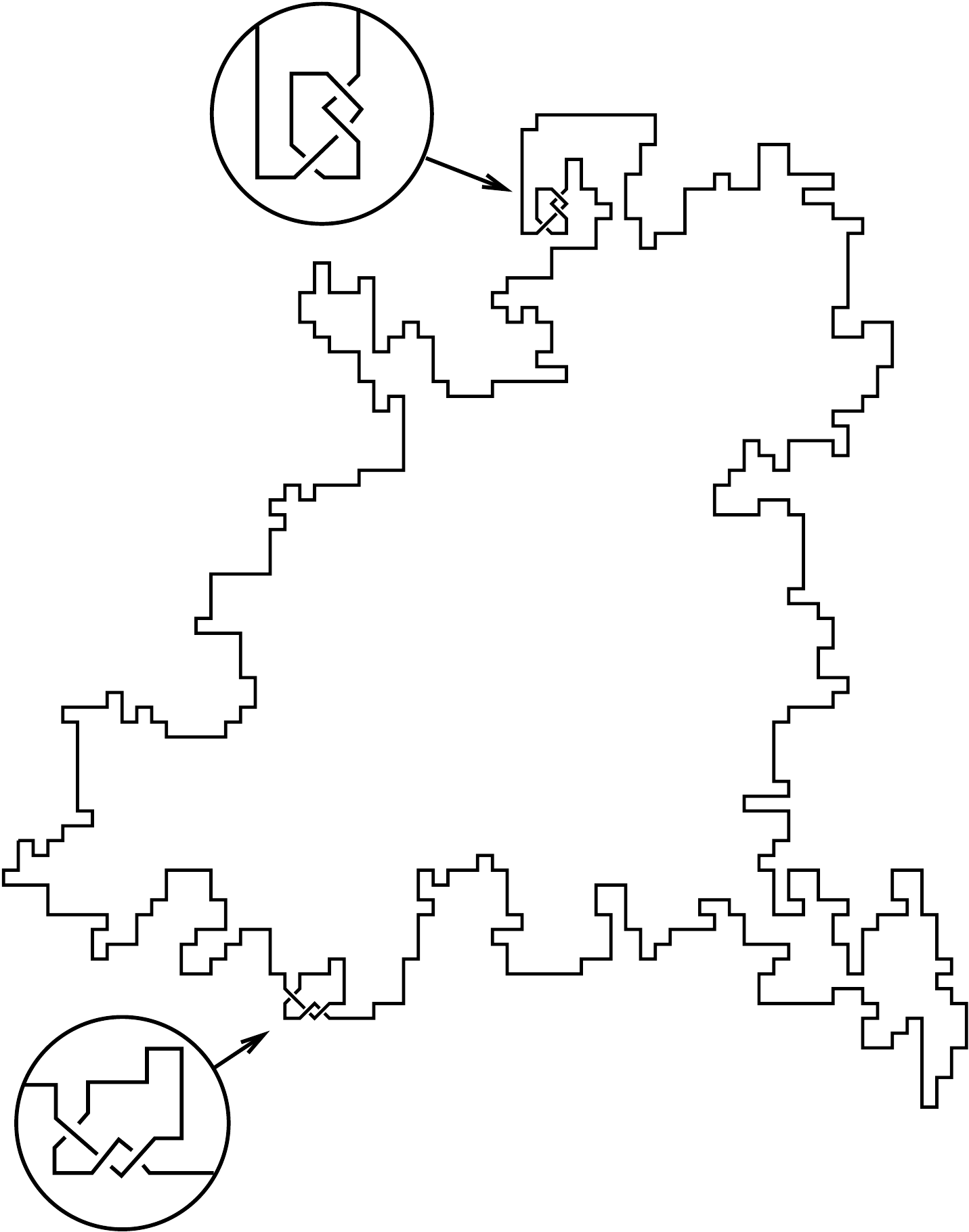}
\caption{Examples of flat knots on the square diagonal
    lattice. The configuration in panel (a) has three crossings and has the topology of a (negative)
    trefoil knot. The one in panel (b)  six crossings and has the topology of a
composite knot $3_1\#3_1$. Note the localization (and the reciprocal separation) of the two prime
knots.} \label{fig:flatknots}
\end{figure}

A definition of knotted arcs and knot size may be much easier to
formulate for \emph{flat knots}
~\citep{Guitter&Orlandini:1999:J-Phys-A}, a simplified model which has
played an important role in the development of ideas concerning
localization and delocalization properties of real knots in
equilibrium
~\citep{Guitter&Orlandini:1999:J-Phys-A,Metzler:2002:PRL,Orlandini:2003:PRE,Hanke:2003:PRE,Orlandini:2004:J-Stat-Phys}.
Flat knots are closed curves in the plane, e.g. in the square lattice,
that can be seen as quasi-2D projections of 3D knots. Physically they
can be realized by adsorbing 3D polymer rings on a strongly attractive
planar surface~\citep{Ercolini:2007:PRL,Rivetti:1996:JMB} or membrane
or by confining the polymer between two close parallel walls (see the
model introduced in~\citep{Michels&Wiegel:1989:J-Phys-A} and
mentioned in Section~\ref{knot_prob_ads}). In these cases the flat
knots can still equilibrate in 2D. Macroscopic realization of flat
knots comes also from experiments in which macroscopic knotted chains
are flattened by gravity onto a vibrating
plane~\citep{Ben-Naim:2001:PRL}. Flat knots have the advantage of
being relatively easy to image by microscopy~\citep{Ercolini:2007:PRL}
and can be studied theoretically either by numerical or analytical
approaches on properly defined two-dimensional models~\citep{Grosberg_Nechaev_JPA_1992}. 
A model of flat knots on a lattice was first proposed 
in~\citep{Guitter&Orlandini:1999:J-Phys-A} where, the authors 
considered configurations embedded in the
square lattice where the bonds lie on lattice edges except in
correspondence of chain crossings (over/under-passes) where the
diagonals of a {\em single} lattice square are used (see
Figure~\ref{fig:flatknots}a).  With this model a numerical
estimate of the entropy and the average extension of flat knots with a
given fixed, topology was carried on. The stochastic sampling was
based on BFACF plus some additional moves corresponding to lattice
versions of Reidemeister moves on knot diagrams. These moves assure a
good mobility within the space of flat knots with a given knot type
that is defined by the initial configuration corresponding to a
projection of a knotted curve in 3-space.

This investigation shows  that the connective constant of flat knots is independent of knot type,
to numerical accuracy, and that
\begin{equation}
\alpha(\tau) = \alpha(\emptyset) + N_{\tau}
\end{equation}
where $N_{\tau}$ is the number of prime components in the knot $\tau$
and $\alpha(\tau)$ is the exponent that enters in a scaling equation
similar to (\ref{scaling_entropy_tau}) where $p_N(\tau)$ is replaced
by the number of $N$-bonds flat knots with a given knot type
$\tau$. By computing the mean squared radius of gyration it was also
shown that the critical exponent $\nu$ of flat knots is, within error
bars, independent of the knot type. Although a detailed study of
tightness was not carried out, in
~\citep{Guitter&Orlandini:1999:J-Phys-A} it was observed that the
knotted regions were typically very small compared to the rest of the
ring (see Figure~\ref{fig:flatknots}b for a typical configuration of
the composite knot $3_1\# 3_1$.) Note that all the results presented
above for flat knots agree with the ones found for real knots in 3D
(see Section~\ref{fixed_knot}) indicating that, at least for some
properties, flat knots are a fair approximation of real knots. As for
real knots the investigations
in~\citep{Guitter&Orlandini:1999:J-Phys-A} refer to the asymptotic
behaviour of the entropy and of the average size of flat knots. These
furnish an indirect measure of the knot size. By exploring the
properties of flat knots in the limit in which they display a minimal
number of crossings it is possible to better characterize the size of the
knotted region. This will be the subject of the next section.

\subsubsection{Size of flat knots}
In the model presented in~\citep{Guitter&Orlandini:1999:J-Phys-A} the number of non essential
crossings were free to fluctuate in the ensemble.  If, on the other hand, the number of overlaps of
the polymer ring with itself is restricted to the minimum compatible with the topology (e.g. 3 for
a $3_1$ knot) the flat knot model enjoys a drastic simplification. With this approximation an
analytic approach to the problem of the size of a flat knot is possible~\citep{Metzler:2002:PRL}.
The idea is to interpret the essential self-overlaps of the chain as vertices of a two-dimensional
polymer network, for which a well developed theory exists (see for example,
~\citep{Duplantier:1989:J-Stat-Phys,Ohno&Binder:1988:J-Phys} and
\citep{Schaefer:1992:Nucl-Phys-B}). By exploiting results coming from this theory it was possible
to discuss the scaling of the length distribution of prime flat knots and predict a \emph{strong
localization regime}~\citep{Metzler:2002:PRL} (i.e. $\langle N_{\tau}\rangle = O(\log(N))$). This
prediction was confirmed by Monte Carlo simulations~\citep{Metzler:2002:PRL}.

The possibility of getting both analytical results and quite sharp
numerical estimates on the localization of flat knots has recently
stimulated an extension of this model that incorporates an attractive
vertex-vertex interaction as a pseudo-potential to mimic poor solvent
conditions. This gives the possibility of looking at the average size
of the knotted regions as a function of the quality of the solvent and
see, in particular if the knot localization property found in the
swollen phase may be affected by the coil-globule ($\theta$)
transition~\citep{DeGennes:1979,Vanderzande:1998}.

\begin{figure}[tbp]
\begin{center}
\includegraphics[width=\WIDTHB]{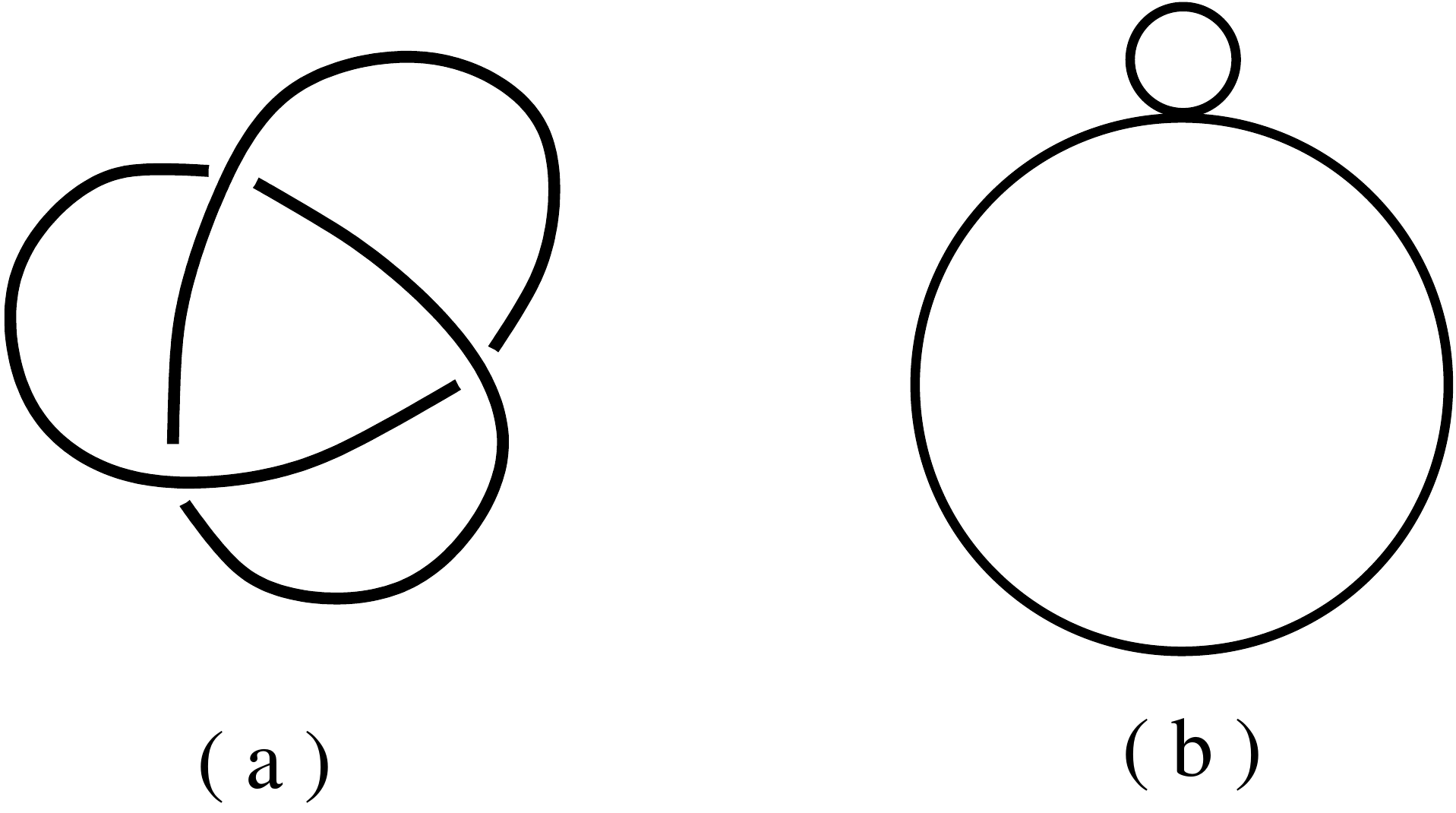}
\caption{Uncontracted (original) flat trefoil configuration (a)
and its figure eight contraction (b).} \label{fig:flat_diag}
\end{center}
\end{figure}

By simulating numerically this \emph{self-attracting flat knots} model
it was found~\citep{Orlandini:2003:PRE,Orlandini:2004:J-Stat-Phys}
that while in the good solvent regime the knots are strongly
localized, consistent with the results of
ref.~\citep{Metzler:2002:PRL}, in the globular state flat knots are
delocalized, (i.e. $\langle N_{\tau}\rangle = O(N)$). Right at the
$\theta$-temperature weak localization prevails with
$t=0.44\pm0.02$. This last result can be explained by extending the
scaling theory of polymer networks of fixed topology at the $\theta$
point~\citep{Duplantier&Saleur:1987:Phys-Rev-Lett} and assuming that
the configurations having the shape of a figure eight dominate the
statistics in that
regime~\citep{Orlandini:2003:PRE,Orlandini:2004:J-Stat-Phys} (see
Figure~\ref{fig:flat_diag}). This assumption, supported by the
numerical data, leads us to speculate that the exact value of $t$ is
$3/7$~\citep{Orlandini:2003:PRE,Orlandini:2004:J-Stat-Phys}. If, on
the other hand, the above assumption is rejected the leading scaling
order of any prime flat knot is the uncontracted configuration (see
Figure~\ref{fig:flat_diag}) that is delocalized, i.e. $t =1$, also at
the $\theta$ point~\citep{Hanke:2003:PRE}.

\subsection{The size of ``real knots''}
\label{realknots}

As anticipated in sections \ref{openknots} and \ref{knotted_arc} the
identification, in a knotted ring, of the shortest ring segment
containing the knot is a difficult and, in general, ambiguous
problem. For this reason the number of studies available on this topic
is still limited and largely restricted to numerical investigations.

One of the earliest attempts to measure the length of the knotted
region for polymer rings at equilibrium was done in
~\citep{Katritch:2000:PRE}. This investigation was carried out with a
Monte Carlo sampling of equilateral random polygons. In this work the
identification of the \emph{minimal knotted arc} was carried out
following procedures (1) and (2) in section ~\ref{knotted_arc}.

For polygons with $N=500$ bonds it was found that the typical length
of the minimal knotted arc were about 6 to 8 for trefoils, about 15
for figure eight knots and about 20 for $5_1$ and $5_2$.  Since this
investigation was done for one value of $N$ it is difficult to put
these results within the context of knot localization. Nevertheless
the general conclusion was that knots were typically tight.

In~\citep{Marcone:2005:J-Phys-A} a first study of knot localization
for lattice polygons in a good solvent regime was carried out by Monte
Carlo simulations using the BFACF algorithm. Since BFACF is ergodic
within each knot class it was possible, for a fixed knot type $\tau$,
to consider several values of $N$ and to explore properly the
statistics of the knotted regions. These regions were detected by
using a variant of the \emph{cutting and closure} schemes outlined in
Section~\ref{knotted_arc} that allows one to minimize the risk of knot
modifications or disentanglements during the closure of the
subwalk. By detecting the minimal knotted arc in polygons having knot
type $3_1$ and with $N$ up to $1500$, it was found
\begin{equation}
\langle m_{\tau}\rangle \sim N^t \label{weak}
\end{equation}
where $t \approx 0.75$ for the good solvent regime.  This result turns
out to be robust with respect to a change of prime knot type. For
example by replacing the $3_1$ with a $4_1$ or $5_1$ or $7_1$ knot,
the scaling behaviour (\ref{weak}) remains valid with the same (within
error bars) estimated value of
$t$~\citep{Marcone:2005:J-Phys-A,Marcone:2007:PRE}.

These results for the good solvent regime are
  qualitatively consistent with a molecular dynamics study by
  Mansfield~\citep{Mansfield:1998:Macromol} who, back in 1998, showed
  maximally-tight knots in a polymer becomes is loosened up by the
  dynamical evolution of the polymer (hence suggesting that
  equilibrated knots are not strongly localized).

At the same time it should be noted that a more
  recent study by Mansfield and Douglas
  \citep{Mansfield&Douglas:2010:J-Chem-Phys} reported a value of
  $t=0.54$ for trefoil knots occurring in self-avoiding rings on a
  cubic lattice. This value is appreciably different from the one
  reported before and prompts for further investigations of the source
  of this discrepancy.  In particular, it is possible that the
  difference reflects the fact that in
  ref.~\citep{Mansfield&Douglas:2010:J-Chem-Phys}, only a subset of the
  possible subportions of a ring are considered as candidate segments
  accommodating the knot. Specifically, considerations are restricted
  only to those ring segments that have both ends at the surface of
  the whole ring so to eliminate topological ambiguities upon
  closure. It is not implausible that introducing this selection
  criterion can results in different statistical properties of the
  shortest knotted segments.

\subsubsection{The size of globular knots}

It is important to stress that the above \emph{direct} approaches to
measuring the size of the minimal knotted arc strongly rely on the
assumption that the closure procedure does not systematically
introduce artifacts in the detection of the arc topology.

This condition ought to be realised in the good solvent regime where
ring configurations are not compact. On the other hand, in situations
such as the poor solvent regime where the whole polymer is quite
compact and strongly geometrically entangled, this condition is not
expected to hold.

To circumvent the problem, Marcone {\em et
  al}~\citep{Marcone:2005:J-Phys-A,Marcone:2007:PRE} recently proposed
a novel strategy to identify the knotted arc of minimal length. The
method exploits the so-called \emph{entropic competition} between two
unlinked, mutually-avoiding, knotted sub-loops obtained by
introducing a \emph{slip ring} in the full ring. The slip ring is small
enough to prevent a complete migration of one sub-loop, or of its
knot, into the other sub-loop (see also~\citep{Edwards:1968:JPA} and
~\citep{Metzler:2002:PRE} for related works). The whole topology of
the system is characterized by the knot types $\tau_1$ and $\tau_2$,
respectively, of the first and the second sub-loop, and by the linking
state between the two loops.

In the formulation of Marcone {\em et al}, the rings in the poor
solvent regime are described by polygons on the cubic lattice with an
effective attraction between non consecutive vertices. This model is
known to be adequate for reproducing the salient features of the
coil-globule transition ($\theta$ point) and the equilibrium
properties of the rings in the compact regime. By using a Monte Carlo
approach based on BFACF moves the overall topology of the system was
kept fixed and it was studied how the sub-rings shrunk or inflated
depending on their topological state. We recall that the sub-rings are
assumed to be unlinked. If a given prime knot $\tau$ is tied in each
loop one can identify the average knot size $\langle m_{\tau}\rangle$
with the length of the smallest loop.

In the good solvent regime the above scheme yields $\langle
m_{\tau}\rangle \sim N^t$ with $t \approx 0.75$. The results are in
good agreement with the findings based on the closure scheme. For poor
solvents the estimate was $t\approx 1$. This indicates that
knots are weakly localized in the good solvent regime but are fully
delocalized in the poor solvent regime. This behaviour was later
confirmed in~\citep{Virnau:2005:J_Am_Chem_Soc} for an off-lattice
model.

Another approach that could be effective either in the swollen or in
the compact regime is the one introduced
in~\citep{Farago:2002:EPL}. The method builds on the idea of
quantifying the degree of tightness of knotted polymers by comparing the
response to mechanical pulling (force-extension relation) of knotted
and unknotted chains.  By simulating rod-bead chains with $N$ (the
number of beads) up to $750$ it was found $t=0.4\pm 0.1$ in the
swollen regime~\citep{Farago:2002:EPL}. This result agrees with
the one in~\citep{Marcone:2005:J-Phys-A,Marcone:2007:PRE} in showing
that knotted polymers in the extended phase are weakly localized
although the estimate of $t$ is considerably smaller.

\subsubsection{The size of adsorbed knots}

A possible link between the theory of flat knots and knot localization
in $3D$ is given by the phenomena of polymer adsorption onto an
impenetrable plane~\citep{DeGennes:1979,Vanderzande:1998} (see section
\ref{knot_prob_ads}). It is known that, in the case of unrestricted
topology, this system exhibits a phase transition between a
\emph{desorbed phase} (zero density of contacts with the surface) and
an \emph{adsorbed phase} where there is finite fraction of monomers at
the surface. This fraction increases with decreasing temperature and
at zero temperature the polymer is practically a fully adsorbed $2D$
object (see Section~\ref{knot_prob_ads}). If, on the other hand, the
topology of the polymer is kept fixed, there will be a certain number
of desorbed monomers.  While the exact number of expected desorbed
monomers cannot be predicted, it is plausible that it is simply
related to the minimal number of crossings necessary to satisfy the
given knot type. At very small temperature, the statistics should be
dominated by fully-adsorbed configurations having the minimal number
of desorbed monomers.

Thus, in adsorption conditions, it is possible to investigate how a
knotted polymer (e.g. in good solvent) crosses over from the weak
localisation regime found at high $T$ to the strong localisation one
corresponding to the occurrence of flat knots. In a recent study this
interplay between topology and adsorption was investigated within a
standard lattice polymer model~\citep{Marcone:2007:PRE-B}.
Specifically, the polymer ring was described by \emph{positive
  polygons} i.e. lattice polygons confined to the half space $z \geq 0$ and
with at least one vertex anchored at the $z=0$ plane. For a given
configuration an energy $\epsilon N_a$ is assigned, where $N_a$ is the
number of vertices in the surface $z=0$. Usually $\epsilon = -1$. The
Monte Carlo sampling was based on BFACF moves while the determination
of knot size was based on the cutting and closure scheme introduced
in~\citep{Marcone:2005:J-Phys-A}. By exploring the whole adsorption
phase diagram for the prime knots $3_1$, $4_1$, $5_1$ and $5_2$, it was
found that knots in the desorbed phase and right at the transition are
weakly localised. Below the transition ($T<T_c$), the knot becomes
more and more localised, reaching a strong localisation regime deep
into the adsorbed phase. This crossover to more localised states is
certainly triggered by the adsorption transition but turns out to be
quite smooth as $T$ decreases. In particular, on the basis of the
collected data, it could not be ruled out that below $T_c$ there is a
continuous variation of the exponent $t$ with the flat knot regime
reached only in the limit $T\to 0$. An alternative scenario would be
the existence of a genuine \emph{localization} transition at some
$T_{loc}<T_c$, where the region $T<T_{loc}$ would correspond to the
strong localization regime. This appealing scenario is however less
probable unless one assumes that the data in
~\citep{Marcone:2007:PRE-B} are affected by strong finite-size
corrections.

With the exception of the above cited study
in~\citep{Marcone:2007:PRE-B} there are no other investigations on the
interplay between topological constraints and polymer adsorption. It
would be interesting, for example, to perform similar studies for
off-lattice models of polymer rings in which self-avoidance and
bending rigidity are taken into account. This would be an interesting
issue to explore given that experimental results showing localization
in adsorbed knots are nowadays available~\citep{Ercolini:2007:PRL}.

The above results indicate that the presence of a topological
constraint introduces a new length scale $R_{\tau}\propto \langle
m\rangle_{\tau}\sim N^{t\nu}$ into the problem that may affect the
scaling properties of knotted rings at equilibrium. In the swollen
phase, since $t\simeq 0.7 < 1$ (weak localization), the new length
scale is subdominant and would not affect the leading behaviour of
scaling laws such as eqs. (\ref{scaling_entropy_tau}) and
(\ref{r2scalknot}). It may have, however, important consequences in
the leading term of the correction to scaling expansion as shown
in~\citep{Marcone:2007:PRE}. In the collapse regime, since $t=1$,
$R_{\tau}$ is comparable to the average extension of the whole polymer
and changes in the scaling theory, already at leading orders may
occur. For example, there are strong numerical indications that
relation (\ref{entr_exp}) is not valid for compact knotted
rings~\citep{ourselves}.

The presence of a new, topological, length scale may also be important
in determining the equilibrium properties of knotted polymer rings
under geometrical confinement or under stretching forces. For example
for polymers under geometrical confinement two crossover regimes may
be considered. The first one occurs when the effective confining size, $D$, is comparable to $R$
though still larger than $R_{\tau}$. In this situation the knot, does
not experience, on average, the confinement and the scaling behaviour
such as eq.~(\ref{scaling2}) should hold. If, on the other hand, $D <
R_{\tau} < R$ then the knotted region will be affected by confinement. In
this case it would be interesting to explore whether the knot would
further localize or instead delocalize, as suggested by the numerical
study of knotting in spherically-confined DNA molecules described in
ref.~\citep{Marenduzzo:2009:Proc-Natl-Acad-Sci-U-S-A:20018693}.

A confinement-induced knot localization is probably at play in the case of
polymer rings confined into slab as $D\to 0$ where we know from the
theory of flat knots that almost $2D$ knots are strongly
localized. Similar non trivial effects may be expected for stretched
rings when the screening length $\xi=1 /f$ becomes comparable to the
average knot size $R_{\tau}$ and a single knot cannot fit in
a single blob of the Pincus argument. With the exception of very
preliminary studies in~\citep{Janse-van-Rensburg:2007:JSTAT} and
~\citep{Sheng:2000:PRE,Sheng:2001:PRE}, these issues are still largely
unexplored and would represent very appealing avenues for future
research.

\section{Summary and perspectives}
\label{summary}

The focus of review was mostly on the salient properties of polymers chains
subject to spatial or topological constraints.

Although these systems are highly interesting  from the theoretical and
applicative points of view several of their key physical
properties are still not fully understood. For example, the exact
value of the metric exponent of infinitely-thin rings with a fixed knot type
is still actively investigated and debated. An increasingly important
avenue is also offered by biological systems, where spatially-confined
long biopolymers (DNA and RNA) are ubiquitous. This poses the pressing
need to understand the extent to which the {in vivo} organization of
these biomolecules relies on ``passive'' physical mechanisms or
``active'' ones involving biological machinery.

The first three Chapters of the review provide a self-contained
account of all the basic concepts and methodologies in polymer physics
and topology that are be necessary to tackle the state-of-the-art problems.

In particular, Sections \ref{sec:2} and \ref{knot_theory} can be
regarded as a primer to the topology of closed curves.
Non-specialists will be progressively introduced to the topic starting
from basic questions such as: what is a knot? What is the handedness
of a given knot? How can we classify knots?

Next, in Section \ref{pol_models}, we offer a didactical survey of the standard
models that are used to characterise and study the kinetics and thermodynamics
of polymers. Since we typically focus on universal, or transferable,
physical properties of polymers subject to various constraints, we
restrict considerations to {\em coarse grained} models. In these
approaches, the polymer is described at a much coarser level then in
atomistic approaches which are, in fact, more tailored to address the
detailed properties of a specific type of polymer. The simplified
nature of the models that we shall review, is not for
the sake of a general and minimalistic statistical mechanical
formulation of the problem; it is often an obligatory choice to keep
the complexity of the problem at a manageable level. Indeed, global
properties, such as polymer self-avoidance and spatial or topological
constraints, are extremely difficult to deal within analytical
approaches and can only be satisfactorily accounted for in extensive
computational approaches.

In Section \ref{Monte_Carlo} we have accordingly reviewed the most advanced stochastic
algorithms for studying the equilibrium properties of confined or
entangled polymers. We note that there are additional methods which
become relevant when non-equilibrium or kinetic effects become
important. These methods typically involve Brownian or molecular
dynamics simulations, or generalisations thereof. While we do report
on selected problems where kinetic effects are important, we do not
review these methods here, but refer the reader to the several
authoritative methodological reviews that are presently available.

After introducing the fundamental theoretical concepts and
methodologies, we first characterize the behaviour of a single polymer
subject to geometric confinement. Specifically, in Sections
\ref{confined_polymers} and \ref{knot_prob_conf} we discuss the
statistical and topological properties of confined polymers and
rings. Particular attention is paid to the so-called knot spectrum --
that is the frequency of occurrence of various knot types -- of
confined polymers, starting from lattice models (for which analytical
results are available) to more complex off-lattice models of flexible
chains. The physical description of such systems is complicated due to
the interplay of various length scales, i.e. the contour length,
persistence length, excluded-volume interaction range, and finally the
confining region size. In general, the effect of confinement is to
make the incidence of knots much more frequent than in unconstrained
cases.

As a matter of fact, quantitative estimates of the level of
topological entanglement computed for standard models of DNA subject
to high confinement (such as the genome of some viria) largely exceeds
the ones detected experimentally. In Section \ref{dna_capsid} we
report on the quest that has been undertaken in recent years to bridge
the gap between the theoretical and experimental data on DNA knotting
and which has been instrumental in advancing the current understanding
of the physics of confined DNA. Besides the open issues regarding DNA
knotting in viria and in the cell, there are a number of outstanding
questions regarding the topology of confined polymers in
general. Among these we largely focused on the case of polymer
confinement in slabs and prisms, which ought to acquire increasing
importance with the advent of nano-fabrication techniques.

Likewise, when confinement is induced by an attractive interaction to
a surface, which we study in Section \ref{knot_prob_ads}, it is still unclear whether
there is a topology-dependent character of the adsorption transition.
As discussed in Section \ref{knot_prob_stress}, little is known about how many and what
knot types occur in stretched polymers. Again the latter may well be
soon extensively characterized experimentally by means of
single-molecule manipulation techniques.

The remainder of the review deals with cases where the entanglement is
not driven by spatial confinement but rather by the proximity of other
chains in concentrated solutions or cases where topological
constraints affect the geometric properties of polymer rings.

The characterization of the equilibrium topological properties of
polymers in concentrated solutions is given in Sections \ref{sec:3rdpart} and \ref{linking}. In these
sections the reader is first introduced to the concept of linked
curves and to the various algorithms that have been designed by
topologists to detect computationally the linked state of polymer
rings. The presented results mostly focus on the linking probability
of two or more rings as a function of the concentration. This topic
has been actively investigated and several exact results are available
for polymer on a lattice. Several major issues, including the impact
that the linking or unlinking has on the configurational properties of
polymers in dense solutions is still largely uncharacterised.

Sections \ref{sec:4thpart} and \ref{topological_constrains} are devoted instead to the case where the topological
state of the polymer(s) is fixed. In this case it is interesting to
understand how and to which extent the topological constraint reverberates on the
geometrical properties of the polymer chains.

These systems are challenging to simulate. In fact, it is typically
necessary to sample the polymer configurational space by using
topology-preserving polymer deformations which are, by necessity,
local. This is reflected in the very long autocorrelation time of the
stochastic Monte-Carlo or Molecular Dynamics simulation. At the same
time, this type of problems is most interesting and connects to a
number of open questions in polymer physics. These range from
fundamental issues such as the determination of the universality class
and critical exponents which characterise the
infinitely-thin ring with
fixed topology (see Section \ref{metr_fixed_knot}), to very applied and highly
interdisciplinary problems, such as the segregation of chromosomes in
eukaryotic and bacterial cells (this problem is reviewed in Section \ref{chromterr}). Indeed, while in the nucleus and in the cell there are enzymes
which allow crossing of DNA strands and hence the changing of the knot
and link states of the genome, the action of these enzymes requires
energy and time so that the fixed topology case is a good starting
point and has been analysed with increasing attention by
biophysicists. The problem of territories and in general of the
structure and 3-dimensional organization of chromatin and bacterial
DNA are fascinating open problems which have just now begun to be
studied by physicists. Preliminary results suggest that confinement,
crowding and topology should all be crucial to the understanding of
these phenomena, and this area seems a particularly promising
playground for the physics and topology of confined and concentrated
polymers.

Finally, Section \ref{size_knots} introduces the study of the ``length'' of a knot
in a single polymer, or equivalently the concept of localisation or
delocalisation of knots. A knot is localised when the portion of the
chain that it occupies scales with a sublinear power of the total
length of the polymer, and delocalised otherwise. While it is
intuitive that stretched knots localise, other results on knot
localisation are quite unexpected. For instance, knots in
unconstrained polymers are weakly localised, and they delocalise below
the theta temperature. Likewise, knots in bacteriophages appear to
delocalise when they are inside the capsid, and this may be at the
origin of the seemingly paradoxical observation that DNA can be
ejected into the bacterial host even when it is highly knotted (see
Sections \ref{dna_capsid} and \ref{realknots}). However, the interplay of knot size with
confinement and concentration is poorly understood to date and will
certainly attract more researchers in the years to come.

In summary, we have provided a survey of the methodology, the current
results and the open problems in the physics and topology of confined
and concentrated polymers. Most of the latest progress in this
interdisciplinary area, which covers vital aspects of biophysics and
technology, depends on the use of advanced theoretical and especially
computational techniques. These have been covered in detail here with
the aim of providing an organised summary of the state-of-the art and
an accessible guide for researchers entering in the field.

\section{Acknowledgements}

We are indebted to Tetsuo Deguchi, Claus Enrst, Alexander Grosberg,
Ken Millett, Eric Rawdon, De Witt Sumners, Luca Tubiana, Peter Virnau,
Stu Whittington and Guillaume Witz for the critical reading of the
manuscript and for suggesting several corrections. We acknowledge
support from the Italian Ministry of Education.

\appendix

\section{Distance between two segments}
\label{app:segdist}

We recall here the procedure necessary to calculate the distance of minimum approach of two
segments with endpoints respectively equal to $\vec{P}_1$, $\vec{P}_2$ and $\vec{Q}_1$,
$\vec{Q}_2$. It is convenient to parametrize the points on the segments as follows:
\begin{equation}
\vec{P}(s) = \vec{P}_1 + s \, \vec{v}\ \ \ \vec{Q}(t) = \vec{Q}_1 + s \, \vec{w}\ \ \
\end{equation}

\noindent where $\vec{v}=\vec{P}_2 - \vec{P}_1$ and $\vec{w}=\vec{Q}_2
- \vec{Q}_1$ and the parameters $s$ and $t$ are limited to the [0,1]
range. The required distance is clearly given by minimizing
$d(s,t)\equiv |\vec{P}(s) - \vec{Q}(t)|$ over the unit square in the
$s,t$ plane.
First it is necessary to calculate the values of $s$ and
$t$ which yield the global minimum of $d$:
\begin{eqnarray}
s^* = {(\vec{v} \cdot \vec{w}) (\vec{w} \cdot \vec{\Delta_1}) -  (\vec{w} \cdot \vec{w}) (\vec{v} \cdot \vec{\Delta_1}) \over (\vec{v} \cdot \vec{v}) (\vec{w} \cdot \vec{w}) -  |(\vec{v} \cdot \vec{w})|^2} \\
t^* =  {(\vec{v} \cdot \vec{v}) (\vec{w} \cdot \vec{\Delta_1}) -  (\vec{v} \cdot \vec{w}) (\vec{v} \cdot \vec{\Delta_1}) \over (\vec{v} \cdot \vec{v}) (\vec{w} \cdot \vec{w}) -  |(\vec{v} \cdot \vec{w})|^2} \\
\end{eqnarray}

\noindent where $\Delta_1 \equiv \vec{P}_1 - \vec{Q}_1$. The points associated to $s^*$ and $t^*$
are the points of minimum approach of the infinite lines which contain the two segments. If $s^*$
and $t^*$ are both in the [0,1] range then the points of closest approach of the lines fall inside
the segments and $d(s^*,t^*)$ is clearly the sought inter-segment distance.  On the contrary, if
$(s^*,t^*)$ falls outside the unit square then the minimum value of $d$ must be searched on the
perimeter of the unit square. This requires setting in turn $s$ and $t$ equal to 0 or 1 and
minimizing $d^2(s^*,t^*)$ over the ``free'' parameter restricted to the [0,1] range.

\section{Local chain moves that preserve topology}
\label{app:topopres}

Knot simplification algorithms usually rely on various geometrical manipulations of a ring that do
not alter the ring topology. A ubiquitous simplification operation consists of picking a chain
node, $i$, at random and displacing it from the current position, $\vec{r}_i$, to a new one,
$\vec{r}_i^\prime$, that is more collinear with the flanking nodes:
\begin{equation}
\vec{r}_i^\prime = \alpha \, \vec{r}_i + {1- \alpha \over 2} (\vec{r}_{i+1} + \vec{r}_{i-1})
\end{equation}
\noindent with $\alpha \in [0,1]$. It must be checked  that during a continuously parametrised
movement of the bond from $\vec{r}_i$ to $\vec{r}^\prime_i$ no chain segment is crossed, as this
might alter the topology.

To check the preservation of topology it is necessary to 
check~\citep{Taylor:2000:Nature:10972297,Koniaris&Muthukumar:1991b} that the two triangular faces with vertices
$\{i-1,i,i^\prime\}$ and $\{i+1,i,i^\prime\}$ do not intersect any of the chain segments that do
not include $i-1, i$ or $i+1$ as endpoints. For the sake of self-containedness we recall here the
procedure for checking whether a segment with endpoints $\vec{P}_1$, $\vec{P}_2$ intersects a
triangular face with vertices $\{\vec{Q}_1, \vec{Q}_2, \vec{Q}_3\}$. First it is calculated the
point of intersection of the face and the infinite line containing the $P_1P_2$ segment and the
infinite plane containing the triangular face.

It is convenient to indicate with $\vec{q}$ the normal to the plane containing the triangular face
and with $\vec{P}(s) = \vec{P}_1 + s \, (\vec{P}_2 - \vec{P}_1)$ the parametrization of the line
containing the segment. The point on the line at the intersection with the plane is identified as
$P(s^*)$ where,
\begin{equation}
s^* = {\vec{q}\cdot (\vec{Q}_1 - \vec{P}_1) \over \vec{q} \cdot (\vec{P}_2 -   \vec{P}_1)}
\end{equation}

\noindent if the value of $s^*$ falls outside the [0,1] range (which
parametrizes the $P_1P_2$ segment) then it means that the segment does not intersect the plane, and hence the face.  However, if $s^*$
is in the [0,1] range, it is necessary to check if the point of
intersection, $\vec{P}(s^*)$ falls within the face. To do so it is
sufficient to write the vector $\vec{P}(s^*)- \vec{Q}_1$ as a linear
combination of the two oriented face edges departing from $Q_1$:
\begin{equation}
\vec{P}(s^*)- \vec{Q}_1 = \alpha \, (\vec{Q}_3- \vec{Q}_1) + \beta \,  (\vec{Q}_2- \vec{Q}_1)
\label{eqn:lincomb}
\end{equation}
\noindent The coefficients $\alpha$ and $\beta$ are conveniently
obtained by solving the two linear equations obtained multiplying both
sides of eq.~(\ref{eqn:lincomb}) by $ (\vec{Q}_3- \vec{Q}_1)$ and
$(\vec{Q}_3- \vec{Q}_1)$. if both $\alpha$ and $\beta$ are in the
[0,1] range it means that the point falls inside the triangular
face. In this case the displacement of the bead cannot be
accomplished without causing the crossing of two chain segments. As
the topology would not be necessarily preserved the bead move must be
rejected.

We remark that the above mentioned procedure is used not only in knot
simplification contexts, but also in topology-preserving Monte Carlo
evolutions of a given conformation. In this case, the $i^{\rm th}$ bead is
displaced randomly or with a crankshaft (local) move and the above
criteria are applied. Notice that this requires that only a single
bead is moved at a time, as the check of topology preservation is not
easily generalizable to the case of simultaneous displacement of two
beads. Also, this method does not preserve the bond lengths.

\end{document}